\documentclass[12pt]{article}        %%%%%%%%%%%%%%%%%%
\usepackage{amsmath}
\usepackage{amssymb}
\usepackage{euscript}

\usepackage{color}    % not mandatory
\usepackage{epsfig}

\usepackage{cite}

\usepackage{alltt}

\usepackage{epic}
\def\Order#1{${\cal O}(#1$)}

\def\Order#1{${\cal O}(#1)$}
\def\Ordex#1{${\cal O}(#1)_{exp}$}
\def\Oeex#1{${\cal O}(#1)_{\rm EEX}$}
\def\Oceex#1{${\cal O}(#1)_{\rm CEEX}$}
\def\KK{${\cal KK}$}
\def\Ordpr#1{${\cal O}(#1)_{prag}$}
\def\born{{\rm Born}}
\def\st{\hbox{}} %% hbox is lower
\def\talpha{\tilde{\alpha}}
\def\tbeta{\tilde{\beta}}
\def\bbeta{\bar{\beta}}
\def\hbeta{\hat{\beta}}

\newcommand{\Reu}{\EuScript{R}}

\newcommand{\Meu}{\EuScript{M}}

\newcommand{\Bmf}{\mathfrak{B}}

\newcommand{\Mmf}{\mathfrak{M}}

\newcommand{\umf}{\mathfrak{u}}

\newcommand{\sfac}{\mathfrak{s}}

\begin{document}                     %%%%%%%%%%%%%%%%%%%%%%

\allowdisplaybreaks

%\endinput

%%%%%%%%%%%%%%%%%%%%%%%%%
% make ceex2-all.ps
%%%%%%%%%%%%%%%%%%%%%%%%%

%%%%%%%%%%%%%%%%%%%%%%%%%%%%%%%%%%%%%%%%%%%%%%%%%%%%%%%%%%%%%%%%%%%%%%%%%%%%%%%%%
%%%%%%%%%%%%%%%%%%%%%%%%%%%%%%%%%%%%%%%%%%%%%%%%%%%%%%%%%%%%%%%%%%%%%%%%%%%%%%%%%
\begin{titlepage}

%%\begin{center}
%%{\huge Semifinal Draft -- \today{} }
%%\end{center}

\begin{flushright}
{\bf  
CERN-TH/2000-087  \\
UTHEP-99-09-01}
\end{flushright}

\vspace{2mm}
\begin{center}{\bf\Large Coherent Exclusive Exponentiation  }\end{center}
\begin{center}{\bf\Large For Precision Monte Carlo Calculations$^{\dag}$ }\end{center}

\vspace{1mm}
\begin{center}
  {\bf   S. Jadach$^{a,b}$,}
  {\bf   B.F.L. Ward$^{a,c}$}
  {\em and}
  {\bf   Z. W\c{a}s$^{b,d}$ }
\\
\vspace{1mm}
{\em $^a$Department of Physics and Astronomy,\\
         The University of Tennessee, Knoxville, TN 37996-1200, USA,}\\
{\em $^b$Institute of Nuclear Physics,
         ul. Kawiory 26a, 30-055 Cracow, Poland,}\\
{\em $^c$SLAC, Stanford University, Stanford, CA 94309, USA,}\\
{\em $^d$CERN, Theory Division, CH-1211 Geneva 23, Switzerland,}\\
\end{center}

\vspace{1mm}
\begin{abstract}
We present the new Coherent Exclusive Exponentiation (CEEX),
the older Exclusive Exponentiation (EEX)
and the semi-analytical Inclusive Exponentiation (IEX)
for the process $e^+e^-\to f\bar{f} +n\gamma$, $f=\mu,\tau,d,u,s,c,b$
with validity for centre of mass energies from $\tau$ lepton threshold to 1TeV,
that is for LEP1, LEP2, SLC, future Linear Colliders, $b,c,\tau$-factories etc.
They are based on Yennie-Frautschi-Suura exponentiation.
In CEEX effects due to photon emission from initial beams and outgoing fermions
are calculated in QED up to second-order, including all interference effects.
Electroweak corrections are included in first-order, at the amplitude level.
Beams can be polarized longitudinally and transversely,
and all spin correlations are incorporated in an exact manner.
EEX is more primitive, lacks initial-final interferences, 
but it is valuable for testing the newer CEEX.
IEX provides us set of a sophisticated semi-analytical formulas
for the total cross section and selected inclusive distributions which are mainly used for
cross-checks of the MC results.
We analyse numerical results at the $Z$-peak 189~GeV and 500~GeV
for simple kinematical cuts (comparisons with IEX) and for realistic experimental cuts.
Physical precision and technical precision are determined for the total cross section
and for the charge asymmetry.
\end{abstract}
\begin{center}
{\it To be Submitted to Phys. Rev. D}
\end{center}

\vspace{1mm}
\footnoterule
\noindent
{\footnotesize
\begin{itemize}
\item[${\dag}$]
Work supported in part by Polish Government grants 
KBN 2P03B08414, %%%<-- Marek
KBN 2P03B14715, %%%<-- Zbyszek
the US DoE contracts DE-FG05-91ER40627 and DE-AC03-76SF00515,
the Maria Sk\l{}odowska-Curie Joint Fund II PAA/DOE-97-316,
and the Polish--French Collaboration within IN2P3 through LAPP Annecy.
\end{itemize}
}
\vspace{1mm}
\begin{flushleft}
{\bf 
  CERN-TH/2000-087  \\
  UTHEP-99-09-01\\
  March 2000}
\end{flushleft}

\end{titlepage}

%%%%%%%%%%%%%%%%%%%%%%%%%%%%%%%%%%%%%%%%%%%%%%%%%%%%%%%%%%%%%%%%%%%%%%%%%%%%%%%%%%%%%%%%%%%%%
%%%%%%%%%%%%%%%%%%%%%%%%%%%%%%%%%%%%%%%%%%%%%%%%%%%%%%%%%%%%%%%%%%%%%%%%%%%%%%%%%%%%%%%%%%%%%
\tableofcontents   % does not work, why???

\newpage
%%%%%%%%%%%%%%%%%%%%%%%%%%%%%%%%%%%%%%%%%%%%%%%%%%%%%%%%%%%%%%%%%%%%%%%%%%%%%%%%%
%%%%%%%%%%%%%%%%%%%%%%%%%%%%%%%%%%%%%%%%%%%%%%%%%%%%%%%%%%%%%%%%%%%%%%%%%%%%%%%%%
%%%%%%%%%%%%%%%%%%%%%%%%%%%%%%%%%%%%%%%%%%%%%%%%%%%%%%%%%%%%%%%%%%%%%%%%%%%%%%%%%

%%%%%%%%%%%%%%%%%%%%%%%%%%%%%%%%%%%%%%%%%%%%%%%%%%%%%%%%%%%%%%%%%%%%%%%%%%%%%%%%%%%%%%%%%%%%%
%%%%%%%%%%%%%%%%%%%%%%%%%%%%%%%%%%%%%%%%%%%%%%%%%%%%%%%%%%%%%%%%%%%%%%%%%%%%%%%%%%%%%%%%%%%%%
\section{Introduction}
%%%%%%%%%%%%%%%%%%%%%%%%%%%%%%%%%%%%%%%%%%%%%%%%%%%%%%%%%%%%%%%%%%%%%%%%%%%%%%%%%%%%%%%%%%%%%
At the end of LEP2 operation the total cross section for 
the process $e^-e^+\to f\bar{f}+n\gamma$
will have to be calculated with the precision $0.2\%-1\%$, depending on event selection.
The arbitrary differential distributions have to be also calculated
with the corresponding precision.
In future linear colliders (LC's) the precision requirement can be even more demanding.
This is especially true for high luminosity linear colliders, like in the case of TESLA.
The above new requirements necessitate development of the new calculational framework 
for the QED corrections and the construction of new dedicated MC programs.
The present work is a part of an effort in this direction.

The main limiting factor preventing us from getting
more precise theoretical predictions for the $e^-e^+\to f\bar{f}+n\gamma$
process is higher-order QED radiative corrections
(QED part of electroweak Standard Model).
In order to achieve the 0.2\% precision tag,
the virtual corrections have to be calculated up to two-three loops and the multiple bremsstrahlung
up to two-three hard photons, 
integrating exactly the multiphoton phase-space for the arbitrary event selection
(phase-space limits).

For any realistic kinematical cuts, one cannot get  
the precise theoretical predictions for $e^-e^+\to f\bar{f}+n\gamma$  
at the above ambitious precision level without
Monte Carlo (MC) event generators.
It is therefore mandatory to formulate perturbative Standard Model (SM) calculations
in a formulation friendly to their use within the Monte Carlo event generator.

Let us stress that the Monte Carlo method is for us nothing more and nothing less
than the numerical integration over the Lorentz invariant phase-space.
It is therefore an exercise in the applied mathematics.
In the present work we shall not, however, elaborate on the methods of the Monte Carlo phase-space
integration and construction of the Monte Carlo event generator.
This is delegated to ref.~\cite{kkcpc:1999} which describes
the new Monte Carlo event generator \KK\ in which the matrix element of the present paper
is implemented and all numerical results presented here are calculated using the
latest version 4.13 of \KK.

In the present work we concentrate on the definition and construction of the matrix element
for the process $e^-e^+\to f\bar{f}$ within Standard Model.
We shall especially address the problem of the higher-order QED corrections.
This work is a continuation of two recent papers
\cite{gps:1998} and \cite{ceex1:1999}.

%%%%%%%%%%%%%%%%%%%%%%%%%%%%%%%%%%%%%%%%%%%%%%%%%%%%%%%%%%%%%%%%%
\subsection{Two types of QED matrix element and exponentiations}
%%%%%%%%%%%%%%%%%%%%%%%%%%%%%%%%%%%%%%%%%%%%%%%%%%%%%%%%%%%%%%%%%
In the \KK\ Monte Carlo and in this paper we use two types of matrix element with two types
of exponentiation: exclusive exponentiation nicknamed EEX and coherent exclusive exponentiation
referred to as CEEX.
Both are termed as ``exclusive'' as opposed to ``inclusive'',
see also the discussion in~\cite{sussex:1989}.
The exclusivity means that the procedure of exponentiation, that is summing up
the infrared (IR) real and virtual contribution, within the standard perturbative scheme
of quantum field theory, is done at the level of the fully differential (multiphoton) cross section,
or even better, at the level of the scattering matrix element (spin amplitudes),
{\em before any phase-space integration over photon momenta is done}.

The other ``inclusive'' exponentiation is an ad hoc procedure of summing up IR corrections
{\em after phase-space integration over photon momenta}, that is for inclusive distributions.
In spite of its weak theoretical basis the inclusive exponentiation is very commonly
done routinely in all semi-analytical approaches 
like that in ref.~\cite{zfitter6:1999}.
In Section \ref{sec:iex} we shall come back to inclusive exponentiation 
and show how to justify it theoretically.

The two exclusive exponentiations EEX and CEEX are well suited for the fully
exclusive Monte Carlo event generators in which four-momenta of all final-state particles
are available.
Historically EEX was formulated for the first time in ref.~\cite{yfs1:1988}
for the initial-state radiation (ISR)
and an improved version was presented in ref.~\cite{yfs2:1990}.
It follows very closely the Yennie-Frautschi-Suura (YFS) exponentiation of 
the classical ref.~\cite{yfs:1961}.
The extension of EEX to the final-state radiation (FSR) was done shortly thereafter~\cite{yfs3-pl:1992},
but it was actually never fully published.
The computer program YFS3, in which EEX for FSR  was implemented, was incorporated in 
KORALZ~\cite{koralz4:1994}
and some numerical results were published in~\cite{yfs3-pl:1992}, without
actually giving details of the QED matrix element.
The present work gives in fact the first full account of the EEX matrix element for ISR and FSR
for the process $e^-e^+\to f\bar{f}+n\gamma, f\neq e$.
This is to be contrasted with the situation for small angle Bhabha scattering
(the well-known LEP-SLC luminosity process) where the EEX type
matrix element was fully documented in refs.~\cite{bhlumi2:1992,bhlumi4:1996,bhlumi-semi:1996}.

CEEX is a new version of the exclusive exponentiation, generally more efficient for calculations
beyond first-order, facilitating inclusion of full spin polarization, narrow resonances 
and any kind of interferences.
Its first version, limited to first-order, was presented in ref.~\cite{ceex1:1999}.
In the present work we extend it to (still incomplete) second-order.

Let us characterize briefly the main features of EEX and CEEX.
EEX is formulated in terms of spin summed/averaged differential distributions,
this is the source of some advantages and disadvantages which may be summarized as follows:
\begin{itemize}
\item
  Differential distributions in practice are given analytically in terms of Mandelstam variables
  and scattering angles, they are therefore easy to inspect by human eye and to check correctness
  of certain important limits like leading-logarithmic and soft limits.
\item
  Analytical representation of the differential distributions allows for analytical 
  phase-space integration and development of the semi-analytical formulas, which are useful
  for cross-check with the MC results.
\item
  Spin effects are difficult to add already at \Order{\alpha^1}, because one is forced to calculate
  radiative corrections to spin density matrices, not an easy task.
\item
  Squaring sums of spin amplitudes from groups of Feynman diagrams leads to many interference terms
  which in the exponentiation procedure are handled analytically and individually.
  Because of that interference terms can be dealt with efficiently in EEX only for simple processes
  dominated by a small number of Feynman diagrams and only up to first-order.
\end{itemize}
CEEX is formulated in terms of spin amplitudes and
this is also the source of some advantages and disadvantages:
\begin{itemize}
\item
  Differential distributions are calculated out of spin amplitudes numerically --
  spin amplitudes are generally simpler/smaller objects, especially beyond \Order{\alpha^1}.
\item
  Since  an analytical representation for differential distributions is not available
  semi-analytical integration over the phase-space is practically impossible.
\item
  Spin effects are added relatively easily, during numerical evaluation of the differential distributions
  out of spin amplitudes. 
  Adding higher-order corrections does no make the treatment of spin polarization more difficult.
\item
  Inclusion of all kind of interference effects (among real photon emissions, many Feynman diagrams etc.)
  comes almost for free -- it is done numerically
  in the process of summing and squaring various contributions to spin amplitudes.
\end{itemize}
As we see CEEX has many advantages over EEX, so why do we keep EEX?
There are important reasons:
\begin{itemize}
\item
  Generally, CEEX is a relatively new invention, the older and more primitive 
  but well established EEX is a useful reference for numerical tests of CEEX.
\item
  EEX is better suited for semi-analytical integration over the phase-space, and can be tested
  with these semi-analytical results.
\item
  In the present \KK\ MC the \Order{\alpha^3} leading logarithmic corrections are available 
  for EEX and are not yet available for CEEX.
\end{itemize}
Summarizing, we see that it make sense to keep EEX as a backup solution even
if we already rely on CEEX as a default and leading solution.

%%%%%%%%%%%%%%%%%%%%%%%%%%%%%%%%%%%%%%%%%%%%%%%
\subsection{Notation, terminology}
%%%%%%%%%%%%%%%%%%%%%%%%%%%%%%%%%%%%%%%%%%%%%%%

%
% Second/third order pragmatic
%
\begin{figure}[ht]
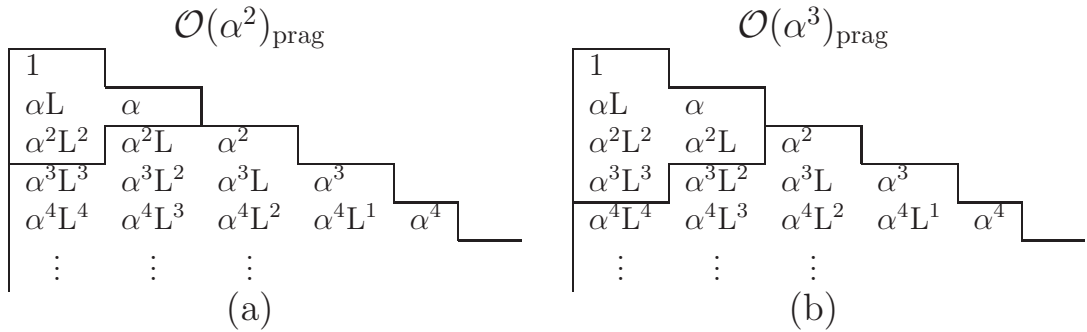

\centering
\begin{tabular}{cc}
%
% --- first table, pragmatic second-order
%
\hbox{
\begin{tabular}{llllll}
& \multicolumn{3}{c}{\large ${\cal O}(\alpha^2)_{{\rm prag}}$}
\\
\cline{1-1}
\multicolumn{1}{|l|}{$1$}
\\ \cline{2-2}
\multicolumn{1}{|l}{$\alpha$L} &
\multicolumn{1}{l|}{$\alpha$}
\\ \cline{2-3}
\multicolumn{1}{|l|}{$\alpha^2$L$^2$} &
\multicolumn{1}{l}{$\alpha^2$L}       &
\multicolumn{1}{l|}{$\alpha^2$}
\\ \cline{1-1} \cline{4-4}
\multicolumn{1}{|l}{$\alpha^3$L$^3$}  &
\multicolumn{1}{l}{$\alpha^3$L$^2$}   &
\multicolumn{1}{l}{$\alpha^3$L}       &
\multicolumn{1}{l|}{$\alpha^3$}
\\ \cline{5-5}
\multicolumn{1}{|l}{$\alpha^4$L$^4$}  &
\multicolumn{1}{l}{$\alpha^4$L$^3$}   &
\multicolumn{1}{l}{$\alpha^4$L$^2$}   &
\multicolumn{1}{l}{$\alpha^4$L$^1$}   &
\multicolumn{1}{l|}{$\alpha^4$}       &
\hbox{\quad}
\\ \cline{6-6}
\multicolumn{1}{|c}{\vdots} &
\multicolumn{1}{c}{\vdots} &
\multicolumn{1}{c}{\vdots}
\\
& \multicolumn{3}{c}{\large (a)}
\end{tabular}}
\quad
%
% --- second table, pragmatic third-order
%
\hbox{
\begin{tabular}{llllll}
& \multicolumn{3}{c}{\large ${\cal O}(\alpha^3)_{{\rm prag}}$}
\\
\cline{1-1}
\multicolumn{1}{|l|}{$1$}
\\ \cline{2-2}
\multicolumn{1}{|l}{$\alpha$L} &
\multicolumn{1}{l|}{$\alpha$}
\\ \cline{3-3}
\multicolumn{1}{|l}{$\alpha^2$L$^2$} &
\multicolumn{1}{l|}{$\alpha^2$L}     &
\multicolumn{1}{|l|}{$\alpha^2$}
\\ \cline{2-2} \cline{4-4}
\multicolumn{1}{|l|}{$\alpha^3$L$^3$} &
\multicolumn{1}{|l}{$\alpha^3$L$^2$}  &
\multicolumn{1}{l}{$\alpha^3$L}       &
\multicolumn{1}{l|}{$\alpha^3$}
\\ \cline{1-1} \cline{5-5}
\multicolumn{1}{|l}{$\alpha^4$L$^4$}  &
\multicolumn{1}{l}{$\alpha^4$L$^3$}   &
\multicolumn{1}{l}{$\alpha^4$L$^2$}   &
\multicolumn{1}{l}{$\alpha^4$L$^1$}   &
\multicolumn{1}{l|}{$\alpha^4$}       &
\hbox{\quad}
\\ \cline{6-6}
\multicolumn{1}{|c}{\vdots} &
\multicolumn{1}{c}{\vdots} &
\multicolumn{1}{c}{\vdots}
\\
& \multicolumn{3}{c}{\large (b)}
\end{tabular}}
\end{tabular}
\caption{\sf
QED perturbative leading and subleading corrections.
Rows represent  corrections in consecutive perturbative orders --
first row is Born contribution.
First column represents leading logarithmic (LL) approximation
and second column depicts next-to-leading (NLL) approximation.
In the figure terms selected for
(a) second and (b) third-order pragmatic expansion
are limited with help of additional line.}
\label{pragma2}
\end{figure}
It is useful to introduce certain notation and terminology already at this stage.
In particular, the most common perturbative calculation (no exponentiation) is ``order-by-order''.
That is all terms beyond a certain order are set to zero.
In Fig.~\ref{pragma2} that means we end at certain row --
at \Order{\alpha^2} we include the first three rows.
Exponentiation is blurring this picture because a certain class of terms is summed up to infinite order
and the meaning of the $r-th$ order exponentiation
is that we truncate to \Order{\alpha^r} the infrared (IR) finite components, the so-called $\beta$'s.
On the other hand, in the leading-logarithmic approximation the focus is on summing up
first the contributions like $\alpha^n L^n$ and later  $\alpha^n L^{n-1}$, that is 
in Fig.~\ref{pragma2} we sum up in column-wise order, neglecting terms far away from the first
column which represents the so-called LL-approximation.
Taking the actual value of $\alpha/<pi \sim 1/400$ and of the big logarithm $L=\ln(s/m_f^2)\sim 10$,
we discover quickly that in Fig.~\ref{pragma2} the limiting line
following the numerical importance of the terms is neither row-wise nor column-wise
but diagonal-wise.
This is why we shall often use \Ordpr{\alpha^r} $r=1,2,3$ approximation,
depicted also in  Fig.~\ref{pragma2}, in which we use (exponentiated or not) \Order{\alpha^r}
calculation in which we use incomplete sub-leading terms, in the sense of the LL approximation.
Note that for the LL approximation we shall never use the strict collinear (zero $p_T$) approximation.
The LL approximation will be done at the level of the differential distributions (or spin amplitudes)
without forcing $p_T=0$ on photons.
Just to give a rough idea,
the precision level of order $0.5-1\%$ corresponds to \Ordpr{\alpha^1},
$0.1-0.5\%$ to \Ordpr{\alpha^2} and going below $0.05\%$ will require \Ordpr{\alpha^3}.
The above is true for the exponentiated calculation.
Lack of exponentiation makes the calculation less precise by a factor $2-5$.
The pure non-logarithmic terms of order \Order{\alpha^2} are negligible  
($<10^{-5}$) for any foreseeable practical application.

%%%%%%%%%%%%%%%%%%%%%%%%%%%%%%%%%%%%%%%%%%%%%%%
\subsection{Outline}
%%%%%%%%%%%%%%%%%%%%%%%%%%%%%%%%%%%%%%%%%%%%%%%
The outline of the paper is the following.
In Section 2 we describe in detail the SM/QED matrix element for the exclusive
exponentiation (EEX) based on the Yennie-Frautschi-Suura (YFS) work of ref.~\cite{yfs:1961},
that is of the type of matrix element defined for the first time in ref.~\cite{yfs1:1988}.
In Section 2 we describe the new 
second-order matrix element with coherent exclusive exponentiation (CEEX),
which is the default matrix element in \KK\ MC.
Its first-order variant was given in~\cite{ceex1:1999}, and is also defined here
for the sake of completeness.
In Section 3 we elaborate on how do we combine the electroweak corrections of 
refs.~\cite{dizet:1989,zfitter6:1999} with the QED corrections within EEX and CEEX.
In Section 4 we discuss the differences between EEX and CEEX.
In Section 5 we integrate analytically over the phase-space for the EEX matrix element
in the case of very simple kinematical cuts.
The resulting analytical results are used in Section 6 where numerical results from
\KK\ MC are presented.
The most important task in Section 6 is, however, 
the determination of the physical and technical precision
for the total cross section and charge asymmetry at the $Z$-peak, LEP2 and 500~GeV.
In particular we discuss the contribution from the initial-final state interference (IFI)
which is included in our new CEEX matrix element (IFI is neglected in EEX).
In the last Section 7 we summarize our work.
In Appendix A we define the Weyl-spinor techniques used in construction of CEEX multi-photon
spin amplitudes.

%%%%%%%%%%%%%%%%%%%%%%%%%%%%%%%%%%%%%%%%%%%%%%%%%%%%%%%%%%%%%%%%%%%%%%%%%%%%%%%%%%%%%%%%%%%%%
%%%%%%%%%%%%%%%%%%%%%%%%%%%%%%%%%%%%%%%%%%%%%%%%%%%%%%%%%%%%%%%%%%%%%%%%%%%%%%%%%%%%%%%%%%%%%
%%%%%%%%%%%%%%%%%%%%%%%%%%%%%%%%%%%%%%%%%%%%%%%%%%%%%%%%%%%%%%%%%%%%%%%%%%%%%%%%%%%%%%%%%%%%%
\section{ Amplitudes for Exclusive Exponentiation}
\label{sec:eex}

%%%%%%%%%%%%%%%%%%%%%%%%%%%%%%%%%%%%%%%%%%%%%%%%%%%%%%%%%%%%%%%%%%%%
%                                                                  %
%  The picture with kinematics for initial + final state emission  %
%                                                                  %
%%%%%%%%%%%%%%%%%%%%%%%%%%%%%%%%%%%%%%%%%%%%%%%%%%%%%%%%%%%%%%%%%%%%
\begin{figure}
\centering
\setlength{\unitlength}{0.08mm}
\begin{picture}(1700,900)
%------definitions------------------------
% horizontal photon line
\newsavebox{\phseg}
\savebox{\phseg}(24,12)[r]{\makebox(24,12)[r]{
\begin{picture}(24,12)
  \put(6,6){\oval(12,12)[t]}
  \put(18,6){\oval(12,12)[b]}
\end{picture}}}
% ------end of definitions -------
%\put(0,0){\framebox(1700,900){}}
%
\thicklines
% initial-state fermion line lower part
\put(200,100){\line(1,0){275}}
\put(475,125){\oval(50,50)[br]}
\put(180,100){\makebox(0,0)[r]{\large $p_2$}}
\put(350, 80){\makebox(0,0)[t]{\large $e^+$}}
\put(250,100){\vector(1,0){100}}
%
% initial-state fermion line upper part
\put(200,800){\line(1,0){275}}
\put(475,775){\oval(50,50)[tr]}
\put(180,800){\makebox(0,0)[r]{\large $p_1$}}
\put(350,820){\makebox(0,0)[b]{\large $e^-$}}
\put(250,800){\vector(1,0){100}}
%
% and vertical middle part
\put(500,125){\line(0,1){650}}
\put(500,450){\circle*{20}}
\put(350,450){\makebox(0,0)[r]{\large $P=p_1+p_2$}}
% photon lines initial state
\multiput(500,750)(24,0){8}{\usebox{\phseg}}
\multiput(500,650)(24,0){8}{\usebox{\phseg}}
\multiput(500,550)(24,0){8}{\usebox{\phseg}}
\multiput(500,250)(24,0){8}{\usebox{\phseg}}
\multiput(500,150)(24,0){8}{\usebox{\phseg}}
\put(740,750){\makebox(0,0)[l]{\large $k_1$}}
\put(740,650){\makebox(0,0)[l]{\large $k_2$}}
\put(740,550){\makebox(0,0)[l]{\large $k_3$}}
\put(740,250){\makebox(0,0)[l]{\large $k_{n-1}$}}
\put(740,150){\makebox(0,0)[l]{\large $k_n$}}
%
% final state fermion line lower part
\put(1225,100){\line(1,0){275}}
\put(1225,125){\oval(50,50)[bl]}
\put(1520,100){\makebox(0,0)[l]{\large $q_2$}}
\put(1400, 80){\makebox(0,0)[t]{\large $\bar{f}$}}
\put(1300,100){\vector(1,0){100}}
%
% final state fermion line upper part
\put(1225,800){\line(1,0){275}}
\put(1225,775){\oval(50,50)[tl]}
\put(1520,800){\makebox(0,0)[l]{\large $q_1$}}
\put(1400,820){\makebox(0,0)[b]{\large $f$}}
\put(1300,800){\vector(1,0){100}}
%
% and vertical middle part
\put(1200,125){\line(0,1){650}}
\put(1200,450){\circle*{20}}
\put(1350,450){\makebox(0,0)[l]{\large $Q=q_1+q_2$}}
% photon lines final state
\multiput(1200,700)(24,0){8}{\usebox{\phseg}}
\multiput(1200,600)(24,0){8}{\usebox{\phseg}}
\multiput(1200,300)(24,0){8}{\usebox{\phseg}}
\multiput(1200,200)(24,0){8}{\usebox{\phseg}}
\put(1440,700){\makebox(0,0)[l]{\large $k'_1$}}
\put(1440,600){\makebox(0,0)[l]{\large $k'_2$}}
\put(1440,300){\makebox(0,0)[l]{\large $k'_{n'-1}$}}
\put(1440,200){\makebox(0,0)[l]{\large $k_{n'}$}}
%
% Z-boson horizontal middle part
\multiput(500,450)(50,0){14}{\line(1,0){25}}
\put(1000,500){\makebox(0,0)[b]{\Large $\gamma$, Z}}
\put(1000,400){\makebox(0,0)[t]{\large $X=P-\sum k_i$}}
\end{picture}
\caption{\it Kinematics of the process with multiple photon emission
from the initial- and final-fermions in the annihilation process.}
\label{kinematics:fig}
\end{figure}
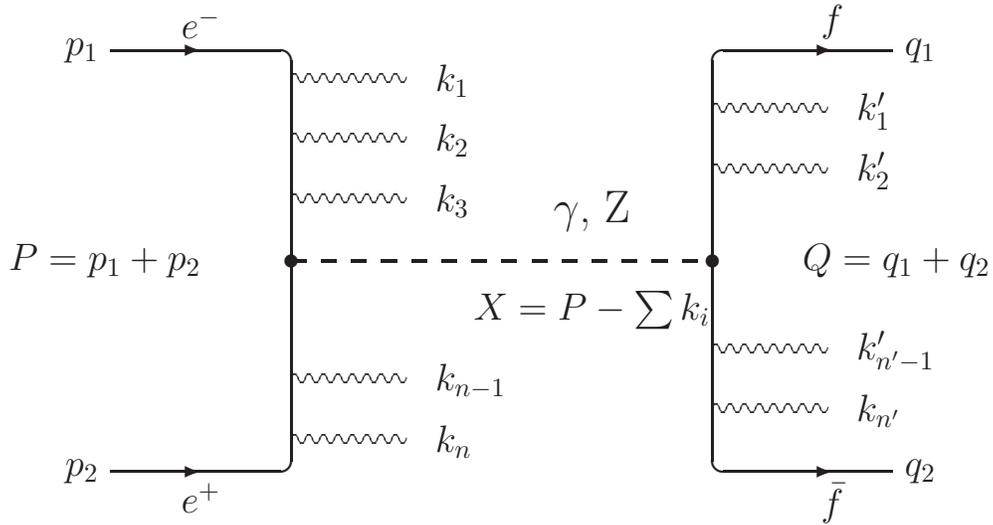

As it was already indicated, the role of the EEX matrix element described in this section
is to provide a testing environment for the new more sophisticated matrix element of the CEEX class,
which will be defined in the next section.

The kinematics of the process $e^-e^+\to f\bar{f}+n\gamma$ is depicted in fig.~\ref{kinematics:fig}.
In the case of the EEX matrix element presented here we neglect the initial-final state interference (IFI).
Consequently, we are allowed in the following
to distinguish among photons emitted from the initial- and final-state fermions.
The four-momentum
\begin{eqnarray}
X = p_1+p_2-\sum_{j=1}^n k_j = q_1 +q_2 +\sum_{l=1}^{n'} k'_l
\end{eqnarray}
of the $s$-channel virtual boson $Z+\gamma$ is then well defined.
Let us denote the rest frame of $X$ as XMS.

%%%%%%%%%%%%%%%%%%%%%%%%%%%%%%%%%%%
\subsection{Master formula}
%%%%%%%%%%%%%%%%%%%%%%%%%%%%%%%%%%%

Denoting Lorentz invariant phase-space by
\begin{equation}
\label{lips}
 d^n{\rm Lips}(P;p_1,p_2,...,p_n)=
   \prod_{j=1}^n  {d^3 p_j\over p^0_j} \;
   \delta^{(4)}\bigg( P -\sum_{j=1}^n p_j \bigg)
\end{equation}
we define for the process
$e^-(p_1)+e^+(p_2) \rightarrow f(q_1)+\bar{f}(q_2)+n\gamma(k_j) +n'\gamma(k'_l)$
the \Order{\alpha^r} total cross-section
\begin{equation}
\label{sigma-eex2}
\sigma^{(r)}_{EEX} =
  \sum_{n=0}^\infty \sum_{n'=0}^\infty {1\over n!}{1\over n'!}
  \int d^{n+n'}{\rm Lips}( p_1+p_2; q_1,q_2, k_1...,k_n, k'_1...,k'_{n'})\;
  \rho^{(r)}_{EEX},\quad r=0,1,2,3
\end{equation}
in terms of the fully differential multiphoton distribution
\begin{equation}
\label{rho-eex2}
\begin{split}
 \rho&^{(r)}_{EEX}(p_1,p_2, q_1,q_2, k_1...,k_n, k'_1...,k'_n) =
  e^{  Y_e(\Omega_I;p_1,p_2) +Y_f(\Omega_F;q_1,q_2) }\\
& \prod_{j=1}^n     \tilde{S}_I(k_j)\;  \bar{\Theta}(\Omega_I;k_j)
  \prod_{l=1}^{n'}  \tilde{S}_F(k'_l)\; \bar{\Theta}(\Omega_F;k'_l)
  \Bigg\{
  \bbeta^{(r)}_0(X,p_1,p_2,q_1,q_2) \\
& +\sum_{j=1}^n    \frac{\bbeta^{(2)}_{1I}(X,p_1,p_2,q_1,q_2,k_j) }{ \tilde{S}_I(k_j) }
  +\sum_{l=1}^{n'} \frac{\bbeta^{(2)}_{1F}(X,p_1,p_2,q_1,q_2,k_l) }{ \tilde{S}_F(k_l) }\\
& +\sum_{n \geq j>k \geq 1}\;\;
       \frac{\bbeta^{(2)}_{2II}(X,p_1,p_2,q_1,q_2,k_j,k_k) }{ \tilde{S}_I(k_j)\tilde{S}_I(k_k)} 
  +\sum_{n' \geq l>m \geq 1}\;\;
       \frac{\bbeta^{(2)}_{2FF}(X,p_1,p_2,q_1,q_2,k_l,k_m) }{  \tilde{S}_F(k_l)\tilde{S}_F(k_m)}\\
& +\sum_{j=1}^n \sum_{l=1}^{n'}
       \frac{\bbeta^{(2)}_{2IF}(X,p_1,p_2,q_1,q_2,k_j,k_l) }{ \tilde{S}_I(k_j)\tilde{S}_F(k_l)}
  +\sum_{n \geq j>k >l \geq 1}\;\;
       \frac{\bbeta^{(3)}_{3III}(X,p_1,p_2,q_1,q_2,k_j,k_k,k_l) }
             { \tilde{S}_I(k_j) \tilde{S}_I(k_k) \tilde{S}_I(k_l) } 
  \Bigg\}.
\end{split}
\end{equation}
Let us explain the notation and physics content in the above expression.
The YFS soft factors for real photons emitted from the initial- and final-state fermions read
\begin{equation}
  \tilde{S}_I(k_j) = -Q_e^2 {\alpha \over 4\pi^2}
                     \Bigg( {p_1\over k_jp_1} -{p_2\over k_jp_2} \Bigg)^2,\qquad
  \tilde{S}_F(k_l)  = - Q_f^2 {\alpha \over 4\pi^2}
                     \Bigg( {q_1\over k_lq_1} -{q_2\over k_lq_2} \Bigg)^2,
\end{equation}
where electric charges of the electron and fermion $f$ are $Q_e$ and $Q_f$.
The $Y$-function in the exponential YFS form factor
is defined as in ref.~\cite{yfs1:1988}:
\begin{equation}
\label{form-factor}
\begin{split}
Y_f(\Omega,p,\bar{p}) 
     \equiv &  2 Q_f^2 \alpha \tilde{B}(\Omega,p,\bar{p})   +2 Q_f^2 \alpha \Re B(p,\bar{p}) \\
     \equiv & -2 Q_f^2 \alpha\;{ 1 \over 8\pi^2} \int {d^3k\over 2k^0} \Theta(\Omega;k)
                       \bigg({p\over kp} - {\bar{p}\over k\bar{p}} \bigg)^2 \\
            & +2 Q_f^2 \alpha \Re \int {d^4k\over k^2} {i\over (2\pi)^3} 
                       \bigg( {2p-k \over 2kp-k^2} -{2\bar{p}-k \over 2k\bar{p}-k^2} \bigg)^2.
\end{split}
\end{equation}
The above form factor is infrared-finite and depends explicitly on the soft
photon domains $\Omega=\Omega_I,\Omega_F$ 
which includes (surrounds) the IR divergence point $k=0$.
We define $\Theta(\Omega;k)=1$ for $k\in\Omega$ and $\Theta(\Omega;k)=0$ for $k\not\in\Omega$.
Contributions from the real photons inside $\Omega$ 
are summed to infinite-order and combined with the analogous virtual contributions forming
the exponential YFS form factor.
In the Monte Carlo we generate photons $k\not\in \Omega$
characterized by the function $\bar{\Theta}(\Omega,k) = 1-\Theta(\Omega,k)$.
We require, as usual, that $\Omega_{I}$ and $\Omega_{F}$ are
small enough (they can be chosen arbitrarily small) such that the total cross
section as defined in eq.~(\ref{rho-eex2}) and any other physically
meaningful  observable do not depend on the actual choice of them,
i.e. $\Omega_{I,F}$ are dummy parameters in the calculation!
If we neglect the initial-final state
interference then we may choose $\Omega_I \neq  \Omega_F$.
Let us define $\Omega_I$ with the $k^0<E_{min}$ condition in the
centre of the mass system of incoming $e^\pm$ beams
and $\Omega_F$ with $k^0<E'_{min}$ in the centre of the mass
of the outgoing fermions $f\bar{f}$.
The two domains differ because the Lorentz frames in which they are defined are different.
The above choice is the easiest for the Monte Carlo generation
but in the later discussion we shall  describe in detail how do
we implement the $\Omega_i=\Omega_F$ option in our Monte Carlo.
The actual YFS form factors for the above choices are well
known \cite{yfs:1961,yfs1:1988,yfs2:1990}:
\begin{equation}
\begin{split}
  Y_e(\Omega_I;p_1,p_2)
 &=   \gamma_e \ln {2E_{min}\over \sqrt{2p_1p_2}}     +{1\over 4}\gamma_e
      +Q_e^2 {\alpha\over\pi} \bigg( -{1\over 2} +{\pi^2\over 3}\bigg),\\
  Y_f(\Omega_F;q_1,q_2)
 &=   \gamma_f \ln {2E_{min}\over \sqrt{2q_1q_2}}     +{1\over 4}\gamma_f
      +Q_f^2 {\alpha\over\pi} \bigg( -{1\over 2} +{\pi^2\over 3}\bigg),
\end{split}
\end{equation}
where
\begin{equation}
\gamma =\gamma_e = 2 Q_e^2 {\alpha\over \pi} \bigg( \ln {2p_1p_2\over m_e^2} -1 \bigg),\quad
\gamma_f         = 2 Q_f^2 {\alpha\over \pi} \bigg( \ln {2q_1q_2\over m_f^2} -1 \bigg).
\end{equation}

%%%%%%%%%%%%%%%%%%%%%%%%%%%%%%%%%%%%%%%%%%%%%%%%%%%%%%%%
\subsection{Pure virtual corrections}
 
The perturbative QED matrix element is located in the $\bbeta$-functions.
The $\bbeta_0$ function is ``proportional'' to the Born
$e^-e^+\rightarrow f\bar{f}$ differential cross section
$d\sigma^\born(s,\vartheta)/d\Omega$ and it contains
(infrared-finite) corrections calculable order by order.
According to our general strategy we shall calculate
$\bbeta_0$ and other $\bbeta$'s in the \Ordpr{\alpha^i}, $i=0,1,2$.

The \Ordpr{\alpha^i} expressions
for $\bbeta_0^{(i)},i=0,1,2$ read%
\footnote{   It may look that we miss pure $(\alpha/\pi)$ term in $\delta^{(1)}_{I,F}$.
             The calculation shows \cite{yfs1:1988} that such a non-logarithmic
             contribution is accidentally equal zero.}
%//////////////////////////////////////////////////
%         beta0 at three orders
%//////////////////////////////////////////////////
\begin{equation}
  \begin{align}
    \label{beta0}
   &\bbeta_0^{(r)}(X,p_1,p_2,q_1,q_2)=
       (1+\delta_I^{(r)} )\; (1+\delta_F^{(r)} )\;
        \frac{1}{4}\sum_{k,l=1,2} {d\sigma^\born \over d\Omega}(X^2,\vartheta_{kl})\\
   &\delta_I^{(0)}=   0,\quad
    \delta_I^{(1)}=                  {1\over 2}  \gamma,\quad
    \delta_I^{(2)}= \delta_I^{(1)} + {1\over 8}  \gamma^2,\quad
    \delta_I^{(3)}= \delta_I^{(2)} + {1\over 48} \gamma^3,\\
   &\delta_F^{(0)}=   0,\quad
    \delta_F^{(1)}=                  {1\over 2}  \gamma_f,\quad
    \delta_F^{(2)}= \delta_I^{(1)} + {1\over 8}  \gamma_f^2,\quad
    \delta_F^{(3)}= \delta_I^{(2)} + {1\over 48} \gamma_f^3,
  \end{align}
\end{equation}
where
\begin{equation}
  \vartheta_{11}= \angle( \vec{p}_1, \vec{q}_1),\;
  \vartheta_{12}= \angle( \vec{p}_1,-\vec{q}_2),\;
  \vartheta_{21}= \angle(-\vec{p}_2, \vec{q}_1),\;
  \vartheta_{22}= \angle(-\vec{p}_2,-\vec{q}_2),\;
\end{equation}
with all 3-vectors taken
in the rest frame of the four-momentum $X$, that is in the frame XMS.

Let us first explain the fact that instead of having a single 
$d\sigma^\born/d\Omega(\vartheta)$ with a single $\vartheta$ we take an average over four $\vartheta_{kl}$.
In fact we could adopt one $\vartheta$, for example 
$\vartheta_0=\angle(\vec{p}_1-\vec{p}_2,\vec{q}_1-\vec{q}_2)$
where all three-momenta are taken in XMS.
The main reason for our apparently more complicated choice 
is related to the implementation of the first and higher-order real photon
contributions in the next subsections.
More precisely, it is well known~\cite{berends-kleiss:1981,mustraal-cpc:1983} 
that the exact single photon ISR matrix element can be cast
as a linear combination of the two $d\sigma^\born/d\Omega(\vartheta_k),k=1,2$ distributions.
The same is true for FSR~\cite{mustraal-cpc:1983}.
(Our implementation of the leading-logarithmic (LL) matrix element for 2 and 3 real photons
will also involve the linear combination of this type.)
It is therefore logical and practical to use a similar solution already for $\bbeta_0$.
One should also keep in mind that in the soft limit, when all photons are soft, then
all four angles $\vartheta_{kl}$ are identical and the averaging over them is a spurious operation anyway.

The reader not familiar with exponentiation may have an even more elementary question:
Why do we have a freedom of defining $\vartheta$ in $d\sigma^\born/d\Omega(\vartheta)$ in first place?
Is this ambiguity dangerous?
These questions are already discussed in refs. \cite{yfs2:1990,bhlumi2:1992}.
The answer is the following:
Strictly speaking the differential cross section $d\sigma^\born(s,\vartheta)/d\Omega$
and $\bbeta_0^{(i)}$ are defined within the two body phase-space.
Later on they are used, however, in eq.~(\ref{rho-eex2}) and in the definitions
of $\bbeta^{(i)},i=1,2,...$ all over the phase-space with additional
soft and/or hard photons.
This requires some kind of extrapolation of $\bbeta_0$ and $d\sigma^\born(s,\vartheta)/d\Omega$
beyond the two body phase-space.
In ref.~\cite{yfs2:1990} this extrapolation was done
using manipulations on the four-momenta and in ref.~\cite{bhlumi2:1992}
it was done as an extrapolation in the Mandelstam variables $s,t,u$.
Here we present another solution which is somewhere in between the previous two ones.
What is really important, however, is that the effect due to change
from one particular choice of extrapolation to another is always,
for the entire calculation, a kind of ``higher-order effect''.
For instance at \Order{\alpha^1} changing the type of extrapolation is an \Order{\alpha^2} effect!
Of course, it is always wise to use some kind of ``smooth''
extrapolation which is able to minimize the higher-order effects.

Another possible question is: 
Why we did not write down the second-order virtual correction 
factor in an {\em additive} way, like for instance
$(1 +\delta_I^{(2)} +\delta_F^{(2)} +\delta^{(1)}_I\delta^{(1)}_F)$?
We have opted for {\em factorized} form because it is generally known that the factorized
form is closer to reality at higher perturbative orders.
Another important reason is that the factorized form is easier for semi-analytical 
integrations over the phase-space in the next section.

%%%%%%%%%%%%%%%%%%%%%%%%%%%%%%%%%%%%%%%%%%%%%%%%%%%%%%%%
\subsection{One real photon with virtual corrections}

The contributions $\bbeta_1^{(2)}$ are needed directly
in eq.~(\ref{rho-eex2}) and the \Ordpr{\alpha^1}
version of $\bbeta_1^{(1)}$ enters indirectly
as a construction element in $\bbeta_2$.
They are constructed from QED distributions with a single
real photon emission and up to one virtual photon contribution.
They are defined separately for initial- and final-state photons
%%%%%%%%%%%%%%%%%%%%%%%%%%%%%%%%%%%%%%%%%%%%%%%%%%%
%   beta1 generic definition
%%%%%%%%%%%%%%%%%%%%%%%%%%%%%%%%%%%%%%%%%%%%%%%%%%%
\begin{equation}
  \begin{split}
    \bbeta_{1I}^{(i)}(X,p_1,p_2,q_1,q_2,k_j)=
    &  D_{1I}^{(i)}(X,p_1,p_2,q_1,q_2,k_j)  -\tilde{S}_I(k_j) \bbeta_0^{(i-1)}(X,p_1,p_2,q_1,q_2),\\
    \bbeta_{1F}^{(i)}(X,p_1,p_2,q_1,q_2,k'_l)=
    &  D_{1F}^{(i)}(X,p_1,p_2,q_1,q_2,k'_l) -\tilde{S}_F(k'_l) \bbeta_0^{(i-1)}(X,p_1,p_2,q_1,q_2),
  \end{split}
\end{equation}
where $i=1,2$.
Let us define first all ingredients for the initial-state contribution.
The single initial-state photon emission differential distribution at \Order{\alpha^r}, $r=1,2,3$, with
the eventual additional up to two-loop virtual correction from the initial- and/or final-state photon reads
%%%%%%%%%%%%%%%%%%%%%%%%%%%%%%%%%%%%%%%%%%%%%%%%%%%
%   1-photon distributions ISR
%%%%%%%%%%%%%%%%%%%%%%%%%%%%%%%%%%%%%%%%%%%%%%%%%%%
\def\halp{\hat{\alpha}}
\def\hbet{\hat{\beta}}
\begin{equation}
\label{single-initial}
  \begin{split}
    D^{(r)}_{1I}(X,&p_1,p_2,q_1,q_2,k_j) =
        Q_e^2 {\alpha \over 4\pi^2}    {2p_1p_2\over (k_jp_1)(k_jp_2)} W_e(\halp_j,\hbet_j)\\
   &\bigg\{  {(1-\halp_j)^2\over 2}  \sum_{r=1,2} {d\sigma^\born \over d\Omega}(X^2,\vartheta_{1r}) 
            +{(1-\hbet_j)^2\over 2}  \sum_{r=1,2} {d\sigma^\born \over d\Omega}(X^2,\vartheta_{2r})
    \bigg\}\\
   &\left(1+\Delta^{(r-1)}_I ( z_j) \right)  (1+\delta^{(r-1)}_F),\\
  \end{split}
\end{equation}
where
\begin{equation}
  \begin{split}
   &\halp_j= {k_jp_2 \over p_1p_2}, \quad \hbet_j= {k_jp_1 \over p_1p_2},\quad
      z_j= (1-\halp_j)(1-\hbet_j),\\
   &\Delta^{(0)}_I(z) \equiv  0,\quad
    \Delta^{(1)}_I(z) \equiv  {1\over 2}\gamma -{1\over 4}\gamma \ln(z),\\
   &\Delta^{(2)}_I(z) \equiv  \Delta^{(1)}_I(z)
        +{1\over 8}\gamma^2 -{1\over 8}\gamma^2 \ln(z) +{1\over 24}\gamma^2 \ln^2(z),\\
   &W_e(a,b) \equiv  1- {m_e^2\over 2p_1p_2}\; {(1-a)(1-b)\over (1-a)^2 + (1-b)^2}\;
                                 \bigg( {a\over b} + {b\over a} \bigg).\\
  \end{split}
\end{equation}
Again the question arises why the averaging over $r$ in $\vartheta_{kr}$ is introduced?
In the case of just one ISR hard photon the averaging trivially disappears because
$\vartheta_{k1}=\vartheta_{k2}$ and in this case
our formula coincides with the exact \Order{\alpha^1} result, 
see~\cite{berends-kleiss:1981,mustraal-np:1983}, as it should.
In the less trivial case of the presence of
the additional hard photons there is an ambiguity in defining $D^{(r)}_{1I}$ which is reflected
in our ``averaging'' procedure; however, it is harmless
i.e. the effect is of \Order{\alpha^2}.,

It is necessary and interesting to check the soft limit.
If in the presence of many additional photons ($n > 1$) we take the soft limit $k_j\rightarrow 0$,
keeping momenta of other photons constant, then $\vartheta_{kr}$ are in general all different.
However, in eq.~(\ref{single-initial}) the sums over $d\sigma^\born / d\Omega$
combine into a simple average over all four angles, as in eq.~(\ref{beta0}) --
in fact the single photon distribution reduces to
\begin{displaymath}
 D^{(2,1)}_{1I}(X,p_1,p_2,q_1,q_2,k_j)\sim
     \tilde{S}_I(k_j)\bbeta_0^{(1,0)}(X,p_1,p_2,q_1,q_2)
\end{displaymath}
and therefore
$\bbeta_{1I}^{(2,1)}(X,p_1,p_2,q_1,q_2,k_j)$ is infrared-finite as required.
The above argument shows
that the extrapolations for $\bbeta_0$ and $\bbeta_1$ have to be of the same type.
If we have opted for another extrapolation in eq.~(\ref{single-initial}),
for example without averaging, with a single angle $\vartheta_{kr}\to\vartheta_{k}$,
then the extrapolation in eqs.~(\ref{beta0}) would need to be changed appropriately.

Another interesting limit is the collinear limit.
If all (possibly hard) photons are collinear to initial- or final-fermions
then all angles $\vartheta_{sr},s,r=1,2$  are identical and equal to
the familiar leading-logarithmic effective scattering angle for the hard process
in the ``reduced frame'' XMS.
This will facilitate introduction of the higher-order LL corrections in the following.

Another remark on eq.~(\ref{single-initial}) is in order:
There are many equivalent ways, modulo term of \Order{m^2/s}, of writing
the single bremsstrahlung spin summed differential distribution \cite{mustraal-np:1983}.
Our choice follows the representation implemented in the Monte Carlo programs YFS2 \cite{yfs2:1990},
KORALZ \cite{koralz4:1994} and MUSTRAAL \cite{mustraal-cpc:1983},
because it minimizes the machine rounding errors
(quite important due to the smallness of electron mass),
and it is explicitly expressed in terms of
Born differential cross sections --
this feature facilitates introduction of electroweak corrections.

The virtual correction $(1+\Delta^{(1)}_I(\halp_j,\hbet_j))$ is
taken in the leading logarithmic approximation
(sufficient for our \Ordpr{\alpha^2} approach) and it agrees
with the corresponding distribution in ref.~\cite{berends-neerver-burgers:1988}.
In the $k_j\rightarrow 0$ limit we have
$\Delta^{(1)}_I(\halp_j,\hbet_j)\rightarrow \delta^{(1)}_I$
as expected, and as required for infrared finiteness of $\bbeta^{(2)}_{1F}$.
The other factor $(1+\delta^{(1)}_F)$ represents the contribution
from the simultaneous emission of the real initial
and the virtual final photon.
We again prefer the factorized form over an additive one
$(1+\Delta^{(1)}_I+\delta^{(1)}_F)$.

The essential ingredients for the \Order{\alpha^r} final-state $\bbeta^{(r)}_{1F},r=1,2$,
is the single final-state photon emission matrix element with
up to one-loop virtual initial- or final-state photon corrections
%%%%%%%%%%%%%%%%%%%%%%%%%%%%%%%%%%%%%%%%%%%%%%%%%%%%%%%%%%%%%%%%
%  1-photon FSR
%%%%%%%%%%%%%%%%%%%%%%%%%%%%%%%%%%%%%%%%%%%%%%%%%%%%%%%%%%%%%%%%
\def\heta{\hat{\eta}}
\def\hzet{\hat{\zeta}}
\begin{equation}
  \label{single-final}
  \begin{split}
    D^{(r)}_{1F}(X,&p_1,p_2,q_1,q_2,k'_l) = 
       Q_f^2 {\alpha \over 4\pi^2} {2q_1q_2\over (k'_lq_1)(k'_lq_2)} W_f(\heta_l,\hzet_l)\\
      &\bigg\{ {(1-\heta_l)^2\over 2} \sum_{r=1,2} {d\sigma^\born \over d\Omega}(X^2,\vartheta_{r1})
              +{(1-\hzet_l)^2\over 2} \sum_{r=1,2} {d\sigma^\born \over d\Omega}(X^2,\vartheta_{r2})
       \bigg\}\\
      &\left(1+\Delta^{(r-1)}_F( z_l ) \right) (1+\delta^{(r-1)}_I)
  \end{split}
\end{equation}
where
%%%%%%%%%%%%%%%%%%%%%%%%%%%%%%%%%%%%%%%%%%%%%%%%%%%%%%%%%%%%%%%%
\begin{equation}
  \begin{split}
  &\eta_l  = {k'_lq_2 \over q_1q_2},\quad
   \zeta_l = {k'_lq_1 \over q_1q_2},\quad
   \heta_l = {\eta_l  \over 1+\eta_l+\zeta_l},\quad
   \hzet_l = {\zeta_l \over 1+\eta_l+\zeta_l},\\
  &z_l=(1-\heta_l)(1-\hzet_l)\\
  &\Delta^{(0)}_F(z) \equiv 0,\quad
   \Delta^{(1)}_F(z) \equiv  {1\over 2}\gamma_f +{1\over 4}\gamma_f \ln(z),\\
  &W_f( a, b) \equiv 1- {m_f^2\over 2q_1q_2}\; {(1- a)(1- b)\over (1- a)^2 + (1- b)^2}\;
                                   \bigg( { a\over b} + { b\over a} \bigg),\\
  \end{split}
\end{equation}
All discussion on the ISR distribution of eq.~(\ref{single-initial}) applies also to the
above FSR distribution.

%%%%%%%%%%%%%%%%%%%%%%%%%%%%%%%%%%%%%%%%%%%%%%%%%%%%%%%%%%%%%%
\subsection{Two real  photons with virtual corrections}
%%%%%%%%%%%%%%%%%%%%%%%%%%%%%%%%%%%%%%%%%%%%%%%%%%%%%%%%%%%%%%
The contributions $\bbeta^{(2)}_{II},\bbeta^{(2)}_{FF}$ and $\bbeta^{(2)}_{IF}$
are related to emission of the real two initial, two final and
one initial and one final photons correspondingly.
They are genuine \Order{\alpha^2} objects because they appear in this order for the first time.
For the same reason they do not include any virtual contributions.
They are defined formally in the usual way
\begin{equation}
  \begin{align}
    \begin{split}
      \bbeta_{II}^{(r)}(X&,p_1,p_2,q_1,q_2,k_j,k_k)
          =D_{II}^{(r)}(X ,p_1,p_2,q_1,q_2,k_j,k_k)\\
          &  -\tilde{S}_I(k_j) \bbeta_{1I}^{(r-1)}(X,p_1,p_2,q_1,q_2,k_k)
             -\tilde{S}_I(k_k) \bbeta_{1I}^{(r-1)}(X,p_1,p_2,q_1,q_2,k_j)\\
          &  -\tilde{S}_I(k_j)\tilde{S}_I(k_k) \bbeta_{0}^{(r-2)}(X,p_1,p_2,q_1,q_2), r=2,3,
    \end{split}\\
    \begin{split}
      \bbeta_{FF}^{(r)}(X&,p_1,p_2,q_1,q_2,k'_l,k'_m)
          =D_{FF}^{(r)}(X ,p_1,p_2,q_1,q_2,k'_l,k'_m)\\
          & -\tilde{S}_F(k'_l) \bbeta_{1F}^{(r-1)}(X,p_1,p_2,q_1,q_2,k'_m)
            -\tilde{S}_F(k'_m) \bbeta_{1F}^{(r-1)}(X,p_1,p_2,q_1,q_2,k'_l)\\
          & -\tilde{S}_F(k'_l)\tilde{S}_F(k'_m) \bbeta_{r-2}^{(r-2)}(X,p_1,p_2,q_1,q_2),r=2,3,
    \end{split}\\
    \begin{split}
      \bbeta_{IF}^{(r)}(X&,p_1,p_2,q_1,q_2,k_j,k'_l)
          =D_{IF}^{(r)}(X ,p_1,p_2,q_1,q_2,k_j,k'_l)\\
          & -\tilde{S}_I(k_j)  \bbeta_{1F}^{(r-1)}(X,p_1,p_2,q_1,q_2,k'_l)
            -\tilde{S}_F(k'_l) \bbeta_{1I}^{(r-1)}(X,p_1,p_2,q_1,q_2,k_j)\\
          & -\tilde{S}_I(k_j)\tilde{S}_F(k'_l) \bbeta_{r-2}^{(0)}(X,p_1,p_2,q_1,q_2),r=2,3.
    \end{split}
   \end{align}
\end{equation}

The new objects in the above expressions are the differential distributions
$D_{II}^{(2)},D_{FF}^{(2)}$ and $D_{IF}^{(2)}$ for double bremsstrahlung.
They are not taken directly from Feynman diagrams
but they are {\em constructed} in such a way that:
\begin{itemize}
\item
  If one photon is hard and one is soft then the single bremsstrahlung expression of 
  eqs.~(\ref{single-initial},\ref{single-final}) are recovered
\item
  If both photons are hard and collinear then
  the proper LL limit, which we know from the double or triple convolution
  of the Altarelli-Parisi kernels, is also recovered.
\end{itemize}
The resulting expressions are rather compact and the LL limit is manifest,
this is not necessarily true for the exact double bremsstrahlung spin amplitudes (see next section).
The method is similar to that of refs.~\cite{yfs2:1990,bhlumi4:1996}.
In the case of ISR we shall also include one-loop virtual corrections
read from the triple convolution of the Altarelli-Parisi kernels, see ref.~\cite{bhlumi4:1996}.

Our construction in the case of the double real ISR reads as follows
%%%%%%%%%%%%%%%%%%%%%%%%%%%%%%%%%%%%%%%%%%%%%%%%%%%%%%%%%%%%%%%%%
%   double ISR
%%%%%%%%%%%%%%%%%%%%%%%%%%%%%%%%%%%%%%%%%%%%%%%%%%%%%%%%%%%%%%%%%
\begin{equation}
\label{double-initial}
  \begin{split}
    D&_{II}^{(2)}(X,p_1,p_2,q_1,q_2,k_1,k_2)\equiv\\
    & Q_e^4\;
     {\alpha \over 4\pi^2}  {2p_1p_2\over (k_1p_1)(k_1p_2)}\;
     {\alpha \over 4\pi^2}  {2p_1p_2\over (k_2p_1)(k_2p_2)}
     W_e(\halp_1,\hbet_1)W_e(\halp_2,\hbet_2)\\
     &\bigg\{ \;
       \Theta(v_1-v_2) 
       \left(1+\Delta^{(r-1)}_{II}( z_1,z_{12} ) \right) (1+\delta^{(r-1)}_F)\\
     &\qquad\bigg[ 
             \chi_2( \halp_1; \halp'_2,\hbet'_2) \sum_{r=1,2} {d\sigma^\born \over d\Omega}(X^2,\vartheta_{1r})
            +\chi_2( \hbet_1; \halp'_2,\hbet'_2) \sum_{r=1,2} {d\sigma^\born \over d\Omega}(X^2,\vartheta_{2r}) 
            \bigg]\\
     &+\Theta(v_2-v_1) 
       \left(1+\Delta^{(r-1)}_{II}( z_2,z_{21} ) \right) (1+\delta^{(r-1)}_F)\\
      &\qquad\bigg[ 
             \chi_2( \halp_2; \halp'_1,\hbet'_1) \sum_{r=1,2} {d\sigma^\born \over d\Omega}(X^2,\vartheta_{1r})
            +\chi_2( \hbet_2; \halp'_1,\hbet'_1) \sum_{r=1,2} {d\sigma^\born \over d\Omega}(X^2,\vartheta_{2r}) 
            \bigg]
      \bigg\},
   \end{split}
\end{equation}
where
%%%%%%%%%%%%%%%%%%%%%%%%%%%%%%%%%%%%%%%%%%%%%%%%%%%%%%%%
\begin{equation}
  \begin{split}
     &  \halp'_1  ={\halp_1 \over 1-\halp_2},\quad \halp'_2={\halp_2 \over 1-\halp_1},\quad
        \hbet'_1  ={\hbet_1 \over 1-\hbet_2},\quad \hbet'_2={\hbet_2 \over 1-\hbet_1},\\
     &  v_i   = \halp_i + \hbet_i,\quad
        z_i   = (1-\halp_i)(1 -\hbet_i),\quad
        z_{ij}= (1-\halp_i-\halp_j)(1 -\hbet_i-\hbet_j),\\
     &  \chi_2(u;a,b) \equiv {1\over 4} (1-u)^2   \big[ (1-a)^2 +(1-b)^2  \big],\\
     &  \Delta^{(0)}_{II}=0,\quad
        \Delta^{(1)}_{II}( z_i,z_{ij} ) 
          = \frac{1}{2}\gamma -\frac{1}{6}\gamma\ln(z_i) -\frac{1}{6}\gamma\ln(z_{ij}).
   \end{split}
\end{equation}
The variables $\halp_i,\hbet_i$ for $i$-th photon are defined as in eq.~(\ref{single-initial}).

In order to understand our construction let us examine how the LL collinear limit is realized
in the exact single bremsstrahlung matrix element of eq.~(\ref{single-initial}).
If the photon carrying the fraction $x_1$ of the beam energy is collinear, let us say, with $p_1$ then
$\halp_1\sim x$, $\hbet_1\sim0$, all four angles are the same $\vartheta_{sr}\to\vartheta^*$
and we recover immediately the correct LL formula
\begin{displaymath}
  \frac{1}{2}(1-\halp_1)^2 \sum_{r=1,2} {d\sigma^\born \over d\Omega}(\vartheta_{1r}) 
 +\frac{1}{2}(1-\hbet_1)^2 \sum_{r=1,2} {d\sigma^\born \over d\Omega}(\vartheta_{2r})
  \to
  \frac{1}{2}(1+(1-x)^2) {d\sigma^\born \over d\Omega}(\vartheta^*).
\end{displaymath}
It is therefore natural to employ for the double emission
the {\em angular dependent} Altarelli-Parisi (AP) factors of the kind
\begin{displaymath}
  \frac{1}{2}[(1-\halp_2)^2 +(1-\hbet_2)^2]\; \frac{1}{2}[(1-\halp_1)^2 +(1-\hbet_1)^2].
\end{displaymath}
The above formula is too simple, however, to reproduce correctly the result of the double
convolution of the AP kernels in the case when both photons are collinear with the same fermion
\begin{displaymath}
  \frac{1}{2}(1+(1-x_1)^2)\; \frac{1}{2}(1+(1-[x_2/(1-x_1)]^2)) 
           {d\sigma^\born \over d\Omega}(\vartheta^*).
\end{displaymath}
where $x'_2=x_1/(1-x_1)$ reflects the loss of energy in the emission cascade due to emission of $k_1$.
In order to match the above cascade limit we construct a better angular dependent AP factor as
\begin{displaymath}
  \frac{1}{2}[(1-\halp_1)^2 +(1-\hbet_1)^2]\; \frac{1}{2}[(1-\halp'_2)^2 +(1-\hbet'_2)^2].
\end{displaymath}
The above fulfils both types of LL collinear limit, when two photons are collinear with
a single beam or each of them follows different beam.
Finally, let us reproduce the limit in which one photon, 
let us say the 1-st, is hard and the other, the 2-nd, is soft, $v_2=\halp_2+\hbet_2\to 0$.
In this case it is logical to split the above double bremsstrahlung angular dependent AP factor
into two pieces
\begin{displaymath}
\begin{split}
  \chi_2(\halp_1;\halp'_2,\hbet'_2)&= \frac{1}{2} (1-\halp_1)^2\; \frac{1}{2}[(1-\halp'_2)^2 +(1-\hbet'_2)^2],\\
  \chi_2(\hbet_1;\halp'_2,\hbet'_2)&= \frac{1}{2} (1-\hbet_1)^2\; \frac{1}{2}[(1-\halp'_2)^2 +(1-\hbet'_2)^2]
\end{split}
\end{displaymath}
and associate each one with the corresponding $d\sigma^\born/d\Omega$, following eq.~(\ref{single-initial}).
The order in the cascade does not matter.
We simply symmetrize over the two orderings in the cascade -- it is essentially Bose-Einstein symmetrization.

The above construction clearly provides the correct limit
$D_{II}^{(2)}(k_1,k_2)\rightarrow \tilde{S}(k_2) D^{(1)}_{1I}(k_2)$ for $v_1=const$ and $v_2\to 0$.
As a consequence
$\bbeta_{II}^{(2)}(X,p_1,p_2,q_1,q_2,k_1,k_2)$ is finite
in the limit of one or both photon momenta tending to zero.

The construction of eq.~(\ref{double-initial}) will be inadequate if both photons are hard and at least
one has high transverse momentum.
It reflects the fact that we do not control fully in EEX the second-order NLL, \Order{\alpha^2L}, contributions.
However, we have known since a long time that the construction of the type 
of eq.~(\ref{double-initial}) agrees rather well with the exact double bremsstrahlung
matrix element calculated using spinor techniques, see \cite{ela-ward:1991}.
For both photons having high transverse momenta there is only about 20\% disagreement
for the approximate and exact results (integrated over the double photon phase-space).
This result is confirmed in the present work by the numerical comparisons 
of EEX and CEEX, where the double bremsstrahlung matrix element is exact.
 
The double final-state bremsstrahlung distribution is
defined/constructed in an analogous way
%%%%%%%%%%%%%%%%%%%%%%%%%%%%%%%%%%%%%%%%%%%%%%%%%%%%%%%%%
%   double final
%%%%%%%%%%%%%%%%%%%%%%%%%%%%%%%%%%%%%%%%%%%%%%%%%%%%%%%%%
\def\hetap{{\hat{\eta}'}}
\def\hzetp{{\hat{\zeta}'}}
\def\ha{\hat{a}}
\def\hb{\hat{b}}
\begin{equation}
  \label{double-final}
  \begin{split}
    &D_{FF}^{(r)}(X,p_1,p_2,q_1,q_2,k'_1,k'_2)=\\
    &Q_f^4
    {\alpha \over 4\pi^2}  {2q_1p_2\over (k'_1q_1)(k'_1p_2)}
    {\alpha \over 4\pi^2}  {2q_1p_2\over (k'_2q_1)(k'_2p_2)}
    W_f(\heta_1,\hzet_1)  W_f(\heta_2,\hzet_2)\\
    &\bigg\{
       \Theta(v'_1-v'_2) 
       \bigg[ \chi_2( \eta_1;  \eta'_2, \zeta'_2)
                    \sum_{r=1,2} {d\sigma^\born \over d\Omega}(X^2,\vartheta_{1r})
             +\chi_2( \zeta_1; \eta'_2, \zeta'_2)
                    \sum_{r=1,2} {d\sigma^\born \over d\Omega}(X^2,\vartheta_{2r})
       \bigg]\\
    & +\Theta(v'_2-v'_1) 
       \bigg[ \chi_2( \eta_2;  \eta'_1, \zeta'_1)
                    \sum_{r=1,2} {d\sigma^\born \over d\Omega}(X^2,\vartheta_{1r})
             +\chi_2( \zeta_2; \eta'_1, \zeta'_1)
                    \sum_{r=1,2} {d\sigma^\born \over d\Omega}(X^2,\vartheta_{2r})
       \bigg]
     \bigg\}\\
     &\left(1+\Delta^{(r-1)}_I ( z_j) \right),
\end{split}
\end{equation}
where
%%%%%%%%%%%%%%%%%%%%%%%%%%%%%%%%
\begin{equation}
  \eta'_1  ={\eta_1  \over 1+\eta_2}, \quad \eta'_2  ={\eta_2  \over 1+\eta_1},\quad
  \zeta'_1 ={\zeta_1 \over 1+\zeta_2},\quad \zeta'_2 ={\zeta_2 \over 1+\zeta_1}.
\end{equation}
The ``primed'' Sudakov variables are here defined differently than in the
ISR case because the fermion momenta $q_{1,2}$ get affected by photon emission.
Virtual corrections are absent because we restrict FSR to \Order{\alpha^2}$_{LL}$.
The above expression is tagged with $r=2,3$ for \Order{\alpha^r}, however,
FSR we implement essentially only in \Order{\alpha^2} and the only correction in \Order{\alpha^3}
is the ISR one-loop correction.
 
The  distribution for one photon from the initial-state and
one photon from the final-state at \Order{\alpha^r} $r=1,2$ we construct as follows
%%%%%%%%%%%%%%%%%%%%%%%%%%%%%%%%%%%%%%%%%%%%%%%%%%%%%%%%%
%   double initial-final
%%%%%%%%%%%%%%%%%%%%%%%%%%%%%%%%%%%%%%%%%%%%%%%%%%%%%%%%%
\begin{equation}
\label{initial-final}
\begin{split}
D^{(r)}_{IF}&(X,p_1,p_2,q_1,q_2,k_j,k'_l)=\\
& Q_e^2 {\alpha \over 4\pi^2} {2p_1p_2\over (k_j p_1)(k_j p_2)}   W_e(\halp_j,\hbet_j)\;\;
  Q_f^2 {\alpha \over 4\pi^2} {2p_1p_2\over (k'_lp_1)(k'_lp_2)}   W_f(\heta_l,\hzet_l)\\
&\bigg\{
   {(1-\halp_j)^2\over 2}
   {(1-\heta_l)^2\over 2}
      {d\sigma^\born \over d\Omega}(X^2,\vartheta_{11})
  +{(1-\halp_j)^2\over 2}
   {(1-\hzet_l)^2\over 2}
      {d\sigma^\born \over d\Omega}(X^2,\vartheta_{12})\\
&+{(1-\hbet_j)^2\over 2}
   {(1-\heta_l)^2\over 2}
      {d\sigma^\born \over d\Omega}(X^2,\vartheta_{21})
  +{(1-\hbet_j)^2\over 2}
   {(1-\hzet_l)^2\over 2}
      {d\sigma^\born \over d\Omega}(X^2,\vartheta_{22})
  \bigg\}\\
& \left(1+\Delta^{(r-1)}_I(z_1)\right) \left(1+\Delta^{(r-1)}_F(z'_2)\right)
\end{split}
\end{equation}
where the variables $\halp_j,\hbet_j,\heta_l,\hzet_l$ and other components 
are defined as in eqs.~(\ref{single-initial},\ref{single-final}).
The above construction is in fact the easiest because two photons cannot be emitted in a cascade
from one line and we fully exploit the four scattering angles in the Born differential
cross sections.
It is trivial to check that all soft and collinear limits are correct.

%%%%%%%%%%%%%%%%%%%%%%%%%%%%%%%%%%%%%%%%%%%%%%%%%%%%%%%%
\subsection{Three real photons}

The differential distribution for of 3 real ISR photons is essentially obtained by the triple
convolution of the AP kernel, for each beam separately and the the two results are
combined with help of additional convolution.
This exercise was done for the collinear sub-generator of BHLUMI~\cite{bhlumi4:1996}
and we exploit here these results.
Even though the collinear limit is of primary importance, we have to be very careful in construction
of the fully differential triple photon distribution to preserve all soft limits: 
when all three photons are soft, when two of them are soft , and only one of them is soft.
In these limits the three-photon differential distribution has
to reproduce smoothly the previously defined Born, single and double
bremsstrahlung distributions times the appropriate soft factor(s).
Otherwise we may have a problem with IR finiteness of 
\begin{equation}
    \begin{split}
      \bbeta&_{III}^{(3)}(X,p_i,q_j,k_1,k_2,k_3)
          =D_{III}^{(r)}(X ,p_i,q_j,k_1,k_2,k_3)\\
          &  -\tilde{S}_I(k_1) \bbeta_{1I}^{(2)}(X,p_i,q_j,k_2,k_3)
             -\tilde{S}_I(k_2) \bbeta_{1I}^{(2)}(X,p_i,q_j,k_1,k_3)
             -\tilde{S}_I(k_3) \bbeta_{1I}^{(2)}(X,p_i,q_j,k_1,k_2)\\
          &  -\tilde{S}_I(k_1)\tilde{S}_I(k_2)  \bbeta_{1I}^{(1)}(X,p_i,q_j,k_3)
             -\tilde{S}_I(k_3)\tilde{S}_I(k_1)  \bbeta_{1I}^{(1)}(X,p_i,q_j,k_2)\\
          &  -\tilde{S}_I(k_2)\tilde{S}_I(k_3)  \bbeta_{1I}^{(1)}(X,p_i,q_j,k_1)
             -\tilde{S}_I(k_1)\tilde{S}_I(k_2)\tilde{S}_I(k_3) \bbeta_{0}^{(0)}(X,p_i,q_j).
    \end{split}
\end{equation}
It is therefore not completely straightforward to turn the strictly collinear expression
for three real photon distributions
of ref.~\cite{bhlumi4:1996} into the fully differential (finite $p_T$) triple photon distribution
which we need.
As in the case of double real ISR the guiding principle is that (i) the hardest photon
decides which of the angles is used in $d\sigma^\born /d\Omega (X^2,\vartheta_{lr})$
and (ii) we have to perform Bose symmetrization, that is sum over all orderings
in a cascade emission of several photons from one beam.
For three real photons there are no virtual corrections.

Our construction in the case of the triple real ISR reads as follows
%%%%%%%%%%%%%%%%%%%%%%%%%%%%%%%%%%%%%%%%%%%%%%%%%%%%%%%%%%%%%%%%%
%   triple ISR
%%%%%%%%%%%%%%%%%%%%%%%%%%%%%%%%%%%%%%%%%%%%%%%%%%%%%%%%%%%%%%%%%
\begin{equation}
\label{triple-initial}
  \begin{split}
   D_{II}^{(3)}(X,p_1,p_2,q_1,q_2,k_1,k_2,k_3) &\equiv
   \prod_{l=1,3}
         Q_e^2\; {\alpha \over 4\pi^2}\;  {2p_1p_2\over (k_l p_1)(k_l p_2)}\; W_e(\halp_l,\hbet_l)\\
     \bigg\{ \;
       \Theta(v_1-v_2) \Theta(v_2-v_3)
       \bigg[
      &\;\;\chi_3( \halp_1; \halp'_2,\hbet'_2, \halp''_3,\hbet''_3) 
                \sum_{r=1,2} {d\sigma^\born \over d\Omega}(X^2,\vartheta_{1r})\\
      &   +\chi_3( \hbet_1; \halp'_2,\hbet'_2, \halp''_3,\hbet''_3) 
                \sum_{r=1,2} {d\sigma^\born \over d\Omega}(X^2,\vartheta_{2r}) 
       \bigg]\\
      \quad +\; {\rm remaining}\; & {\rm five}\; {\rm permutations}\; {\rm of}\; (1,2,3) \quad
      \bigg\},
   \end{split}
\end{equation}
where
\begin{equation}
  \begin{split}
  &\chi_3( u_1; a_2,b_2, a_3,b_3) 
     \equiv {1\over 8} (1-u_1)^2 \big[ (1-a_2)^2 +(1-b_2)^2  \big] \big[ (1-a_3)^2 +(1-b_3)^2  \big],\\
  &  \halp''_3  ={\halp_3 \over 1-\halp_1-\halp_2},\quad 
     \hbet''_3  ={\hbet_3 \over 1-\hbet_1-\hbet_2},\quad 
  \end{split}
\end{equation}

In most cases such an approach should be enough; however, in some 
special cases with two hard photons explicitly tagged it may not be sufficient. 
We have programmed and run special tests (unpublished)
relying on the up to 3 hard-photon ISR amplitudes
\cite{erw-private:1997} constructed
with the methods similar to these in ref.\cite{erw:1994},
in order to get additional confidence in the approximate real emission
distrubutions presented in this Section.

%%%%%%%%%%%%%%%%%%%%%%%%%%%%%%%%%%%%%%%%%%%%%%%%%%%%%%%%%%%%%%%%%%%%%%%%%%%%%%%%%%%%%%%%%%%
%%%%%%%%%%%%%%%%%%%%%%%%%%%%%%%%%%%%%%%%%%%%%%%%%%%%%%%%%%%%%%%%%%%%%%%%%%%%%%%%%%%%%%%%%%%
%%%%%%%%%%%%%%%%%%%%%%%%%%%%%%%%%%%%%%%%%%%%%%%%%%%%%%%%%%%%%%%%%%%%%%%%%%%%%%%%%%%%%%%%%%%

%%%%%%%%%%%%%%%%%%%%%%%%%
% make ceex2-all-ps
%%%%%%%%%%%%%%%%%%%%%%%%%

%%%%%%%%%%%%%%%%%%%%%%%%%%%%%%%%%%%%%%%%%%%%%%%%%%%%%%%%%%%%%%%%%%%%%%%%%%%%%%%%%%%%%%%%%%%%%
%%%%%%%%%%%%%%%%%%%%%%%%%%%%%%%%%%%%%%%%%%%%%%%%%%%%%%%%%%%%%%%%%%%%%%%%%%%%%%%%%%%%%%%%%%%%%
%%%%%%%%%%%%%%%%%%%%%%%%%%%%%%%%%%%%%%%%%%%%%%%%%%%%%%%%%%%%%%%%%%%%%%%%%%%%%%%%%%%%%%%%%%%%%
\newpage
\section{Amplitudes for Coherent Exclusive Exponentiation}
\label{sec:ceex}

The Coherent Exclusive Exponentiation (CEEX) was introduced for the first time in
ref.~\cite{ceex1:1999}.
It is deeply rooted in the Yennie-Frautschi-Suura (YFS) exponentiation~\cite{yfs:1961}.
It applies in particular to processes with narrow resonances where
it is related also to works of Greco et.al.~\cite{greco:1975,greco:1980}.
The exponentiation procedure, that is a reorganisation of the QED perturbative series
such that infrared (IR) divergences  are summed up to infinite-order
is done at the spin-amplitude level for both real and virtual IR singularities.
This is to be contrasted with traditional YFS exponentiation, 
on which our EEX is based,
where isolating the {\em real} IR divergences is done for squared spin-summed spin amplitudes,
that is for differential distributions and spin density matrices%
\footnote{ The realization of EEX for spin density matrices is an obvious generalisation
  of the EEX/YFS exponentiation which, however, was never fully implemented in practice.}.

Our calculations of the spin amplitudes for fermion pair production
in electron positron scattering is done
with the help of the powerful Weyl spinor (WS) techniques.
There are several variants of WS techniques.
We have opted for the method of Kleiss and Stirling 
(KS)~\cite{kleiss-stirling:1985,kleiss-stirling:1986},
which we found the best suited for our CEEX.
In ref.~\cite{gps:1998} the KS spinor technique for massless and massive fermions
was reviewed and appended with the rules
for controlling their complex  phases, or equivalently, the fermion rest frame (all three axes)
in which the fermion spin is quantised
-- this is a critical point if we want to control fully the spin density matrix of the fermions.
This fermion rest frame we call the GPS frame and the rule for finding it we call the GPS rule.
For the sake of completeness we include definitions of the KS spinors, photon polarization vectors,
and our GPS rules in Appendix~A.

The very interesting feature of CEEX is that,
although it is formulated entirely in terms of the spin-amplitudes,
the IR cancellations in CEEX occur for the integrated cross sections (probabilities), as usual;
in practice they are realised {\em numerically}.
There is no contradiction in the above statement. 
In order to avoid any confusion on this point, we shall provide the new detailed proof
of IR cancellations in CEEX scheme in one of the following subsections.

%%%%%%%%%%%%%%%%%%%%%%%%%%%%%%%%%%%%%%%%%%%%%%%%%%%%%%%%%%%%%%%%
\subsection{Master formula}
%%%%%%%%%%%%%%%%%%%%%%%%%%%%%%%%%%%%%%%%%%%%%%%%%%%%%%%%%%%%%%%%

Defining the Lorentz invariant phase-space as
%////////////////////////////////////////////////
\begin{equation}
\label{eq:lips}
\int d{\rm Lips}_n(P;p_1,p_2,...,p_n) 
    = \int (2\pi)^4\delta(P-\sum_{i=1}^n p_i) \prod_{i=1}^n \frac{ d^3 p}{(2\pi)^3 2p^0_i}
\end{equation}
we write the CEEX total cross section for the process
%//////////////////////////////////////////////////////
\begin{equation}
e^-(p_a) +e^+(p_b) 
  \to f(p_c) +\bar{f}(p_d) +\gamma(k_1) +\gamma(k_2)+...+\gamma(k_n), n=0,1,2,...,\infty
\end{equation}
with polarized beams
and decays of unstable final fermions being sensitive to fermion spin polarizations,
following refs.~\cite{ceex1:1999}, as follows:
%//////////////////////////////////////////////////////
\begin{equation}
  \label{eq:sigma-ceex2}
%  \begin{split}
  \sigma^{(r)} =  {1\over {\rm flux}(s)}
  \sum_{n=0}^\infty 
  \int d{\rm Lips}_{n+2} ( p_a+p_b; p_c,p_d, k_1,\dots,k_n)\;
  \rho^{(r)}_{\rm CEEX}  ( p_a,p_b, p_c,p_d, k_1,\dots,k_n)
%  \end{split}
\end{equation}
where, in the CMS ${\rm flux}(s)\simeq 2s$,
%///////////////////////////////////////////////////////
\begin{equation}
  \label{eq:rho-ceex2}
  \begin{split}
  \rho^{(r)}_{\rm CEEX} &( p_a,p_b, p_c,p_d, k_1,k_2,\dots,k_n)=
  {1\over n!} e^{Y(\Omega;p_a,...,p_d)}\;\bar{\Theta}(\Omega)\;
    \sum_{\sigma_i=\pm 1}\;
    \sum_{\lambda_i,\bar{\lambda}_i=\pm 1}\;
\\&
    \sum_{i,j,l,m=0}^3\;
        \hat{\varepsilon}^i_a                  \hat{\varepsilon}^j_b\;
        \sigma^i_{\lambda_a \bar{\lambda}_a}   \sigma^j_{\lambda_b \bar{\lambda}_b}
    \Mmf^{(r)}_n 
    \left(\st^{p}_{\lambda} \st^{k_1}_{\sigma_1} \st^{k_2}_{\sigma_2}
                                           \dots \st^{k_n}_{\sigma_n} \right)
    \left[
    \Mmf^{(r)}_n 
    \left(\st^{p}_{\bar{\lambda}} \st^{k_1}_{\sigma_1} \st^{k_2}_{\sigma_2}
                                                 \dots \st^{k_n}_{\sigma_n} \right)
    \right]^\star
        \sigma^l_{\bar{\lambda}_c \lambda_c }   \sigma^m_{\bar{\lambda}_d \lambda_d }
        \hat{h}^l_c                             \hat{h}^m_d,
  \end{split}
\end{equation}
%%------------
and assuming domination of the $s$-channel exchanges, including resonances,
the complete set of spin amplitudes
for emission of $n$ photons we define in \Oceex{\alpha^r} $r=0,1,2$ as follows:
%///////////////////////////////////////////////////
%     CEEX  O(alf2)
%///////////////////////////////////////////////////
\begin{equation}
  \label{eq:ceex-master}
  \begin{align}
  &\Mmf^{(0)}_n\left( \st^{p}_{\lambda} \st^{k_1}_{\sigma_1}
                                   \dots \st^{k_n}_{\sigma_n}  \right)
   =\!\!\!\sum_{\wp\in\{I,F\}^n}\;  
     \prod_{i=1}^n \; \sfac^{\{\wp_i\}}_{[i]}\;
      \hbeta^{(0)}_0 \left( \st^{p}_{\lambda}; X_\wp \right),
\\
  &\Mmf^{(1)}_n\left( \st^{p}_{\lambda} \st^{k_1}_{\sigma_1}
                                   \dots \st^{k_n}_{\sigma_n}  \right)
   =\!\!\!\sum_{\wp\in\{I,F\}^n}\;  
     \prod_{i=1}^n \; \sfac^{\{\wp_i\}}_{[i]}\;
     \left\{ \hbeta^{(1)}_0 \left( \st^{p}_{\lambda}; X_\wp \right)
            +\sum_{j=1}^n 
             {\hbeta^{(1)}_{1\{\wp_j\}} \left( \st^{p}_{\lambda} \st^{k_j}_{\sigma_j} ; X_\wp \right)
                                          \over \sfac^{\{\wp_j\}}_{[j]} }\;
    \right\},
\\ \nonumber
  &\Mmf^{(2)}_n\left( \st^{p}_{\lambda} \st^{k_1}_{\sigma_1}
                                   \dots \st^{k_n}_{\sigma_n}  \right)
   =
\\
  \label{eq:ceex-master2}
   &=\!\!\!\sum_{\wp\in\{I,F\}^n}\;  
    \prod_{i=1}^n \; \sfac^{\{\wp_i\}}_{[i]}\;
    \left\{  \hbeta^{(2)}_0 \left( \st^{p}_{\lambda}; X_\wp \right)
            +\sum_{j=1}^n 
             {\hbeta^{(2)}_{1\{\wp_j\}} \left( \st^{p}_{\lambda} \st^{k_j}_{\sigma_j} ; X_\wp \right)
                                          \over \sfac^{\{\wp_j\}}_{[j]} }\;
             +\sum_{1\leq j<l\leq n}\;\!\!\!\!
              {\hbeta^{(2)}_{2\{\wp_j\wp_l\}}
                   \left( \st^{p}_{\lambda} \st^{k_j}_{\sigma_j}  \st^{k_l}_{\sigma_l} ;X_\wp \right)
                               \over \sfac^{\{\wp_j\}}_{[j]} \sfac^{(\wp_l)}_{[l]} }\;
    \right\},
  \end{align}
\end{equation}
%%------------
In the following subsections we shall explain all basic notation,
then in the next section we shall discuss in detail the IR structure in CEEX,
effectively deriving all the above formulas.
At \Order{\alpha^r} we have to provide
for functions $\hbeta^{(r)}_{k}, k=0,1,...,r$ from Feynman diagrams, which are
infrared-finite by construction~\cite{yfs:1961}.
Their actual precise definitions will be given in the following.
We shall define/calculate them explicitly up to \Order{\alpha^2}.

%%%%%%%%%%%%%%%%%%%%%%%%%%%%%%%%%%%%%%%%%%%%%%%%%%%%%%%%%%%%%%%%
\subsubsection{Spin notation}
%%%%%%%%%%%%%%%%%%%%%%%%%%%%%%%%%%%%%%%%%%%%%%%%%%%%%%%%%%%%%%%%
In order to shorten our many formulas, we use a compact collective notations
\begin{displaymath}
    \left(  \st^{p}_{\lambda} \right) =
    \left(  \st^{p_a}_{\lambda_a} \st^{p_b}_{\lambda_b}
            \st^{p_c}_{\lambda_c} \st^{p_d}_{\lambda_d} \right)
\end{displaymath}
for fermion four-momenta $p_A,A=a,b,c,d$ (i.e., $p_1=p_a, p_2=p_b, q_1=p_c, q_2=p_d$)
and helicities $\lambda_A,A=a,b,c,d$.
For $k=1,2,3$,  $\sigma^k$ are Pauli matrices and
$\sigma^0_{\lambda,\mu} = \delta_{\lambda,\mu}$ is the unit matrix.
The components $\hat{\varepsilon}^j_1, \hat{\varepsilon}^k_2, j,k=1,2,3$ 
are the components of the conventional spin polarization vectors 
of $e^-$ and $e^+$ respectively, defined in the so-called GPS fermion rest frames
(see Appendix A and ref.~\cite{gps:1998} for the exact definition of these frames).
We define $\hat{\varepsilon}^0_A=1$
in a non-standard way  (i.e. $p_A\cdot \hat{\varepsilon}_A=m_e, A=a,b$).
The {\em polarimeter} vectors $\hat{h}_C$ are similarly defined
in the appropriate GPS rest frames of the final unstable fermions  ($p_C\cdot \hat{h}_C=m_f, C=c,d$).
Note that, in general, $\hat{h}_C$ may
depend in a non-trivial way on the momenta of all decay products,
see refs.~\cite{jadach-was:1984,tauola2.4:1993,jadach:1985,gps:1998} for details.
We did not introduce polarimeter vectors for bremsstrahlung photons,
i.e. we take advantage of the fact that luckily all high-energy experiments are completely blind
to photon spin polarizations. 

%%%%%%%%%%%%%%%%%%%%%%%%%%%%%%%%%%%%%%%%%%%%%%%%%%%%%%%%%%%%%%%%
\subsubsection{IR regulators and YFS form-factor}
%%%%%%%%%%%%%%%%%%%%%%%%%%%%%%%%%%%%%%%%%%%%%%%%%%%%%%%%%%%%%%%%
Here we introduce/explain
our notation for IR integration limits for real photons
in eqs.~(\ref{eq:sigma-ceex2}) and (\ref{eq:rho-ceex2}) and in the following sections.
In general, the factor $\bar{\Theta}(\Omega)$ in eq.~(\ref{eq:sigma-ceex2})
defines the infrared (IR) integration limits for all real photons.
More precisely for a single photon, $\Omega$ is the domain surrounding the IR divergence point $k=0$,
which is in fact {\em excluded} from the MC phase-space.
In CEEX there is no real distinction among ISR and FSR photons,
$\Omega$ is therefore necessarily the same for all photons.
We define a characteristic function $\Theta(\Omega,k)$
of the IR domain $\Omega$ as
    $\Theta(\Omega,k)=1$  for $k\in\Omega$ 
and $\Theta(\Omega,k)=0$ for $k\not\in \Omega$.
The characteristic function for the part of the phase-space {\em included} in the MC integration 
for a single real photon is
$\bar{\Theta}(\Omega,k) = 1-\Theta(\Omega,k)$.
The analogous characteristic function 
for {\em all} real photons is, of course, the following product
%///////////////////////////////
\begin{equation}
  \label{eq:barOmega}
  \bar{\Theta}(\Omega) = \prod_{i=1}^n \bar{\Theta}(\Omega,k).
\end{equation}
In the present calculation corresponding
to the \KK\ Monte Carlo  program we opt for $\Omega$ defined traditionally with the photon energy cut
condition $k^0<E_{\min}$.

The YFS form factor~\cite{yfs:1961} for  $\Omega$ defined  with the condition $k^0<E_{\min}$ reads
%//////////////////////////////////////////////////
\begin{equation}
  \label{eq:YFS-ffactor}
  \begin{split}
   &Y(\Omega;p_a,...,p_d)
  =   Q_e^2   Y_\Omega(p_a,p_b)  +Q_f^2   Y_\Omega(p_c,p_d)\\
&\qquad\qquad
     +Q_e Q_f Y_\Omega(p_a,p_c)  +Q_e Q_f Y_\Omega(p_b,p_d) 
     -Q_e Q_f Y_\Omega(p_a,p_d)  -Q_e Q_f Y_\Omega(p_b,p_c).
  \end{split}
\end{equation}
where
\begin{equation}
\label{eq:form-factor}
\begin{split}
Y_\Omega(p,q) 
     \equiv &  2 \alpha \tilde{B}(\Omega,p,q)   +2 Q_f^2 \alpha \Re B(p,q) \\
     \equiv & -2 \alpha\;{ 1 \over 8\pi^2} \int {d^3k\over k^0} \Theta(\Omega;k)
                       \bigg({p\over kp} - {q\over kq} \bigg)^2 \\
            & +2 \alpha \Re \int {d^4k\over k^2} {i\over (2\pi)^3} 
                       \bigg( {2p-k \over 2kp-k^2} -{2q-k \over 2kq-k^2} \bigg)^2
\end{split}
\end{equation}
is given analytically in terms of logarithms and Spence functions.
As we see, the above YFS form factor includes terms due to the initial-final state interference (IFI).
The above form-factor will be derived in the following.
The additional contribution to the YFS form-factor due to the narrow $Z$-resonance will be discussed
in detail separately.

%%%%%%%%%%%%%%%%%%%%%%%%%%%%%%%%%%%%%%%%%%%%%%%%%%%%%%%%%%%%%%%%
\subsubsection{Partitions and $\sfac$-factors}
%%%%%%%%%%%%%%%%%%%%%%%%%%%%%%%%%%%%%%%%%%%%%%%%%%%%%%%%%%%%%%%%
The {\em coherent} sum is taken over the set $\{ \wp \} =\{ I,F\}^n $ of all $2^n$ partitions --
the single partition $\wp$ is defined as a vector $(\wp_1,\wp_2,\dots, \wp_n)$
where $\wp_i=I$ for an ISR photon and  $\wp_F=F$ for a FSR photon,
see the analogous construction in refs.~\cite{greco:1975,greco:1980}.
The set of all partitions is explicitly the following
%//////////////////////////////////////////////////
\begin{displaymath}
\{\wp\}=\{
(I,I,I,\dots,I), (F,I,I,\dots,I),
(I,F,I,\dots,I), (F,F,I,\dots,I),\dots
(F,F,F,\dots,F) \}.
\end{displaymath}
The $s$-channel four-momentum in the (possibly) resonant $s$-channel  propagator is
$X_\wp = p_a+p_b -\sum\limits_{ \wp_i = I} k_i.$

The soft (eikonal) amplitude factors 
$\sfac^{(\omega)}_{[i]}, \omega=I,F$, are complex numbers
and they are defined as follows
%//////////////////////////////////////////////////
%            CEEX soft factors
%//////////////////////////////////////////////////
\begin{equation}
  \begin{align}
  \label{eq:soft-fac-isr}
  & \sfac^{\{I\}}_{[i]} \equiv \sfac^{\{I\}}_{\sigma_i}(k_i) =\;\;\;
    -eQ_e{b_\sigma(k,p_a) \over 2k_ip_a} +eQ_e{b_\sigma(k_i,p_b) \over 2k_ip_b},\;\;
    \left|\sfac^{\{I\}}_{[i]}\right|^2 = 
             -\frac{e^2Q_e^2}{2} \bigg( {p_a\over k_ip_a} - {p_b\over k_i p_b}  \bigg)^2,\\
  \label{eq:soft-fac-fsr}
  & \sfac^{\{F\}}_{[i]} \equiv \sfac^{\{F\}}_{\sigma_i}(k_i) =
            +eQ_f{b_\sigma(k_i,p_c) \over 2kp_c} -eQ_f{b_\sigma(k_i,p_d) \over 2k_ip_d},\;
    \left|\sfac^{\{F\}}_{[i]}\right|^2 = 
            -\frac{e^2Q_f^2}{2} \bigg( {p_c\over k_ip_c} - {p_d\over k_i p_d}  \bigg)^2,\\
  \label{b-sigma}
  & b_\sigma(k,p)
    =  \sqrt{2}\; { \bar{u}_\sigma(k) \not\!p \; \umf_\sigma(\zeta)
                    \over      \bar{u}_{-\sigma}(k) \umf_\sigma(\zeta) }
    =  \sqrt{2}\; \sqrt{ {2\zeta p \over 2\zeta k} } s_\sigma(k,\hat{p}),
    \end{align}
\end{equation}
see also Appendix A for more details.
As indicated above, the moduli squared of the CEEX soft factors 
coincide up to a normalization constant with the corresponding 
EEX real photon soft factors $\tilde{S}(k_i)$.

%%%%%%%%%%%%%%%%%%%%%%%%%%%%%%%%%%%%%%%%%%%%%%%%%%%%%%%%%%%%%%%%
\subsubsection{Born}
%%%%%%%%%%%%%%%%%%%%%%%%%%%%%%%%%%%%%%%%%%%%%%%%%%%%%%%%%%%%%%%%
The simplest IR-finite $\hbeta$-function
$\hbeta^{(0)}_{0}$ is just the Born spin amplitude 
times a certain kinematical factor (see the next subsection)
\begin{equation}
\hbeta^{(0)}_{0}\left(\st^{p}_{\lambda} ; X \right)
       = \Bmf \left( \st^{p}_{\lambda}; X \right)\;\;
           { X^2  \over (p_c+p_d)^2 }.
\end{equation}
The Born spin amplitude $\Bmf \left( \st^{p}_{\lambda}; X \right)$
is a basic building block in the construction of all of our spin amplitudes --
let us define it already at this point.
The many equivalent notations for $\Bmf$  will be introduced for flexibility --
in view of its role as a basic building block
in the calculation of the multi-bremsstrahlung amplitudes.
Using  Feynman rules and our basic massive spinors
with definite GPS helicities of Appendix A,
Born spin amplitudes for%
\footnote{ For the moment we require $f\neq e$.}
the $e^-(p_a)e^+(p_b)\to f(p_c)\bar{f}(p_d)$
process are given by
%/////////////////////////////////////////////
%      Born spin amplitudes
%/////////////////////////////////////////////
\begin{equation}
  \label{eq:born-def}
  \begin{split}
     &\Bmf\left(\st^{p}_{\lambda}; X \right)=
      \Bmf\left(  \st^{p_a}_{\lambda_a} \st^{p_b}_{\lambda_b}
                  \st^{p_c}_{\lambda_c} \st^{p_d}_{\lambda_d}; X \right)=
      \Bmf\left[  \st^{p_b}_{\lambda_b} \st^{p_a}_{\lambda_a}\right]
          \left[  \st^{p_c}_{\lambda_c} \st^{p_d}_{\lambda_d}\right]\!(X)=
      \Bmf_{[bc][cd]}(X)=
\\  &\qquad\qquad
      =ie^2 \sum_{B=\gamma,Z} \Pi^{\mu\nu}_B(X)\; (G^{B}_{e,\mu})_{[ba]}\; (G^{B}_{f,\nu})_{[cd]}\; H_B
      =\sum_{B=\gamma,Z} \Bmf^B_{[bc][cd]}(X),
\\
     &(G^{B}_{e,\mu})_{[ba]} \equiv \bar{v}(p_b,\lambda_b) G^{B}_{e,\mu} u(p_a,\lambda_a),\;\;
      (G^{B}_{f,\mu})_{[cd]} \equiv \bar{u}(p_c,\lambda_c) G^{B}_{f,\mu} v(p_d,\lambda_d),
\\
     &G^{B}_{e,\mu} = \gamma_\mu \sum_{\lambda=\pm} \omega_\lambda g^{B,e}_\lambda,\quad
      G^{B}_{f,\mu} = \gamma_\mu \sum_{\lambda=\pm} \omega_\lambda g^{B,f}_\lambda,\quad
      \omega_\lambda = {1\over 2}(1+\lambda\gamma_5),
\\
     &\Pi^{\mu\nu}_B(X) = { g^{\mu\nu} \over X^2 - {M_{B}}^2 +i\Gamma_{B} {X^2 / M_{B}} },
  \end{split}
\end{equation}
where $g^{B,f}_\lambda$ are the usual chiral ($\lambda=+1,-1=R,L$) coupling constants
of the vector boson $B=\gamma,Z$ to fermion $f$ in units of the elementary charge $e$.
If not specified otherwise, the {\em ``hook function''} $H_B$ is trivial $H_\gamma=H_Z=1$.
It will be used to introduce special effects into Born spin amplitudes,
like running coupling constants or an additional form-factor due to a narrow resonance.

Spinor products are reorganized with the help of
the Chisholm identity, see eq~(\ref{Chisholm}) in the Appendix A,
which applies assuming that electron spinors are massless,
and the inner product of eq.~(\ref{inner-massive}), also in the Appendix A:
%//////////////////////////////////////////////////////////////
%      Born spin amplitudes
%      PropZet =    1d0/DCMPLX(svar-MZ**2, GammZ*svar/MZ)
%//////////////////////////////////////////////////////////////
\begin{equation}
  \label{born}
    \Bmf^B_{[ba][cd]}(X)
 = 2 ie^2 \frac{ 
            \delta_{\lambda_a, -\lambda_b}
            \big[\;
                 g^{B,e}_{ \lambda_a} g^{B,f}_{-\lambda_a}\;
                 T_{ \lambda_c \lambda_a}\; T'_{\lambda_b \lambda_d}
                +g^{B,e}_{ \lambda_a} g^{B,f}_{ \lambda_a}\;
                 U'_{ \lambda_c \lambda_b}\; U_{\lambda_a \lambda_d}
            \big]
         }
         { X^2 - {M_{B}}^2 +i\Gamma_{B} X^2 /M_{B} },
\end{equation}
where
%//////////////////////////////////////////////////
\begin{equation}
  \begin{split}
  T_{ \lambda_c \lambda_a} =& \bar{u}(p_c, \lambda_c)  u(p_a, \lambda_a)
                            =S(p_c, m_c, \lambda_c,    p_a,   0,    \lambda_a),\\
  T'_{\lambda_b \lambda_d} =& \bar{v}(p_b, \lambda_b)  v(p_d,  \lambda_d)
                            =S(p_b,   0,  -\lambda_b,  p_d,-m_d,   -\lambda_d ),\\
  U'_{\lambda_c \lambda_b} =& \bar{u}(p_c, \lambda_c)  v(p_b,-\lambda_b)
                            =S(p_c, m_c,\lambda_c,     p_b,   0,     \lambda_b),\\
  U_{ \lambda_a \lambda_d} =& \bar{u}(p_a,-\lambda_a)  v(p_d, \lambda_d)
                            =S(p_a,   0,  -\lambda_a,  p_d,-m_d,    -\lambda_d).
  \end{split}
\end{equation}
Note that the use of the Chisholm identity is a technical detail
which should not obscure the generality of our approach.
What we need in practice is {\em any} numerical method of evaluation 
of the Born spin amplitudes defined in eq.~(\ref{eq:born-def}), and Chisholm identity
is just one possibility.

%%%%%%%%%%%%%%%%%%%%%%%%%%%%%%%%%%%%%%%%%%%%%%%%%%%%%%%%%%%%%%%%
\subsubsection{Off-space extrapolation}
%%%%%%%%%%%%%%%%%%%%%%%%%%%%%%%%%%%%%%%%%%%%%%%%%%%%%%%%%%%%%%%%
In eq.~(\ref{eq:ceex-master})
Born spin amplitudes are obviously used for $p_i$
which {\em do not necessarily obey} the four-momentum conservation $p_a+p_b=p_c+p_d$.
In the exclusive exponentiation
this is natural and necessary because, in the presence of the bremsstrahlung photons,
the relation $X=p_a+p_b=p_c+p_d$ may not hold.
In eq.~(\ref{eq:ceex-master}) only fermion momenta enter as an argument of the Born
spin amplitudes. 
Photon momenta play only an indirect role, they disturb fermion momenta through energy and momentum
conservation (sometimes referred to as a ``recoil effect'').
The self-suggesting questions are: Is this acceptable? Is this dangerous?
Can this be avoided?
The clear answer is: 
It is unavoidable and natural feature of the exclusive exponentiation that
certain scattering matrix elements originally defined within $n$-body
phase-space are in fact used in the phase-space with more particles.
Let us call it {\em off-space extrapolation}, analogously to off-shell extrapolation%
\footnote{In the off-shell case particles do not obey $p^2=m^2$, here we also modify
  the dimension of the phase-space.}.
It surely makes sense, and in principle is not dangerous, provided it is done with a little bit of care.

A technical remark: In the actual calculations of the multiphoton
spin amplitudes fermion momenta $p_i$ in eq.~(\ref{born}) may be replaced, 
and occasionally will be replaced, by the momentum $k$ of one of the photons.
This will be  due to purely technical reasons 
(specific to the method of calculating multiphoton spin amplitudes).
In such a case, the spinor into which $k$ enters as an argument is always understood to be massless.

%%%%%%%%%%%%%%%%%%%%%%%%%%%%%%%%%%%%%%%%%%%%%%%%%%%%%%%%%%%%%%%%
\subsubsection{Pseudo-flux factor}
%%%%%%%%%%%%%%%%%%%%%%%%%%%%%%%%%%%%%%%%%%%%%%%%%%%%%%%%%%%%%%%%
One demonstration of the ``off-space extrapolation'' is the presence of
the auxiliary factor  $F = X_\wp / (p_c+p_d)^2$.
In the framework of CEEX, its presence is not really mandatory
and it disappears in the ``in-space'' situation  $p_a+p_b=p_c+p_d$.
In other words, the $F$-factor does not affect the soft limit; 
it really matters if at least one hard FSR photon is present.
It is not related to narrow resonances, 
but rather to the leading-logarithmic (LL) structure of the higher-orders.
Nevertheless, the $F$-factor is useful, because it is already implicitly present in the
photon emission matrix element at \Order{\alpha^1} and in all higher-orders, as can be seen
in the LL approximation.
It is therefore natural to include it at the early stage, already in the \Order{\alpha^0} exponentiation.
If we do not include it at the \Order{\alpha^0} then it will be included order by order anyway.
However, in such a case, the convergence of perturbative expansion will be deteriorated.
As we shall see below, the introduction of the $F$-factor will slightly complicate
the higher-order exponentiation and construction of the $\hbeta$ functions,
but the gain is worth the effort.
Furthermore, the $F$-factor has also been always present in the ``crude distribution'' 
in the YFS-type Monte Carlo generators, see for instance ref.~\cite{yfs2:1990},
so it also improves the variance of the MC weight, especially for \Oceex{\alpha^0}.

%%%%%%%%%%%%%%%%%%%%%%%%%%%%%%%%%%%%%%%%%%%%%%%%%%%%%%%%%%%%%%%%%%%%%%%%%%%%%%%%%%%%%%%%%%%%%
\subsection{IR structure in CEEX}

Let us discuss in detail the origin of the \Oceex{\alpha^r}
expressions eqs.~(\ref{eq:sigma-ceex2}-\ref{eq:rho-ceex2})
and the mechanism of the IR cancellations.
Our real starting point is the infinite order perturbative expression for
the total cross section given by the standard quantum-mechanical expression
of the type ``matrix element squared modulus times phase-space''
%//////////////////////////////////////////////////
%               The very beginning
%//////////////////////////////////////////////////
\begin{equation}
\label{eq:master}
  \sigma^{(\infty)} = 
  \sum_{n=0}^\infty {1\over n!}
  \int d\tau_{n} ( p_a+p_b ;\; p_c,p_d,\; k_1,\dots,k_n)\;
  \frac{1}{4}
  \sum_{\lambda,\sigma_i,\dots,\sigma_n =\pm}\;
  \left|
    \Meu_n 
    \left(  \st^{p}_{\lambda}
            \st^{k_1}_{\sigma_1} 
            \st^{k_2}_{\sigma_2}
            \dots
            \st^{k_n}_{\sigma_n}
    \right)
  \right|^2,
\end{equation}
where $d\tau_{n}$ is the respective $n\gamma+2f$ Lorentz invariant phase space,
and $\Meu_n$ are the corresponding spin amplitudes.
To simplify the discussion we take the unpolarized case, without narrow resonances.

%%%%%%%%%%%%%%%%%%%%%%%%%%%%%%%%%%%%%%%%%%%%%%%%%%%%%%%%%%%%%%
\subsubsection{IR virtual factorization to infinite-order}

According to the Yennie-Frautschi-Suura fundamental factorization theorem~\cite{yfs:1961},
all virtual IR corrections can be re-located into an exponential form-factor%
\footnote{ 
  In the LL approximation it is, of course, the doubly-logarithmic Sudakov form-factor.}
order by order and in infinite order
%//////////////////////////////////////////////////
\begin{equation}
  \Meu^{(\infty)}_n =e^{\alpha B_4(p_a,p_b,p_c,p_d)} \Mmf^{(\infty)}_n.
\end{equation}
As the convergence of the perturbative series is questionable,
the above equation is in practice treated as a symbolic representation of the order-by-order relation which
at \Order{\alpha^r} reads
%//////////////////////////////////////////////////
\begin{equation}
  \Meu^{(r)}_n = \sum_{l=0}^{r-n} { (\alpha B_4)^{r-l} \over (r-l)!}  \Mmf^{[l+n]}_n,\quad (n\leq r),
\end{equation}
where the index $l$ is the number of loops in $\Mmf^{[l+n]}_n$.
The above identity is quite powerful because $\Mmf^{[l+n]}_n$ are 
not only free of the {\em virtual} IR-divergences, but are also universal:
they are the same in every  perturbative order $r$ --
for example for one photon, the one-loop (IR-subtracted) component, $\Mmf^{(1)}_1$, 
is the same in the fifth-order and, let us say, in the second order,
where it appears for the first time.
The above identity can also be reformulated as follows
%//////////////////////////////////////////////////
%
%//////////////////////////////////////////////////
\begin{equation}
  \label{eq:virtual-subtraction}
  \Mmf^{(r)}_n = \sum_{l=0}^{r-n}  \Mmf^{[l+n]}_n
   =  \left[ e^{-\alpha B_4(p_a,p_b,p_c,p_d)} \Meu^{(r)}_n \right]\Big|_{{\cal O}(\alpha^r) },
\end{equation}
where, $\Meu^{(r)}_n$ has to be calculated from Feynman diagrams in at least%
\footnote{ The  use of  
  $\Mmf^{(r+m)}_n$ at \Order{\alpha^{(r+m)}}, $m>0$ will yield the same result --
  this is another way of stating the universality property.}
\Order{\alpha^r}.
The above steps are exactly the same as in~\cite{yfs:1961}.

The YFS form-factor $B_4$ for 
$e^-(p_a)+e^+(p_b) \to f(p_c)+\bar{f}(p_d)+n\gamma$ reads
%%%%%%%%%%%%%%%%%%%%%%%%%%%%%%%%%%%%%%%%%%%%
\begin{equation}
  \label{eq:B4}
  \begin{split}
   &\alpha B_4(p_a,p_b,p_c,p_d)
   =\int {d^4k\over k^2 -m_\gamma^2 +i\epsilon}\; {i\over (2\pi)^3}\;
                 \left| J_I(k)  -J_F(k) \right|^2,\\
  &J_I = eQ_e(\hat{J}_a(k) - \hat{J}_b(k)),\;\;
   J_F = eQ_f(\hat{J}_c(k) - \hat{J}_d(k)),\;
   \hat{J}_f^\mu(k) = { 2p_f^\mu +k^\mu \over k^2 +2k\cdot p_f +i\epsilon}.
  \end{split}
\end{equation}
Using the identity $(\sum_k Z_kJ_k)^2 = -\sum_{i>k} Z_iZ_k (J_i-J_k)^2$,
valid for $\sum Z_k=0$,
where $Z_k$ is the charge or minus charge of the particle in the initial- 
or final-state respectively,
we may cast (see ref.~\cite{yfs:1961})  $B_4$ into a sum of the simpler dipole components
%//////////////////////////////////////////////////
\begin{equation}
  \begin{align}
  \label{eq:b-virtual}
    \begin{split}
     &B_4(p_a,p_b,p_c,p_d) =  Q_e^2   B_2(p_a,p_b)  +Q_f^2   B_2(p_c,p_b)\\
&\qquad
      +Q_e Q_f B_2(p_a,p_c)  +Q_e Q_f B_2(p_b,p_d) 
      -Q_e Q_f B_2(p_a,p_d)  -Q_e Q_f B_2(p_b,p_c),
  \end{split}\\
& B_2(p_i,p_j) \equiv 
    \int {d^4k\over k^2 -m_\gamma^2 +i\epsilon}\; {i\over (2\pi)^3}\;
         \left( \hat{J}(p_i,k)- \hat{J}(p_j,k)\right)^2.
  \end{align}
\end{equation}
In the above we assume that IR singularities are regularized with a finite
photon mass $m_\gamma$ which enters into all $B_2$'s and implicitly into $\sfac$-factors 
(and in the real photon phase-space integrals, see the following discussion).

%%%%%%%%%%%%%%%%%%%%%%%%%%%%%%%%%%%%%%%%%%%%%%%%%%%%%%%%%%%%%%
\subsubsection{IR real factorization to infinite-order}

The next step is isolation of the {\em real} IR singularities and it is worth to elaborate 
on this point because here the CEEX method differs in essential details
from the original YFS method~\cite{yfs:1961}.
We use again results of the basic analysis of real IR singularities of ref.~\cite{yfs:1961},
the essential difference is that we do not square the amplitudes immediately --
it is done numerically at the later stage.
The validity of the whole basic analysis of the IR cancellations
in ref.~\cite{yfs:1961} remains, however, useful
because it is done in terms of the currents
%//////////////////////////////////////////////////
\begin{equation}
        j_f^\mu(k) = {2p_f^\mu \over 2p_f\cdot k}, f=a,b,c,d.
\end{equation}
The above currents are simply related to our $\sfac$-factors:
%//////////////////////////////////////////////////
\begin{equation}
  \begin{split}
   &\sfac^{\{I\}}_\sigma(k) = const \times Q_e (j_{a}-j_{b})\cdot \epsilon_\sigma(\beta),\\
   &\sfac^{\{F\}}_\sigma(k) = const \times Q_f (j_{c}-j_{d})\cdot \epsilon_\sigma(\beta).
  \end{split}
\end{equation}
It is important to remember that
the whole structure of the real IR divergences is entirely controlled by 
the squares of the currents $|j(k)|^2$, for $j=j_{a}-j_{b}$ or $j=j_{c}-j_{d}$,
independently whether we prefer to work with the amplitudes or their squares,
because only the squares $|j(k)|^2$ are IR divergent and
the other contractions do not matter (as was already stressed in ref.~\cite{yfs:1961}).
Similarly, if we express spin amplitudes in terms of $\sfac$-factors, 
only the squares $|\sfac(k)|^2$ are IR divergent
and not the interference terms like $\Re\{\sfac(k)(\dots)^*\}$.

Having the above in mind we may proceed using results of ref.~\cite{yfs:1961}
and we see that for instance the most IR divergent part of $\Meu_n$
is proportional to the products of $n$ $\sfac$-factors
%//////////////////////////////////////////////////
%
%//////////////////////////////////////////////////
\begin{equation}
  \begin{split}
  \Mmf_n &\left( \st^{p}_{\lambda}
                      \st^{k_1}_{\sigma_1} 
                      \st^{k_2}_{\sigma_2}\dots
                      \st^{k_n}_{\sigma_n}
               \right)
  \sim
  \hbeta_0 \left(  \st^{p}_{\lambda} ; X \right)
             \sfac_{\sigma_1}(k_1)
             \sfac_{\sigma_2}(k_2)\dots
             \sfac_{\sigma_n}(k_n)
  \end{split}
\end{equation}
where the function $\hbeta_0$ is not IR divergent any more, 
and we assumed for the moment the absence of the
narrow resonances, using the sum of ISR and FSR $\sfac$-factors%
\footnote{In the non-resonant case we may set $X=p_a+p_b$, for example.}
%//////////////////////////////////////////////////
\begin{equation}
        \sfac_{\sigma}(k) \equiv \sfac_{\sigma}^{\{F\}}(k) + \sfac_{\sigma}^{\{I\}}(k).
\end{equation}

However, there are also non-leading IR singularities.
Suppressing inessential spin indices the whole real IR structure is revealed in the following
decomposition \cite{yfs:1961}:
%//////////////////////////////////////////////////
%
%//////////////////////////////////////////////////
\begin{equation}
  \begin{split}
  \label{eq:beta-decomposition}
    \Mmf&^{(\infty)}_n(k_1,k_2,k_3,...,k_n) = 
                       \hbeta_0 \prod_{s=1}^n \sfac(k_s)
    +\sum_{j=1}^n      \hbeta_1(k_j)    \prod_{s\neq j}  \sfac(k_s)\\
   &+\sum_{j_1>j_2}    \hbeta_2(k_{j_1},k_{j_2}) \prod_{s\neq j_1,j_2}  \sfac(k_s)
    +\sum_{j_1>j_2>j_3}\hbeta_2(k_{j_1},k_{j_2},k_{j_3}) \prod_{s\neq j_1,j_2,j_3}  \sfac(k_s)+...\\
   &+\sum_{j=1}^n      \hbeta_{n-1} (k_1,...k_{j-1},k_{j+1},...,k_n) \sfac(k_j)
                      +\hbeta_n (k_1,k_2,k_3,...,k_n)
  \end{split}
\end{equation}
where functions $\hbeta_i$ are IR free and include finite loop corrections to infinite-order.
Let us stress that these functions $\hbeta_i$ are {\em genuinely new objects}.
They were not used and even not considered in ref.~\cite{yfs:1961}.

%%%%%%%%%%%%%%%%%%%%%%%%%%%%%%%%%%%%%%%%%%%%%%%%%%%%%%%%%%%%%%
\subsubsection{Finite-order  $\hbeta$'s}

The decomposition of eq.~(\ref{eq:beta-decomposition}) has also its order-by-order representation,
which at \Order{\alpha^r}, $r=n+l$, reads as follows:
%//////////////////////////////////////////////////
%
%//////////////////////////////////////////////////
\begin{equation}
  \label{eq:beta-decomposition2}
  \begin{split}
    \Mmf&^{(n+l)}_n(k_1,k_2,k_3,...,k_n)
                      =\hbeta^{(l)}_0 \prod_{s=1}^n \sfac(k_s)
    +\sum_{j=1}^n      \hbeta^{(1+l)}_1(k_j)    \prod_{s\neq j}  \sfac(k_s)\\
   &+\sum_{j_1<j_2}    \hbeta^{(2+l)}_2(k_{j_1},k_{j_2}) \prod_{s\neq j_1,j_2}  \sfac(k_s)
    +\sum_{j_1<j_2<j_3}\hbeta^{(3+l)}_2(k_{j_1},k_{j_2},k_{j_3}) \prod_{s\neq j_1,j_2,j_3}  \sfac(k_s)+...\\
   &+\sum_{j=1}^n      \hbeta^{(n-1+l)}_{n-1} (k_1,...k_{j-1},k_{j+1},...,k_n) \sfac(k_j)
                      +\hbeta^{(n+l)}_n (k_1,k_2,k_3,...,k_n)\\
  =& \prod_{s=1}^n \sfac(k_s) \Bigg\{
             \hbeta^{(l)}_0
    +\sum_{j=1}^n      {\hbeta^{(1+l)}_1(k_j) \over \sfac(k_j) }
    +\sum_{j_1<j_2}    {\hbeta^{(2+l)}_2(k_{j_1},k_{j_2}) \over \sfac(k_{j_1})\sfac(k_{j_2}) }
    +\sum_{j_1<j_2<j_3}{\hbeta^{(3+l)}_2(k_{j_1},k_{j_2},k_{j_3}) \over \sfac(k_{j_1})\sfac(k_{j_2}\sfac(k_{j_3}) }\\
   &\qquad
    +\sum_{j=1}^n      {\hbeta^{(n-1+l)}_{n-1} (k_1,...k_{j-1},k_{j+1},...,k_n) \over \prod\limits_{s\neq j} \sfac(k_s) }
                      +{\hbeta^{(n+l)}_n (k_1,k_2,k_3,...,k_n)  \over \prod\limits_{s} \sfac(k_s) }
    \Bigg\}.
  \end{split}
\end{equation}
The new functions $\hbeta^{(n+l)}_n (k_1,k_2,k_3,...,k_n)$ contain up to $l$-loop corrections,
and are not only completely IR-finite, but are also universal:
for instance the $\hbeta^{(2)}_1(k)$, which appears for the first time in decomposition of 
$\Mmf^{(2)}_1(k)$, is functionally the same when decomposing $\Mmf^{(3)}_2(k_1,k_2)$ or
any higher-order $\Mmf^{(n+l)}_n$.
This feature is essential for reversing the relations of eq.~(\ref{eq:beta-decomposition2}),
that is for practical order-by-order calculations of $\hbeta^{(n+l)}_n$
from $\Mmf^{(r)}_n$, obtained directly from the Feynman rules:
%//////////////////////////////////////////////////
%
%//////////////////////////////////////////////////
\begin{equation}
  \label{eq:BetaRecursive}
  \begin{split}
   &\hbeta^{(l)}_0                = \Mmf^{(l)}_0,\\
   &\hbeta^{(1+l)}_1(k_1)         = \Mmf^{(1+l)}_1(k_1) -\hbeta^{(l)}_0  \sfac(k_1),\\
   &\hbeta^{(2+l)}_2(k_1,k_2)     = \Mmf^{(2+l)}_2(k_1,k_2) 
            -\hbeta^{(1+l)}_1(k_1)\sfac(k_2)-\hbeta^{(1+l)}_1(k_2)\sfac(k_1)
            -\hbeta^{(l)}_0  \sfac(k_1)\sfac(k_2),\\
   &\hbeta^{(3+l)}_3(k_1,k_2,k_3) = \Mmf^{(3+l)}_3(k_1,k_2,k_3)\\
   &\qquad  -\hbeta^{(2+l)}_2(k_1,k_2)\sfac(k_3)-\hbeta^{(2+l)}_2(k_1,k_3)\sfac(k_2)-\hbeta^{(2+l)}_2(k_2,k_3)\sfac(k_1)\\
   &\qquad  -\hbeta^{(1+l)}_1(k_1)\sfac(k_2)\sfac(k_3)-\hbeta^{(1+l)}_1(k_2)\sfac(k_1)\sfac(k_3)
                                                     -\hbeta^{(1+l)}_1(k_3)\sfac(k_1)\sfac(k_2)\\
   &\qquad  -\hbeta^{(l)}_0  \sfac(k_1)\sfac(k_2)\sfac(k_3),\dots,\\
%%%%%
   &\hbeta^{(n+l)}_n(k_1,...,k_n) = \Mmf^{(n+l)}_n(k_1,...,k_n)
            -\sum_{j=1}^n \hbeta^{(n-1+l)}_{n-1}(k_1,...k_{j-1},k_{j+1},...,k_n) \sfac(k_j)\\
   &\qquad  -\sum_{j_1<j_2}\hbeta^{(n-2+l)}_{n-2}(k_1,...k_{j_1-1},k_{j_1+1},...k_{j_2-1},k_{j_2+1},...,k_n) 
                              \sfac(k_{j_1})\sfac(k_{j_2}) ...\\
   &\qquad  -\sum_{j_1<j_2}\hbeta^{(1+l)}_2(k_{j_1},k_{j_2}) \prod_{s\neq j_1,j_2}  \!\!\!\sfac(k_s)
            -\sum_{j=1}^n  \hbeta^{(1+l)}_1(k_j)             \prod_{s\neq j}  \sfac(k_s)
                          -\hbeta^{(l)}_0 \prod_{s=1}^n \sfac(k_s).
  \end{split}
\end{equation}
The above set of equations is a recursive rule, i.e., higher-order $\hbeta$'s
are constructed in terms of lower-order ones.
In practical calculations we do not go to infinite-order but we stop at some \Order{\alpha^r}
and the above set of equations is truncated  for $\hbeta^{(n+l)}_n$ 
by the requirement $n+l\leq r$.
The above truncation is harmless from the point of view of IR cancellations because
we omit higher-order $\hbeta$'s which are IR-finite.
As a consequence of the above fixed-order truncation
eq.~(\ref{eq:beta-decomposition}) takes the following form:
%//////////////////////////////////////////////////
%
%//////////////////////////////////////////////////
\begin{equation}
  \label{eq:beta-trunc}
  \begin{split}
    \Mmf&_n^{(r)}(k_1,k_2,k_3,...,k_n) = \\
  =& \prod_{s=1}^n \sfac(k_s) \Bigg\{
             \hbeta^{(r)}_0
    +\sum_{j=1}^n      {\hbeta^{(r)}_1(k_j) \over \sfac(k_j) }
    +\sum_{j_1<j_2}    {\hbeta^{(r)}_2(k_{j_1},k_{j_2}) \over \sfac(k_{j_1})\sfac(k_{j_2}) }
    +\sum_{j_1<j_2<j_3}{\hbeta^{(r)}_2(k_{j_1},k_{j_2},k_{j_3}) \over \sfac(k_{j_1})\sfac(k_{j_2}\sfac(k_{j_3}) }\\
   &\qquad\qquad
    +\sum_{j_1<j_2<...<j_r}{\hbeta^{(r)}_r(k_{j_1},k_{j_2},...,k_{j_r}) 
              \over \sfac(k_{j_1})\sfac(k_{j_2})...\sfac(k_{j_r}) }
   \Bigg\},
  \end{split}
\end{equation}
where, contrary to eq.~(\ref{eq:beta-decomposition2}),
we now allow only for $r<n$; in such a case the sum has $r+1$ terms instead of $n$.

The above formula represents the general finite-order \Ordex{\alpha^r} case
while for $r=0$ only the first term survives, and
in our \Order{\alpha^2} case there are three terms.
The CEEX spin amplitudes in our master formula eq.~(\ref{eq:ceex-master}) represent
the cases of $r=0,1,2$.

Just to give an explicit example, in the recursive calculation of $\hbeta$'s
in \Order{\alpha^3} we would need to calculate 
$\hbeta^{(l)}_0, l=0,1,2,3$, $\;\hbeta^{(1+l)}_1, l=0,1,2$, $\;\hbeta^{(2+l)}_2, l=0,1$
and $\hbeta^{(3)}_3$.
In the present work, at \Order{\alpha^r}, $r=0,1,2$,  we shall employ
the following set of recursive definitions
based on eqs.~(\ref{eq:BetaRecursive})
%//////////////////////////////////////////////////
\begin{equation}
  \label{eq:BetRecursive2}
  \begin{split}
   &    \hbeta^{(l)}_0\left(\st^{p}_{\lambda}\right) 
    =    \Mmf^{(l)}_0\left(\st^{p}_{\lambda}\right),\; l=0,1,2,\\
   &  \hbeta^{(1+l)}_1\left(\st^{p}_{\lambda}\st^{k_1}_{\sigma_1}\right)      
    =  \Mmf^{(1+l)}_1\left(\st^{p}_{\lambda}\st^{k_1}_{\sigma_1}\right) 
       -\hbeta^{(l)}_0\left(\st^{p}_{\lambda}\right)  \sfac_{\sigma_1}(k_1),\;l=0,1,\\
   &    \hbeta^{(2)}_2\left(\st^{p}_{\lambda}\st^{k_1}_{\sigma_1}\st^{k_2}_{\sigma_2}\right)    
    =    \Mmf^{(2)}_2\left(\st^{p}_{\lambda}\st^{k_1}_{\sigma_1}\st^{k_2}_{\sigma_2}\right)
       -\hbeta^{(1)}_1\left(\st^{p}_{\lambda}\st^{k_1}_{\sigma_1}\right) \sfac_{\sigma_2}(k_2)  
       -\hbeta^{(1)}_1\left(\st^{p}_{\lambda}\st^{k_2}_{\sigma_2}\right) \sfac_{\sigma_1}(k_1)
       -\hbeta^{(0)}_0\left(\st^{p}_{\lambda}\right)  \sfac_{\sigma_1}(k_1)
                                                         \sfac_{\sigma_2}(k_2),\\
  \end{split}
\end{equation}
where the $\Mmf$-amplitude is given by eq.~(\ref{eq:virtual-subtraction}).
Here we restored spin indices but we still specialize to the non-resonant case,
and our $\hbeta$'s do not have the partition
dependent $X_\wp$ argument as in $\hbeta$'s 
of eqs.~(\ref{eq:ceex-master}-\ref{eq:ceex-master2}).
We shall provide a definition for $\hbeta$'s in the resonant case in the following section~\ref{sec:resobetas}.

%%%%%%%%%%%%%%%%%%%%%%%%%%%%%%%%%%%%%%%%%%%%%%%%%%%%%%%%%%%%%%
\subsubsection{IR cancellations in CEEX}

At fixed-order \Oceex{\alpha^r}, and remembering that $|\exp(B_4)|^2= \exp(2\Re B_4)$,
we have obtained
%//////////////////////////////////////////////////
%               r-th order
%//////////////////////////////////////////////////
\begin{equation}
  \label{eq:master-bis}
  \sigma^{(r)} = 
  \sum_{n=0}^\infty {1\over n!}
  \int d\tau_{n} ( p_1+p_2 ;\; p_3,p_4,\; k_1,\dots,k_n)\;
  e^{2\alpha\Re B_4(p_a,...,p_d)}
  {1\over 4}\sum_{\rm spin} \left| \Mmf^{(r)}_n \left( k_1, k_2, \dots k_n \right) \right|^2,
\end{equation}
where $\Mmf^{(r)}_n$ is defined in eq.~(\ref{eq:beta-trunc}) 
and we factorize out the $\sfac$-factors
%//////////////////////////////////////////////////
\begin{equation}
  \begin{split}
    {1\over 4}\sum_{\rm spin}
    \left| \Mmf_n^{(r)}(k_1,k_2,k_3,...,k_n) \right|^2 
 &= d_{n}(k_1,k_2,k_3,...,k_n)\; 
    \prod_{s=1}^n |\sfac(k_s)|^2 ,  \\
    d_{n}(k_1,k_2,k_3,...,k_n) = \Bigg| &\hbeta^{(r)}_0
    +\sum_{j=1}^n      {\hbeta^{(r)}_1(k_j) \over \sfac(k_j) }
    +\sum_{j_1<j_2}    {\hbeta^{(r)}_2(k_{j_1},k_{j_2}) \over \sfac(k_{j_1})\sfac(k_{j_2}) }
    +\sum_{j_1<j_2<j_3}{\hbeta^{(r)}_2(k_{j_1},k_{j_2},k_{j_3}) 
                        \over \sfac(k_{j_1})\sfac(k_{j_2}\sfac(k_{j_3}) }\\
 &  +\sum_{j_1<j_2<...<j_r}{\hbeta^{(r)}_r(k_{j_1},k_{j_2},...,k_{j_r}) 
              \over \sfac(k_{j_1})\sfac(k_{j_2})...\sfac(k_{j_r}) }
   \Bigg|^2,
  \end{split}
\end{equation}
In the above the function $d_n(k_1,k_2,k_3,...,k_n)$ is IR-finite 
and we are allowed set $m_\gamma \to 0$ in it.
Apart from $2\alpha\Re B_4$ 
the IR regulator  $m_\gamma$ still remains in all $\sfac(k_i)$-factors
and in the lower phase-space boundary of all real photons in $\int d^3k/2k^0$.

The above total cross section is perfectly IR-finite,
as can be checked with a little bit of effort
by {\em analytical} partial differentiation%
\footnote{ This method of validating IR-finiteness was noticed by 
  G. Burgers~\cite{burgers-private:1990}.
  The classical method of ref.~\protect\cite{yfs:1961} relies on the techniques of the
  Melin transform, which could be also used here.}
with respect the photon mass
%//////////////////////////////////////////////////
%
%//////////////////////////////////////////////////
\begin{equation}
  \label{eq:intermediate}
  \begin{split}
    &\frac{\partial}{\partial m_\gamma} \sigma^{(r)}=\\
    &= \sum_{n=0}^\infty {1\over n!}
    \int d\tau_{n} (P ;\; p_3,p_4,\; k_1,\dots,k_n)\;
    e^{2\alpha\Re B_4 }
   \frac{\partial}{\partial m_\gamma} \{ 2\alpha\Re B_4 \}
    {1\over 4}\sum_{\rm spin} \left| \Mmf^{(r)}_n \left( k_1, k_2, \dots k_n \right) \right|^2\\
    &+\sum_{n=1}^\infty {1\over n!}
    \sum_{s=1}^n
    \int d\tau_{n-1} (P ;\; p_3,p_4,\; k_1,\dots,k_{s-1},k_{s+1},\dots,k_n)\;
    e^{2\alpha\Re B_4 }\\
   &\qquad\qquad\qquad\times
   \frac{\partial}{\partial m_\gamma}\left\{ \int {d^3k_s \over 2k^0_s}  |\sfac(k_s)|^2 \right\}
    \prod_{j\neq s} |\sfac(k_j)|^2 \;
    d_{n}(k_1,k_2,...,k_s,...,k_n)
  \end{split}
\end{equation}
It is now necessary to notice that
\begin{displaymath}
   \frac{\partial}{\partial m_\gamma}\left\{ \int {d^3k_s \over 2k^0_s}  |\sfac(k_s)|^2 \right\}
\end{displaymath}
is a $\delta$-like function concentrated at $k_s=0$ and we may therefore use the limit
\begin{displaymath}
 d_{n}(k_1,...,k_s,...,k_n)\to d_{n}(k_1,k_2,...,k_{s-1},0,k_{s+1},...,k_n)
   \equiv d_{n-1}(k_1,k_2,...,k_{s-1},k_{s+1},...,k_n)
\end{displaymath}
The above helps us to notice that all terms in $\sum_{s=1}^n$ are identical and we may sum them up,
(after formally renaming the photon integration variables in the second integral) and
rewrite eq.~(\ref{eq:intermediate}) as follows
%//////////////////////////////////////////////////
%
%//////////////////////////////////////////////////
\begin{equation}
  \begin{split}
    \frac{\partial}{\partial m_\gamma} \sigma^{(r)}
    = \sum_{n=0}^\infty {1\over n!}
    \int d\tau_{n}& (P ;\; p_3,p_4,\; k_1,\dots,k_n)\;
    e^{2\alpha\Re B_4 }
    {1\over 4}\sum_{\rm spin} \left| \Mmf^{(r)}_n \left( k_1, k_2, \dots k_n \right) \right|^2\\
   &\times\frac{\partial}{\partial m_\gamma}
       \left\{ 2\alpha\Re B_4 + \int {d^3k_s \over 2k^0_s}  |\sfac(k_s)|^2 \right\}=0,
  \end{split}
\end{equation}
where the independence on $m_\gamma$ of the sum of the 1-photon real and virtual integrals
is due to the usual cancellation of the IR-divergences in the YFS scheme, shown explicitly
many times.

The integral of eqs.~(\ref{eq:master}) and (\ref{eq:master-bis}) are perfectly implementable
in the Monte Carlo form, with small $m_\gamma$ being the IR regulator,
using a method very similar to that in ref.~\cite{yfs2:1990}.
Traditionally, however, the lower boundary on the real soft photons is defined
using the energy cut condition $k^0>\varepsilon \sqrt{s}/2$ in the laboratory frame.
The practical advantage of such a cut is the lower photon multiplicity in the MC simulation,
and consequently a faster computer program%
\footnote{ The disadvantage of the cut $k^0>\varepsilon \sqrt{s}/2$ is that in the MC 
  it has to be implemented in {\em different} reference frames for ISR and for FSR -- this costs
  the additional delicate procedure of bringing these two boundaries together,
  see ref.~\cite{kkcpc:1999} and/or discussion in the analogous 
  $t$-channel case in ref.~\cite{bhlumi2:1992}.}.
If the above energy cut on the photon energy is adopted, then the real soft-photon
integral between the lower LIPS boundary defined by $m_\gamma$ and that
defined by $\varepsilon$ can be evaluated by hand and summed up rigorously
(the only approximation is $m_\gamma/m_e \to 0$)
in the following.

%%%%%%%%%%%%%%%%%%%%%%%%%%%%%%%%%%%%%%%%%%%%%%%%%%%%%%%%%%%%%%
\subsubsection{Explicit IR boundary for real photons}

A general notation for the IR domain $\Omega$ was already introduced,
see eq.~(\ref{eq:barOmega}).
Let us now exclude the $\Omega$ domain from the real photon phase space (integrate out analytically).
Splitting the real photon integration phase space we rewrite
the eq.~(\ref{eq:master-bis}) as follows
\begin{equation}
  \begin{split}
    \sigma^{(r)}=&
    \sum_{n=0}^\infty {1\over n!}
    \prod_{j=1}^n \left\{ 
     \int  {d^3k_j \over 2k^0_j} |\sfac(k_j)|^2 \Theta(\Omega,k_j)
    +\int  {d^3k_j \over 2k^0_j} |\sfac(k_j)|^2 \bar{\Theta}(\Omega,k_j)
    \right\}\; \\
   &\int d\tau_{2} (P -\sum_{j=1}^n k_j;\; p_3,p_4)\;
    e^{2\alpha\Re B_4 }
    d_{n}(k_1,k_2,...,k_n).
  \end{split}
\end{equation}
After expanding the binomial product into $2^n$ terms let us consider
for instance the sum of all $({n\atop 1})=n$ terms in which one photon is in $\Omega$
and the other ones are not:
\begin{equation}
  \begin{split}
  {1\over n!} &\sum_{s=1}^n 
  \int  {d^3k_s \over 2k^0_s} |\sfac(k_s)|^2 \Theta(\Omega,k_s)
  \prod_{j\neq s}^n \int  {d^3k_j \over 2k^0_j} |\sfac(k_j)|^2 \bar{\Theta}(\Omega,k_j)\\
  &\qquad
    \int d\tau_{2} (P -\sum_{j=1}^n k_j;\; p_3,p_4)\;
    e^{2\alpha\Re B_4 }
    d_{n}(k_1,k_2,...,k_{s-1},0,k_{s+1},...,k_n)\\
 =& {1\over n!} \left({n\atop 1}\right)
    \int  {d^3k \over 2k^0} |\sfac(k)|^2 \Theta(\Omega,k)\\
  &\qquad
    \int d\tau_{n+1} (P;\; p_3,p_4, k_1,k_2,...,k_{n-1})
    \prod_{j=1}^{n-1}\bar{\Theta}(\Omega,k_j) |\sfac(k_j)|^2 \;
    d_{n-1}(k_1,k_2,...,k_{n-1})
  \end{split}
\end{equation}
A similar summation is performed for the $({n\atop s})$ terms where $s$ photons are in $\Omega$
giving rise to
\begin{equation}
  \begin{split}
  &\sigma^{(r)}=
    \sum_{n=0}^\infty {1\over n!} \sum_{s=0}^n
    \left({n\atop s}\right)
    \left( \int  {d^3k \over 2k^0} |\sfac(k)|^2 \Theta(\Omega,k) \right)^s\\
  &\quad \int d\tau_{2+n-s} (P;\; p_3,p_4, k_1,k_2,...,k_{s})
    \prod_{j=1}^{n-s} \left\{ |\sfac(k_j)|^2 \bar{\Theta}(\Omega,k_j) \right\}\;
    e^{2\alpha\Re B_4 }
    d_{n-s}(k_1,k_2,...,k_{n-s})\\
  &=\sum_{n=0}^\infty {1\over n!}
   \int d\tau_{2+n} (P;\; p_3,p_4, k_1,k_2,...,k_{n})
    \exp\left( \int  {d^3k_j \over 2k^0_j} |\sfac(k_j)|^2 \Theta(\Omega,k_j) \right)
    e^{ 2\alpha\Re B_4(p_1,...,p_4) }\\
  &\qquad\qquad\times
    \prod_{j=1}^{n} \left\{ |\sfac(k_j)|^2 \bar{\Theta}(\Omega,k_j) \right\}\;
    d_{n}(k_1,k_2,...,k_{n}).
  \end{split}
\end{equation}
The additional overall exponential factor contains the well known function
%//////////////////////////////////////////////////
\begin{equation}
  \begin{split}
   &\tilde{B}_4(p_1,...,p_4)=
     \int  {d^3k_j \over 2k^0_j} |\sfac(k_j)|^2 \Theta(\Omega,k_j)
   =  Q_e^2   \tilde{B}_2(p_1,p_2)  +Q_f^2   \tilde{B}_2(p_3,p_4)\\
&\qquad\qquad
    +Q_e Q_f \tilde{B}_2(p_1,p_3)  +Q_e Q_f \tilde{B}_2(p_2,p_4) 
    -Q_e Q_f \tilde{B}_2(p_1,p_4)  -Q_e Q_f \tilde{B}_2(p_2,p_3),
\\& \tilde{B}_2(p,q) 
     \equiv -\int {d^3k\over 2k^0}\; \Theta(\Omega,k_j)\;
             \bigg( j_p(k)-j_q(k)  \bigg)^2
     \equiv  \int {d^3k\over 2k^0}\; \Theta(\Omega,k_j)\;
             \frac{(-1)}{8\pi^2} \bigg( {p\over kp} - {q\over k q}  \bigg)^2,
  \end{split}
\end{equation}
which forms together with $2\alpha\Re B_4(p_1,...,p_4)$
the conventional YFS form-factor 
%//////////////////////////////////////////////////
\begin{equation}
 Y(\Omega;p_1,...,p_4) = 2\alpha \tilde{B}_4(p_1,...,p_4) + 2\alpha \Re B_4(p_1,...,p_4)
\end{equation}
in our master eqs.~(\ref{eq:sigma-ceex2},\ref{eq:rho-ceex2}).
The dependence on $m_\gamma$ in $Y$ cancels out. 
Photon mass gets effectively replaced by  the size of $\Omega$ in its role of the IR regulator.
The YFS form-factor $Y$ can be decomposed into six dipole components,
see eq.~(\ref{eq:YFS-ffactor})
and can be calculated analytically in terms of logs and Spence functions,
see refs.~\cite{bhwide:1997,yfsww:1996,yfsww:1998}
keeping all fermion masses exactly.

As already indicated, 
in the MC with the YFS exponentiation it would be possible to do without $\Omega$ (declare it as empty)
and rely uniquely on the IR regularization with a small photon mass $m_\gamma$ only~\cite{ceex1:1999}.
In such a case the formulas~(\ref{eq:form-factor}) for YFS form factor
would include only the second virtual photon integral part.

For the sake of the completeness of the discussion it is necessary
to examine once again the IR cancellations in the total cross section
with $\Omega$ as the new IR-regulator
%%%%%%%%%%%%%%%%%%%%%%%%%%%%%%%%%%%%%%%%%%%%%%%%%%%%%%
\begin{equation}
  \label{eq:sigtot}
  \begin{split}
   \sigma^{(r)}=
   \sum_{n=0}^\infty {1\over n!}
   \int d\tau_{2+n} &(P;\; p_3,p_4, k_1,k_2,...,k_{n})
    \prod_{j=1}^{n} \left\{ |\sfac(k_j)|^2 \bar{\Theta}(\Omega,k_j) \right\}\;\\
  &\times
    e^{ \tilde{B}_4(\Omega;p_1,...,p_4) +2\alpha\Re B_4(p_1,...,p_4) }\;
    d_{n}(k_1,k_2,...,k_{n}).
  \end{split}
\end{equation}
IR-finiteness of the total cross section now simply translates into independence 
on the $\Omega$ domain, (assuming, as usual, that the size of $\Omega$ is very small)
\begin{equation}
  {\delta \over \delta \Omega} \sigma^{(r)}= 0.
\end{equation}
The proof can be done along the same lines as the previous one for the photon mass. 
Let us assume that we want to vary $\Omega \to \Omega'=\Omega+\delta\Omega$,
that is $\bar\Omega'=\bar\Omega-\delta\Omega$.
Note that $\Omega'$ can be much bigger or smaller than of $\Omega$,
the only requirement is that both are very small%
\footnote{$\delta\Omega$ does not need to be infinitesimal with respect to $\Omega$.}
We proceed as follows
%%%%%%%%%%%%%%%%%%%%%%%%%%%%%%%%%%%%%%%%%%%%%%%%%%%%%%
\begin{equation}
  \begin{split}
   \sigma^{(r)}
  &=\sum_{n=0}^\infty {1\over n!}\; \prod_{j=1}^{n} 
   \left\{
     \int{d^3k_j\over k_j^0} |\sfac(k_j)|^2  \bar{\Theta}(\Omega',k_j)\;
    +\int{d^3k_j\over k_j^0} |\sfac(k_j)|^2 {\Theta}(\delta\Omega,k_j)\;
   \right\}\\
  &\times
   \int d\tau_{2} (P-\sum k_j ;\; p_3,p_4)\;
    e^{ \tilde{B}_4(\Omega;p_1,...,p_4) +2\alpha\Re B_4(p_1,...,p_4) }\;
    d_{n}(k_1,k_2,...,k_{n})\\
  &=\sum_{n=0}^\infty {1\over n!}
    \sum_{s=0}^n \left( n\atop s \right)
    \left\{ \int{d^3k\over k^0} |\sfac(k)|^2 {\Theta}(\delta\Omega,k) \right\}^s
   \int d\tau_{2+n-s} (P;\; p_3,p_4,k_1,...,k_{n-s})\;\\
  &\times
    \prod_{j=1}^{n-s} \left\{ |\sfac(k_j)|^2 \bar{\Theta}(\Omega',k_j) \right\}\;
    e^{ \tilde{B}_4(\Omega;p_1,...,p_4) +2\alpha\Re B_4(p_1,...,p_4) }\;
    d_{n-s}(k_1,k_2,...,k_{n-s})\\
  &=\sum_{n=0}^\infty {1\over n!}
   \int d\tau_{2+n} (P;\; p_3,p_4,k_1,...,k_{n})\;
    e^{ \int{d^3k\over k^0} |\sfac(k)|^2 {\Theta}(\delta\Omega,k) 
       +\tilde{B}_4(\Omega;p_1,...,p_4) +2\alpha\Re B_4(p_1,...,p_4)}\\
  &\times
    \prod_{j=1}^{n} \left\{ |\sfac(k_j)|^2 \bar{\Theta}(\Omega',k_j) \right\}\;
    d_{n}(k_1,k_2,...,k_{n}).
  \end{split}
\end{equation}
recovering the same expression as (\ref{eq:sigtot}),
but with $\Omega'$ instead of $\Omega$.

%%%%%%%%%%%%%%%%%%%%%%%%%%%%%%%%%%%%%%%%%%%%%%%%%%%%%%
%%%%%%%%%%%%%%%%%%%%%%%%%%%%%%%%%%%%%%%%%%%%%%%%%%%%%%
\subsection{Narrow neutral resonance in CEEX}
%%%%%%%%%%%%%%%%%%%%%%%%%%%%%%%%%%%%%%%%%%%%%%%%%%%%%%

The main new feature of CEEX in comparison with EEX is that the separation
of the IR real singularities is done at the spin amplitude level
and after squaring and spin summing them (numerically) the higher order terms are retained
while in CEEX they are truncated.
For more detailed discussion of see section \ref{sec:relation},
where we show explicitly the relations among $\hbeta's$ of EEX and $\bbeta$'s of EEX.
Keeping the above in mind, we still have at least three possible versions of CEEX.
In the following we shall describe them, concentrating mostly on the third one
which is designed for the neutral
resonances%
\footnote{ The case of exponentiation for charged resonances like $W^\pm$ resonances
is not yet covered in the literature.}
and which is the principal version implemented in the \KK\ Monte Carlo.
Let us stress immediately that the resonance may be arbitrarily narrow.
However, our approach works without any modification
for any value of the resonance width.

%%%%%%%%%%%%%%%%%%%%%%%%%%%%%%%%%%%%%%%%%%%%%%%%%%%%%%
\subsubsection{General discussion}
%%%%%%%%%%%%%%%%%%%%%%%%%%%%%%%%%%%%%%%%%%%%%%%%%%%%%%
We believe that CEEX is the only workable technique for treatment of narrow resonances
in the exclusive MC.
To understand the essential difference among three possible formulations of CEEX
it is enough to limit the discussion to the simplest case of the \Order{\alpha^0}.
The three possible options are:
\begin{itemize}
\item[(A)] Version for the non-resonant Born without partitions:
%//////////////////////////////////////////////////
  \begin{equation}
  \label{M0a}
    \Mmf^{(0)}_n 
    \left( \st^{p}_{\lambda} 
           \st^{k_1}_{\sigma_1} \st^{k_2}_{\sigma_2} \dots \st^{k_n}_{\sigma_n}\right) 
   =\prod_{i=1}^n \; 
    \big(  \sfac_{\sigma_i}^{I}(k_i)+\sfac_{\sigma_i}^{F}(k_i) \big)\;
    \Bmf_{[ba][cd]}\;
  \end{equation}
\item[(B)] Version for the non-resonant Born with partitions:
%//////////////////////////////////////////////////
  \begin{equation}
  \label{M0b}
    \Mmf^{(0)}_n 
    \left( \st^{p}_{\lambda} 
           \st^{k_1}_{\sigma_1} \st^{k_2}_{\sigma_2} \dots \st^{k_n}_{\sigma_n}\right) 
   = \sum_{\wp\in\{I,F\}^n}\;
    \prod_{i=1}^n \; \sfac_{\sigma_i}^{\{\wp_i\}}(k_i)\;
    \Bmf_{[ba][cd]} (X_\wp)\;
  \end{equation}
\item[(C)] Version for the resonant Born:
%//////////////////////////////////////////////////
  \begin{equation}
    \label{M0c}
    \Mmf^{(0)}_n 
    \left( \st^{p}_{\lambda} 
           \st^{k_1}_{\sigma_1} \st^{k_2}_{\sigma_2} \dots \st^{k_n}_{\sigma_n}\right) 
    = \sum_{\wp\in\{I,F\}^n}\;
    \prod_{i=1}^n \; \sfac_{\sigma_i}^{\{\wp_i\}}(k_i)\;
    { X^2_{\wp} \over (p_3+p_4)^2 }\;
    \sum_{R=\gamma,Z}
    \Bmf^B_{[ba][cd]} (X_\wp)\;
    e^{\alpha B_4^R(X_\wp)}
  \end{equation}
\end{itemize}
Let us immediately define the additional form-factor for the $Z$ resonance (case (C))
%//////////////////////////////////////////////////
\begin{equation}
  \label{eq:B4Z}
    \alpha B_4^Z(X) =
    \int {d^4k\over k^2 -m_\gamma^2 +i\epsilon}\; {i\over (2\pi)^3}\;
           {J_I}_\mu(k) (J_F^\mu(k))^*\;
           \left(     { (X)^2 -\bar{M}^2 \over (X-k )^2 -\bar{M}^2 }  -1 \right),
\end{equation}
where $\bar{M}^2 = M_Z^2 -iM_Z\Gamma_Z$, the currents $J^\mu$ are defined in  (\ref{eq:B4}),
while for the non-resonant part we have $B_4^\gamma(X)=0$.
The $B_4^Z(X)$ form-factor sums up to infinite order the virtual $\alpha \ln(\Gamma_Z/M_Z)$
contributions 
-- we postpone discussion of its origin and importance to the latter part of this section.

Coming back to the more elementary level we see that the case (B) becomes (A)
if we can neglect the partition dependence of the four momentum in the Born:
$\Bmf_{[ba][cd]}(X_\wp) \to \Bmf_{[ba][cd]}(P)$, where $P=p_a+p_b$ or $P=p_c+p_d$
or any other choice which does not depend on momenta of the individual photons.
This is thanks to the identity:
%//////////////////////////////////////////////////
\begin{equation}
    \prod_{i=1}^n\; (\sfac_{\sigma}^{\{F\}}(k_i) + \sfac_{\sigma}^{\{I\}}(k_i))
    \equiv
    \sum_{\wp\in\{I,F\}}\;
    \prod_{i=1}^n \; \sfac_{\sigma_i}^{\{\wp_i\}}(k_i)\;
\end{equation}
Only case (C) is efficient for the resonant process, so obviously (A) and (B) are limited to
non-resonant processes.
The immediate question is: which of them is better?
If (A) is not summing higher order much better than (B),
then it has the clear advantage of being simpler
-- summation over partitions makes
the computer code more complicated and adds heavily to the consumption of CPU time.
The answer is that, although we did not investigate quantitatively the differences
between (A) and (B), 
we think that (B) sums up the LL higher orders more efficiently than (A) and is therefore better,
even if there is no resonance.
In our case, since we want to cover the resonant process anyway, it is a natural choice to use (B)
for the non-resonant background component
of the spin amplitudes (off-shell $\gamma$ exchange) even if it is not vital.
Once summation over partitions is in place, it is the easiest to use it for the non-resonant background
as well. 
The additional bonus of better higher order convergence provides an extra justification.
{\em Summarizing, if (C) is implemented then (B) comes for free.}

Having discussed the differences among the three options 
let us now concentrate on option (C) for the resonant process,
remembering that for the non-resonant background component it becomes automatically (B).
First of all, for the narrow neutral resonance (the $Z$ boson in our case) the emission of the photons
in the production and the decay processes are well separated by a long time interval,
and are therefore completely independent and uncorrelated.
In the perturbative QED this simple physical fact is reflected
in a certain specific class of cancellations among the ISR and FSR photons on one hand
and among virtual and real corrections on the other hand.
For the inclusive observables like the total cross section or charge asymmetry
the effects of the ISR-FSR interference (IFI) in the non-resonant case are
of order $\alpha/\pi$, typically up to $1\%$,
as can be seen from many example of the explicit \Order{\alpha^1} calculations.
The IFI effect will be of order $(\alpha/\pi)(E_{\max}/E_{beam})$, when the experimental
cut on photon energy is $E_{\max}$.
Note that the IFI effect is not directly enhanced by the big mass-logarithms like $\ln s/m_e^2\sim 20$.
For the resonant process the IFI effects in the inclusive observables are
of order $(\alpha/\pi)(\Gamma/M)$ and are therefore often negligible
on the scale of the experimental error.
One has to remember, however, that the additional suppression factor $\Gamma/M$
disappears if the experimental cut on photon energy is of order of the resonance width,
$E_{\max}/E_{beam}\sim \Gamma/M$,
and for an even stronger cut $E_{\max}<\Gamma$ the IFI effect becomes
of order $(\alpha/\pi)(E_{\max}/\Gamma)$.

If $\Gamma/M$ is extremely small, like for the $\tau$ lepton, the IFI cancellation
can be taken for granted and the photon emission interference between production and decay can be
neglected whatsoever.
In the case of the $Z$ boson close to the $Z$ resonance (LEP1) the IFI effect is detectable experimentally
but it is small enough that it can be omitted in the Monte Carlo programs
used for correcting for the detector acceptance only.
In this case KORALZ/YFS3 \cite{koralz4:1994} with the EEX matrix element was the acceptable solution.

The most convenient solution is the universal Monte Carlo in which IFI is included,
which can evaluate IFI effects near the resonance, far from the resonance, for inclusive quantities
and for strong energy cuts $E_{\max}\sim\Gamma$.
This is exactly what our CEEX offers.

%%%%%%%%%%%%%%%%%%%%%%%%%%%%%%%%%%%%%%%%%%%%%%%%%%%%%%%%%%%%%%%%%%%%
\subsubsection{Derivation of the resonance formfactor}
%%%%%%%%%%%%%%%%%%%%%%%%%%%%%%%%%%%%%%%%%%%%%%%%%%%%%%%%%%%%%%%%%%%%

As we have already pointed out (following refs.~\cite{greco:1975,greco:1980}),
in the presence of narrow resonances
it is not enough to sum up coherently the real emissions, 
taking properly into account energy shift in the resonance propagator (only due to ISR photons).
It is also necessary to do the same for the virtual emission, 
and also sum them up to infinite-order --
this is why the resonance form factor $\exp(B_4^Z)$ has to be included,
see eqs.~(\ref{M0c}) and (\ref{eq:B4Z}).
In the following we shall derive eq.~(\ref{eq:B4Z}) for $B_4^Z$ and show analytically that the
IFI cancellations do really work, as expected, to infinite order.

Let us write again the formula for standard YFS function in eq.~(\ref{eq:B4}) in a slightly
modified notation
\begin{equation}
\begin{split}
 &B_4(p_a,...,p_d)=
  \int {d^4k\over k^2 -m_\gamma^2 +i\epsilon} {i\over (2\pi)^3} S(k),
\\
 &S(k) = S_I(k) +S_F(k) +S_{Int}(k),\;\;
\\
 &S_I(k) = |J_I(k)|^2,\;\;
  S_F(k) = |J_F(k)|^2,\;\;
  S_{Int}(k) = -2\Re(J_I(k)\cdot J^*_F(k))
\end{split}
\end{equation}
In the presence of the narrow resonance,
the YFS factorization of the virtual IR contributions has to take into account
the dependence of the scalar part of the resonance propagator on photon energies
of order $\Gamma$
(the numerator treated in soft photon approximation as usual).
The relevant integrals with $n$ virtual photons look as follows:
\begin{equation}
\label{eq:virsoft}
  I= (P^2 -\bar{M}^2)
    \sum_{n=0}^\infty  {1\over n!}  \sum_{\wp\in{\cal P}_n}
    \prod_{i=0}^n
    \int  {i\over (2\pi)^3} {d^4 k_{i} \over k_{i}^2 -m_\gamma^2 } S_{\wp_i}(k_i)
    {1\over P_\wp^2 -\bar{M}^2},
\end{equation}
where $\bar{M}^2 = \bar{M}^2 -iM\Gamma$,
and ${\cal P}_n$ is set of all $3^n$ 
partitions $(\wp_1,\wp_2,...,\wp_n)$ with $\wp_i=I,F,Int$,
and $P_\wp\equiv P-\sum\limits_{\wp_i=Int} k_i$
includes only momenta of photons in $S_{Int}$ and not of photons in $S_I$ or $S_F$.
The $(P^2 -\bar{M}^2)$ factor is conventional, to make the integral dimensionless.
We shall show that the above integral factorizes into the conventional YFS formfactor 
(dependent on the photon mass $m_\gamma$)
and the additional non-IR factor due to the resonance $R=Z$
\begin{equation}
  I= \exp(B^R_4(m_\gamma,s,\bar{M})) = \exp(B_4(m_\gamma,s) + \Delta B^R_4(s,\bar{M})).
\end{equation}
Our aim is to find  the analytical form of the the additional function $\Delta B^R_4$.
In the current calculation we use the following
approximate formula, also used by Greco et.al.~\cite{greco:1975,greco:1980},
%%%%%%%%%%%%%%%%%%%%%%%%%%%%%%%%%%%%%%%%%%%%%%%%%%%%%
\begin{equation}
   \label{eq:FormGreco}
   \alpha\Delta B_4^R(s') = -2Q_eQ_f {\alpha\over\pi} 
      \ln\bigg({t\over u}\bigg)
      \ln\bigg({ \bar{M}^2-s \over \bar{M}^2}\bigg)
    =-{1\over 2}\gamma_{Int}\ln\bigg({ \bar{M}^2-s \over \bar{M}^2}\bigg).
\end{equation}
In the following:
\begin{itemize}
\item
  We shall derive the above approximate result and
\item
  show explicitly that the above approximate virtual interference part of the formfactor
  cancels exactly with the corresponding real interference contributions.
\end{itemize}

Since soft virtual photons entering into $S_I$ and $S_F$ in eq.~(\ref{eq:virsoft})
do not enter the resonance propagator,
we may therefore factorize and sum up the contributions with  $S_I$ and $S_F$:
\begin{equation}
  \begin{split}
  I&=
   \sum_{n_1=0}^\infty {1\over n_1!} \prod_{i_1=0}^{n_1}
   \int {i\over (2\pi)^3} {d^4 k_{i_1} \over k_{i_1}^2 -m_\gamma^2 }  S_I(k_{i_1})\;
   \sum_{n_2=0}^\infty {1\over n_2!} \prod_{i_2=0}^{n_2}
   \int {i\over (2\pi)^3} {d^4 k_{i_2} \over k_{i_2}^2 -m_\gamma^2 }  S_F(k_{i_2})
\\&\times
   \sum_{n_3=0}^\infty {1\over n_3!} \prod_{i_3=0}^{n_3}
   \int {i\over (2\pi)^3} {d^4 k_{i_3} \over k_{i_3}^2 -m_\gamma^2 }  S_{Int}(k_{i_3})
    {1\over (P-\sum_{j=1}^{n_3} k_j)^2 -\bar{M}^2}
\\&\equiv
   e^{ \alpha B_I + \alpha B_F}
   \sum_{n=0}^\infty {1\over n!}
   \prod_{i=0}^n
   \int  {i\over (2\pi)^3} {d^4 k_i \over k_i^2 -m_\gamma^2 } S_{Iin}(k_i)
    {1\over (P-\sum_{j=1}^{n} k_j)^2 -\bar{M}^2}
 \end{split}
\end{equation}
Now we neglect the quadratic terms in photon energies ${\cal O}( k_ik_j)$
\begin{equation}
  \begin{split}
  &{1\over (P-\sum_{j=1}^{n} k_j)^2 -\bar{M}^2}
  \simeq {1\over P^2 -2P\sum_{j=1}^{n} k_j -\bar{M}^2}
  =  {1\over P^2-\bar{M}^2}\;   {1\over 1 -\sum_{j=1}^{n} {2P k_j \over P^2-\bar{M}^2} }
\\&
  \simeq {1\over P^2-\bar{M}^2}\; \prod_{j=1}^{n} {1\over 1 - {2P k_j \over P^2-\bar{M}^2} }
  \simeq {1\over P^2-\bar{M}^2}\; \prod_{j=1}^{n} {P^2-\bar{M}^2 \over (P-k_j)^2-\bar{M}^2 }
  \end{split}
\end{equation}
and this leads to
\begin{equation}
\label{eq:kicka}
  \begin{split}
   &I=    e^{ \alpha B_I + \alpha B_F}
    \exp\Bigg( 
         \int  {i\over (2\pi)^3} {d^4 k \over k_i^2 -m_\gamma^2 +i\epsilon} S_{Iin}(k)  
               {P^2-\bar{M}^2 \over (P-k)^2-\bar{M}^2 } \Bigg)
     = e^{ B_4(m_\gamma) + \Delta B^R_4(\Gamma)}
\\&
     \Delta B^R_4(\Gamma)
     = \int  {i\over (2\pi)^3} {d^4 k \over k^2 } S_{Iin}(k)  
               \Bigg( {P^2-\bar{M}^2 \over (P-k)^2-\bar{M}^2 } -1\Bigg)
  \end{split}
\end{equation}
How solid is the above ``derivation''?
Strictly speaking it is justified in the limit 
where we follow Yennie, Frautschi and Suura in ref.~\cite{yfs:1961}
and express the $k\to0$ emission amplitude as
\begin{displaymath}
  \Meu\to   {1\over k}\bigg(  \varepsilon_1+{\cal O}(k/\bar{M}) 
                             +{k \over \Gamma_Z} (\varepsilon_2+{\cal O}(k/\bar{M}))
                      \bigg),
\end{displaymath}
where $\varepsilon_{1,2}$ are constants independent of $k$,
so that
\begin{displaymath}
  \big|{2P k_j / (P^2-\bar{M}^2)} \big| \ll 1,
\end{displaymath}
that is if photon energy is below the resonance width.
This restriction is thus entirely analogous to the usual YFS expansion into IR-singular part and the rest.
We note that Greco et.al. in refs.~\cite{greco:1975,greco:1980} have also pointed out  that the result
for $\Delta B^R_4(\Gamma)$ in eq.~(\ref{eq:kicka}) follows from the YFS expansion; 
here shall show how this happens in detail.

The best situation would be to have a more precise evaluation of the integral of eq.~(\ref{eq:virsoft})
(the integral is probably calculable analytically).
For the moment, however, following refs.~\cite{greco:1975,greco:1980}:
we choose an easier ``pragmatic'' approach based on the fact
that the virtual and real contributions from IFI
for photons $E_\gamma>\Gamma$ {\em do cancel}, 
as a consequence  of the time separation between production and decay,
and we shall check that the above cancellation really works.
In this way we trade analytical evaluation of the more difficult
multiphoton virtual integral for
an easier evaluation of the multiphoton real integral.

%%%%%%%%%%%%%%%%%%%%%%%%%%%%%%%%%%%%%%%%%%%%%%%%%%%%%%%%%%%%%%%%%%%%%%%%%%%%%%%%%
\subsubsection{Cancellation of the virtual formfactor with the real emissions}

Let us therefore examine analytically 
the real multi-photon emission contribution from the IFI%
\footnote{ In the practical CEEX calculation the contribution from IFI is evaluated
  numerically, inside the MC program.}.
The starting point is the integral in which the {\em total} photon energy
$K=\sum _{j=1}^n k_j$ is kept below
$E_{\max}=v_{\max}\sqrt{s}$, where $\Gamma < E_{\max} << \sqrt{s}$:
\begin{equation}
  \begin{split}
&
  \sigma=
  \sum_{n=0}^\infty {1\over n!}
  \int \prod_{i=1}^n {d^3 k_i \over 2k_i^0}
  \sum_{\sigma_1...\sigma_n}
  \bigg| 
         \sum_{\wp\in\{I,F \}^n}
         \prod_{j=1}^n \sfac_{[j] \{\wp_j\} }\;
         {1\over X_\wp^2 -\bar{M}^2 }\;
         e^{\alpha B_4^R(X_{\wp})}
  \bigg|^2
  \Theta( E_{\max} - \sum _{j=1}^n k_j)
\\&
 =\sum_{n=0}^\infty {1\over n!}
  \int\limits_{K^0<v\sqrt{s}} \prod_{i=1}^n {d^3 k_i \over 2k_i^0}
  \sum_{\sigma_1...\sigma_n}
         \sum_{\wp,\wp'\in\{I,F \}^n}
         \prod_{j=1}^n \sfac_{[j] \{\wp_j\} }\; \sfac^*_{[j] \{\wp'_j\} }\;
         {e^{\alpha B_4^R(X_\wp)}    \over X_\wp^2    -\bar{M}^2 }
   \Bigg({e^{\alpha B_4^R(X_{\wp'})} \over X_{\wp'}^2 -\bar{M}^2 }\Bigg)^*
\\&
  =\sum_{n=0}^\infty {1\over n!} 
  \int\limits_{K^0<v\sqrt{s}} \prod_{i=1}^n {d^3 k_i \over 2k_i^0}
  \sum_{\wp,\in\{I^2,F^2,IF,FI \}^n}
  \prod_{\wp_j=I^2}   \tilde{S}_I(k_j)
  \prod_{\wp_j=F^2}   \tilde{S}_F(k_j)
\\&\qquad
  \prod_{\wp_j=IF}    \tilde{S}_{Int}(k_j)\;
  \prod_{\wp_j=FI}    \tilde{S}_{Int}(k_j)\;
               {e^{\alpha B_4^R(P-K_I-K_{IF})}\over (P-K_I-K_{IF})^2 -\bar{M}^2 }\;
        \Bigg( {e^{\alpha B_4^R(P-K_I-K_{FI})}\over (P-K_I-K_{FI})^2 -\bar{M}^2 } \Bigg)^*
  \end{split}
\end{equation}
where the Born amplitude we have simplified to the level of the scalar part of the resonance propagator 
and we denote
\begin{equation}
  \begin{split}
   &\tilde{S}_I(k_j) = \sum_{\sigma_j} |\sfac_{[j]}^{\{0\} }|^2,
\quad
    \tilde{S}_F(k_j) = \sum_{\sigma_j} |\sfac_{[j]}^{\{1\} }|^2,
\\&
    \tilde{S}_{Int}(k_j)=\sum_{\sigma_j} \sfac_{[j]}^{\{0\} }\; (\sfac_{[j] \{1\} })^*
                        =\sum_{\sigma_j} \sfac_{[j]}^{\{1\} }\; (\sfac_{[j] \{0\} })^*,
\\&
    K_{I^2}   = \sum_{\wp_j=I^2 } k_j,\quad
    K_{F^2}   = \sum_{\wp_j=F^2 } k_j,\quad
\\&
    K_{IF} = \sum_{\wp_j=IF} k_j,\quad
    K_{FI} = \sum_{\wp_j=FI} k_j,\quad
    K = K_{I^2}+K_{F^2}+K_{IF}+K_{FI}
  \end{split}
\end{equation}
As we see, the product of two sums,
each  over $2^n$ partitions $\wp,\wp'\in\{I,F \}^n$,
is now replaced by the single sum over $4^n$ partitions $\wp,\in\{I^2,F^2,IF,FI \}^n$,
where $IF,FI$ represent the interference terms.

Keeping track of the dependence of the propagators on $K_{I^2}$,  $K_{IF}$ and $K_{FI}$,
the summation over the number of photons can be reorganised, leading us back to 
the following factorized formula
\begin{equation}
  \begin{split}
 &\sigma(v_{\max})=
  \sum_{n_1=0}^\infty {1\over n_1!}
  \int \prod_{i_1=1}^{n_1} {d^3 k_{i_1} \over 2k_{i_1}^0}\; 2\tilde{S}_I(k_{i_1})\;
  \sum_{n_2=0}^\infty {1\over n_2!}
  \int \prod_{i_2=1}^{n_2} {d^3 k_{i_2} \over 2k_{i_2}^0}\; 2\tilde{S}_F(k_{i_2})\;
\\&\qquad
  \sum_{n_3=0}^\infty {1\over n_3!}
  \int \prod_{i_3=1}^{n_3} {d^3 k_{i_2} \over 2k_{i_3}^0}\; 2\tilde{S}_{Int}(k_{i_3})\;
               {e^{\alpha B_4^R(P-K_{I^2}-K_{IF})}\over (P-K_{I^2}-K_{IF})^2 -\bar{M}^2 }\;
\\&\qquad
  \sum_{n_4=0}^\infty {1\over n_4!}
  \int \prod_{i_4=1}^{n_3} {d^3 k_{i_4} \over 2k_{i_4}^0}\; 2\tilde{S}_{Int}(k_{i_4})\;
        \Bigg( {e^{\alpha B_4^R(P-K_{I^2}-K_{FI})}\over (P-K_{I^2}-K_{FI})^2 -\bar{M}^2 } \Bigg)^*
\\&\qquad
  \Theta( E_{\max} -K_{I^2}-K_{F^2}-K_{IF}-K_{FI}),
  \end{split}
\end{equation}
where $K_{I^2} = \sum_{i_1} k_{i_1}$,  
$K_{F^2} = \sum_{i_2} k_{i_2}$, $K_{IF} = \sum_{i_3} k_{i_3}$ and  $K_{FI} = \sum_{i_4} k_{i_4}$.
The sums over the pure initial and final state contributions,
and the interference contributions are now well factorized and
can be performed analytically.
As a first step we integrate and sum up contributions from the very
soft photons below $\varepsilon \sqrt{s}$, similarly to what was shown in ref.~\cite{yfs2:1990}:
\begin{equation}
  \begin{split}
 &\sigma(v_{\max})=
  \int_0^{E_{\max}} dE\; \delta( E -E_I-E_F-E_{Int})
  \int_0^{E_{\max}} dE_I\; dE_F\; dE_{IF}\; dE_{FI}\;
\\&\quad
  \sum_{n_1=0}^\infty {1\over n_1!}
  \prod_{i_1=1}^{n_1} \int_{k^0_{i_1}> \varepsilon E } 
  {d^3 k_{i_1} \over 2k_{i_1}^0} 2\tilde{S}_I(k_{i_1})
  e^{2\alpha \tilde{B}_I(\varepsilon E) +2\alpha \Re B_I}
  \delta(E_I-\sum_{i_1} k_{i_1})
\\&\quad
  \sum_{n_2=0}^\infty {1\over n_2!}
  \prod_{i_2=1}^{n_2} \int_{k^0_{i_2}> \varepsilon E} 
  {d^3 k_{i_2} \over 2k_{i_2}^0} 2\tilde{S}_F(k_{i_2})\;
  e^{2\alpha \tilde{B}_F(\varepsilon E) +2\alpha \Re B_F}
  \delta(E_F-\sum_{i_2} k_{i_2})
\\&\quad
  \sum_{n_3=0}^\infty {1\over n_3!}
  \prod_{i_3=1}^{n_3} \int_{k^0_{i_3}> \varepsilon E } 
  {d^3 k_{i_2} \over 2k_{i_3}^0} 2\tilde{S}_{Int}(k_{i_3})\;
  {e^{\alpha \Delta B_4^R(P-K_{I^2}-K_{IF})}\over (P-K_{I^2}-K_{IF})^2 -\bar{M}^2 }\;
  e^{\alpha \tilde{B}_{Int}(\varepsilon E) +\alpha \Re B_{Int}}
\\&\quad
  \sum_{n_4=0}^\infty {1\over n_4!}
  \prod_{i_3=1}^{n_4} \int_{k^0_{i_3}> \varepsilon E } 
  {d^3 k_{i_2} \over 2k_{i_3}^0} 2\tilde{S}_{Int}(k_{i_3})\;
  \Bigg( {e^{\alpha \Delta B_4^R(P-K_{I^2}-K_{FI})}\over (P-K_{I^2}-K_{FI})^2 -\bar{M}^2 } \Bigg)^*
  e^{\alpha \tilde{B}_{Int}(\varepsilon E) +\alpha \Re B_{Int}}
\\&\qquad
  e^{2\alpha \Re \Delta B_4^R }\;
  \delta(E_{Int}-\sum_{i_3} k_{i_3}),
  \end{split}
\end{equation}
where $E={\sqrt{s}\over 2}$.
The integration over photon momenta can be performed without any approximation
leading to the following result
\begin{equation}
  \begin{split}
 &\sigma(v_{\max})=
  \int_0^{v_{\max}} dv\; \delta( v -v_I-v_F-v_{IF}-v_{FI})
\\&\quad
  \int dv_I\;
  F(\gamma_I) \gamma_I v_I^{ \gamma_I -1}\;
  e^{2\alpha \tilde{B}_I(E) +2\alpha \Re B_I}
  \int dv_F\;
  F(\gamma_F) \gamma_F v_F^{ \gamma_F -1}\;
  e^{2\alpha \tilde{B}_F(E) +2\alpha \Re B_F}
\\&
  \int dv_{IF}\;
  F\big({\gamma_{Int}\over 2}\big) {1\over 2}\gamma_{Int} v_{IF}^{ {1\over 2}\gamma_{IF} -1}\;
  \bigg( {e^{\alpha \Delta B_4^R(s(1-v_I)(1-v_{IF}))} \over s(1-v_I)(1-v_{IF}) -\bar{M}^2 } \bigg)\;
  e^{\alpha \tilde{B}_{Int}(E) +\alpha \Re B_{Int} }
\\&
  \int dv_{FI}\;
  F\big({\gamma_{Int}\over 2}\big) {1\over 2}\gamma_{Int} v_{FI}^{ {1\over 2}\gamma_{FI} -1}\;
  \bigg( {e^{\alpha \Delta B_4^R(s(1-v_I)(1-v_{FI}))} \over s(1-v_I)(1-v_{FI}) -\bar{M}^2 } \bigg)^*\;
  e^{\alpha \tilde{B}_{Int}(E) +\alpha \Re B_{Int} }
  \end{split}
\end{equation}
in which is explicitly free of any IR divergences.

The essential question is whether we have perfect 
cancellations of the $\ln(\Gamma/M_Z)$ terms in the interference subintegral
\begin{equation}
  \begin{split}
  I_{Int}= \Re \int_0^{v_{max}-v_I-v_F-v_{FI}} dv_{IF}\;
  F\Big({\gamma_{IF}\over 2}\Big) {1\over 2}\gamma_{IF} 
  v_{IF}^{ {1\over 2}\gamma_{Int} -1}\;
  {  e^{\alpha\Delta B_4^R(s'(1-v_{IF})) } \over s'(1-v_{IF}) -\bar{M}^2 }\;
  \end{split}
\end{equation}
We omit from consideration the constant IR-finite factor 
$e^{\alpha \tilde{B}_{Int}(E) +\alpha \Re B_{Int} }$
because it does not depend on resonance parameters.
The bulk of the integral comes from the neighbourhood of $v_{IF}=0$
and the integrand is $\sim 1/v^2$ at large $v$ due to the resonance;
we can therefore extend the integration
limit to $\int_0^{\infty} dv_{Int}$ at the expense of an error of
${\cal O} \big( {\Gamma\over M_Z} \big)$.
One possible evaluation method is to use the standard techniques of the complex functions.
First, we reformulate the integral as an integral 
over the discontinuity $C_1$ along the real axis%
\footnote{ We have also pulled out of the integral the $e^{\alpha\Delta B_4^R}$ factor,
  because the most of integral comes from the neighbourhood of the singularity at $v_{IF}=0$.}
\begin{equation}
  \begin{split}
  I_{Int}= 
  F\Big({\gamma_{IF}\over 2}\Big)
  e^{\alpha\Delta B_4^R(s') }
  { 1\over i\sin(\pi {1\over 2}\gamma_{Int}) }
   \int_{C_1} dz\;
   {1\over 2}\gamma_{Int} (-z)^{ {1\over 2}\gamma_{Int} -1}\;
        {1 \over s'-\bar{M}^2  -s'z }\;
  \end{split}
\end{equation}
Since the contour can be closed in a standard way with the big circle,
the integral is given by the value of the residue at $z=1-\bar{M}^2/s'$.
\begin{equation}
  \begin{split}
 &I_{Int}= 
  F\Big({\gamma_{IF}\over 2}\Big)
  e^{\alpha\Delta B_4^R(s') }\;
  { \pi{1\over 2}\gamma_{Int} \over \sin(\pi {1\over 2}\gamma_{Int}) }\;
   \Bigg({\bar{M}^2 -s'\over s'} \Bigg)^{\gamma_{Int} -1}\;
   { 1 \over s'}
\\&
  ={ 1 \over \bar{M}^2-s'}
    F\Big({\gamma_{Int}\over 2}\Big)
  { \pi{1\over 2}\gamma_{Int} \over \sin(\pi {1\over 2}\gamma_{Int}) }\;
  e^{\alpha\Delta B_4^R(s') }\;
   \Bigg({\bar{M}^2 -s'\over s'} \Bigg)^{{1\over 2}\gamma_{Int}}\;
\\&
  ={ 1 \over \bar{M}^2-s'}
   (1+{\cal O}(\gamma_{Int}))
  \end{split}
\end{equation}
The above is true because
\begin{equation}
  \begin{split}
    \alpha\Delta B_4^R(s') = -2Q_eQ_f {\alpha\over\pi} 
      \ln\bigg({t\over u}\bigg)
      \ln\bigg({ \bar{M}^2-s' \over \bar{M}^2}\bigg)
    =-{1\over 2}\gamma_{Int}\ln\bigg({ \bar{M}^2-s' \over \bar{M}^2}\bigg)
  \end{split}
\end{equation}
We have therefore proven
the full cancellation of the dependence on the resonance parameters
for the integrated cross section.

%%%%%%%%%%%%%%%%%%%%%%%%%%%%%%%%%%%%%%%%%%%%%%%%%%%%%%%%%%%%%%%%%%%%
\subsubsection{Definitions of $\hbeta$'s with partitions}
\label{sec:resobetas}
%%%%%%%%%%%%%%%%%%%%%%%%%%%%%%%%%%%%%%%%%%%%%%%%%%%%%%%%%%%%%%%%%%%%

The \Order{\alpha^r}, $r=0,1,2$,
$\hbeta$-functions  for the variant of the CEEX with summation over the partitions,
as in eqs.~(\ref{eq:ceex-master}-\ref{eq:ceex-master2}), are derived
with the recursive relations of eqs.~(\ref{eq:BetaRecursive})
(similar to those of eqs.~(\ref{eq:BetRecursive2})).
The only additional complication is that we must keep track of the indices which
say whether an external real photon is of ISR or FSR type and of the total photon momentum
after emission of the ISR photons 
(the one which enters resonance propagator, if such a resonance is present)
%///////////////////////////////////////////////////////////////////////////
\begin{equation}
  \label{eq:DefBet}
  \begin{split}
   &  \hbeta^{(l)}_0\left(\st^{p}_{\lambda}; P \right) 
      =\Mmf^{(l)}_0\left(\st^{p}_{\lambda}; P \right),\; l=0,1,2,
\\&
      \hbeta^{(1+l)}_{1\{I\}}\left(\st^{p}_{\lambda}\st^{k_1}_{\sigma_1}; P-k_1 \right)      
      =\Mmf^{(1+l)}_{1\{I\}}\left(\st^{p}_{\lambda}\st^{k_1}_{\sigma_1}; P-k_1 \right) 
       -\hbeta^{(l)}_0\left(\st^{p}_{\lambda}; P-k_1 \right) \sfac^{\{I\}}_{\sigma_1}(k_1),\;l=0,1,
\\&
      \hbeta^{(1+l)}_{1\{F\}}\left(\st^{p}_{\lambda}\st^{k_1}_{\sigma_1}; P \right)      
      =\Mmf^{(1+l)}_{1\{F\}}\left(\st^{p}_{\lambda}\st^{k_1}_{\sigma_1}; P \right) 
       -\hbeta^{(l)}_0\left(\st^{p}_{\lambda}; P \right)  \sfac^{\{F\}}_{\sigma_1}(k_1),\;l=0,1,
\\&
       \hbeta^{(2)}_{2\{\omega_1,\omega_2\}}
           \left(\st^{p}_{\lambda}\st^{k_1}_{\sigma_1}\st^{k_2}_{\sigma_2}; X_\omega \right)    
    =    \Mmf^{(2)}_{2\{\omega_1,\omega_2\}}
           \left(\st^{p}_{\lambda}\st^{k_1}_{\sigma_1}\st^{k_2}_{\sigma_2}; X_\omega \right)
\\&\quad
       -\hbeta^{(1)}_{1\{\omega_1\}}\left(\st^{p}_{\lambda}\st^{k_1}_{\sigma_1}; X_\omega \right) 
                                                              \sfac^{\{\omega_2\}}_{\sigma_2}(k_2)  
       -\hbeta^{(1)}_{1\{\omega_2\}}\left(\st^{p}_{\lambda}\st^{k_2}_{\sigma_2}; X_\omega \right) 
                                                              \sfac^{\{\omega_1\}}_{\sigma_1}(k_1)
       -\hbeta^{(0)}_0\left(\st^{p}_{\lambda}; X_\omega \right)  
                       \sfac^{\{\omega_1\}}_{\sigma_1}(k_1) \sfac^{\{\omega_2\}}_{\sigma_2}(k_2),\\
  \end{split}
\end{equation}
where  $X_\omega=P-\sum\limits_{\omega_i=I} k_i$, $P=p_a+p_b$.
Introduction of the partition index $\omega_i$ defining whether a photon belongs to ISR or FSR
is in a sense not such a deep and great complication --
it is now just another (third) attribute of the photon like its helicity.

Let us look closer into the structure of the term like
$\hbeta^{(1)}_{1\{\omega_1\}}\left(\st^{p}_{\lambda}\st^{k_1}_{\sigma_1}; X_\omega \right) 
                                                     \sfac^{\{\omega_2\}}_{\sigma_2}(k_2)$.
For example $\omega_1=F$ and $\omega_2=I$ it reads
$\hbeta^{(1)}_{1\{F\}}\left(\st^{p}_{\lambda}\st^{k_1}_{\sigma_1}; P-k_2 \right) 
                                                     \sfac^{\{I\}}_{\sigma_2}(k_2)$,
that is the total shift in $X$ in $\hbeta^{(1)}$ depends 
not only on the type $\omega_1$ of ``its own photon'' but also on the type $\omega_2$
of the photon in $\sfac^{\{\omega_2\}}$ factor which multiplies it!

The $\Mmf$-amplitude in eq.~(\ref{eq:DefBet}) is given essentially by eq.~(\ref{eq:virtual-subtraction})
with the formfactor including the resonance part (if present)
%//////////////////////////////////////////////////
\begin{equation}
  \label{eq:VirtSubt}
  \Mmf^{(r)R}_{n\{\omega\}}
           \left(\st^{p}_{\lambda}\st^{k_1}_{\sigma_1}...\st^{k_n}_{\sigma_n}; X_\omega \right)
   =\left[ e^{-\alpha B_4 -\alpha B_4^R(X_\omega) } 
           \Meu^{(r)R}_{n\{\omega\}}
           \left(\st^{p}_{\lambda}\st^{k_1}_{\sigma_1}...\st^{k_n}_{\sigma_n}; X_\omega \right)
    \right] \Big|_{{\cal O}(\alpha^r) },
\end{equation}
As we see the type $R=\gamma,Z$ of the ``resonance'' formfactor $B_4^R$ has to be adjusted 
to the type of the component in $\Meu^{(r)R}$ 
(we have temporarily introduced an explicit  index $R$ into $\Meu$ and $\Mmf$ and $\gamma$
is essentially a ``resonance'' with the zero width).

%%%%%%%%%%%%%%%%%%%%%%%%%%%%%%%%%%%%%%%%%%%%%%%%%%%%%%%%%%%%%%%%%%%%%%%%%%%%%%%%%%%%%%%%%%%%%
%%%%%%%%%%%%%%%%%%%%%%%%%%%%%%%%%%%%%%%%%%%%%%%%%%%%%%%%%%%%%%%%%%%%%%%%%%%%%%%%%%%%%%%%%%%%%
\subsection{Virtual corrections, no real photons}
%%%%%%%%%%%%%%%%%%%%%%%%%%%%%%%%%%%%%%%%%%%%%%%%%%%%%%%%%%%%%%%%%%%%%%%%%%%%%%%%%%%%%%%%%%%%%
We now start to accumulate the actual formulas for the $\hbeta$-functions entering the CEEX
amplitudes of in eqs.~(\ref{eq:ceex-master}-\ref{eq:ceex-master2})
with the case of no real photons and up to two virtual photons.
The ``raw material'' are the $\Meu$-amplitudes from Feynman diagrams
which are turned into  $\hbeta$-functions using the recursive
relations of eqs.~(\ref{eq:DefBet}).

%%%%%%%%%%%%%%%%%%%%%%%%%%%%%%%%%%%%%%%%%%%%%%%%%%%%%%
\subsubsection{Photonic corrections}
%%%%%%%%%%%%%%%%%%%%%%%%%%%%%%%%%%%%%%%%%%%%%%%%%%%%%%

%====================================================================================
%------------------------------------------------------------------------------------
\begin{figure}[h]
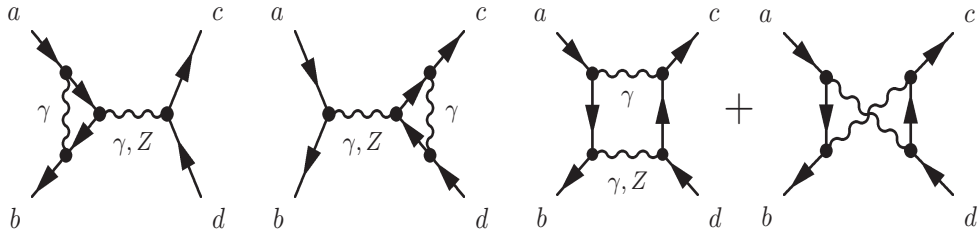

\centering
\setlength{\unitlength}{1mm}
\begin{picture}(120,30)
%\put(0,0){\framebox( 100,25){ }}
\put( -5, 0){\makebox(0,0)[lb]{\epsfig{file=feyn-IvirF1.eps,width=30mm,height=30mm}}}
\put( 30, 0){\makebox(0,0)[lb]{\epsfig{file=feyn-FvirF1.eps,width=30mm,height=30mm}}}
\put( 65, 0){\makebox(0,0)[lb]{\epsfig{file=feyn-Box.eps,width=60mm,height=30mm}}}
\end{picture}
\caption{\sf First order diagrams.}
\label{fig:IvirF}
\end{figure}
%====================================================================================
Let us start with the simple case of 
\Order{\alpha^1} spin amplitudes with one virtual and zero real photon
coming directly from Feynman diagrams, which will be used to obtain
the first order $\hbeta^{(1)}_0$.
The relevant spin amplitudes are
%//////////////////////////////////////////////////////////////
%      First-order, 0-photon
%         F1ini = DCMPLX(0.5d0*m_Alfpi*Qe*Qe)*( BVR_CnuA(svar,m1,m2) -DCMPLX(1d0) )
%//////////////////////////////////////////////////////////////
\begin{equation}
\label{one-virtual}
    \Meu^{(1)}_0 \left(\st^{p}_{\lambda};X \right)
  = \Bmf \left(\st^{p}_{\lambda} ; X \right)
   \left[1+ Q_e^2 F_1(s,m_e,m_\gamma) \right]
   \left[1+ Q_f^2 F_1(s,m_f,m_\gamma) \right]
   +\Meu^{(1)}_{\rm box} \left(\st^{p}_{\lambda} ; X \right),
\end{equation}
where $F_1$ is the standard electric form-factor regularized with a photon mass, 
see fig.~\ref{fig:IvirF}.
We omit, for the moment, the magnetic form-factor $F_2$; this is justified
for light final fermions. It will be restored in the future.
In $F_1$ we keep  the exact final fermion mass.
If not stated otherwise, the four-momentum conservation $p_a+p_b=p_c+p_d$ holds.

In the present work we use spin amplitudes for $\gamma$-$\gamma$ and $\gamma$-$Z$-boxes
in the small mass approximation $m_e^2/s\to 0, m_f^2/s\to 0$,
see fig.~\ref{fig:IvirF},
following refs.~\cite{was:1987,brown:1984},
%//////////////////////////////////////////////////////////////
%      Boxes
%//////////////////////////////////////////////////////////////
\begin{equation}
  \label{boxy}
  \begin{split}
    \Meu^{(1)}_{\rm Box} \left(\st^{p}_{\lambda} ; X \right)
  &= 2 ie^2 \sum_{B=\gamma,Z}
    \frac{ 
                 g^{B,e}_{ \lambda_a} g^{B,f}_{-\lambda_a}\;
                 T_{ \lambda_c \lambda_a} T'_{\lambda_b \lambda_d}
                +g^{B,e}_{ \lambda_a} g^{B,f}_{ \lambda_a}\;
                 U'_{ \lambda_c \lambda_b} U_{\lambda_a \lambda_d}
         }
         { X^2 - {M_{B}}^2 +i\Gamma_{B} X^2 /M_{B} }\;
         \delta_{\lambda_a, -\lambda_b}
         \delta_{\lambda_c, -\lambda_d}\\
& \frac{\alpha}{\pi} Q_eQ_f
  \left[
         \delta_{\lambda_a,  \lambda_c}\;
         f_{\rm BDP}(\bar{M}^2_B,m_\gamma,s,t,u)
        -\delta_{\lambda_a, -\lambda_c}\;
         f_{\rm BDP}(\bar{M}^2_B,m_\gamma,s,u,t)
  \right],
  \end{split}
\end{equation}
where
%//////////////////////////////////////////////////////////////
\begin{equation}
  \begin{split}
&f_{\rm BDP}(\bar{M}^2_B,m_\gamma,s,u,t)
 =         \ln\left( {t\over u} \right) \ln\left( {m_\gamma^2\over (tu)^{1/2}} \right)
          -2\ln\left( {t\over u} \right) \ln\left( {\bar{M}^2_Z-s \over \bar{M}^2_Z } \right)
\\ &
          +{\rm Li}_2\left( { \bar{M}^2_Z+u \over \bar{M}^2_Z} \right)
          -{\rm Li}_2\left( { \bar{M}^2_Z+t \over \bar{M}^2_Z} \right)
\\ &
          +{(\bar{M}^2_Z-s)(u-t-\bar{M}^2_Z)\over u^2} \left\{
                   \ln\left( {-t\over s} \right) \ln\left( { \bar{M}^2_Z-s \over \bar{M}^2_Z} \right)
                  +{\rm Li}_2\left( {\bar{M}^2_Z+t \over \bar{M}^2_Z} \right)  
                  -{\rm Li}_2\left( {\bar{M}^2_Z-s \over \bar{M}^2_Z} \right)  \right\}
\\ &
          +{(\bar{M}^2_Z-s)(\bar{M}^2_Z-s)\over u s}\; 
                   \ln\left( { \bar{M}^2_Z-s \over \bar{M}^2_Z} \right) 
          +{\bar{M}^2_Z-s \over u}  \ln\left( {-t\over \bar{M}^2_Z}\right),
  \end{split}
\end{equation}
$\bar{M}^2_Z = M_Z^2-iM_Z\Gamma_Z$, $\bar{M}^2_\gamma=m_\gamma^2$,
and the function $f_{\rm BDP}$ is that of eq.~(11) of ref.~\cite{brown:1984}.
The standard Mandelstam variables $s,t$ and $u$ are defined as usual:
$s=(p_a+p_b)^2,\; t=(p_a-p_c)^2,\; t=(p_a-p_d)^2$.
Since in the rest of our calculation we do not use $m_f^2/s\to 0$,
we therefore intend to replace the above box spin amplitudes with
the finite-mass results%
\footnote{ For the $\gamma$-$\gamma$ box we use the spin amplitudes with the exact final fermion mass.
  It seems, however, that the $\gamma$-$Z$ box for 
  the heavy fermion is missing in the literature.}
that were given in ref.~\cite{koralb:1985}.)

Now using eq.~(\ref{eq:VirtSubt}) we determine
%/////////////////////////////////////////////////////////////////////////////////////////
\begin{equation}
  \label{beta01}
  \hbeta^{(1)}_0 \left(  \st^{p}_{\lambda} ; X \right) 
  =\Bmf \left(  \st^{p}_{\lambda} ; X \right)
  \left( 1+ \delta^{(1)e}_{Virt}(s) \right)
  \left( 1+ \delta^{(1)f}_{Virt}(s) \right)
  +\Reu_{\rm Box}^{(1)} \left(\st^{p}_{\lambda}; X \right)
\end{equation}
where 
%//////////////////////////////////////////////////////////////
\begin{equation}
  \begin{split}
   &\delta^{(1)e}_{Virt}(s)=Q_e^2 F_1(s,m_e,m_\gamma)-Q_e^2 \alpha B_2(p_a,p_b,m_\gamma)
   = Q_e^2 {\alpha\over \pi} {1\over 2} \bar{L}_e,
\\
   &\delta^{(1)f}_{Virt}(s)=Q_f^2 F_1(s,m_f,m_\gamma)-Q_f^2 \alpha B_2(p_c,p_d,m_\gamma)
   = Q_f^2 {\alpha\over \pi} {1\over 2} \bar{L}_f,
\\
   &\bar{L}_e=\ln\bigg({s\over m^2_e}\bigg)+i\pi-1,\;\;
    \bar{L}_f=\ln\bigg({s\over m^2_f}\bigg)+i\pi-1.
  \end{split}
\end{equation}
Note that we departed in eq.~(\ref{beta01}) from the strict \Order{\alpha^1} by retaining
the $\delta^{(1)e}_{Virt}(s)\delta^{(1)f}_{Virt}(s)$ term, i.e.,
replacing the ``additive'' form
  $ 1+ \delta^{(1)e}_{Virt}(s) +\delta^{(1)f}_{Virt}(s)$
with the  ``factorized'' form
  $( 1+ \delta^{(1)e}_{Virt}(s) )( 1+ \delta^{(1)f}_{Virt}(s) )$.
The above does not need really much justification --
it is obviously closer to the reality of the higher-orders, so 
the ``factorized'' form is preferable.
The only question is whether the above method does not disturb IR-cancellations.
It does not, as it is seen from the definitions of 
$\delta^{(1)e}_{Virt}(s)$ and $\delta^{(1)f}_{Virt}(s)$.

%====================================================================================
%------------------------------------------------------------------------------------
\begin{figure}[ht]
\centering
\setlength{\unitlength}{1mm}
\begin{picture}(160,80)
%%\put(0,0){\framebox( 160,80){ }}
\put( 0, 0){\makebox(0,0)[lb]{\epsfig{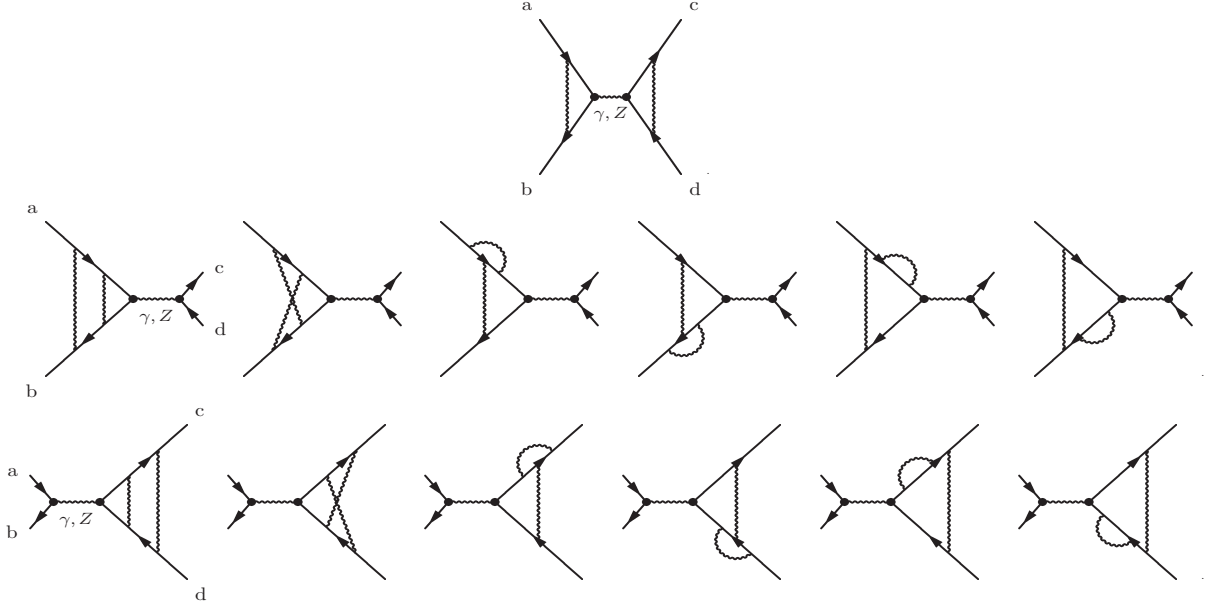}}}
\end{picture}
\caption{\sf Second order vertex diagrams.}
\label{fig:2v}
\end{figure}
%====================================================================================
The IR-subtraction  in $\Meu^{(1)}_{\rm Box}$ 
using eq.~(\ref{eq:VirtSubt}) at \Order{\alpha^1} leads to the IR-finite $\Reu_{\rm Box}$.
The above subtraction is equivalent to the following substitution 
%//////////////////////////////////////////////////////////////
%      IR finite of Boxes
%//////////////////////////////////////////////////////////////
\begin{equation}
 f_{\rm BDP}(\bar{M}^2_B,m_\gamma,s,t,u) 
   \to
 f_{\rm BDP}(\bar{M}^2_B,m_\gamma,s,t,u)  - f_{\rm IR}(m_\gamma,t,u),
\end{equation}
where
%//////////////////////////////////////////////////////////////
\begin{equation}
    f_{\rm IR}(m_\gamma,t,u)
          = \frac{2}{\pi} B_2(p_a,p_c,m_\gamma) -\frac{2}{\pi} B_2(p_a,p_d,m_\gamma)
          = \ln\left( \frac{t}{u} \right) 
            \ln\left( \frac{m_\gamma^2}{\sqrt{tu}} \right)
           +\frac{1}{2} \ln\left( \frac{t}{u} \right),
\end{equation}
and the additional resonance factor $\exp\big(-\alpha B_4^Z(s)\big)$ in eq.~(\ref{eq:VirtSubt})
induces the additional subtraction in the $\gamma$-Z box part:
%//////////////////////////////////////////////////////////////
\begin{equation}
  f_{\rm BDP}(s,t,u) \to f_{\rm BDP}(s,t,u) -\alpha B_4^Z(s),
\end{equation}
see eq.~(\ref{eq:FormGreco}) for the definition of $\alpha B_4^Z$.

Our \Order{\alpha^2} expressions for $\hbeta^{(2)}_0$ are still incomplete.
We base them on the graphs depicted in fig.~\ref{fig:2v}.
(In fig.~\ref{fig:2v} we omitted some trivial transpositions of the diagrams.)
Following again the eq.~(\ref{eq:VirtSubt}), we obtain
%/////////////////////////////////////////////////////////////////////////////////////////
\begin{equation}
    \hbeta^{(2)}_0 \left(  \st^{p}_{\lambda} ; X \right) 
    = \Bmf \left(  \st^{p}_{\lambda}; X \right)
      \left(1+ \delta^{(2)e}_{Virt}(s,m_e)  \right)
      \left(1+ \delta^{(2)f}_{Virt}(s,m_f)  \right)
      +\Reu_{\rm Box}^{(2)} \left(\st^{p}_{\lambda}; X \right)
\end{equation}

%====================================================================================
%------------------------------------------------------------------------------------
\begin{figure}[ht]
\centering
\setlength{\unitlength}{1mm}
\begin{picture}(120,80)
%%\put(0,0){\framebox( 120,80){ }}
\put( 0, 0){\makebox(0,0)[lb]{\epsfig{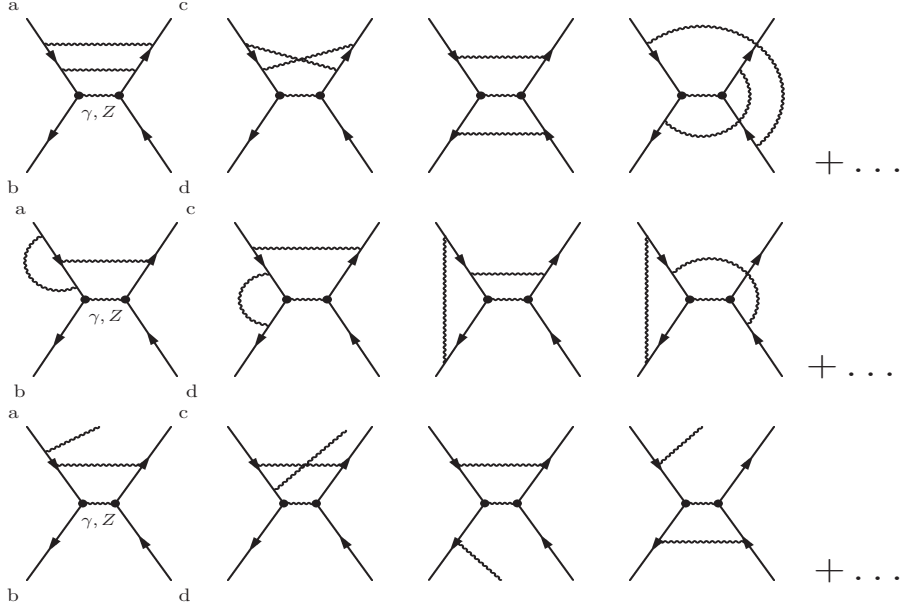}}}
\end{picture}
\caption{\sf Missing second order diagrams.}
\label{fig:missing}
\end{figure}
%====================================================================================
In the present calculation we neglect 
the two-loop double-box contributions in $\Reu_{\rm Box}^{(2)}$,
depicted in the first row in fig.~\ref{fig:missing}
and vertex-box type of diagrams, see examples of diagrams
in the second row  in fig.~\ref{fig:missing}
\footnote{ In fact the two-loop double-box contributions 
  became known recently~\cite{Smirnov:1999wz},
  so there is a chance to include it in the future.}.
In fact we keep only the first order box contribution $\Reu_{\rm Box}^{(1)}$
in our incomplete \Order{\alpha^2} type matrix element.

Two remarks: in spite of the  temporary lack of the above contribution
we are not stuck because what we neglect is IR-finite!
This statement is not so trivial as it may look because
in the calculation without exponentiation neglecting such contributions
would violate IR cancellations, 
and correcting for such a violation would be rather complicated and physically dangerous.
Secondly, what we neglect is expected to be numerically small, of \Order{\alpha^2L^1}
and therefore it does not make much harm to our overall physical precision.

Coming back to the \Order{\alpha^2} corrections to the electric form factor
from the diagrams in fig.~\ref{fig:2v},
they are well known since they were calculated in 
refs.~\cite{Barbieri:1972,burgers:1985,BBVN:1986,berends-neerver-burgers:1988}
and they contribute as follows
%//////////////////////////////////////////////////////////////
\begin{equation}
  \begin{split}
    \delta^{(2)e}_{Virt}(s,m_e) = \delta^{(1)e}_{Virt}(s)+
    {\left(\alpha\over \pi\right)}^2 \left(
          {\bar{L_e}^2 \over 8}
          +\bar{L_e} \left( {3\over 32} -{3\over 4}\zeta_2 +{3\over 2}\zeta_3 \right) \right),
\\
    \delta^{(2)f}_{Virt}(s,m_f) = \delta^{(1)f}_{Virt}(s)+
    {\left(\alpha\over \pi\right)}^2 \left(
          {\bar{L_f}^2 \over 8}
          +\bar{L_f} \left( {3\over 32} -{3\over 4}\zeta_2 +{3\over 2}\zeta_3 \right) \right),
  \end{split}
\end{equation}
In the above we kept terms of \Order{\alpha^2L^2} and \Order{\alpha^2L^1},
and neglected the known~\cite{BBVN:1986,Z-physics-at-lep-1:89}
negligible terms of \Order{\alpha^2L^0}.

%%%%%%%%%%%%%%%%%%%%%%%%%%%%%%%%%%%%%%%%%%%%%%%%%%%%%%
\subsubsection{Electroweak corrections}
%%%%%%%%%%%%%%%%%%%%%%%%%%%%%%%%%%%%%%%%%%%%%%%%%%%%%%

In the not so interesting case of the absence of the electroweak (EW) corrections
the couplings of two neutral bosons $\gamma$ and $Z$ are defined in a conventional way:
%//////////////////////////////////////////////////////////////
\begin{equation}
  \begin{split}
    &G^{Z,f}_{\lambda}     = g^{Z,f}_{V} -\lambda g^{Z,f}_{A} (?),\;\;\;
     G^{\gamma,f}_{\lambda}= g^{Z,f}_{V},\; \lambda=+,-=R,L,
\\
    &g^{\gamma,e}_{V} =Q_e=1,\; g^{\gamma}_{V,f} =Q_f,\; 
     g^{\gamma,e}_{A} =0,\; g^{\gamma}_{A,f} =0,\;
\\
    &g^{Z,e}_{V}  = {2 T^3_e -4 Q_e \sin^2\theta_W \over 16 \sin^2\theta_W \cos^2\theta_W},\;
     g^{Z,f}_{V}  = {2 T^3_f -4 Q_f \sin^2\theta_W \over 16 \sin^2\theta_W \cos^2\theta_W},\;
\\
    &g^{Z,e}_{A}  = {2 T^3_e                        \over 16 \sin^2\theta_W \cos^2\theta_W},\;
     g^{Z,f}_{A}  = {2 T^3_f                        \over 16 \sin^2\theta_W \cos^2\theta_W},\;
  \end{split}
\end{equation}
where $T^3_f$ is the isospin of the left-handed component of the fermion
($T^3_d=-1/2,\; T^3_e=-1/2$).

The actual implementation of EW corrections is practically the same as in KORALZ \cite{koralz4:1994}.
It goes as follows:
The $\gamma$ and $Z$-propagators are multiplied
by the corresponding hook-functions (scalar form-factors) due to vacuum polarizations 
%//////////////////////////////////////////////////////////////
\begin{equation}
         H_{\gamma} \to H_{\gamma} \times {1   \over 2-\Pi_\gamma},\;\;
         H_{Z} \to H_{Z} \times
                16\sin^2\theta_W \cos^2\theta_W \;
                {G_{\mu} M_Z^2 \over \alpha_{_{\rm QED}}  8\pi \sqrt{2}} \;
                \rho_{\rm EW}.
\end{equation}
In addition the vector couplings of the $Z$ get multiplied by extra form factors.
First of all we replace
%//////////////////////////////////////////////////////////////
\begin{equation}
  \begin{split}
    &g^{Z,e}_{V}  = {2 T^3_e -4 Q_e \sin^2\theta_W  \over 16\sin^2\theta_W \cos^2\theta_W}
     => {2 T^3_e -4 Q_e \sin^2\theta_W F^{e}_{EW}(s) \over 16\sin^2\theta_W \cos^2\theta_W}
\\
    &g^{Z,f}_{V}  = {2 T^3_f -4 Q_f \sin^2\theta_W  \over 16\sin^2\theta_W \cos^2\theta_W}
     => {2 T^3_f -4 Q_f \sin^2\theta_W F^{f}_{EW}(s) \over 16\sin^2\theta_W \cos^2\theta_W}
  \end{split}
\end{equation}
where  $F^{e}_{EW}(s)$ and $F^{f}_{EW}(s)$ are electroweak form factors provided
by the DIZET package~\cite{dizet:1989}, which is part the ZFITTER semianalytical code
\cite{zfitter6:1999} and correspond to electroweak vertex corrections.

The electroweak box diagrams require more complicated treatment.
In the Born spin amplitudes we have essentially two products of the coupling constants
%//////////////////////////////////////////////////////////////
\begin{equation}
  \begin{split}
    &g^{Z,e}_{\lambda} g^{Z,f}_{-\lambda}
      =(g^{Z,e}_{V} -\lambda g^{Z,e}_{A})(g^{Z,f}_{V} +\lambda g^{Z,f}_{A})
      =         g^{Z,e}_{V} g^{Z,f}_{V}   -\lambda g^{Z,e}_{A} g^{Z,f}_{V} 
       +\lambda g^{Z,e}_{V} g^{Z,f}_{A}           -g^{Z,e}_{A} g^{Z,f}_{A},
\\
    &g^{Z,e}_{\lambda} g^{Z,f}_{\lambda}
      =(g^{Z,e}_{V} -\lambda g^{Z,e}_{A})(g^{Z,f}_{V} -\lambda g^{Z,f}_{A})
      =         g^{Z,e}_{V} g^{Z,f}_{V}   -\lambda g^{Z,e}_{A} g^{Z,f}_{V} 
       -\lambda g^{Z,e}_{V} g^{Z,f}_{A}           +g^{Z,e}_{A} g^{Z,f}_{A}.
  \end{split}
\end{equation}
In the above the following modification is done for the doubly-vector component
%//////////////////////////////////////////////////////////////
\begin{equation}
g^{Z,e}_{V} g^{Z,f}_{V}
=>  {  4 T^3_e T^3_f  
      -8 T^3_e Q_f  F^{f}_{EW}(s)   
      -8 T^3_f Q_e  F^{e}_{EW}(s)  
      +16  Q_f Q_f  F^{ef}_{EW}(s,t)
      \over (16\sin^2\theta_W \cos^2\theta_W)^2 },
\end{equation}
where the new form factor $F^{ef}_{EW}(s,t)$ corresponds to electroweak
boxes and is angle dependent.
The Born spin amplitudes modified in the above way are used also
in the case of the presence of the single and multiple real photons,
see next sections.

%%%%%%%%%%%%%%%%%%%%%%%%%%%%%%%%%%%%%%%%%%%%%%%%%%%%%%%%%%%%%%%%%%%%%%%%%%%%%%%%%%%%%%%%%%%%%
%%%%%%%%%%%%%%%%%%%%%%%%%%%%%%%%%%%%%%%%%%%%%%%%%%%%%%%%%%%%%%%%%%%%%%%%%%%%%%%%%%%%%%%%%%%%%
\subsection{One real photon}
%%%%%%%%%%%%%%%%%%%%%%%%%%%%%%%%%%%%%%%%%%%%%%%%%%%%%%%%%%%%%%%%%%%%%%%%%%%%%%%%%%%%%%%%%%%%%
The discussion of the $\hbeta_1$ tensors corresponding to emission of a single real photon
we start with the tree level case (zero virtual photons).
The starting point is the well known \Order{\alpha^1} split amplitude for the single
bremsstrahlung which we shall reconsider
separately first in the case of the emission from the initial state beams (ISR)
and later for emission from the final state fermions (FSR).
This will be the ``raw material'' for obtaining $\hbeta^{(0)}_1$
using eqs.~(\ref{eq:DefBet}).

%====================================================================================
%------------------------------------------------------------------------------------
\begin{figure}[h]
\centering
\setlength{\unitlength}{1mm}
\begin{picture}(100,50)
%\put(0,0){\framebox( 100,25){ }}
\put( -5, 0){\makebox(0,0)[lb]{\epsfig{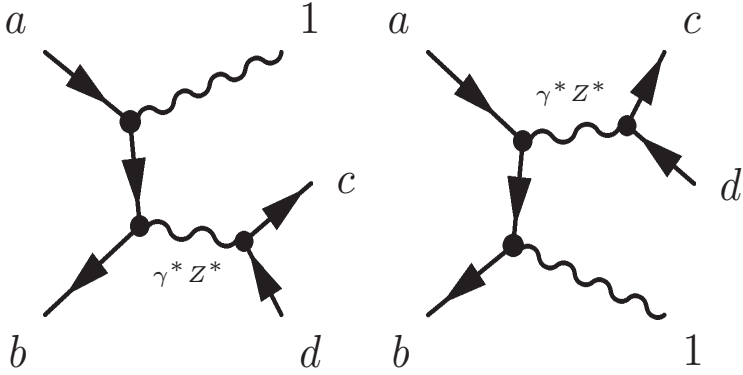}}}
\end{picture}
\caption{\sf ISR diagrams.}
\label{fig:bremI}
\end{figure}
%====================================================================================
The first-order, 1-photon, ISR matrix element from the Feynman diagrams
depicted in fig.~\ref{fig:bremI} reads
%//////////////////////////////////////////////////
%       pure ISR 1-photon from Feynman diags
%//////////////////////////////////////////////////
\begin{equation}
  \label{isr-feynman}
  \begin{split}
    \Meu_{1\{I\}}\left( \st^{p}_{\lambda} \st^{k_1}_{\sigma_1} \right)=
   & eQ_e \;
    \bar{v}(p_b,\lambda_b)\; \mathbf{M}_1\;
         {\not\!{p_a}+m-{\not\!k_1} \over -2k_1p_a} \not\!{\epsilon}^\star_{\sigma_1}(k_1)\;
    u(p_a,\lambda_a)\\
   +&eQ_e \;
    \bar{v}(p_b,\lambda_b)
         \not\!{\epsilon}^\star_{\sigma_1}(k_1)\; {-{\not\!p_b}+m+\not\!{k_1} \over -2k_1p_b} \;
         \mathbf{M}_{\{I\}}\;
    u(p_a,\lambda_a),
  \end{split}
\end{equation}
where 
\begin{equation}
 \mathbf{M}_{\{I\}} =ie^2 \sum_{B = \gamma,Z} \Pi^{\mu\nu}_B(X)\; 
                              G^{B}_{e,\mu}\; (G^{B}_{f,\nu})_{[cd]},
\end{equation}
is the annihilation scattering spinor matrix, including final-state spinors.
The above expression we split into soft IR parts%
\footnote{ This kind of separation was already  exploited
in ref.~\protect\cite{erw:1994}.
  We thank E. Richter-W\c{a}s for attracting our attention to this method.
  }
proportional to $(\not\!{p} \pm m)$
and non-IR parts proportional to ${\not\!k_1}$.
Employing the completeness relations of eq.~(\ref{transition-defs}) in the Appendix A
to those parts we obtain:
%//////////////////////////////////////////////////
%               ISR
%//////////////////////////////////////////////////
\begin{equation}
  \begin{split}
    \Meu_{1\{I\}}
    \left( \st^{p}_{\lambda} \st^{k_1}_{\sigma_1}
    \right)=
   &-{eQ_e\over 2k_1p_a}\; \sum_\rho
     \Bmf\left[ \st^{p_b}_{\lambda_b}  \st^{p_a}_{\rho_a}\right]\st_{[cd]}
          U\left[ \st^{p_a}_{\rho_a}  \st^{k_1}_{\sigma_1}  \st^{p_a}_{\lambda_a} \right]
    +{eQ_e\over 2k_1p_b}\; \sum_\rho
           V\left[ \st^{p_b}_{\lambda_b} \st^{k_1}_{\sigma_1} \st^{p_b}_{\rho_b}  \right]
      \Bmf\left[ \st^{p_b}_{\rho_b}  \st^{p_a}_{\lambda_a} \right]\st_{[cd]}\\
   +&{eQ_e\over 2k_1p_a}\; \sum_\rho
     \Bmf\left[\st^{p_b}_{\lambda_b}  \st^{k_1}_{\rho}  \right]\st_{[cd]}
          U\left[\st^{k_1}_{\rho}        \st^{k_1}_{\sigma_1}  \st^{p_a}_{\lambda_a} \right]
    -{eQ_e\over 2k_1p_b}\; \sum_\rho
           V\left[ \st^{p_b}_{\lambda_b} \st^{k_1}_{\sigma_1}  \st^{k_1}_{\rho}      \right]
      \Bmf\left[ \st^{k_1}_{\rho}  \st^{p_a}_{\lambda_a}\right]\st_{[cd]}.
  \end{split}
\end{equation}
The summation in the first two terms gets eliminated due to the diagonality
property of $U$ and $V$, see eq.~(\ref{diagonality}) in the Appendix A, and leads to
%//////////////////////////////////////////////////
%            again ISR
%//////////////////////////////////////////////////
\begin{equation}
  \label{first-order-isr}
  \begin{split}
   &\Meu^{1\{I\}}
      \left( \st^{p}_{\lambda} \st^{k_1}_{\sigma_1} \right)
   =\sfac^{\{I\}}_{\sigma_1}(k_1) 
    \Bmf\left[\st^{p}_{\lambda} \right]
   +r_{\{I\}} \left(\st^{p}_{\lambda} \st^{k_1}_{\sigma_1} \right),\\
%----------
  & r_{\{I\}} \left( \st^{p}_{\lambda} \st^{k_1}_{\sigma_1} \right)=
    +{eQ_e\over 2k_1p_a}\; \sum_\rho
     \Bmf\left[ \st^{p_b}_{\lambda_b} \st^{k_1}_{\rho}   \right]\st_{[cd]}
        U\left[   \st^{k_1}_{\rho} \st^{k_1}_{\sigma_1}  \st^{p_a}_{\lambda_a}   \right]
    -{eQ_e\over 2k_1p_b}\; \sum_\rho
        V\left[   \st^{p_b}_{\lambda_b} \st^{k_1}_{\sigma_1} \st^{k_1}_{\rho}   \right]
     \Bmf\left[ \st^{k_1}_{\rho}    \st^{p_a}_{\lambda_a} \right]\st_{[cd]},\\
%----------
  & \sfac^{\{I\}}_{\sigma_1}(k_1) =
     -eQ_e{b_{\sigma_1}(k_1,p_a) \over 2k_1p_a} +eQ_e{b_{\sigma_1}(k_1,p_b) \over 2k_1p_b},\quad
    \end{split}
\end{equation}
The soft part is now clearly separated and the remaining non-IR part,
necessary for the CEEX, is obtained.

%====================================================================================
%------------------------------------------------------------------------------------
\begin{figure}[h]
\centering
\setlength{\unitlength}{1mm}
\begin{picture}(100,50)
%\put(0,0){\framebox( 50,25){ }}
\put( -5, 0){\makebox(0,0)[lb]{\epsfig{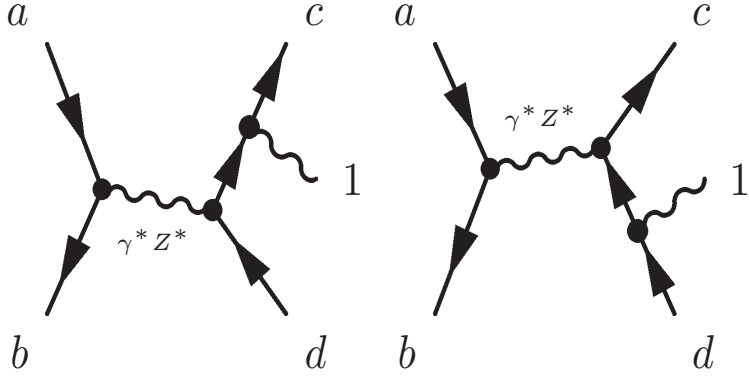}}}
\end{picture}
\caption{\sf FSR diagrams.}
\label{fig:bremF}
\end{figure}
%====================================================================================
The case of final-state one real photon emission (FSR), see fig.~\ref{fig:bremF},
can be analysed in a similar way.
The first-order FSR, 1-photon, matrix element is
%%//////////////////////////////////////////////////
%       pure FSR 1-photon from Feynman diags
%//////////////////////////////////////////////////
\begin{equation}
  \label{fsr-feynman}
  \begin{split}
    \Meu_{1\{F\}}\left( \st^{p}_{\lambda} \st^{k_1}_{\sigma_1} \right)
   &=eQ_f\;
    \bar{u}(p_c,\lambda_c)
         \not\!{\epsilon}^\star_{\sigma_1}(k_1)\; {\not\!{p_c}+m+{\not\!k_1} \over 2k_1p_c}\;
         \mathbf{M}_0\;
    v(p_d,\lambda_d)\\
   &+eQ_f\;
    \bar{u}(p_c,\lambda_c)\;
         \mathbf{M}_{\{F\}}\;
         {-{\not\!p_d}+m-\not\!{k_1} \over 2k_1p_d}\; \not\!{\epsilon}^\star_{\sigma_1}(k_1)\;
    v(p_d,\lambda_d),\\
  \end{split}
\end{equation}
where 
\begin{equation}
 \mathbf{M}_{\{F\}} =ie^2 \sum_{B = \gamma,Z} \Pi^{\mu\nu}_B(X)\; 
                            (G^{B}_{e,\mu})_{[ba]}\;  G^{B}_{f,\nu},
\end{equation}
is spinor matrix for annihilation scattering,including initial spinors.
Similarly, the expansion into soft and non-IR parts for the FSR 
spin amplitudes is done in the way completely analogous to the ISR case
%//////////////////////////////////////////////////
%               FSR
%//////////////////////////////////////////////////
\begin{equation}
  \begin{split}
   &\Meu_{1\{F\}}
    \left(\st^{p}_{\lambda} \st^{k_1}_{\sigma_1} 
    \right)
   =\sfac^{\{F\}}_{\sigma_1}(k_1) 
    \Bmf\left( \st^{p}_{\lambda} \right)
   +r_{\{F\}} \left( \st^{p}_{\lambda}  \st^{k_1}_{\sigma_1} \right),\\
%-----------
  & r_{\{F\}} \left( \st^{p}_{\lambda} \st^{k_1}_{\sigma_1} \right) =
   {eQ_f\over 2k_1p_c}\; \sum_\rho
     U\left[    \st^{p_c}_{\lambda_c}  \st^{k_1}_{\sigma_1}   \st^{k_1}_{\rho} \right]
     \Bmf\st_{[ba]}\left[ \st^{k_1}_{\rho} \st^{p_d}_{\lambda_d}     \right]
    -{eQ_f\over 2k_1p_d}\; \sum_\rho 
      \Bmf\st_{[ba]}\left[ \st^{p_c}_{\lambda_c}\st^{k_1}_{\rho} \right]
      V\left[   \st^{k_1}_{\rho}  \st^{k_1}_{\sigma_1}  \st^{p_d}_{\lambda_d} \right],\\
%------
  & \sfac^{\{F\}}_{\sigma_1}(k_1) =
            eQ_f{b_{\sigma_1}(k_1,p_c) \over 2k_1p_c} -eQ_f{b_{\sigma_1}(k_1,p_d) \over 2k_1p_d}.\\
    \end{split}
\end{equation}

For the purpose of the 
following discussion of the remaining non-IR terms it is useful to introduce
an even more compact tensor notation:
\begin{equation}
   U\left[\st^{p_f}_{\lambda_f}  \st^{k_i}_{\sigma_i} \st^{k_j}_{\sigma_j} \right]
   \equiv U_{[f,i,j]},\quad
   \Bmf\left[ \st^{p_b}_{\lambda_b}\st^{p_a}_{\lambda_a} \right]
       \left[ \st^{p_c}_{\lambda_c}\st^{p_d}_{\lambda_d} \right]
   \equiv \Bmf_{[ba][cd]},
\end{equation}
etc.
For the ``primed'' indices we understand contractions, for instance
\begin{equation}
  U_{[a,i,j']} V_{[j',j,b]} \equiv \sum_{\sigma'_j=\pm}
  U\left[\st^{p_a}_{\lambda_a}  \st^{k_i}_{\sigma_i} \st^{k_j}_{\sigma'_j} \right]
  V\left[\st^{k_j}_{\sigma'_j}  \st^{k_j}_{\sigma_j} \st^{p_b}_{\lambda_b} \right].
\end{equation}

Using the above notation,
the complete \Order{\alpha^1} spin amplitude for 1-photon ISR+FSR,
coming directly from Feynman diagrams, with the explicit split into 
IR and non-IR parts, ISR and FSR parts, reads
%//////////////////////////////////////////////////
%          ISR+FSR, 1-photon
%//////////////////////////////////////////////////
\begin{equation}
  \label{one-photon}
  \begin{split}
   &\Mmf^{(1)}_1\left(\st^{p}_{\lambda} \st^{k_1}_{\sigma_1}\right)
  = \Mmf^{(1)}_{1\{I\}}\left(\st^{p}_{\lambda} \st^{k_1}_{\sigma_1}\right)(P-k_1) 
   +\Mmf^{(1)}_{1\{F\}}\left(\st^{p}_{\lambda} \st^{k_1}_{\sigma_1}\right)(P)
\\&\qquad
  =   \sfac^{\{I\}}_{[1]}\; \Bmf \left( \st^{p}_{\lambda}; P-k_1 \right)
    + r_{\{I\}} \left(\st^{p}_{\lambda} \st^{k_1}_{\sigma_1}; P-k_1 \right)
    + \sfac^{\{F\}}_{[1]}\; \Bmf \left( \st^{p}_{\lambda}; P     \right)
    + r_{\{F\}} \left(\st^{p}_{\lambda} \st^{k_1}_{\sigma_1}; P   \right),
\\&
   r_{\{I\}} \left( \st^{p}_{\lambda} \st^{k_1}_{\sigma_1} ;X \right)
 ={ eQ_e \over 2k p_a} \Bmf_{[b1'cd]}(X)\; U_{[1'1a]}
 -{ eQ_e \over 2k p_b} V_{[b11']}\; \Bmf_{[1'acd]}(X)
\\&
   r_{\{F\}} \left( \st^{p}_{\lambda} \st^{k_1}_{\sigma_1}; X \right)
 ={eQ_f \over 2k p_c} U_{[c11']}\; \Bmf_{[ba1'd]}(X)
 -{eQ_f \over 2k p_d} \Bmf_{[bac1']}(X)\; V_{[1'1d]}
  \end{split}
\end{equation}
In the lowest-order the Born spin amplitudes $\Bmf$
are defined in eq.~(\ref{born}), and we show explicitly as an argument
the four-momentum $X$ which enters the propagator of the
$s$-channel exchange.
Note that the formulas here differ by an overall sign from those of 
ref.~\cite{ceex1:1999}

%%%%%%%%%%%%%%%%%%%%%%%%%%%%%%%%%%%%%%%%%%%%%%%%%%%%%%
%%%%%%%%%%%%%%%%%%%%%%%%%%%%%%%%%%%%%%%%%%%%%%%%%%%%%%
\subsubsection{First- and second-order $\hbeta_1$}
%%%%%%%%%%%%%%%%%%%%%%%%%%%%%%%%%%%%%%%%%%%%%%%%%%%%%%

Now we employ the tree level, \Order{\alpha^1} variant of eqs.~(\ref{eq:DefBet})
getting the following results:
\begin{equation}
  \label{beta11}
  \begin{split}
    &\hbeta^{(1)}_{1\{I\}}\left( \st^{p}_{\lambda} \st^{k_1}_{\sigma_1}; P-k_1 \right)
         \equiv r_{\{I\}}\left( \st^{p}_{\lambda} \st^{k_1}_{\sigma_1}; P-k_1 \right)\;
\\
    &\hbeta^{(1)}_{1\{F\}}\left( \st^{p}_{\lambda} \st^{k_1}_{\sigma_1}; P \right)
         \equiv r_{\{F\}}\left( \st^{p}_{\lambda} \st^{k_1}_{\sigma_1}; P \right)\;
     +\left( { (p_c+p_d+k_1)^2  \over (p_c+p_d)^2} - 1 \right)
      \Bmf\left(\st^{p}_{\lambda}; X \right),
  \end{split}
\end{equation}
The ``context dependent'' reduced total momentum $X$
(the total four-momentum in the resonance propagator, if present) 
is in the above definition uniquely defined as
$X=P-k_1$ in the case of ISR, and $X=P$ in the case of FSR.
In the general context of the CEEX amplitude of eqs.~(\ref{eq:ceex-master}-\ref{eq:ceex-master2}),
that is in presence of the additional ``spectator'' ISR photons in a given term,
$X$  is also defined quite unambiguously:
$X_\wp$ includes not only $k_1$ but also all additional ISR momenta in the process.
For the pseudo-flux factor there is some ambiguity, however. 
In the presence of the additional ``spectator'' ISR photons it can be defined either as
$(p_a+p_b-k_1)^2/(p_a+p_b)^2$ or $(p_c+p_d+k_1)^2/(p_c+p_d)^2$.
We are free to choose any of them and we opted for the second choice 
(it seems to lead to more stable MC weights).

The one-loop level, \Order{\alpha^2} case of $\hbeta^{(2)}_1$ is quite interesting
because this is the first time that we deal with the nontrivial case 
of the simultaneous emission of virtual and real photons.
It is therefore instructive to write the formal definitions of $\hbeta^{(2)}_1$
following eqs.~(\ref{eq:VirtSubt}) and (\ref{eq:DefBet}) in this particular case:
%/////////////////////////////////////////////////////////////////////////////////////////
\begin{equation}
  \begin{align}
  \label{beta12def1}
    \Mmf^{(2)}_{1\{\omega\}}\left( \st^{p}_{\lambda} \st^{k_1}_{\sigma_1}; X_\omega \right) 
   &=\left\{ e^{-\alpha B_4 -\alpha B_4^R(X_\omega)} 
    \Meu^{(2)}_{1\{\omega\}}\left( \st^{p}_{\lambda} \st^{k_1}_{\sigma_1}; X_\omega \right)
      \right\}\bigg|_{{\cal O}(\alpha^2)},\; \omega=I,F,\; R=\gamma,Z,
\\ \label{beta12def2}
    \hbeta^{(2)}_{1\{I\}}\left( \st^{p}_{\lambda} \st^{k_1}_{\sigma_1}; P-k_1 \right) 
 & = \Mmf^{(2)}_{1\{I\}}\left( \st^{p}_{\lambda} \st^{k_1}_{\sigma_1}; P-k_1 \right) 
         -\sfac^{\{I\}}_{\sigma_1}(k_1) \;\hbeta^{(1)}_0\left(\st^{p}_{\lambda}; P-k_1 \right)\; 
\\ \nonumber
    \hbeta^{(2)}_{1\{F\}}\left( \st^{p}_{\lambda} \st^{k_1}_{\sigma_1}; P \right) 
 & = \Mmf^{(2)}_{1\{F\}}\left( \st^{p}_{\lambda} \st^{k_1}_{\sigma_1}; P \right) 
         -\sfac^{\{F\}}_{\sigma}(k_1) \;\hbeta^{(1)}_0\left(\st^{p}_{\lambda}; P   \right).
  \end{align}
\end{equation}

%====================================================================================
%------------------------------------------------------------------------------------
\begin{figure}[h!]
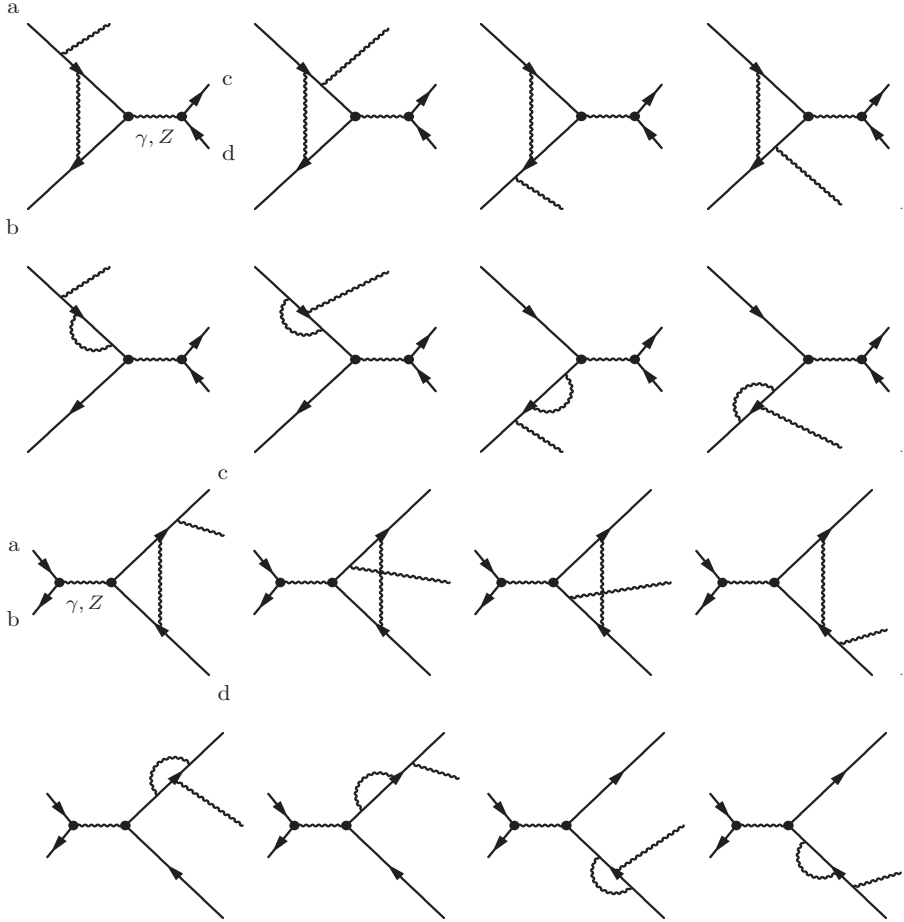

\centering
\setlength{\unitlength}{1mm}
\begin{picture}(120,120)
%%\put(0,0){\framebox( 100,100){ }}
\put( 0,61){\makebox(0,0)[lb]{\epsfig{file=feyn-1r1vI.eps,width=120mm,height=60mm}}}
\put( 0,-1){\makebox(0,0)[lb]{\epsfig{file=feyn-1r1vF.eps,width=120mm,height=60mm}}}
\end{picture}
\caption{\sf One-loop corrections to single bremsstrahlung.}
\label{fig:VirReal}
\end{figure}
%====================================================================================
What is presently available from the Feynman diagrams? 
For the moment we have at our disposal the amplitudes corresponding to vertex-like diagrams in
fig.~(\ref{fig:VirReal}),
and we miss diagrams of the ``5-box'' type shown in the third (bottom) row in fig.~\ref{fig:missing}.
More precisely, after applying the IR virtual subtraction of eq.~(\ref{beta12def1})
we expand in the number of loops, keeping track of the initial/final state attachment 
of the {\em virtual} photon:
%/////////////////////////////////////////////////////////////////////////////////////////
\begin{equation}
  \begin{split}
     &\Mmf^{(2)}_{1\{\omega\}}\left(\st^{p}_{\lambda} \st^{k_1}_{\sigma_1}; X \right)
    = \Mmf^{(1)}_{1\{\omega\}}\left(\st^{p}_{\lambda} \st^{k_1}_{\sigma_1}; X \right)
    +\alpha Q_e^2   \Mmf^{[1]}_{1\{\omega\}, I^2} \left(\st^{p}_{\lambda}\st^{k_1}_{\sigma_1}; X \right)
    +\alpha Q_f^2   \Mmf^{[1]}_{1\{\omega\}, F^2} \left(\st^{p}_{\lambda}\st^{k_1}_{\sigma_1}; X \right)
\\&\qquad\qquad\qquad
    +\alpha Q_e Q_f \Mmf^{[1]}_{1\{\omega\}, Box5}\left(\st^{p}_{\lambda}\st^{k_1}_{\sigma_1}; X \right).
  \end{split}
\end{equation}
In the above expression the first term describes the already discussed tree level single bremsstrahlung,
the next two correspond to vertex-like diagrams in fig.~\ref{fig:VirReal},
and the last one represents the ``5-box'' type diagrams in the third row of fig.~\ref{fig:missing}.
In the present version we temporarily omit from the calculation  the 
contribution to $\hbeta^{(2)}_1$ from the last, ``5-box'' term which looks as follows:
%/////////////////////////////////////////////////////////////////////////////////////////
\begin{equation}
  \hbeta^{(2)}_{1\{\omega\}, Box5}\left(\st^{p}_{\lambda}\st^{k_1}_{\sigma_1}; X \right)
  =\alpha Q_e Q_f \Mmf^{[1]}_{1\{\omega\}, Box5}\left(\st^{p}_{\lambda}\st^{k_1}_{\sigma_1}; X \right)
  -\sfac^{\{I\}}_{[1]}\; \Reu_{\rm Box}^{(1)} \left(\st^{p}_{\lambda}; X \right)
  -\sfac^{\{F\}}_{[1]}\; \Reu_{\rm Box}^{(1)} \left(\st^{p}_{\lambda}; X \right).
\end{equation}
As we see the trivial IR-part, which we remove, is proportional to the ordinary box contributions
already discussed before.
We expect the above to contribute in the integrated cross section to be at most of \Order{\alpha^2 L^1},
and in the resonance scattering it will be suppressed by the additional $\Gamma/M$ factor.

Limiting ourselves to the pure ``vertex-like'' diagrams of fig.~\ref{fig:VirReal},
for one real ISR $(\omega=I)$ photon 
we obtain from the Feynman rules the following \Order{Q_e^2\alpha^2} result
%%%%%%%%%%%%%%%%%%%%%%%%%%%%%%%%%%%%%%
\begin{equation}
  \label{beta12isr}
  \begin{split}
    \hbeta^{(2)}_{1\{I\}}\left( \st^{p}_{\lambda} \st^{k_1}_{\sigma_1}; X \right)
     &\equiv 
      r_{\{I\}}\left( \st^{p}_{\lambda} \st^{k_1}_{\sigma_1}; X \right)\;
          \left(1+\delta^{(1)e}_{Virt}(s) +\rho^{(2)e}_{Virt}(s,\talpha_1,\tbeta_1)\right)\;
          \left(1+\delta^{(1)f}_{Virt}(s)\right)
\\
    &+\Bmf\left(\st^{p}_{\lambda}; X \right)\;
          \sfac^{\{I\}}_{\sigma_1}(k_1)\;
          \rho^{(2)e}_{Virt}(s,\talpha,\tbeta)
  \end{split}
\end{equation}
where 
%%%%%%%%%%%%%%%%%%%%%%%%%%%%%%%%%%%%%%
\begin{equation}
  \begin{split}
   &\rho^{(2)e}_{Virt}(s,\talpha,\tbeta) = 
        {\talpha\over\pi} Q_e^2\; {1\over 2} (V(s,\talpha,\tbeta)+V(s,\tbeta,\talpha)),
\\
   &V(s,\talpha,\tbeta)
    =  \ln(\talpha) \ln(1-\tbeta)
\\ &\qquad\qquad
      +{\rm Li}_2(\talpha)
      -{1 \over 2} \ln^2(1-\talpha)
      +{3 \over 2} \ln(1-\talpha)
      +{1 \over 2} {\talpha (1-\talpha)\over (1+(1-\talpha)^2)}
  \end{split}
\end{equation}
and we use Sudakov variables
%%%%%%%%%%%%%%%%%%%%%%%%%%%%%%%%%%%%%%
\begin{equation}
    \talpha_i = {2k_i p_b \over 2p_a p_b},\;\;
    \tbeta_i  = {2k_i p_a \over 2p_a p_b}.
\end{equation}
Let us make a number of observations concerning eq.~(\ref{beta12isr}):
\begin{itemize}
\item
  The terms of \Order{\alpha^4} like $|\sfac^{\{I\}}_{\sigma}\rho^{(2)e}_{Virt}|^2$
  in the cross section, although beyond
  \Order{\alpha^2}, are not rejected, as it would be the case in the ordinary  \Order{\alpha^2}
  calculation without exponentiation.
  They are included in the process of numerical evaluation of the differential cross sections
  out of spin amplitudes.
  (It is essential that they are IR-finite.)
\item
  The term  $r_{\{I\}} \delta^{(1)e}_{Virt}$ contributes \Order{\alpha^2L^2} to the integrated
  cross section -- one $L^1$ is explicit (from the virtual photon) and another $L^1$ is from the 
  integration over the  angle of the real photon.
\item
  The term $\sim\ln(\talpha) \ln(1-\tbeta)$ contributes a correction of \Order{\alpha^2L^2} 
  to the integrated cross section with the double logarithm $L^2$ resulting directly from the integration
  over the angle of the real photon:
  \begin{displaymath}
    \int {dk^3\over k^0}\; \Re[ \rho^{(2)e}_{Virt}(k) \{\hbeta_0 \sfac^{\{I\}}_{\sigma}(k)\}^*)]
    \sim Q_e^2 \alpha^2  \int_{m_e^2/s}\; {d\talpha\over\talpha} \ln(\talpha)
    \sim Q_e^2 \alpha^2  \ln^2 {s\over m_e^2}
  \end{displaymath}
\item
  The other terms in $\hbeta^{(2)}_{1\{I\}}$ contribute at most \Order{\alpha^2L^1}.
\item
  The FSR virtual corrections are included multiplicatively through
  the factor $(1+\delta^{(1)f}_{Virt}(s))$ and not additively
  like  $(1+\delta^{(1)e}_{Virt}(s)+\delta^{(1)f}_{Virt}(s))$.
  This is our deliberate choice.
\item
  The subleading term $\talpha (1-\talpha)/(1+(1-\talpha)^2)$ has in fact a more complicated
  spin structure than that of the Born amplitude (it should be restored in future).
  The unpolarized integrated cross section is however correct in \Order{\alpha^2L^1}.
\end{itemize}

The analogous \Order{Q_f^2\alpha^2} contribution for one real FSR $(\omega=0)$ photon is
%%%%%%%%%%%%%%%%%%%%%%%%%%%%%%%%%%%%%%
\begin{equation}
  \label{beta12fsr}
  \begin{split}
    \hbeta^{(2)}_{1\{F\}}\left( \st^{p}_{\lambda} \st^{k}_{\sigma}; X \right)
     &\equiv 
      r_{\{F\}}\left( \st^{p}_{\lambda} \st^{k}_{\sigma}; X \right)\;
          \left(1+\delta^{(1)e}_{Virt}(s)\right)\;
          \left(1+\delta^{(1)f}_{Virt}(s) +\rho^{(2)f}_{Virt}(s,\talpha',\tbeta')\right)
\\
    &+\Bmf\left(\st^{p}_{\lambda}; X \right)\;
          \sfac^{\{F\}}_{\sigma}(k)\;
          \rho^{(2)f}_{Virt}(s,\talpha',\tbeta')
\\
    &+\Bmf\left(\st^{p}_{\lambda}; X \right)
          \sfac^{\{F\}}_{\sigma}(k)\;
          \left(1+\delta^{(1)e}_{Virt}(s)\right)\;\left(1+\delta^{(1)f}_{Virt}(s)\right)\;
          \left( 1 -{ (p_c+p_d+k)^2  \over (p_c+p_d)^2} \right)
  \end{split}
\end{equation}
where 
%%%%%%%%%%%%%%%%%%%%%%%%%%%%%%%%%%%%%%
\begin{equation}
  \begin{split}
   &\rho^{(2)f}_{Virt}(s,\talpha',\tbeta') = 
        {\talpha\over\pi} Q_f^2\; {1\over 4} \bar{L}_f (\ln(1-\talpha'')+\ln(1-\tbeta'')),
\\&
    \talpha'= {2k p_d \over 2p_c p_d},\;\;
    \tbeta' = {2k p_c \over 2p_c p_d},\;\;
    \talpha''= { \talpha' \over 1+\talpha'+\tbeta'},\;\;
    \tbeta'' = { \tbeta'  \over 1+\talpha'+\tbeta'}.\;\;
  \end{split}
\end{equation}
In the above FSR amplitudes we keep only the LL part averaged over the photon angles,
similarly as in EEX.
This corresponds to present status of our CEEX amplitudes implemented in \KK\ MC version 4.13,
and we expect this to be improved in the future.

%%%%%%%%%%%%%%%%%%%%%%%%%%%%%%%%%%%%%%%%%%%%%%%%%%%%%%%%%%%%%%%%%%%%%%%%%%%%%%%%%%%%%%%%%%%%%
%%%%%%%%%%%%%%%%%%%%%%%%%%%%%%%%%%%%%%%%%%%%%%%%%%%%%%%%%%%%%%%%%%%%%%%%%%%%%%%%%%%%%%%%%%%%%
\subsection{2-real  photons}
%%%%%%%%%%%%%%%%%%%%%%%%%%%%%%%%%%%%%%%%%%%%%%%%%%%%%%%%%%%%%%%%%%%%%%%%%%%%%%%%%%%%%%%%%%%%%

In the \Order{\alpha^2} contributions from two real photons are completely at
the tree level, without virtual corrections
(in the future \Order{\alpha^3} version we shall include virtual corrections to double
bremsstrahlung in the LL approximation).
The double bremsstrahlung is considered in three separate cases, two ISR photons,
two FSR photons and one ISR plus one FSR photon.
The corresponding spin amplitudes will be given without any approximation, in particular
we do not use the small mass approximation $m_f/\sqrt{s}<<1$.
The main problems to be solved will be
\begin{itemize}
\item[(a)] to write all spin amplitudes in a form easy for numerical evaluation,
  that is in terms of $U$ and $V$ matrices,
\item[(b)] to extract $\hbeta_2$ functions by means of removing IR-singular parts.
\end{itemize}

%%%%%%%%%%%%%%%%%%%%%%%%%%%%%%%%%%%%%%%%%%%%%%%%%%%%
\subsubsection{2-real ISR  photons}
%%%%%%%%%%%%%%%%%%%%%%%%%%%%%%%%%%%%%%%%%%%%%%%%%%%%

%====================================================================================
%------------------------------------------------------------------------------------
\begin{figure}[h]
\centering
\setlength{\unitlength}{1mm}
\begin{picture}(125,50)
%\put(0,0){\framebox( 50,25){ }}
\put( -5, 0){\makebox(0,0)[lb]{\epsfig{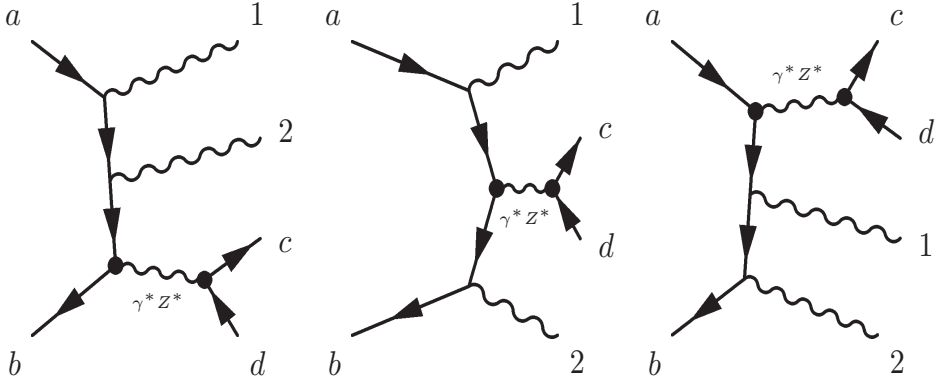}}}
\end{picture}
\caption{\sf Feynman diagrams of the ISR double bremsstrahlung.}
\label{fig:bremII}
\end{figure}
%====================================================================================
The second-order, two-photon, ISR matrix element from the Feynman rules, see Fig.~\ref{fig:bremII},
reads as follows
%//////////////////////////////////////////////////
%       pure ISR 2-photon from Feynman diags
%//////////////////////////////////////////////////
\begin{equation}
  \label{isr2-feynman}
  \begin{split}
    \Meu^{(2)}_{2\{II\}} &\left( \st^{p_a}_{\lambda_a} \st^{p_b}_{\lambda_b} 
                            \st^{k_1}_{\sigma_1}  \st^{k_2}_{\sigma_2}; P-k_1-k_2 \right)
    =ie^2 \sum_{B=\gamma,Z} \Pi^{\mu\nu}_B(P-k_1-k_2)\; (G^{B}_{f,\nu}\;)_{[cd]}\;
     (eQ_e)^2 \; \bar{v}(p_b,\lambda_b)\; \Bigg\{
\\
   &G^{B}_{e,\mu}\;
    {(\not\!{p_a}+m)-{\not\!k_1}-{\not\!k_2} \over-2k_1p_a-2k_2p_a +2k_1k_2}\; 
    \not\!{\epsilon}^\star_{\sigma_1}(k_1)\;
    {(\not\!{p_a}+m)-{\not\!k_2} \over -2k_2 p_a}\; 
    \not\!{\epsilon}^\star_{\sigma_2}(k_2)\;
\\
  +&\not\!{\epsilon}^\star_{\sigma_1}(k_1)\; 
    {(-{\not\!p_b}+m)+\not\!{k_1} \over -2k_1 p_b} \;
    \not\!{\epsilon}^\star_{\sigma_2}(k_2)\; 
    {(-{\not\!p_b}+m)+\not\!{k_1}+\not\!{k_2} \over -2k_1p_b-2k_2p_b+2k_1k_2} \;
    G^{B}_{e,\mu}\;
\\
  +&\not\!{\epsilon}^\star_{\sigma_1}(k_1)\; 
    {(-{\not\!p_b}+m)+\not\!{k_1} \over -2k_1 p_b} \;
    G^{B}_{e,\mu}\;
    {(\not\!{p_a}+m)-{\not\!k_2} \over -2k_2 p_a}\; 
    \not\!{\epsilon}^\star_{\sigma_2}(k_2)\;
    +(1 \leftrightarrow 2)  \Bigg\} u(p_a,\lambda_a).
  \end{split}
\end{equation}
We shall use eq~(\ref{eq:DefBet}) which in this case reads
%%%%%%%%%%%%%%%%%%%%%%%%%%%%%%%%%%%%%%%%%%%%%%%%%%%%
\begin{equation}
  \label{eq:beta2II}
  \begin{split}
      &\hbeta^{(2)}_{2\{II\}}
           \left(\st^{p}_{\lambda}\st^{k_1}_{\sigma_1}\st^{k_2}_{\sigma_2}; P-k_1-k_2 \right)    
    =    \Mmf^{(2)}_{2\{II\}}
           \left(\st^{p}_{\lambda}\st^{k_1}_{\sigma_1}\st^{k_2}_{\sigma_2}; P-k_1-k_2 \right)
       -\hbeta^{(1)}_{1\{I\}}\left(\st^{p}_{\lambda}\st^{k_1}_{\sigma_1}; P-k_1-k_2 \right) 
                                                              \sfac^{\{I\}}_{\sigma_2}(k_2)  
\\&\qquad
       -\hbeta^{(1)}_{1\{I\}}\left(\st^{p}_{\lambda}\st^{k_2}_{\sigma_2}; P-k_1-k_2 \right) 
                                                              \sfac^{\{I\}}_{\sigma_1}(k_1)
       -\hbeta^{(0)}_0\left(\st^{p}_{\lambda}; P-k_1-k_2 \right)  
                       \sfac^{\{I\}}_{\sigma_1}(k_1) \sfac^{\{I\}}_{\sigma_2}(k_2).
  \end{split}
\end{equation}
We shall proceed similarly as in 1-photon case, we shall isolate from the above expression
the group of terms containing two factors $(\not\!{p}+m)$,
then the group containing single factor $(\not\!{p}+m)$ 
and finally the rest.
Such a split represents almost exactly the split in eq.~(\ref{eq:beta-trunc})
into contribution with two $\sfac$-factors (double IR singularity), 
with single $\sfac$-factor (single IR singularity) and the IR-finite remnant 
$\hbeta^{(2)}_2$ which is our primary goal.
In other words, we decompose $\Mmf^{(2)}_{2\{II\}}$ into several terms/parts, as described above,
and we apply the IR-subtraction of eq.~(\ref{eq:beta2II}) term-by-term.

Let us discuss first the doubly IR-singular part proportional to two factors $(\not\!{p}+m)$.
To simplify maximally the discussion let us neglect for the moment $2k_1k_2$
in the propagator. 
Using the completeness relations of eq.~(\ref{transition-defs})
and the diagonality property of eq.~(\ref{diagonality}) in Appendix A, we can factorize soft
factors exactly and completely
%//////////////////////////////////////////////////
%       pure ISR 2-photon from Feynman diags
%//////////////////////////////////////////////////
\begin{equation}
  \label{eq:doubleIR}
  \begin{split}
    (eQ_e)^2 \; 
    \bar{v}(p_b,\lambda_b)\; &\Bigg\{
    G^{B}_{e,\mu}\;
    {(\not\!{p_a}+m)\over 2k_1p_a+2k_2p_a}\; \not\!{\epsilon}^\star_{\sigma_1}(k_1)\;
    {(\not\!{p_a}+m)\over 2k_2 p_a}\;        \not\!{\epsilon}^\star_{\sigma_2}(k_2)\;
\\
  &+\not\!{\epsilon}^\star_{\sigma_1}(k_1)\; {(-{\not\!p_b}+m)\over 2k_1 p_b} \;
    \not\!{\epsilon}^\star_{\sigma_2}(k_2)\; {(-{\not\!p_b}+m)\over 2k_1p_b+2k_2p_b} \;
    G^{B}_{e,\mu}\;
\\
  &+\not\!{\epsilon}^\star_{\sigma_1}(k_1)\; {(-{\not\!p_b}+m)\over 2k_1 p_b} \;
    G^{B}_{e,\mu}\;
   {(\not\!{p_a}+m)\over 2k_2 p_a}\;         \not\!{\epsilon}^\star_{\sigma_2}(k_2)\;
    +(1 \leftrightarrow 2)  \Bigg\} u(p_a,\lambda_a)
\\
  =(G^{B}_{e,\mu})_{[ba]}\;
     (eQ_e)^2 &\Bigg\{
      {b_{\sigma_1}(k_1,p_a)\over 2k_1p_a +2k_2p_a}\;
      {b_{\sigma_2}(k_2,p_a)\over 2k_2 p_a}\;
     +{b_{\sigma_1}(k_1,p_b)\over 2k_1 p_b} \;
      {b_{\sigma_2}(k_2,p_b)\over 2k_1p_b +2k_2p_b} \;\\
    &-{b_{\sigma_1}(k_1,p_b)\over 2k_1 p_b}
      {b_{\sigma_2}(k_2,p_a)\over 2k_2 p_a}\;
     +(1 \leftrightarrow 2) \Bigg\}
\\
   =(G^{B}_{e,\mu})_{[ba]}\; & \sfac^{\{I\}}_{\sigma_1}(k_1) \sfac^{\{I\}}_{\sigma_2}(k_2),
  \end{split}
\end{equation}
where the identity 
%//////////////////////////////////////////////////
\begin{equation}
 {1\over 2k_1p_a +2k_2p_a}\; {1\over 2k_1p_a}
+{1\over 2k_1p_a +2k_2p_a}\; {1\over 2k_2p_a} = {1\over 2k_1p_a}\;{1\over 2k_2p_a}
\end{equation}
was instrumental.

If we restore the terms $2k_1k_2$ in the propagator
the corresponding analog of (\ref{eq:doubleIR}) $\Meu_{2\{II\}}^{\rm DoubleIR}$
leads to
%//////////////////////////////////////////////////
\begin{equation}
  \begin{split}
  &\hbeta^{(2){\rm Double}}_{2\{II\}}\left[\st^{p}_{\lambda}  \st^{k_1}_{\sigma_1} \st^{k_2}_{\sigma_2} \right]= 
       \Meu_{2\{II\}}^{\rm DoubleIR}\left[\st^{p}_{\lambda}  \st^{k_1}_{\sigma_1} \st^{k_2}_{\sigma_2} \right]
   -\sfac^{\{I\}}_{\sigma_1}(k_1)\; \sfac^{\{I\}}_{\sigma_2}(k_2)\;\Bmf\left[\st^{p}_{\lambda} \right]
\\
  &\qquad\qquad=
    \bigg(\sfac^{(a)}_{[1]} \sfac^{(a)}_{[2]} \Delta_a
         +\sfac^{(b)}_{[1]} \sfac^{(b)}_{[2]} \Delta_b \bigg)
    \Bmf\left[\st^{p}_{\lambda} \right],
\\
  &\sfac^{(a)}_{\sigma_i}(k_i)\equiv \sfac^{(a)}_{[i]} = - eQ_e {b_{\sigma_1}(k_i,p_a)\over 2k_i p_a},\;\;
   \sfac^{(b)}_{\sigma_i}(k_i)\equiv \sfac^{(b)}_{[i]} = + eQ_e {b_{\sigma_1}(k_i,p_b)\over 2k_i p_b},\;\;
\\
  &\sfac^{\{I\}}_{\sigma_i}(k_i)\equiv \sfac^{(a)}_{[i]}+\sfac^{(b)}_{[i]} 
                              \equiv \sfac^{(a)}_{\sigma_i}(k_i) +\sfac^{(b)}_{\sigma_i}(k_i),
\\
  &\Delta_f={2k_1p_f +2k_2p_f \over 2k_1p_f +2k_2p_f \mp 2k_1k_2} -1
               = {\pm 2k_1k_2\over 2k_1p_f +2k_2p_f \mp 2k_1k_2},\; f=a,b,c,d
  \end{split}
\end{equation}
and the upper sign should be taken for $f=a,b$.
Obviously $\hbeta^{(2){\rm Double}}$ is IR-finite because of the $\Delta_f$ factor.
In the above we have introduced a more compact notation for $\sfac$-factors.
In addition from now on we shall use the following shorthand notation
%//////////////////////////////////////////////////
\begin{equation}
  r_{if}=2k_i\cdot p_f,\;\;  r_{ij}=2k_i\cdot k_j,\;\; f=a,b,c,d,\;\;i,j,=1,2,...n.
\end{equation}

The next class of terms which we are going to consider carefully is the one in which we sum terms 
with a single $(\not\!{p}+m)$, more precisely, 
let us include terms, which may lead to a single IR singularity (if $k_1 << k_2$ 
or $k_2 << k_1$),
that is with $(\not\!{p}+m)$ next to a spinor, at the end of the fermion line:
%//////////////////////////////////////////////////
%       single (p+m)
%//////////////////////////////////////////////////
\begin{equation}
  \begin{split}
   \Meu_{2\{II\}}^{\rm SingleIR} \left[\st^{p}_{\lambda}  \st^{k_1}_{\sigma_1} \st^{k_2}_{\sigma_2} \right]
  =ie^2 \sum_{B=\gamma,Z} &\Pi^{\mu\nu}_B(X)\; (G^{B}_{f,\nu}\;)_{[cd]}\;
\\
   \times(eQ_e)^2 \; \bar{v}(p_b,\lambda_b)\;
  &\Bigg\{
   G^{B}_{e,\mu}\;
   {-{\not\!k_1}-{\not\!k_2} \over -r_{1a}-r_{2a} +r_{12}}\; 
   \not\!{\epsilon}^\star_{\sigma_1}(k_1)\;
   {(\not\!{p_a}+m)\over -r_{2a}}\; 
   \not\!{\epsilon}^\star_{\sigma_2}(k_2)\;
\\
  +&\not\!{\epsilon}^\star_{\sigma_1}(k_1)\; 
   {(-{\not\!p_b}+m)\over -r_{1b}} \;
    \not\!{\epsilon}^\star_{\sigma_2}(k_2)\; 
   {\not\!{k_1}+\not\!{k_2} \over -r_{1b}-r_{2b}+r_{12}} \;
    G^{B}_{e,\mu}\;
\\
  +&\not\!{\epsilon}^\star_{\sigma_1}(k_1)\; 
   {(-{\not\!p_b}+m)\over -r_{1b}} \; G^{B}_{e,\mu}\;
   {-{\not\!k_2} \over -r_{2a}}\; \not\!{\epsilon}^\star_{\sigma_2}(k_2)\;\\
  +&\not\!{\epsilon}^\star_{\sigma_1}(k_1)\; 
   {\not\!{k_1} \over -r_{1b}}\; 
   G^{B}_{e,\mu}\;
   {(\not\!{p_a}+m)\over -r_{2a}}\; \not\!{\epsilon}^\star_{\sigma_2}(k_2)\;
    +(1 \leftrightarrow 2)  
   \Bigg\} u(p_a,\lambda_a).
  \end{split}
\end{equation}

Using the compact notation, already introduced when (re)calculating single bremsstrahlung,
we express $\Meu_{2\{II\}}^{\rm SingleIR}$
in a form friendly for numerical evaluation,
that is in terms of $U$ and $V$ matrices,
%//////////////////////////////////////////////////
%       single (p+m)
%//////////////////////////////////////////////////
\begin{equation}
  \begin{split}
   &\Meu_{2\{II\}}^{\rm SingleIR}\left(\st^{p}_{\lambda} \st^{k_1}_{\sigma_1}\st^{k_2}_{\sigma_2}\right) =
\\
   & =eQ_e\; {-\Bmf_{[b1'][cd]} U_{[1'1a]}-\Bmf_{[b2'][cd]} U_{[2'1a]} \over -r_{1a}-r_{2a} +r_{12}}\;
                 \sfac^{(a)}_{[2]}
     +eQ_e\; \sfac^{(b)}_{[1]}\; 
              { V_{[b22']}\Bmf_{[2'a][cd]} +V_{[b21']}\Bmf_{[1'a][cd]} \over -r_{1a}-r_{2a} +r_{12}}\;
\\
   & -eQ_e\; \sfac^{(b)}_{[1]}\;         \Bmf_{[b2'][cd]}\;    {U_{[2'2a]} \over -r_{2a} }
     +eQ_e\; {V_{[b11']}\over -r_{1b}}\;  \Bmf_{[1'a][cd]}\;    \sfac^{(a)}_{[2]}
    +(1 \leftrightarrow 2)  
  \end{split}
\end{equation}
On the other hand the single-IR part to be eliminated is
%//////////////////////////////////////////////////
\begin{equation}
  \begin{split}
     \hbeta^{(1)}_{1(1)[1]} \sfac^{\{I\}}_{[2]} 
   &+\hbeta^{(1)}_{1(1)[2]} \sfac^{\{I\}}_{[1]}
    = r^{\{I\}}_{[1]}  \sfac^{\{I\}}_{[2]}
    + r^{\{I\}}_{[2]}  \sfac^{\{I\}}_{[1]}\\
   &= \left( eQ_e \Bmf_{[b1'][cd]}\; {U_{[1'1a]}\over r_{1a}} 
            -eQ_e {V_{[b11']}\over r_{1a}} \Bmf_{[1'a][cd]}\;  \right)\sfac^{\{I\}}_{[2]}
     +(1 \leftrightarrow 2).
  \end{split}
\end{equation}
Altogether we get
%//////////////////////////////////////////////////
\begin{equation}
  \begin{split}
   &\hbeta_{2\{II\}}^{\rm Single}  \left(\st^{p}_{\lambda} \st^{k_1}_{\sigma_1}\st^{k_2}_{\sigma_2}\right) =
     \Meu_{2\{II\}}^{\rm SingleIR}\left(\st^{p}_{\lambda} \st^{k_1}_{\sigma_1}\st^{k_2}_{\sigma_2}\right)
    -\hbeta^{(1)}_{1(1)[1]} \sfac^{\{I\}}_{[2]} 
    -\hbeta^{(1)}_{1(1)[2]} \sfac^{\{I\}}_{[1]}
\\
   & =-eQ_e\; \Bmf_{[b2'][cd]}\;  {U_{[2'1a]} \over -r_{1a}-r_{2a} +r_{12}}\; \sfac^{(a)}_{[2]}
      +eQ_e\; \sfac^{(b)}_{[1]}\; {V_{[b21']} \over -r_{1a}-r_{2a} +r_{12}}\; \Bmf_{[1'a][cd]}
\\
   &\quad -eQ_e\;\Bmf_{[b1'][cd]}\; 
     \left({ U_{[1'1a]}\over -r_{1a}-r_{2a}+r_{12}} -{ U_{[1'1a]}\over -r_{1a}}\right) 
              \sfac^{(a)}_{[2]}      
\\
   &\quad +eQ_e\; \sfac^{(b)}_{[1]}\; 
     \left({V_{[b22']} \over -r_{1a}-r_{2a} +r_{12}} -{ V_{[b22']}\over -r_{2b}} \right) \Bmf_{[2'a][cd]}\;
     +(1 \leftrightarrow 2).
  \end{split}
\end{equation}
It is rather straightforward to see that the above is IR-finite.

Finally, we have to include all remaining terms from eq.~(\ref{eq:doubleIR}) which have not
yet included in $\hbeta_{2\{II\}}$.
They are IR-finite (in the case of only soft photon energy) and they read
%//////////////////////////////////////////////////
\begin{equation}
  \begin{split}
   \hbeta_{2\{II\}}^{\rm Rest}
     \left( \st^{p}_{\lambda} \st^{k_1}_{\sigma_1}  \st^{k_2}_{\sigma_2} \right)&
  =ie^2 \sum_{B=\gamma,Z} \Pi^{\mu\nu}_B(X)\; (G^{B}_{f,\nu}\;)_{[cd]}\;
   (eQ_e)^2 \; \bar{v}(p_b,\lambda_b)\; \Bigg\{
\\
   &G^{B}_{e,\mu}\;
   {(\not\!{p_a}+m)-{\not\!k_1}-{\not\!k_2} \over -r_{1a}-r_{2a} +r_{12}}\; 
   \not\!{\epsilon}^\star_{\sigma_1}(k_1)\;
   {-{\not\!k_2} \over - r_{2a}}\; 
   \not\!{\epsilon}^\star_{\sigma_2}(k_2)\;
\\
  +&\not\!{\epsilon}^\star_{\sigma_1}(k_1)\; 
   {\not\!{k_1} \over -r_{1b}} \;
    \not\!{\epsilon}^\star_{\sigma_2}(k_2)\; 
   {(-{\not\!p_b}+m)+\not\!{k_1}+\not\!{k_2} \over -r_{1b}-r_{2b}+r_{12}} \;
    G^{B}_{e,\mu}\;
\\
  +&\not\!{\epsilon}^\star_{\sigma_1}(k_1)\; 
   {\not\!{k_1} \over -r_{1b}} \;
    G^{B}_{e,\mu}\;
   {-{\not\!k_2} \over -r_{2a}}\; 
   \not\!{\epsilon}^\star_{\sigma_2}(k_2)\;
    +(1 \leftrightarrow 2)  \Bigg\} u(p_a,\lambda_a),
  \end{split}
\end{equation}
Using tensor notation in the fermion helicity indices 
the above can be expressed in terms of $U$ and $V$ matrices as follows
%//////////////////////////////////////////////////
\begin{equation}
  \begin{split}
  \hbeta_{2\{II\}}^{\rm Rest}\left( \st^{p}_{\lambda} \st^{k_1}_{\sigma_1}  \st^{k_2}_{\sigma_2} \right)
   =&(eQ_e)^2\;
   { \Bmf_{[ba'][cd]} U_{[a'12'']} -\Bmf_{[b1'][cd]} U_{[1'12'']} -\Bmf_{[b2'][cd]} U_{[2'12'']} 
             \over -r_{1a}-r_{2a} +r_{12}}\; 
   {-U_{[2''2a]} \over -r_{2a}}\; 
\\
  +&(eQ_e)^2\; { V_{[b11'']} \over -r_{1b}} \;\;
   {-V_{[1''2b']}\Bmf_{[b'a][cd]} +V_{[1''21']}\Bmf_{[1'a][cd]} +V_{[1''22']}\Bmf_{[2'a][cd]}
             \over -r_{1b}-r_{2b}+r_{12}} \;
\\
  +&(eQ_e)^2\;   {V_{[b11']} \over -r_{1b}} \; \Bmf_{[1'2'][cd]}\; {-U_{[2'2a]} \over -r_{2a}}\; 
    +(1 \leftrightarrow 2).
  \end{split}
\end{equation}
The total ISR $\hbeta_{2\{II\}}$ is the sum of the three
%//////////////////////////////////////////////////
\begin{equation}
  \hbeta_{2\{II\}}\left( \st^{p}_{\lambda} \st^{k_1}_{\sigma_1}  \st^{k_2}_{\sigma_2} \right)=
  \hbeta_{2\{II\}}^{\rm Double}\left( \st^{p}_{\lambda} \st^{k_1}_{\sigma_1}  \st^{k_2}_{\sigma_2} \right)
 +\hbeta_{2\{II\}}^{\rm Single}\left( \st^{p}_{\lambda} \st^{k_1}_{\sigma_1}  \st^{k_2}_{\sigma_2} \right)
 +\hbeta_{2\{II\}}^{\rm Rest}  \left( \st^{p}_{\lambda}\st^{k_1}_{\sigma_1}  \st^{k_2}_{\sigma_2} \right).
\end{equation}

%%%%%%%%%%%%%%%%%%%%%%%%%%%%%%%%%%%%%%%%%%%%%%%%%%%%
%%%%%%%%%%%%%%%%%%%%%%%%%%%%%%%%%%%%%%%%%%%%%%%%%%%%
\subsubsection{2-real FSR  photons}
%%%%%%%%%%%%%%%%%%%%%%%%%%%%%%%%%%%%%%%%%%%%%%%%%%%%

The case of final-state double real photon emission 
can be analysed in a similar way.
The second-order FSR, two-photon, matrix element is
%%//////////////////////////////////////////////////
%       pure FSR 2-photon from Feynman diags
%//////////////////////////////////////////////////
\begin{equation}
  \label{fsr2-feynman}
  \begin{split}
    \Meu^{(2)}_{2\{FF\}} &
      \left( \st^{p}_{\lambda} \st^{k_1}_{\sigma_1} \st^{k_2}_{\sigma_2};P \right)
    =ie^2 \sum_{B=\gamma,Z} \Pi^{\mu\nu}_B(P)\; (G^B_{e,\mu})_{[ba]}\;
     (eQ_f)^2\; \bar{u}(p_c,\lambda_c)\Bigg\{\\
     &\not\!{\epsilon}^\star_{[1]}\; {(\not\!{p_c}+m)+{\not\!k_1} \over 2k_1p_c}\;
      \not\!{\epsilon}^\star_{[2]}\; 
            {(\not\!{p_c}+m)+{\not\!k_1}+{\not\!k_2} \over 2k_1p_c+2k_2p_c +2k_1k_2}\;
      G^{B}_{f,\nu}\;
\\
     &+G^{B}_{f,\nu}\;
       {(-{\not\!p_d}+m)-\not\!{k_1}-\not\!{k_2} \over 2k_1p_d+2k_2p_d +2k_1k_2}\; 
               \not\!{\epsilon}^\star_{[1]}\;
       {(-{\not\!p_d}+m)-\not\!{k_2} \over 2k_2p_d }\; \not\!{\epsilon}^\star_{[2]}\;
\\
     &+\not\!{\epsilon}^\star_{[1]}\; {(\not\!{p_c}+m)+{\not\!k_1} \over 2k_1p_c}\;
      G^{B}_{f,\nu}\;
      {(-{\not\!p_d}+m)-\not\!{k_2} \over 2k_2p_d}\; \not\!{\epsilon}^\star_{[2]}\;
      +(1 \leftrightarrow 2) \Bigg\} v(p_d,\lambda_d),
  \end{split}
\end{equation}
Similarly, the expansion into soft and non-IR parts for FSR 
spin amplitudes is done in the way completely analogous to the ISR case.
The subtraction formula is now
%%%%%%%%%%%%%%%%%%%%%%%%%%%%%%%%%%%%%%%%%%%%%%%%%%%%
\begin{equation}
  \label{eq:beta2FF}
  \begin{split}
      &\hbeta^{(2)}_{2\{FF\}}
           \left(\st^{p}_{\lambda}\st^{k_1}_{\sigma_1}\st^{k_2}_{\sigma_2}; P \right)    
    =    \Mmf^{(2)}_{2\{FF\}}
           \left(\st^{p}_{\lambda}\st^{k_1}_{\sigma_1}\st^{k_2}_{\sigma_2}; P \right)
       -\hbeta^{(1)}_{1\{F\}}\left(\st^{p}_{\lambda}\st^{k_1}_{\sigma_1}; P \right) 
                                                     \sfac^{\{F\}}_{\sigma_2}(k_2)  
\\&\qquad\qquad
       -\hbeta^{(1)}_{1\{F\}}\left(\st^{p}_{\lambda}\st^{k_2}_{\sigma_2}; P \right) 
                                                     \sfac^{\{F\}}_{\sigma_1}(k_1)
       -\hbeta^{(0)}_0\left(\st^{p}_{\lambda}; P \right)  
                       \sfac^{\{F\}}_{\sigma_1}(k_1) \sfac^{\{F\}}_{\sigma_2}(k_2).
  \end{split}
\end{equation}

First we obtain the contribution from terms with two $({\not\!p}-m)$ factors
%//////////////////////////////////////////////////
\begin{equation}
  \begin{split}
  &\hbeta^{(2){\rm Double}}_{2\{FF\}}\left[\st^{p}_{\lambda}  \st^{k_1}_{\sigma_1} \st^{k_2}_{\sigma_2} \right]= 
       \Meu_{2\{FF\}}^{\rm DoubleIR}\left[\st^{p}_{\lambda}  \st^{k_1}_{\sigma_1} \st^{k_2}_{\sigma_2} \right]
   -\sfac^{\{F\}}_{[1]} \sfac^{\{F\}}_{[2]}
         \Bmf_{[ba][cd]} { (p_c+p_d+k_1+k_2)^2 \over (p_c+p_d)^2}
\\
  &\qquad=
    \bigg(\Delta_c \sfac^{(c)}_{[1]} \sfac^{(c)}_{[2]}
         +\Delta_d \sfac^{(d)}_{[1]} \sfac^{(d)}_{[2]}\bigg)
    \Bmf_{[ba][cd]}
\\
  &\qquad -\sfac^{\{F\}}_{[1]} \sfac^{\{F\}}_{[2]}
          \Bmf_{\rm F} \left[\st^{p_b}_{\lambda_b}  \st^{p_a}_{\lambda_a} \right]
          \left( { (p_c+p_d+k_1+k_2)^2 \over (p_c+p_d)^2} -1 \right)
\\
  &\sfac^{(c)}_{\sigma_i}(k_i)\equiv \sfac^{(c)}_{[i]} =+eQ_f {b_{\sigma_i}(k_i,p_c)\over r_{ic}},\;\;
   \sfac^{(d)}_{\sigma_i}(k_i)\equiv \sfac^{(d)}_{[i]} =-eQ_f {b_{\sigma_i}(k_i,p_d)\over r_{id}},\;\;
\\
  &\sfac^{\{F\}}_{\sigma_i}(k_i) \equiv \sfac^{(c)}_{\sigma_i}(k_i) +\sfac^{(d)}_{\sigma_i}(k_i)
                               \equiv \sfac^{(c)}_{[i]} + \sfac^{(d)}_{[i]},
  \end{split}
\end{equation}
and is explicitly IR-finite.
The second group of terms with only one $({\not\!p}-m)$ factor at the end of the fermion line is
%%//////////////////////////////////////////////////
%       FSR 2-photon single (p-m)
%//////////////////////////////////////////////////
\begin{equation}
  \begin{split}
    \Meu&_{2\{FF\}}^{\rm SingleIR}
    \left( \st^{p}_{\lambda} \st^{k_1}_{\sigma_1} \st^{k_2}_{\sigma_2} \right)
    =ie^2 \sum_{B=\gamma,Z} \Pi^{\mu\nu}_B(X)\; (G^B_{e,\mu})_{[ba]}\;
     (eQ_f)^2\;\bar{u}(p_c,\lambda_c)\Bigg\{
\\
     &\not\!{\epsilon}^{\star}_{[1]}\; {(\not\!{p_c}+m) \over r_{1c}}\;
      \not\!{\epsilon}^{\star}_{[2]}\; 
            {{\not\!k_1}+{\not\!k_2} \over r_{1c}+r_{2c} +r_{12}}\;
       G^{B}_{f,\nu}\;
      +G^{B}_{f,\nu}\;
       {-\not\!{k_1}-\not\!{k_2} \over r_{1d}+r_{2d} +r_{12}}\; 
               \not\!{\epsilon}^{\star}_{[1]}\;
       {(-{\not\!p_d}+m) \over r_{2d} }\; \not\!{\epsilon}^{\star}_{[2]}\;
\\
     +&\not\!{\epsilon}^{\star}_{[1]}\; {(\not\!{p_c}+m)\over r_{1c}}\;
      G^{B}_{f,\nu}\;
      {-\not\!{k_2} \over r_{2d}}\; \not\!{\epsilon}^{\star}_{[2]}\;
      +\not\!{\epsilon}^{\star}_{[1]}\; {{\not\!k_1} \over r_{1c}}\;
      G^{B}_{f,\nu}\;
      {(-{\not\!p_d}+m)\over r_{2d}}\; \not\!{\epsilon}^{\star}_{[2]}\;
      +(1 \leftrightarrow 2) \Bigg\} v(p_d,\lambda_d),
  \end{split}
\end{equation}
and it translates in the matrix notation (in fermion spin indices) into
%//////////////////////////////////////////////////
\begin{equation}
  \begin{split}
    \Meu&_{2\{FF\}}^{\rm SingleIR}
    \left( \st^{p}_{\lambda} \st^{k_1}_{\sigma_1} \st^{k_2}_{\sigma_2} \right)=
\\
  =&eQ_f\;\sfac^{(c)}_{[1]}\; { U_{[c21']} \over r_{1c}+r_{2c} +r_{12}}\; \Bmf_{[ba][1'd]}\;
   +eQ_f\;\sfac^{(c)}_{[1]}\; { U_{[c22']} \over r_{1c}+r_{2c} +r_{12}}\; \Bmf_{[ba][2'd]}\;
\\
  +&eQ_f\;\Bmf_{[ba][c1']}\;   {-V_{[1'1d]} \over r_{1d}+r_{2d} +r_{12}}\; \sfac^{(d)}_{[2]}\;
   +eQ_f\;\Bmf_{[ba][c2']}\;   {-V_{[2'1d]} \over r_{1d}+r_{2d} +r_{12}}\; \sfac^{(d)}_{[2]}\;
\\
  +&eQ_f\;\sfac^{(c)}_{[1]}\; \Bmf_{[ba][c2']}\; {-V_{[2'2d]} \over r_{2d}}\;
  +eQ_f\;{U_{[c11']} \over r_{1c}}\; \Bmf_{[ba][1'd]}\; \sfac^{(d)}_{[2]}\;
  +(1 \leftrightarrow 2),
  \end{split}
\end{equation}
On the other hand the single-IR part to be eliminated is
%//////////////////////////////////////////////////
\begin{equation}
  \begin{split}
    \hbeta^{(1)}_{1(0)[1]}& \sfac^{\{F\}}_{[2]}     +\hbeta^{(1)}_{1(0)[2]} \sfac^{\{F\}}_{[1]}
    =       r^{\{F\}}_{[1]}  \sfac^{\{F\}}_{[2]}        + r^{\{F\}}_{[2]} \sfac^{\{F\}}_{[1]}\\
   =& \left( +eQ_e  \Bmf_{[ba][1'd]}\;         {U_{[c11']}\over r_{1c}} 
             -eQ_e  {V_{[1'1d]}\over r_{1d}}  \Bmf_{[ba][c1']}\;   \right)\sfac^{\{F\}}_{[2]}
     +(1 \leftrightarrow 2)\\
   & -\Bmf_{[ba][cd]}\;  \left( { (p_c+p_d+k_1)^2 \over (p_c+p_d)^2} -1 \right) 
                         \sfac^{\{F\}}_{[1]} \sfac^{\{F\}}_{[2]} 
       +(1 \leftrightarrow 2)
  \end{split}
\end{equation}
Altogether we get
%//////////////////////////////////////////////////
\begin{equation}
  \begin{split}
   \hbeta_{2\{FF\}}^{\rm Single}& \left(\st^{p}_{\lambda} \st^{k_1}_{\sigma_1}\st^{k_2}_{\sigma_2}\right) =
    \Meu_{2\{FF\}}^{\rm SingleIR}\left(\st^{p}_{\lambda} \st^{k_1}_{\sigma_1}\st^{k_2}_{\sigma_2}\right)
    -\hbeta^{(1)}_{1(0)}\left(\st^{p}_{\lambda} \st^{k_1}_{\sigma_1}\right) 
                              \sfac^{\{F\}}\left[\st^{k_2}_{\sigma_2}\right]
    -\hbeta^{(1)}_{1(0)}\left(\st^{p}_{\lambda} \st^{k_2}_{\sigma_2}\right) 
                              \sfac^{\{F\}}\left[\st^{k_1}_{\sigma_1}\right]
\\
    =&eQ_f\;\sfac^{(c)}_{[1]}\;\Bigg\{
      \left( {U_{[c22']}\over r_{2c}+r_{1c}+r_{12}} -{ U_{[c22']} \over r_{2c} }\right)\; \Bmf_{[ba][2'd]}\; 
    + {U_{[c21']} \over r_{2c}+r_{1c}+r_{12} }\; \Bmf_{[ba][1'd]}\; \Bigg\}
\\
    +&eQ_f\; \Bigg\{\Bmf_{[ba][c1']}\;
      \left({-V_{[1'1d]}\over r_{1d} +r_{2d} +r_{12}} -{-V_{[1'1d]} \over r_{1d} }\right)\;       
    +{-V_{[2'1d]} \over r_{1d} +r_{2d} +r_{12}}\;  \Bmf_{[ba][c2']}
   \;\Bigg\}\sfac^{(d)}_{[2]}
\\
    &+\Bmf_{[ba][cd]}\;  \left( { (p_c+p_d+k_1)^2 \over (p_c+p_d)^2} -1 \right) 
                         \sfac^{\{F\}}_{[1]} \sfac^{\{F\}}_{[2]} 
     +(1 \leftrightarrow 2),
  \end{split}
\end{equation}
Finally we include the remaining terms in eq.~(\ref{fsr2-feynman})
%%//////////////////////////////////////////////////
%       pure FSR 2-photon Rest
%//////////////////////////////////////////////////
\begin{equation}
  \begin{split}
    \Meu_{2\{FF\}}^{\rm Rest}
      \left( \st^{p}_{\lambda} \st^{k_1}_{\sigma_1} \st^{k_2}_{\sigma_2} \right)
      =ie^2 \sum_{B=\gamma,Z} &\Pi^{\mu\nu}_B(X)\; (G^B_{e,\mu})_{[ba]}\;
      (eQ_f)^2\;\bar{u}(p_c,\lambda_c)\Bigg\{
\\
     &\not\!{\epsilon}^{\star}_{[1]}\; {{\not\!k_1} \over r_{1c}}\;
      \not\!{\epsilon}^{\star}_{[2]}\; 
            {(\not\!{p_c}+m)+{\not\!k_1}+{\not\!k_2} \over r_{1c}+r_{2c} +r_{12}}\;
      G^{B}_{f,\nu}\;
\\
     &+G^{B}_{f,\nu}\;
       {(-{\not\!p_d}+m)-\not\!{k_1}-\not\!{k_2} \over r_{1d}+r_{2d} +r_{12}}\; 
               \not\!{\epsilon}^{\star}_{[1]}\;
       {-\not\!{k_2} \over r_{2d} }\; \not\!{\epsilon}^{\star}_{[2]}\;
\\
     &+\not\!{\epsilon}^{\star}_{[1]}\; {{\not\!k_1} \over r_{1c}}\;
      G^{B}_{f,\nu}\;
      {-\not\!{k_2} \over r_{2d}}\; \not\!{\epsilon}^{\star}_{[2]}\;
      +(1 \leftrightarrow 2) \Bigg\} v(p_d,\lambda_d),
  \end{split}
\end{equation}
which in the programmable matrix notation looks as follows
%%//////////////////////////////////////////////////
%       pure FSR 2-photon Rest
%//////////////////////////////////////////////////
\begin{equation}
  \begin{split}
    \hbeta_{2\{FF\}}^{\rm Rest}&\left(\st^{p}_{\lambda} \st^{k_1}_{\sigma_1} \st^{k_2}_{\sigma_2} \right)
   = \Meu_{2\{FF\}}^{\rm Rest} \left(\st^{p}_{\lambda} \st^{k_1}_{\sigma_1} \st^{k_2}_{\sigma_2} \right)=
\\
    =&(eQ_f)^2\; {U_{[c11'']}\over r_{1c}}\;\;
      {U_{[1''2c']}\Bmf_{[ba][c'd]} +U_{[1''21']}\Bmf_{[ba][1'd]} +U_{[1''22']}\Bmf_{[ba][2'd]}
              \over r_{1c}+r_{2c} +r_{12}}\;
\\
    +&(eQ_f)^2\; 
      {-\Bmf_{[ba][cd']}V_{[d'12'']} -\Bmf_{[ba][c1']}V_{[1'12'']} -\Bmf_{[ba][c2']}V_{[2'12'']} 
           \over r_{1d}+r_{2d} +r_{12}}\;\; {-V_{[2''2d]} \over r_{2d} }\;
\\
    +&(eQ_f)^2\;{U_{[c11']}\over r_{1c}}\; \Bmf_{[ba][1'2']}\;  {-V_{[2'2d]} \over r_{2d}}\;
      +(1 \leftrightarrow 2)
  \end{split}
\end{equation}
The total contribution from double FSR real photon emission is
%//////////////////////////////////////////////////
\begin{equation}
    \hbeta_{2\{FF\}}             \left(\st^{p}_{\lambda} \st^{k_1}_{\sigma_1} \st^{k_2}_{\sigma_2} \right)
   =\hbeta_{2\{FF\}}^{\rm Double}\left(\st^{p}_{\lambda} \st^{k_1}_{\sigma_1} \st^{k_2}_{\sigma_2} \right)
   +\hbeta_{2\{FF\}}^{\rm Single}\left(\st^{p}_{\lambda} \st^{k_1}_{\sigma_1} \st^{k_2}_{\sigma_2} \right)
   +\hbeta_{2\{FF\}}^{\rm Rest}  \left(\st^{p}_{\lambda} \st^{k_1}_{\sigma_1} \st^{k_2}_{\sigma_2} \right).
\end{equation}

%%%%%%%%%%%%%%%%%%%%%%%%%%%%%%%%%%%%%%%%%%%%%%%%%%%%
%%%%%%%%%%%%%%%%%%%%%%%%%%%%%%%%%%%%%%%%%%%%%%%%%%%%
\subsubsection{1-real ISR  and 1-real FSR photon}
%%%%%%%%%%%%%%%%%%%%%%%%%%%%%%%%%%%%%%%%%%%%%%%%%%%%

As we have seen in the previous cases of double real emission
most complications are due to simultaneous emission from one fermion ``leg''.
The case of one real ISR and one real FSR photon is easier
because there is at most one photon on one leg:
%/////////////////////////////////////////////
%    1 ISR + 1 FSR
%/////////////////////////////////////////////
\begin{equation}
  \begin{split}
    \Meu^{(2)}_{2\{IF\}} &
      \left( \st^{p_a}_{\lambda_a} \st^{p_b}_{\lambda_b} \st^{p_c}_{\lambda_c} \st^{p_d}_{\lambda_d} 
             \st^{k_1}_{\sigma_1} \st^{k_2}_{\sigma_2};P-k_1 \right)
  = ie^2 \sum_{B=\gamma,Z} \Pi^{\mu\nu}_B(P-k_1)\;\;
\\
   & eQ_e \bar{v}(p_b,\lambda_b)\left(  
         G^{B}_{e,\mu}  {\not\!{p_a}+m-{\not\!k_1} \over -2k_1p_a} \not\!{\epsilon}^\star_{[1]}
        +\not\!{\epsilon}^\star_{[1]}\; {-{\not\!p_b}+m+\not\!{k_1} \over -2k_1p_b}\; G^{B}_{e,\mu}
    \right) u(p_a,\lambda_a)
\\
   &eQ_f\bar{u}(p_c,\lambda_c)\left(  
         G^{B}_{f,\nu}\; {-{\not\!p_d}+m-\not\!{k_2} \over 2k_2p_d}\; \not\!{\epsilon}^\star_{[2]}\;
        +\not\!{\epsilon}^\star_{[2]}\; {\not\!{p_c}+m+{\not\!k_2} \over 2k_2p_c}\; G^{B}_{f,\nu}
     \right) v(p_d,\lambda_d)
  \end{split}
\end{equation}
and the subtraction formula is now
%%%%%%%%%%%%%%%%%%%%%%%%%%%%%%%%%%%%%%%%%%%%%%%%%%%%
\begin{equation}
  \label{eq:beta2IF}
  \begin{split}
      &\hbeta^{(2)}_{2\{IF\}}
           \left(\st^{p}_{\lambda}\st^{k_1}_{\sigma_1}\st^{k_2}_{\sigma_2}; P-k_1 \right)    
    =   \Mmf^{(2)}_{2\{IF\}}
           \left(\st^{p}_{\lambda}\st^{k_1}_{\sigma_1}\st^{k_2}_{\sigma_2}; P-k_1 \right)
       -\hbeta^{(1)}_{1\{I\}}\left(\st^{p}_{\lambda}\st^{k_1}_{\sigma_1}; P-k_1 \right) 
                                                     \sfac^{\{F\}}_{\sigma_2}(k_2)  
\\&\qquad\qquad
       -\hbeta^{(1)}_{1\{F\}}\left(\st^{p}_{\lambda}\st^{k_2}_{\sigma_2}; P-k_1 \right) 
                                                     \sfac^{\{I\}}_{\sigma_1}(k_1)
       -\hbeta^{(0)}_0\left(\st^{p}_{\lambda}; P-k_1 \right)  
                       \sfac^{\{I\}}_{\sigma_1}(k_1) \sfac^{\{F\}}_{\sigma_2}(k_2).
  \end{split}
\end{equation}

The simplicity of this contribution is manifest in the fact that
$\hbeta_{2\{IF\}}$ is obtained by simple subtraction (omission) of
all terms proportional to one or two $(\not\!p-m)$ factors
%/////////////////////////////////////////////
\begin{equation}
  \begin{split}
    \hbeta_{2\{IF\}} &
      \left( \st^{p}_{\lambda}  \st^{k_1}_{\sigma_1} \st^{k_2}_{\sigma_2};X \right)
  = ie^2 \sum_{B=\gamma,Z} \Pi^{\mu\nu}_B(X)\;\;
\\
   & eQ_e \bar{v}(p_b,\lambda_b)\left(  
         G^{B}_{e,\mu}  {-{\not\!k_1} \over -r_{1a}} \not\!{\epsilon}^{\star}_{[1]}
        +\not\!{\epsilon}^{\star}_{[1]}\; { \not\!{k_1} \over -r_{1b}}\; G^{B}_{e,\mu}
    \right) u(p_a,\lambda_a)
\\
   &eQ_f\bar{u}(p_c,\lambda_c)\left(  
         G^{B}_{f,\nu}\; {-\not\!{k_2} \over r_{2d}}\; \not\!{\epsilon}^{\star}_{[2]}\;
        +\not\!{\epsilon}^{\star}_{[2]}\; {{\not\!k_2} \over r_{2c}}\; G^{B}_{f,\nu}
     \right) v(p_d,\lambda_d).
  \end{split}
\end{equation}
In the computation-friendly matrix notation it reads
%/////////////////////////////////////////////
\begin{equation}
  \begin{split}
    &\hbeta_{2\{IF\}}
      \left( \st^{p}_{\lambda} \st^{k_1}_{\sigma_1} \st^{k_2}_{\sigma_2};X \right)
  = ie^2 \sum_{B=\gamma,Z} \Pi^{\mu\nu}_B(X)\;\;eQ_e eQ_f
\\ &\times
     \left(  (G^{B}_{e,\mu})_{[b1']}\;       {-U_{[1'1a]} \over -r_{1a}}\;
                 +{V_{[b11']} \over -r_{1b}}\;   (G^{B}_{e,\mu})_{[1'a]} \right)
     \left(  (G^{B}_{f,\nu})_{[c2']}\;       {-V_{[2'2d]} \over r_{2d}}\;
                 +{U_{[c22']} \over r_{2c}}\;   (G^{B}_{f,\nu})_{[2'd]} \right)
\\
  & =eQ_e eQ_f\Bigg(
    \Bmf_{[b1'][c2']}(X) {-U_{[1'1a]}\over -r_{1a}}\;  {-V_{[2'2d]}\over r_{2d}}\;
    +{ U_{[c22']}\over r_{2c}}\; \Bmf_{[b1'][2'd]}(X) {-U_{[1'1a]}\over -r_{1a}}\;
\\&\qquad\qquad\qquad
    +{ V_{[b11']}\over -r_{1b}}\; \Bmf_{[1'a][c2']}(X) {-V_{[2'2d]}\over r_{2d}}\;
    +{ V_{[b11']}\over -r_{1b}}\; { U_{[c22']}\over r_{2c}} \Bmf_{[1'a][2'd]}(X) \Bigg)
  \end{split}
\end{equation}

%%%%%%%%%%%%%%%%%%%%%%%%%%%%%%%%%%%%%%%%%%%%%%%%%%%%%%%%%%%%%%%%%%%%%%%%%%%%%%%%%%%%%%%%%%%%%
%%%%%%%%%%%%%%%%%%%%%%%%%%%%%%%%%%%%%%%%%%%%%%%%%%%%%%%%%%%%%%%%%%%%%%%%%%%%%%%%%%%%%%%%%%%%%
\section{Relations between CEEX and EEX}
%%%%%%%%%%%%%%%%%%%%%%%%%%%%%%%%%%%%%%%%%%%%%%%%%%%%%%%%%%%%%%%%%%%%%%%%%%%%%%%%%%%%%%%%%%%%%
Having shown the CEEX and EEX schemes in a detail,
we would like to compare certain important/interesting fearures of both schemes
in a more detail.
In particular we would like to show how the two examples of the EEX 
scheme can be obtained as a limiting case of CEEX,
and to show the exact relation among $\bbeta$'s of EEX and $\hbeta$'s of CEEX.
From these considerations it will be clear that the CEEX scheme is more general
than the EEX scheme.

%%%%%%%%%%%%%%%%%%%%%%%%%%%%%%%%%%%%%%%%%%%%%%%%%%%%%%%%%%%%%%
\subsection{Neglecting partition dependence}
%%%%%%%%%%%%%%%%%%%%%%%%%%%%%%%%%%%%%%%%%%%%%%%%%%%%%%%%%%%%%%
Let us first examine the interesting limit of CEEX in which 
which we drop the dependence on the partition index $X_\wp \to P$, where $P=p_a+p_b$, for example.
Note that it is not in the EEX class.
In this limit, in the simplest  case of the \Order{\alpha^0} exponentiation we have:
\begin{equation}
    \sum\limits_{\wp\in{\cal P}}
    e^{\alpha {B_4^\star(X_\wp})}
    { X_\wp^2  \over s_{cd} }\;
    \Bmf \big(  \st^{p}_{\lambda}; X_\wp \big)
    \prod\limits_{i=1}^n \sfac^{\{\wp_i\}}_{[i]}
    \Longrightarrow
    e^{\alpha B_4}
    \Bmf \big(\st^{p}_{\lambda};P \big)
    \prod\limits_{i=1}^n \big( \sfac^{\{0\}}_{[i]} + \sfac^{\{1\}}_{[i]} \big),
\end{equation}
because of the identity
\begin{displaymath}
    \sum\limits_{\wp\in{\cal P}}
    \prod\limits_{i=1}^n \sfac^{\{\wp_i\}}_{[i]}
    \equiv
    \prod\limits_{i=1}^n \big( \sfac^{\{0\}}_{[i]} + \sfac^{\{1\}}_{[i]} \big).
\end{displaymath}
The relevance, advantages and disadvanteges of this scenario were already discussed
in Section ???
Is this realized in \KK MC?????
Note that in the above transition we keep the ISR$\otimes$FSR interference contribution.

%%%%%%%%%%%%%%%%%%%%%%%%%%%%%%%%%%%%%%%%%%%%%%%%%%%%%%%%%%%%%%
\subsection{Neglecting IFI}
%%%%%%%%%%%%%%%%%%%%%%%%%%%%%%%%%%%%%%%%%%%%%%%%%%%%%%%%%%%%%%
The second important case we would like to discuss is the case of the
very narrow resonances, when the ISR$\otimes$FSR interference contribution
to any physicaal observable is so small that it can be neglected whatsoever.
This corresponds to a well defined limit in the CEEX scheme.
In this limit, in the simplest  case of the \Order{\alpha^0} exponentiation we have:
\begin{equation}
\begin{split}
|\Meu^{(0)}_n|^2 
 &= \sum\limits_{\wp\in{\cal P}}
    \sum\limits_{\wp'\in{\cal P}}
    e^{\alpha  B_4^\star(X_\wp)}
    e^{\alpha (B_4^\star(X_{\wp'}))^*}
    \Bmf\big(\st^{p}_{\lambda}; X_\wp \big)
    \Bmf\big(\st^{p}_{\lambda}; X_{\wp'} \big)^*
    \prod\limits_{i=1}^n  \sfac^{\{\wp_i\}}_{[i]}
    \prod\limits_{j=1}^n {\sfac^{\{\wp_i'\}}_{[j]}}\st^*
\\
  & \Longrightarrow
    e^{2\alpha \Re B_2(p_a,p_b)}
    e^{2\alpha \Re B_2(p_c,p_d)}
    \sum\limits_{\wp\in{\cal P}}
    \big|\Bmf \big(  \st^{p}_{\lambda}; X_\wp \big)\big|^2
    \prod\limits_{i=1}^n \big| \sfac^{\{\wp_i\}}_{[i]}\big|^2.
\end{split}
\end{equation}
What we did in the above transition,
we have neglected the ISR$\otimes$FSR interferences entirely
by dropping non-diagonal terms $\wp \neq \wp'$ in the double sum over partitions,
and we have replaced the resonance-type form-factor by the sum of the
traditional YFS formfactors for ISR and FSR (no interferenence).
In this way we have got the \Oeex{\alpha^0} which at this order
is identical to \Oceex{\alpha^0}.
At the \Oceex{\alpha^r} $r=1,2$, in order to get from
\Oceex{\alpha^r} to \Oeex{\alpha^r} we have in addition to truncate
$\hbeta$'s  down to $\bbeta$', as will be shown in the next subsection.

The \Oeex{\alpha^r} $r=1,2$ neglecting the ISR$\otimes$FSR interferences
was used in YFS2/3~\cite{yfs2:1990,yfs3-pl:1992} of KORALZ~\cite{koralz4:1994}
and it is well justified close to $Z$ resonance position at LEP1, 
see also relevant numerical results in the next Section.
At LEP2 the above approximation cannot be justified any more.

%%%%%%%%%%%%%%%%%%%%%%%%%%%%%%%%%%%%%%%%%%%%%%%%%%%%%%%%%%%%%%%%%%%%%%%%
\subsection{Relation among  $\bbeta$'s for EEX and $\hbeta$'s of CEEX}
%%%%%%%%%%%%%%%%%%%%%%%%%%%%%%%%%%%%%%%%%%%%%%%%%%%%%%%%%%%%%%%%%%%%%%%%
\label{sec:relation}

For the sake of completeness of the discussion,
it is necessary to find out the relation of the  $\beta$'s defined at the amplitude level
to the older EEX/YFS $\bar{\bbeta}$'s defined at the level of the differential distributions.
Let us suppress all spin indices, understanding that for every term like
$|...|^2$ or $\Re[A B^*]$ the appropriate spin sum/average is done.
The traditional $\bbeta$'s of the YFS scheme at the \Order{\alpha^2} level are
%//////////////////////////////////////////////////
\begin{equation}
  \begin{split}
    &\bbeta^{(l)}_0    = \Big|\Mmf^{(l)}_0\Big|^2_{(\alpha^l)},\; l=0,1,2,\\
    &\bbeta^{(l)}_1(k) = \Big|\Mmf^{(l)}_1(k)\Big|^2_{(\alpha^{l+1})} 
              - \bbeta^{(l)}_0 |\sfac(k)|^2,\; l=0,1,\\
    &\bbeta^{(2)}_2(k_1,k_2) = \Big|\Mmf^{(2)}_1(k_1,k_2)\Big|^2 
              - \bbeta^{(1)}_1(k_1) |\sfac(k_2)|^2
              - \bbeta^{(1)}_1(k_2) |\sfac(k_1)|^2
              - \bbeta^{(0)}_0 |\sfac(k_1)| |\sfac(k_2)|^2,
  \end{split}
\end{equation}
where subscript $|_{(\alpha^{r})}$ means truncation to \Order{\alpha^r}.
Now  for each $\Mmf^{(n+l)}_n$we substitute its expansion 
in terms of $\hbeta$'s according to eq.~(\ref{eq:beta-decomposition2}) getting
the following relation
%//////////////////////////////////////////////////
\begin{equation}
  \begin{split}
    &\bbeta^{(l)}_0    = |\hbeta^{(l)}_0|^2_{(\alpha^l)},\; l=0,1,2,\\
    &\bbeta^{(l)}_1(k) = |\hbeta^{(l)}_1(k)|^2 
                       +2\Re[\hbeta^{(l)}_0 (\hbeta^{(l)}_1(k))^*]_{(\alpha^{l+1})},\; l=0,1,\\
    &\bbeta^{(2)}_2(k_1,k_2) =|\hbeta^{(2)}_2(k_1,k_2)|^2
            +2\Re[\hbeta^{(1)}_1(k_1)\sfac(k_2)\{\hbeta^{(1)}_1(k_2)\sfac(k_1)\}^*]\\
    &\qquad +2\Re[\hbeta^{(2)}_2(k_1,k_2) \{\hbeta^{(1)}_1(k_1)\sfac(k_2)
                              \hbeta^{(1)}_1(k_2)\sfac(k_1) +\hbeta^{(0)}_1 \sfac(k_1)\sfac(k_2)\}^*],
  \end{split}
\end{equation}
As we see, the relation is not completely trivial; there are some extra terms on the RHS,
which are all IR-finite.
From the above exercise it is obvious that $\bbeta$'s are generally 
more complicated objects than the $\hbeta$'s and that for example
the inclusion of the spin density matrix formalism into the $\bbeta$'s would be a quite nontrivial exercise 
-- the great advantage of the CEEX scheme is that this is done numerically.
It is also seen that in the $\bbeta_0$ and $\bbeta_1$ some higher-order
virtual terms are {\em unnecessarily} truncated, which probably is worsening perturbative convergence
of the EEX/YFS scheme in comparison with CEEX.
The above formula shows in a most clear and clean way the difference between the EEX and CEEX
exponentiation schemes.

%%%%%%%%%%%%%%%%%%%%%%%%%%%%%%%%%%%%%%%%%%%%%%%%%%%%%%%%%%%%%%%%%%%%%%%%%%%%%%%%%%%%%%%%%%%
%%%%%%%%%%%%%%%%%%%%%%%%%%%%%%%%%%%%%%%%%%%%%%%%%%%%%%%%%%%%%%%%%%%%%%%%%%%%%%%%%%%%%%%%%%%
%%%%%%%%%%%%%%%%%%%%%%%%%%%%%%%%%%%%%%%%%%%%%%%%%%%%%%%%%%%%%%%%%%%%%%%%%%%%%%%%%%%%%%%%%%%

%%%%%%%%%%%%%%%%%%%%%%%%%
% make ceex2-all-ps
%%%%%%%%%%%%%%%%%%%%%%%%%

%%%%%%%%%%%%%%%%%%%%%%%%%%%%%%%%%%%%%%%%%%%%%%%%%%%%%%%%%%%%%%%%%%%%%%%%%%%%%%%%%%%%%%%%%%%%%
%%%%%%%%%%%%%%%%%%%%%%%%%%%%%%%%%%%%%%%%%%%%%%%%%%%%%%%%%%%%%%%%%%%%%%%%%%%%%%%%%%%%%%%%%%%%%
\section{Semi-analytical approach}
\label{sec:semi}
In this section we shall present results of semi-analytical
calculations which reproduce in the \Ordpr{\alpha^2}, or even in the \Ordpr{\alpha^3},
certain selected results, mainly integrated cross-sections, 
of the Monte Carlo calculation.

The semi-analytical approach in which one integrates over the phase space analytically,
often leaving the last one- or two-dimensional integrations for numerical treatment
(usually non Monte Carlo), is the oldest one.
Four decades ago there were no computers powerful enough
even to dream about the numerical integration over the complete multiphoton phase space.
Even now, in spite of proliferation of the MC programs,
the non Monte Carlo semianalytical programs are still very popular and useful,
especially programs used to calculate the total cross section and charge asymmetry
for the fermion-pair final state,
near the $Z$ resonance, like ZFITTER and TOPAZ0~\cite{zfitter6:1999, topaz0:1998}.
Semi-analytical programs have certain advantages over the MC programs 
--  they are generally faster in terms
of computer CPU time and are therefore better suited for fitting input parameters of
the Standard Model, like the Higgs mass%
\footnote{ 
  It is definitely possible to fit input SM parameters with help of the Monte Carlo
  event generators, as it is currently done in the $W$-mass measurement in LEP2.}.
Nevertheless, semi-analytical calculations have also two long-standing important disadvantages:
\begin{itemize}
\item[(a)]
  They are able to provide predictions for rather very primitive or absent
  experimental cut-offs. In practice they always have to be used in combination with
  the MC event generators anyway. The MC is used to remove or ``straighten''
  the real experimental cuts to be closer to these which
  are practically implementable in the semianalytical calculation.
  Obviously this introduces additional systematic errors.
\item[(b)] They are prohibitively complicated beyond the three-body final state, 
  that is they are relatively easy up to \Order{\alpha^1},
  (single photon emission in the fermion pair production). 
  In the collinear approximation (structure function method)
  one is able to add the effects due to emission of the second photon and further photons;
  however, this can improve the precision only within the leading logarithmic scope
  and makes the introduction of the realistic cuts even harder.
\end{itemize}
The other important role of the semianalytical calculations is to provide
the numerical tests of the Monte Carlo programs.
They are typically used to check technical correctness
of the phase space integration,
the so-called technical precision,
and also correctness of the implementation of the SM matrix element.
In the following we shall see examples of both kinds of tests.
In particular, we shall see the test of the technical precision
of \KK MC at the $2\cdot 10^{-4}$ level based of the semianalytical formula
obtained in this section, in the case of a single kinematical cut on the
total energy of all photons.

The role of semianalytical calculations as a test of the Monte Carlo programs
is important but limited.
One can easily imagine the situation in which the numerical problem shows up
not for simple kinematical cuts, for which the ``calibration'' with help
of the semianalytical program has been done,
but for more realistic and complicated cuts.
It is therefore always true that the {\em ultimate test} of the MC calculation
is always the comparison of two MC programs,
because it can be done for arbitrary cut-offs%
\footnote{
  This kind of test was for instance done for the first modern \Order{\alpha^1}
  Monte Carlo event generator MUSTRAAL of ref.~\cite{mustraal-cpc:1983,mustraal-np:1983},
  with the very high precision at that time of 1\%.}.
This recipe may look often prohibitively expensive in the effort required to realize it;
however, at the sub-permille precision level the amount of work required
for the realization of the semianalytical formulas and testing the semianalytical program
is probably comparable.
So altogether, for the sub-permille precision prediction the approach
with two independent MC programs seems to be the most economical one.
In any case the semi-analytical calculations will be always very useful
especially when the precision requirements are not excessive and we do not deal
with the observables involving complicated experimental cuts.

%%%%%%%%%%%%%%%%%%%%%%%%%%%%%%%%%%%%%%%%%%%%%%%%
\subsection{Inclusive exponentiation, IEX}
\label{sec:iex}

An important ingredient in many semi-analytical calculations is the ``exponentiation''.
The meaning and the technique of exponentiation in the context of the semi-analytical calculations
is however different from the exclusive exponentiation discussed in most of this paper.
Let us elaborate more on what the exponentiation really means in the semianalytical approach.

As already discussed in ref.~\cite{sussex:1989}, in the typical semianalytical approach
one is practicing what we call an
``ad-hoc exponentiation'' or ``naive exponentiation'' procedure
in which one takes the QED finite-order, let us say \Order{\alpha} or
\Order{\alpha^2}, analytical result for a certain
one- or two-dimensional {\em inclusive} distribution and this result 
is ``improved'' by hand,
in such a way that the soft limit (the limit in which hard photons are eliminated)
in the resulting distribution conforms to the Yennie-Frautschi-Suura work \cite{yfs:1961}.
The well known examples of the ad-hoc exponentiation
are presented in refs.~\cite{jackson-sharre:1975,kuraev-fadin:1985} 
and later in ref.~\cite{third-order:1991} for the initial-state
bremsstrahlung in $e^+e^-$ annihilation.
There are also many other examples of the ad-hoc exponentiation, including calculations for
the deep inelastic and Bhabha scattering processes.
The ad-hoc exponentiation procedure may get improved gradually by taking
into account higher-order effects.
For example, the \Order{\alpha} procedure of ref.~\cite{jackson-sharre:1975}
was extended to \Order{\alpha^2} in ref.~\cite{kuraev-fadin:1985}
and later to \Order{\alpha^3} in ref.~\cite{third-order:1991}.
The problem is that the procedure of ad-hoc exponentiation is essentially
rather art than science -- one may regard it at best 
as a ``by hand interpolation'' between two kinds of analytical formulas -- 
one valid in the soft photon limit and another
in the hard photon limit, see ref.~\cite{sussex:1989} for more detailed discussion
on this interpretation.
Since this approach is not systematic, 
it is therefore difficult to estimate the uncertainty of the results
and it has to be ``reinvented'' for each perturbative order
and for each inclusive observable again and again.

The self-suggesting question is therefore:
{\em Is there any better and more systematic way of re-formulating the 
ad-hoc exponentiation for any inclusive distributions in the analytical form?}
I would be also desirable to find a direct connection
to the {\em exclusive exponentiation} YFS exponentiation (of the EEX or CEEX type) 
which is discussed and implemented in this work.
There is an obvious hint in which direction to go.
If all soft photons are soft, then we know exactly the analytical formula
for the multi-photon phase space integral, 
\begin{equation}
\begin{split}
f(\gamma,V) &=
 e^{\gamma\ln\varepsilon}
\sum_{n=0}^\infty {1\over n!}
\prod_{i=1}^n \;
\int\limits_{k_0 > 2\varepsilon\sqrt{s}/2}\;
      {d^3k_i \over k^0_i}
      \tilde{S}(p_1;p_2;k_i)\;
\delta\bigg( V -
     {1\over s} (p_1+p_2)\cdot \bigg( \sum_{i=1}^n k_i  \bigg)
      \bigg)
\nonumber \\
& = {e^{-C\gamma} \over \Gamma(1+\gamma)} \gamma V^{\gamma-1}
  = F(\gamma) \gamma V^{\gamma-1},\;\; C=0.57721566...,
\label{soft-integral}
\end{split}
\end{equation}
already obtained in the original YFS paper~\cite{yfs:1961}.
So why not include hard photons in the game?
Let us therefore define the ``YFS inclusive exponentiation''
as a result of the analytical phase space integration of the distributions
of the YFS exclusive exponentiation:
\\
\framebox{\parbox{5.5cm}{\em YFS inclusive exponentiation} } $\equiv$
\framebox{\parbox{5cm}{\em Analytical integration of YFS multiphoton integrals. }  }
\\
In this way we have got
a clear and clean connection between the YFS exponentiation (as implemented in the Monte Carlo)
and the ``YFS inclusive exponentiation'' --
the connection is simply the analytical phase-space integration!
As a result, we do not need any obscure ``recipes'' any more and what we only need to know
is how to integrate (analytically) the phase-space.

The above looks promising but 
the YFS exclusive exponentiation involves non-trivial integrals over
the multiphoton phase-space, this is why it is
implemented in the form of the Monte Carlo integration/simulation numerical program.
The relevant integrals over $n$ real photon phase space
do not seem at first sight to be treatable analytically at all.

The situation is not so hopeless, however, as it may look at first sight,
and in the following we shall present the solution.
We start, as promised, from the full YFS expression, the same as in the
Monte Carlo and are able to do ``the impossible'' ---
that is to perform the phase-space integral analytically.
We calculate the phase-space integral approximately.
We shall do it, however, in such a way
that in the approximate method becomes {\em exact} in the soft photon limit,
The soft photon contributions are there integrated exactly 
and only the remaining ``non-infrared''
contribution will be calculate using approximate methods, 
typically the leading-logarithmic (LL) collinear approximation.
The LL approximations in non-IR parts may concern both the phase-space parametrization
and the matrix element%
\footnote{ 
  Let us note that the LL evaluation of the
  phase-space integral was already employed to some extent
  in the original  YFS work \cite{yfs:1961}.
  At that time, because of lack of fast computers, it was the only accessible method.}.

The profit from the above approach is two-fold: 
Contrary to the traditional ad-hoc exponentiation we gain,
for a given exponentiated inclusive distribution,
a clear and clean connection between the YFS exponentiation (as implemented in the Monte Carlo)
and the ``YFS inclusive exponentiation'',
we do not need any obscure ``recipes'' any more.
It means that the resulting inclusive exponentiation (IEX)
is now a systematic order-by-order procedure, this feature is simply inherited
from the exclusive YFS exponentiation.

As already stressed, the inclusive YFS exponentiation will never
fully replace the YFS Monte Carlo because
it is possible to deal analytically with only a very limited
number of the distributions, without cuts or with very
simple cuts, while in the Monte Carlo one may calculate
an arbitrary distribution in the presence of the most complicated cuts.

In the following subsections we shall show explicitly the analytical
integrations leading to \Ordpr{\alpha^2} IEX results.
We shall typically compare the Monte Carlo with EEX matrix element and
IEX formulas, both in the \Ordpr{\alpha^2} class.
Their difference will be then necessarily  of \Ordpr{\alpha^3},
i.e. up to factor 10 smaller -- quite a strong test of both calculations.
On one occasion, we shall go to a more difficult level of the \Ordpr{\alpha^3},
in which case the difference between MC and IEX is of order \Ordpr{\alpha^4}.

Finally, let us note that the set of IEX formulas presented in this section
was used over many years as a basic test of the precision of the YFS2~\cite{yfs2:1990}
and KORALZ/YFS3\cite{yfs3-pl:1992,koralz4:1994} programs.
Some of the IEX results were already shown in ref.~\cite{yfs2:1990} and \cite{sussex:1989}.
Most of them are, however, shown here for the first time.
In the mean time the analogous set of IEX results was obtained and published
for the $t$-channel dominated process~\cite{bhlumi-semi:1996}.
In fact ref.~\cite{bhlumi-semi:1996} shows an even more sophisticated case 
in the \Ordpr{\alpha^3} class than the \Ordpr{\alpha^2} results presented here.
Using the experience of ref.~\cite{bhlumi-semi:1996} it would be definitely possible,
for the $s$-channel process of this paper, to upgrade systematically the calculation
of this section to \Ordpr{\alpha^3}, both for ISR and FSR.

%%%%%%%%%%%%%%%%%%%%%%%%%%%%%%%%%%%%%%%%%%%%%%%%%%%%%%%%%%%%%%%%%%%%%%%%%%%
\subsection{Semi-analytical formulas for ISR}
%%%%%%%%%%%%%%%%%%%%%%%%%%%%%%%%%%%%%%%%%%%%%%%%%%%%%%%%%%%%%%%%%%%%%%%%%%%
We shall start the construction in IEX expressions with the ISR case,
first showing the basic techniques working out the example 
with the \Order{\alpha^0} EEX matrix element.
In this case the multiphoton differential distribution is just the Born 
cross section times real soft-factors.
While for the other IEX formulas the
phase space will be  integrated basically in the \Ordpr{\alpha^2},
the case of the of \Oeex{\alpha^0} we shall make more effort and do it
in the \Ordpr{\alpha^3}, like in  ref.~\cite{bhlumi-semi:1996}.
Let us attract attention of the reader to the fact that we have the matrix element
in the \Oeex{\alpha^0} and the phase space integration is in \Ordpr{\alpha^2}
or \Ordpr{\alpha^3}.
There is no contradiction in this, as we shall see in the following.

%%%%%%%%%%%%%%%%%%%%%%%%%%%%%%%%%%%%%%%%%%%%%%%%%%%%%%%%%%%%%%%%%%%%%%%%%%%
\subsubsection{Baseline high precision results for \Oeex{\alpha^{(0)}}}
%%%%%%%%%%%%%%%%%%%%%%%%%%%%%%%%%%%%%%%%%%%%%%%%%%%%%%%%%%%%%%%%%%%%%%%%%%%
The complete \Ordpr{\alpha^2} calculation/exponentiation  
according to the rules layed down in the beginning of this Section will be rather involved,
let us therefore illustrate our calculational methods with the simplest possible example.
Even this simple example features some nontrivial technical features
and we shall therefore present two versions of the calculation.
 
The basic example discussed in the following is the
\Order{\alpha^0} initial-state YFS inclusive exponentiation.
In the master equation (\ref{rho-eex2}) we set the charge of
the final fermion to zero, $Q_f=0$
and we replace the sum of $\bbeta$'s with the \Order{\alpha^0}
version of $\bbeta_0$ proportional to the Born differential cross section
\begin{equation}
\bbeta_0^{(0)}(q_1,q_2)= {2\over \beta_f}\;
          {d\sigma^\born \over d\Omega}
             \Big((q_1+q_2)^2,\vartheta \Big),\quad
\beta_f=\big(1-4m_f^2/(q_1+q_2)^2\big)^{1/2},
\label{beta-zero-zero}
\end{equation}
where the normalization is such that
\begin{equation}
\int {d^3q_1\over q_1^0} {d^3q_2\over q_2^0}\;
   \delta^{(4)}(X-q_1-q_2)\;
    \bbeta_0^{(0)}(q_1,q_2)
   = \sigma^\born\Big((q_1+q_2)^2\Big).
\end{equation}
The initial-state \Order{\alpha^0} YFS formula reads
\begin{equation}
\begin{split}
\sigma_0&= \sum_{n=0}^\infty {1\over n!}\;
   \int
   {d^3q_1\over q_1^0} {d^q_2\over q_2^0}\;
   \prod_{i=1}^n\;
   \int\;    {d^3 k_i \over k_i^0}
     \tilde{S}_I(k_i)
     \Theta\Big(k^0_i - {1\over 2} \varepsilon \sqrt{s}\Big)
\\
&\qquad\qquad
   \delta^{(4)}(p_1+p_2-q_1-q_2-\sum_j k_j)\;
   e^{Y_I(\varepsilon)}\;
   \bbeta_0^{(0)} \Big( (q_1+q_2)^2,\vartheta_0 \Big)
\end{split}
\end{equation}
Integration over the final-state fermion 2-body phase space  is done trivially leading to
\begin{equation}
\sigma_0= \int_0^1 dv\;
\sigma^\born (s(1-v)) e^{\delta_{YFS}}\;
\rho_0(v)
\end{equation}
where the essential multiphoton integral
\begin{equation}
\rho_0(v) = e^{\gamma\ln\varepsilon}
   \sum_{n=0}^\infty  {1\over n!}\;
   \int {d^3 q_1 \over q_1^0} {d^3q_2 \over q_2^0}
   \prod_{i=1}^n\;
   \int\limits_{k^0_i > \varepsilon \sqrt{s}/2 }\;
   {d^3 k_i \over k_i^0}
   \tilde{S}_I(k_i)\;
   \delta\Big(1 - v - {1\over s} (p_1+p_2-\sum_j k_j)^2 \Big)
\end{equation}
is the main object of our interest.
Note that we have split
$Y_I(\varepsilon)= \gamma\ln\varepsilon + \delta_{YFS}$.
 
The QED matrix element, beyond the soft photon integral,
is in this simplified case totally absent.
The inclusive YFS exponentiation, as defined above,
amounts to calculating analytically
the multiphoton phase-space integral for $\rho_0(v)$.
As explained above, we shall do it in \Ordpr{\alpha^2},
but we shall keep the proper soft limit undestroyed.
Let us note first that in the soft limit $v \rightarrow 0$
the function $\rho_0(v)$ coincides with the soft integral of eq.~(\ref{soft-integral}), i.e.
$\rho (v) \rightarrow  f(\gamma,v) $.
Since the most singular part in this limit is known
we isolated it and we expect the \Ordpr{\alpha^2} result to be in form
\begin{equation}
  \rho_0(v)= f(\gamma,v)  (1 + v\gamma f_1(v))
\end{equation}
where $f_1(v)$ is nonsingular. How does one find the function $f_1(v)\;$?
Let us inspect the difference
\begin{equation}
\begin{split}
d_0(v) &= \rho(v)-f(\gamma,v)\\
       &= {1\over 2!}
 \int {d^3 k_1 \over k_1^0} \tilde{S}_I(k_1)
 \int {d^3 k_2 \over k_2^0} \tilde{S}_I(k_2)
\\
\qquad &
 \Bigg[
   \delta\Big(1 - v - {1\over s} (p_1+p_2-\sum_j k_j)^2 \Big)
   -\delta\bigg( v -
     {1\over s} (p_1+p_2)\cdot \bigg( \sum_{i=1}^n k_i  \bigg)
      \bigg)
 \Bigg]
\end{split}
\end{equation}
This new object has rather interesting properties.
First of all, the \Order{\alpha^1} integrals cancel exactly and
the first nontrivial integral is of \Order{\alpha^2}.
This second-order integral is not, however, infrared divergent!
According to our rules we are therefore allowed,
without any danger of spoiling the soft limit,
to calculate it in the leading-logarithmic approximation.
 
Let us present now our first of two methods of calculating $\rho_0(v)$.
In the LL approximation we replace collinear singularities in
the photon angle $\vartheta_\gamma=0,\pi$ by $\delta$-like peaks
\begin{equation}
\begin{split}
 \int {d^3 k_i \over k_i^0} \tilde{S}_I(k_i)
&=
 { \alpha \over 2\pi^2 }
  \int\limits_0^1 {dx_i \over x_i}\;
  \int\limits_{-1}^1  dc_i {s_i^2 \over (1-\beta_e^2c_i^2)^2 }
  \int\limits_0^{2\pi} d\phi_i
\\
&\longrightarrow
  \int\limits_0^1 {dx_i \over x_i}\;
  \int dc_i\;
  \Bigg[
   {1\over 2} \gamma \delta(c_i-1)
  +{1\over 2} \gamma \delta(c_i+1)
   \Bigg]
\end{split}
\end{equation}
where
$$
\beta_e= (1-4m_e^2/s)^{1/2},\;
c_i=\cos\theta_i,\; s_i=\sin\theta_i,\; i=1,2,
$$
and using the above LL substitution we get
\begin{equation}
d_0(v)=
  \lim_{\varepsilon \rightarrow 0}
{ \gamma^2\over 4}
\int\limits_\varepsilon^1 {dx_1 \over x_1}
\int\limits_\varepsilon^1 {dx_2 \over x_2}
\Big[ \delta\Big(v-(1-x_1)(1-x_2)\Big)
     -\delta(v-x_1-x_2)\Big]
\end{equation}
Two immediate remarks are in order: Out of four terms in the product
$[\delta(c_1-1)+\delta(c_1+1)][\delta(c_2-1)\delta(c_2+1)]$
only two contribute, these with
two anticollinear photons, $c_1=1,c_2=-1$ and $c_1=-1,c_2=1$.
The result of integration depends critically on the careful
regularization and for this reason we keep
explicitly the $\varepsilon$ infrared regulator.
A quick calculation gives a zero value for the integral.
The very similar phenomenon is present in the
calculation of $f(\gamma,v)$ where a naive calculation
up to second-order gives the $v^{\gamma-1}$ factor only.
The remaining $F(\gamma)=1 -{\pi^2\over 12}\gamma^2+...$
factor comes from careful consideration of the
$k^0>\varepsilon \sqrt{s}/2$ condition for two photons.
With our proper regularization we obtain
\begin{equation}
d_0(v)
= -{\gamma^2 \over 4}\;  {\ln(1-v) \over v},
\end{equation}
which is finite in the $v\rightarrow 0$ limit.
 
Now comes the second calculation method which will be often employed in the following.
In this variant we take into account the influence of additional soft photons
(in addition to the two hard ones).
They do not change the second-order result but provide the proper infrared
regulation replacing the former $\varepsilon$ regulator.
The LL treatment of the phase-space will be a little bit different.
Starting from eq.~(\ref{soft-integral}) we split (in CMS frame) the photon
integration into its forward and backward hemisphere parts
$$
\int {d^3k\over k^0} =
 \int\limits_{\theta>\pi/2}\; {d^3k\over k^0}
+\int\limits_{\theta<\pi/2}\; {d^3k\over k^0}
$$
and after changing the summation order we get
\begin{equation}
\begin{split}
f(\gamma,v) &=
e^{\gamma\ln\varepsilon}
\sum_n \sum_{n'} {1\over n!} {1\over n'!}
\prod_{i=1}^n\;    \int\limits_{\theta_i>\pi/2}
          {d^3 k_i^+ \over k_i^{+0} }
          \tilde{S}_I(k_i^+)
          \Theta\bigg(k_i^{+0} -\varepsilon {\sqrt{s}\over 2}\bigg)
\\
&\prod_{j=1}^{n'}\; \int\limits_{\theta_j<\pi/2}
          {d^3 k_j^- \over k_j^{-0} }
          \tilde{S}_I(k_j^-)
          \Theta\bigg(k_j^{-0} -\varepsilon {\sqrt{s}\over 2}\bigg)\;
\delta\Big( v- {2\over s} P\cdot (K^+ + K^-)  \Big)
\end{split}
\end{equation}
where
$P=p_1+p_2,\; K^+= \sum\limits_{i=1}^n k_i^+,\;
              K^-= \sum\limits_{j=1}^{n'} k^-_j$.
The above sum of integrals factorizes into two sums.
Each of the sums can be evaluated exactly
leading to the following simple convolution
\begin{equation}
f(\gamma,v)=\int dv_+ dv_-
\delta(v-v_-  -v_+) f\Big({\gamma\over 2},v_+\Big)
                    f\Big({\gamma\over 2}, v_-\Big),\;
\end{equation}
This identity holds for the integration result anyway, but we have also
obtained it through the direct phase-space integration.
So far all calculations were exact and we only reorganized
the phase-space integration which will be useful in the next step.
Let us consider the $d_0(v)$ difference again
\begin{equation}
\begin{split}
d_0(v) &=
e^{\gamma\ln\varepsilon}
\sum_n \sum_{n'} {1\over n!} {1\over n'!}
\prod_{i=1}^n\;    \int\limits_{\theta_i>\pi/2}
          {d^3 k_i^+ \over k_i^{+0} }
          \tilde{S}_I(k_i^+)
          \Theta\bigg(k_i^{+0} -\varepsilon {\sqrt{s}\over 2}\bigg)
\\
&\prod_{j=1}^{n'}\; \int\limits_{\theta_j<\pi/2}
          {d^3 k_j^- \over k_j^{-0} }
          \tilde{S}_I(k_j^-)
          \Theta\Big(k_j^{-0} -\varepsilon {\sqrt{s}\over 2}\Big)
\Bigg[
\delta\Big( v - 1 +{1\over s} (P -K^+ -K^- )^2  \Big)
\\
&\qquad\qquad -\delta\Big( v- {2\over s} P\cdot (K^+ + K^-)  \Big)
\Bigg].
\end{split}
\end{equation}
As before, the whole integral is finite in $v \rightarrow 0$
limit and it gets the first non-zero contribution in the second-order.
From the previous exercises we know that the essential
second-order leading-logarithmic contribution comes from
two anticollinear photons -- this is why we divided
photon phase-space into two hemispheres.
Now, the LL approximation is realized by substituting in the first $\delta$
$$
K^{\pm\mu} = (K^{\pm 0},0,0,\pm |K^{\pm 0}|\;).
$$
Note that, contrary to the previous calculation, we did not modify the $\tilde{S}$ factors,
we did not introduce collinear $\delta$'s in the photon angular distribution
and we keep infinite numbers of photons.
In spite of the apparent increase of the complication level,
the integral reduces to a nice form
\begin{equation}
f(\gamma,v)=\int dv_+ dv_-
\Big[
  \delta(v-v_-  -v_+  +v_+ v_-)
- \delta(v-v_-  -v_+)
\Big]
      f\bigg({\gamma\over 2}, v_+\bigg)
      f\bigg({\gamma\over 2}, v_-\bigg),\;
\end{equation}
which is calculable analytically!
Neglecting terms \Order{\gamma^3} we obtain
\begin{equation}
d_0(v)
= - { e^{-C\gamma}\over \Gamma(1+\gamma) }
    \gamma v^{\gamma-1}\;{1\over 4}\; \gamma \ln (1-v).
\end{equation}
Note that since in the present variant of the calculation
we have treated soft photons more friendly we recovered the proper soft factor
$f(\gamma,v)$ as a factor in the solution.
 
%/////////////////////////////////////////////////////////////////////////////////////////
%-----------------------------------------------------------------------------------------
\begin{table}[!ht]
\centering
\setlength{\unitlength}{0.1mm}
\begin{picture}(800,800)
\put(   500,600){\makebox(0,0)[b]{\Large $\bbeta^{(2)}_{0,I}\otimes\bbeta^{(2)}_{0,F}$ }}
\put(   400,  0){\makebox(0,0)[b]{\epsfig{file=flat_bt0xbt0.eps,width=80mm,height=80mm}}}
\end{picture}
%%%%%%%%%%%%%%%%%%%%%%%%%%%%%%%%%%%%%%%%%%%%%%%%
\caption{\small\sf
 The comparison between \KK MC and IEX \Ordpr{\alpha^2} formula
 of eq.~(\protect\ref{eq:IEX-bt0xbt0}) for the constant Born cross section. (200GeV?)
 The difference between \KK MC in EEX mode and semianalytical formula divided by Born
 is plotted with dotted line, 
 as a function of the $v_{\max}$ cutoff on the total energy of all ISR and FSR photons.
 We also include the integrated cross section divided by Born,
 and multiplied by factor $10^{-2}$, as dots for IEX and a line for MC.
}
\label{fig:flat-bt0xbt0}
\end{table}
%-----------------------------------------------------------------------------------------
Summarizing, the \Ordpr{\alpha^2} phase-space integration result is
\begin{equation}
\rho_0(v)
= { e^{-C\gamma}\over \Gamma(1+\gamma) }
  \gamma v^{\gamma-1} \Big(1 -{1\over 4} \gamma \ln (1-v) \Big)
\end{equation}
and the corresponding cross section reads
\begin{equation}
\sigma_0(v_{max}) =
 e^{\delta_{YFS}}\;
 { e^{-C\gamma}\over \Gamma(1+\gamma) }
\int\limits_0^{v_{max}} dv\;
\sigma^\born (s(1-v))
  \gamma v^{\gamma-1} \Big(1 -{1\over 4} \gamma \ln (1-v) \Big)
\label{sigma-zero}
\end{equation}
The above integration methods provide us with the \Ordpr{\alpha^2} phase-space integration 
result for any of the $\bbeta_0$ contributions as listed in eqs.~(\ref{beta0}).
For example the contribution from $\bbeta_0^{(2)}$ reads
%%%
\begin{equation}
\label{eq:IEX-beta02}
\begin{split}
\sigma_0^{(2)} &= \int\limits_0^1 dv\; \sigma^\born\big(s(1-v)\big) \rho_0^{(2)},
\\
  \rho_0^{(2)} &= F(\gamma)
  e^{\delta_{YFS}}\; \gamma v^{\gamma-1} (1+\delta^{(1)}_I+\delta_I^{(2)} )
  \Big(1  -{1\over 4} \gamma \ln (1-v) \Big)
\end{split}
\end{equation}
From now one we shall not restrict ourselves to \Ordpr{\alpha^0} EEX matrix elements,
but rather consider the complete EEX-class \Ordpr{\alpha^2} EEX matrix elements 
as defined in Section \ref{sec:eex}.

%/////////////////////////////////////////////////////////////////////////////////////////
%-----------------------------------------------------------------------------------------
\begin{figure}[!ht]
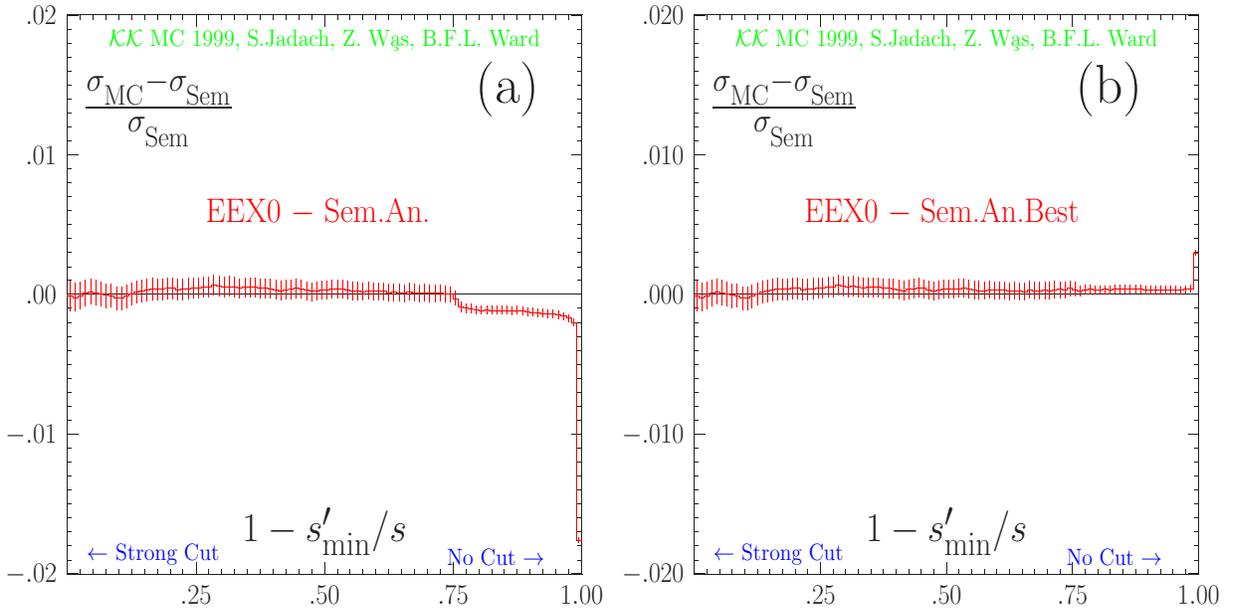

\centering
\setlength{\unitlength}{0.1mm}
\begin{picture}(1600,800)
%\put( 0,0){\framebox( 1600,800){ }}
\put( 650,650){\makebox(0,0)[b]{\LARGE (a)}}
\put(1450,650){\makebox(0,0)[b]{\LARGE (b)}}
\put(  -20, 0){\makebox(0,0)[lb]{\epsfig{file=chi-mcan-O0dif-vmax1.eps,width=80mm,height=80mm}}}
\put(  800, 0){\makebox(0,0)[lb]{\epsfig{file=chi-mcan-O0tech-vmax1.eps,width=80mm,height=80mm}}}
\end{picture}
%----------------------------------------
\caption{\small\sf 
 The comparison between \KK MC and IEX \Ordpr{\alpha^2} formula
 of eq.~(\protect\ref{eq:IEX-bt0xbt0}) for the $s$-dependent Born cross section at 189GeV.
 The difference between \KK MC in EEX mode and IEX formula divided by IEX
 is plotted as a function of the $v_{\max}$ cutoff on the total energy of all ISR and FSR photons.
}
\label{fig:bt0xbt0-prag2-3}
\end{figure}
%----------------------------------------------------------------------------------------

The practical significance of IEX formula of eq.~\ref{eq:IEX-beta02} is rather important.
The biggest terms neglected in it are of \Order{\gamma^3} and \Order{\alpha\gamma} and
we expect them to stay below 0.1\%.
(This will be true provided they are no extra enhancement factors, see discussion below.)
In other words we expect for the $\bbeta_0^{(2)}$ contribution
in the EEX matrix element in Section \ref{sec:eex} the result of the Monte Carlo phase-space
integration will agree with the formula
(\ref{eq:IEX-beta02}) to within about 0.1\% for an arbitrary cut $v_{max}$.

Let us check the above conjecture with the numerical exercise.
In the numerical test we shall already include at this moment not only
ISR $\bbeta_0^{(2)}$ contribution of eq.~(\ref{eq:IEX-beta02}),
see also Table~\ref{table-initial},
but also the analogous FSR $\bbeta_0^{(2)}$ 
contribution which will be calculated%
\footnote{
  We could present results of the numerical tests (which we have done) for ISR alone.
  They look however very much the same like as simultaneous ISR and FSR
  so we decided not to present their figures.}
in the next sub-section, see eq.(\ref{eq:IEX-FSR}) and Table~\ref{table-final}.
We consider the total cross section with the cut on the total photon energy
defined by $v_{max}$ as follows
%/////////////////////////////////////////////////////////////////////////
\begin{equation}
\label{eq:IEX-bt0xbt0}
    \sigma^{(2)}_{\bbeta^{(2)}_0\otimes\bbeta^{(2)}_0}
   = \int_0^{v_{\max}} dv\;
     \int_0^{v/(1-u_{\max})} du\;
     \sigma^f_{\rm Born}\big(s(1-u)(1-v)\big)\;
     \rho^{(2)}_{I\bbeta^{(2)}_0}(v)\; 
     \rho^{(2)}_{F\bbeta^{(2)}_0}(u)
\end{equation}
In order to get a clearer picture about the magnitude of the discrepancy between
EEX MC and IEX formula we use the artificially flat Born cross-section
$\sigma^f_{\rm Born}(s(1-u)(1-v))\to \sigma^f_{\rm Born}(s)$ in both.
Results of the comparison are presented in Figure~\ref{fig:flat-bt0xbt0}.
Following our expectation the difference is well below 0.1\%
for the entire range of the photon energy cutoff $v_{\max}$.

The situation does not look as good when we switch-on the $s$-dependence in Born cross-section.
In Figure~\ref{fig:bt0xbt0-prag2-3}(a) we see the relevant comparison.
At the CMS energy of 189GeV the position of the Z radiative return is at $v=0.75$
and we clearly see a worsening there with respect to the previous case in Figure~\ref{fig:flat-bt0xbt0}
where the discrepancy is now almost 0.2\% (0.4\% in terms of $\sigma_{\rm Born}$).
The situation is even more dramatic in the last bin which corresponds to 
$v_{\max}=1-4m_\mu^2/s$ and here the discrepancy among \Ordpr{\alpha^2} IEX
and \Ordpr{\alpha^2} MC EEX is -2\% of the total cross section,
that is -7\% in terms of the Born cross-section!
This is, of course due to the $Z$ resonance and $1/s$ behaviour of the Born cross-section at very
low $s$ 
(especially for the case of the $\mu$ channel shown in Figure~\ref{fig:bt0xbt0-prag2-3}).
In order to be sure that the above effect is not due to some technical problem in the MC integration
we had to improve our IEX formula and upgrade the analytical phase space integration for ISR
to the level of \Ordpr{\alpha^3}.
The comparison with the \Ordpr{\alpha^3} IEX for the same EEX \Ordpr{\alpha^2} MC
we show in Figure~\ref{fig:bt0xbt0-prag2-3}(b).
Now the difference is reduced to less than 0.1\% everywhere, an in the last bin it is reduced
from -2\% to +0.2\%, as expected.
The additional terms of the \Order{L^1\alpha^2} and \Order{L^3\alpha^3}
are shown in Table~\ref{table-initial} at the end of this Section.
We do not show the details on the phase space integration which provide these two additional terms.
The method is generally rather 
similar to the one used in this Section and in ref.~\cite{bhlumi-semi:1996}.

%%%%%%%%%%%%%%%%%%%%%%%%%%%%%%%%%%%%%%%%%%%%%%%%%%%%%%%%%%
\subsubsection{Beta-bar-one, $\bbeta_1$}
%%%%%%%%%%%%%%%%%%%%%%%%%%%%%%%%%%%%%%%%%%%%%%%%%%%%%%%%%%

In the following step our aim is to calculate analytically the ISR contribution
to the total cross section from $\bbeta^{(2)}_{1I}$ as given by
\begin{equation}
\begin{split}
& \sigma=
\sum_{n=0}^\infty  {1\over n!}
\int {d^3q_1\over q_1^0} {d^3q_2\over q_2^0}
\int \prod_{j=1}^n  {d^3k_j\over k^0_j}
                   \tilde{S}_I(p_1,p_2;k_j) (1-\Theta(\Omega_I;k_j))
\\& \qquad
e^{Y(\Omega_I)}
\sum_{j=1}^n
    \bbeta^{(2)}_{1I}(X,p_1,p_2,q_1,q_2,k_j)/\tilde{S}_I(k_j)
\delta^{(4)}\bigg( p_1+p_2 -q_1-q_2 -\sum_{j=1}^n k_j \bigg)
\\& =
\sum_{n=0}^\infty  {1\over n!}
    \int {d^3q_1\over q_1^0} {d^3q_2\over q_2^0}
    \int\prod_{j=1}^n  {d^3k_j\over k^0_j} \tilde{S}_I(p_1,p_2;k_j) 
    (1-\Theta(\Omega_I;k_j)) e^{Y(\Omega_I)}
\\& \qquad
    {d^3k\over k^0} (1-\Theta(\Omega_I;k))
    \delta^{(4)} \bigg( p_1+p_2 -q_1-q_2 -k-\sum_{j=1}^n k_j \bigg)
    \bbeta^{(2)}_{1I}(q_1+q_2,p_1,p_2,q_1,q_2,k).
\end{split}
\end{equation}
We start again from the EEX \Ordpr{\alpha^2} matrix element for
the initial-state bremsstrahlung and we shall perform
the  phase-space integration also in \Ordpr{\alpha^2}.
We integrate first over final-state fermion four-momenta
\begin{equation}
  \begin{split}
&   \int {d^3q_1\over q_1^0} {d^3q_2\over q_2^0}
    \delta^{(4)}\Big(  p_1+p_2 -q_1-q_2 -k   \Big)
    \bbeta^{(2)}_{1I}(q_1+q_2,p_1,p_2,q_1,q_2,k)
\\& \qquad
    = B_1^{(2)}(p_1,p_2,k)\; \sigma^\born((q_1+q_2)^2)
  \end{split}
\end{equation}
where
\begin{equation}
  \begin{split}
    B_1^{(2)}(p_1,p_2,k) =&
    {\alpha \over 4\pi^2} {2p_1p_2\over (kp_1)(kp_2)} W_e(\halp,\hbet)
    (1+\Delta^{(1)}_I(\halp,\hbet))
    {1\over 2} \Big\{ (1-\halp)^2 +(1-\hbet)^2  \Big\}
\\&
    -\tilde{S}_I(p_1,p_2,k) (1+\delta_I^{(1)})
  \end{split}
\end{equation}
and obtain
\begin{equation}
  \begin{split}
    \sigma &= \int\limits_0^1 dv
    e^{\delta_{YFS} + \gamma\ln\varepsilon }
    \sum_{n=0}^\infty  {1\over n!}
    \int\limits_{k^0_j>\varepsilon {\sqrt{s}\over 2}}\;
    \prod_{j=1}^n  {d^3k_j\over k^0_j} \tilde{S}_I(p_1,p_2;k_j)\;
    \int\limits_{k^0_j>\varepsilon {\sqrt{s}\over 2}}\;
    {d^3k\over k^0}
\\& \qquad
    B_1^{(2)}(p_1,p_2,k)\;
    \sigma^\born(s(1-v))\;
    \delta\Big(v-{1\over s} (P-\sum_j k_j -k)^2\Big)
\\&
    \equiv\int\limits_0^1 dv \rho_1^{(2)}(v) \sigma^\born(s(1-v))
  \end{split}
\end{equation}
In the calculation of $\rho_1^{(2)}$ we could follow
the first of the methods employed for $\bbeta_0$.
Let us describe it briefly without going into details of the calculation.
We calculate the first two non-zero integrals \Order{\alpha} and \Order{\alpha^2}.
The first one \Order{\alpha} has to be calculated
keeping both the leading \Order{L\alpha} and the subleading term \Order{L^0\alpha}.
This can be done following the well known \Order{\alpha} 
analytical calculations \cite{berends-kleiss:1981}.
The \Order{\alpha^2} integral with two real photons
can be treated in LL approximation, i.e. keeping only \Order{L^2\alpha^2} terms.
This can be done introducing collinear peaks
in the photon angles as demonstrated in the case of $\bbeta_0$.
Both integrals are connected due to infrared regulation with $\varepsilon$.
The first one is proportional to
$e^{\gamma\ln\varepsilon} \simeq 1+\gamma\ln\varepsilon$
and the term $\gamma\ln\varepsilon$ from the first one
cancels the infrared divergence in the second (independently of the LL approximation).
As in the case of $\bbeta_0$ one has to pay attention
to the subtle ``edge effects'' in the $\varepsilon$ regularization%
\footnote{
  Generally, the calculation for $\bbeta_1$ is more difficult
  than for $\bbeta_0$ and $\bbeta_2$ because this is
  the only case in \Order{\alpha^2} where we deal with simultaneous
  real and virtual photon emission.
}.
 
Let us describe in detail the second method in which soft photons provide the
convenient infrared regulation.
The main \Order{\alpha} contribution comes from the
configuration in which we have $k^0\simeq v\sqrt{s}/2$ and one or more soft photons.
This part has to be calculated exactly in \Order{\alpha}. 
We split, as before,
\begin{equation}
  \rho_1^{(2)}(v)  = f_1^{(2)}(v) + d_1^{(2)}(v)
\end{equation}
in such a way that $d_1^{(2)}(v)$ is vanishing in \Order{\alpha} -- 
it can be therefore calculated in second-order LL while $f_1^{(2)}(v)$ is simple enough
to be calculated exactly in the \Order{\alpha}.
We define
%%%
\begin{equation}
  \begin{split}
    &f_1^{(2)}(v) =
    e^{\delta_{YFS}} \int {d^3k\over k^0}
    e^{\gamma\ln\varepsilon}
    \sum_{n=0}^\infty  {1\over n!}
    \int\limits_{k^0_j>\varepsilon {\sqrt{s}\over 2}}\;
    \prod_{j=1}^n  {d^3k_j\over k^0_j}
                   \tilde{S}_I(k_j)\;
\\ \qquad
    &\delta\Big(v-{2\over s} P\cdot(\sum_j k_j +k)\Big) B_1^{(1)}(p_1,p_2,k)
    =e^{\delta_{YFS}} \int {d^3k\over k^0}
    f\Big(\gamma,v-{2\over s}P\cdot k\Big)
    B_1^{(1)}(p_1,p_2,k)
  \end{split}
\end{equation}
where
\begin{equation}
  B_1^{(1)}(p_1,p_2,k)=
     {\alpha \over 4\pi^2} {2p_1p_2\over (kp_1)(kp_2)} W_e(\halp,\hbet)
     {1\over 2}  \Big\{ (1-\halp)^2 +(1-\hbet)^2  \Big\} - \tilde{S}_I(p_1,p_2,k).
\end{equation}
The remarkable feature of  $f_1^{(2)}$ is that we could integrate over spectator photons exactly.
Note that the  $\varepsilon$ regulator has disappeared from the $k-$integral.
In the next step we integrate {\em exactly} over photon angles following 
the old \Order{\alpha} calculations
and we are left with the single integral over the photon energy $x=2k^0/\sqrt{s}$, 
with the strongest singularity $(v-x)^{\gamma-1}$ being regularized nicely by soft photons
\begin{equation}
  \begin{split}
    f_1^{(2)}(v) &= e^{\delta_{YFS}} F(\gamma)
    \int\limits_0^v dx\;
    \gamma (v-x)^{\gamma-1}\;
    \gamma \Big[ -1 +{1\over 2}x \Big]
\\&=
    e^{\delta_{YFS}} F(\gamma)\;
    \gamma v^{\gamma} \;
    \Big[ -1 +{1\over 2}v -{1\over 2}\gamma v \Big]
    +{\cal O}(\gamma^3).
  \end{split}
\end{equation}

%/////////////////////////////////////////////////////////////////////////////////////////
%-----------------------------------------------------------------------------------------
\begin{figure}[!ht]
\centering
\setlength{\unitlength}{0.1mm}
\begin{picture}(1600,800)
%\put( 0,0){\framebox( 1600,800){ }}
\put( 400,750){\makebox(0,0)[t]{\Large (a)\qquad $\bbeta^{(2)}_{1I}\otimes\bbeta^{(2)}_{0F}$}}
\put(1250,750){\makebox(0,0)[t]{\Large (b)\qquad $\bbeta^{(2)}_{0I}\otimes\bbeta^{(2)}_{1F}$}}
\put(  -20, 0){\makebox(0,0)[lb]{\epsfig{file=flat_bt1xbt0.eps,width=80mm,height=80mm}}}
\put(  800, 0){\makebox(0,0)[lb]{\epsfig{file=flat_bt0xbt1.eps,width=80mm,height=80mm}}}
\end{picture}
%----------------------------------------
\caption{\small\sf 
 The comparison between \KK MC and IEX \Ordpr{\alpha^2} formulas for the
 integrated cross section as a function of the cut-off parameter $v_{\max}$
 on ISR and FSR photons.
 Presented are \KK MC (solid line) and IEX (line with open circles) results for:
 (a) ISR $\bbeta^{(2)}_{1I}$ and the FSR $\bbeta^{(2)}_{0F}$,
 (b) FSR $\bbeta^{(2)}_{1F}$ and the ISR $\bbeta^{(2)}_{0I}$,
 multiplied by factor 0.1 (in order to fit into the scale).
 The difference between \KK MC and IEX is shown as a dotted curve.
 Center of the mass energy is 189GeV. Final state fermion is muon.
}
\label{fig:bt1xbt0-prag2-3}
\end{figure}
%----------------------------------------------------------------------------------------
 
Now we shall calculate the remaining part $d_1^{(2)}$ of $\rho_1^{(2)}$. 
Since it vanishes at \Order{\alpha} we may calculate it in LL approximation.
Although strictly speaking it is not necessary,
we treat photons gently (as in $\bbeta_0$ example)
such that we do not use the crude collinear approximation.
As before, we split the photon angular integration into
forward and backward hemispheres and we integrate
immediately over the final fermion momenta
\begin{equation}
  \begin{split}
    & d_1^{(2)}(v) =
    e^{\gamma\ln\varepsilon}
    \sum_n \sum_{n'} {1\over n!} {1\over n'!}\;
    2 \int\limits_{\theta<\pi/2}
          {d^3 k \over k^{0} }\;
          e^{\delta_{YFS}}
\\&
    \prod_{i=1}^n\;    \int\limits_{\theta_i>\pi/2}
    {d^3 k_i^+ \over k_i^{+0} }
    \tilde{S}_I(k_i^+)
    \Theta\bigg(k_i^{+0} -\varepsilon {\sqrt{s}\over 2}\bigg)
    \prod_{j=1}^{n'}\; \int\limits_{\theta_j<\pi/2}
          {d^3 k_j^- \over k_j^{-0} }
          \tilde{S}_I(k_j^-)
          \Theta\Big(k_j^{-0} -\varepsilon {\sqrt{s}\over 2}\Big)
\\&
    \Bigg\{ \Bigg[ \delta\Big( v - 1 +{(P -k  -K^+ -K^-  )^2\over s}  \Big)
\\& \qquad
    -\delta\Big( v - {2  P\cdot (k+ K^+ + K^-)\over s}  \Big)
    \Bigg]
    B_1^{(1)}(p_1,p_2,k)
\\& \qquad
    +\delta\Big( v - 1 +{1\over s} (P -k  -K^+ -K^-  )^2  \Big)
    [B_1^{(2)}(p_1,p_2,k)-B_1^{(1)}(p_1,p_2,k)]
    \Bigg\}.
  \end{split}
\end{equation}
Using the collinear replacement $K^{\pm} = (K^{\pm 0},0,0,\pm |K^{\pm 0}|\;)$
in $\delta$'s allows us to integrate over spectator multiple photons
\begin{equation}
  \begin{split}
    & d_1^{(2)}(v) =
    \int\limits_0^1 dv_+
    \int\limits_0^1 dv_-
    \int\limits_{\theta<\pi/2}
          {d^3 k \over k^{0} }\;
          e^{\delta_{YFS}}\;
          f\bigg( {\gamma\over 2},v_+ \bigg)
          f\bigg( {\gamma\over 2},v_- \bigg)
\\& \qquad
    \Bigg\{\Bigg[
     \delta\Big( v - 1  +(1 -x -v_+)(1- v_-)  \Big)
    -\delta\Big( v - x  -v_+  -v_-  \Big) \Bigg]
    B_1^{(1)}(p_1,p_2,k)
\\& \qquad
    + \delta\Big( v - 1  +(1 -x -v_+)(1- v_-)  \Big)
    [B_1^{(2)}(p_1,p_2,k)-B_1^{(1)}(p_1,p_2,k)]
\Bigg\}.
  \end{split}
\end{equation}
where $x=2 k^0/\sqrt{s}$ and the other notation is the same as in $\bbeta_0$ case.
Integration over photon angles leads to
\begin{equation}
  \begin{split}
    d_1^{(2)}(v) &=
    \int\limits_0^1 dv_+
    \int\limits_0^1 dv_-
    \int\limits_0^1 dx\;
    e^{\delta_{YFS}}\;
    f\bigg( {\gamma\over 2},v_+ \bigg)
    f\bigg( {\gamma\over 2},v_- \bigg)
\\&
    \Bigg\{ \Bigg[
    \delta\Big( v - 1  +(1 -x -v_+)(1- v_-)  \Big)
    -\delta\Big( v - x  -v_+  -v_-  \Big) \Bigg]
    \gamma b_1(x)
\\& \qquad
    + \delta\Big( v - 1  +(1 -x -v_+)(1- v_-)  \Big)
    \gamma^2 b_2(x)
    \Bigg\},
\\
    b_1(x) &=  -1 +{1\over 2} x,\quad
    b_2(x)  =  -1 +{1\over 2} x -{1\over 8} [1+(1-x)^2] {\ln(1-x)\over x}
  \end{split}
\end{equation}

Let us show very briefly the calculation of the part
proportional to the difference of $\delta$'s which is somewhat more tricky.
We convolute first $b_1$ with photons
in the same hemisphere and next with photons from opposite hemisphere
\begin{equation}
  \begin{split}
    &d_{1A}^{(2)}(v)= \int dV dv_-
      \Bigg[  \delta\Big( v - 1  +(1 -V)(1- v_-)  \Big)
             -\delta\Big( v - V  -v_-  \Big) \Bigg]
\\&\qquad
    e^{\delta_{YFS}}\;
    f\bigg( {\gamma\over 2},v_- \bigg)
    \int dx d v_+
    \delta\Big( V - x  -v_+   \Big)
    f\bigg( {\gamma\over 2},v_+ \bigg)
    \gamma b_1(x)
\\&
    = e^{\delta_{YFS}}\;
    F^2\bigg( {\gamma\over 2} \bigg)
    \gamma v^{\gamma-1}
    \int\limits_0^1 dy\; y^{ {1\over 2}\gamma}
               (1-y)^{ {1\over 2}\gamma-1}
   \bigg\{ (1-vy)^{ -{1\over 2}\gamma} -1 \bigg\}
          \left[ -1 +{vy\over 2} \bigg(1-{\gamma\over 2}\bigg) \right]
\\&  
   = e^{\delta_{YFS}}\;
   F(\gamma)\; \gamma v^{\gamma}
   \left( -1+ {1\over 2} v      \right)
   \left( -{1\over 2} \gamma\ln (1-v) \right)
   +{\cal O}(\gamma^3)
  \end{split}
\end{equation}
The remaining part of $d_{1}^{(2)} $ is easier to calculate
because it is explicitly of \Order{\gamma^2}
\begin{equation}
    d_{1B}^{(2)}(v) = e^{\delta_{YFS}}\;
    F(\gamma)\; \gamma v^{\gamma}
    \gamma \bigg\{
    {1\over 2} \left( -1+ {1\over 2} v \right)
    -{1\over 8} \left( 1+(1-v)^2 \right)
    {\ln (1-v)\over v}
    \bigg\}
    +{\cal O}(\gamma^3).
\end{equation}
The  contribution from the initial-state $\bbeta_1$ 
with \Ordpr{\alpha^2} QED matrix element and with
\Ordpr{\alpha^2} analytical integration over the multiphoton phase-space reads
\begin{equation}
  \begin{split}
    \rho_{1}^{(2)}(v) =& e^{\delta_{YFS}}\;
    F(\gamma)\; \gamma v^{\gamma-1}
    \bigg\{ {1\over 2}
     \left( -1+ {1\over 2} v \right)
\\&\qquad
   +\gamma
   \left[ -{1\over 2} v -{1\over 4} v^2
     +{1\over 8}
     \left( -1 + 3(1-v)^2 \right) \ln (1-v) \right] \bigg\}
   +{\cal O}(\gamma^3).
  \end{split}
\end{equation}
The contribution with \Ordpr{\alpha^1} QED matrix element and with analytical 
\Ordpr{\alpha^2} multiphoton phase-space integration
is obtained by retaining only $d_{1A}^{(2)}$ and it reads
\begin{equation}
  \begin{split}
    \rho_{1}^{(1)}(v)
    =& e^{\delta_{YFS}}\;
    F(\gamma)\; \gamma v^{\gamma-1}
    \bigg\{ {1\over 2} \left( -1+ {1\over 2} v \right)
\\&\qquad
   +\gamma \left[
     -{1\over 2} v^2 -{1\over 2}
     \left( -1 +{1\over 2} v \right) \ln (1-v) \right] \bigg\}
   +{\cal O}(\gamma^3).
  \end{split}
\end{equation}

%%%%%%%%%%%%%%%%%%%%%%%%%%%%%%%%%%%%%%%%%%%%%%%%%% 
\subsubsection{Beta-bar-two, $\bbeta_2$}
%%%%%%%%%%%%%%%%%%%%%%%%%%%%%%%%%%%%%%%%%%%%%%%%%% 

In the following step our aim is to calculate analytically  the contribution
to the total cross section from $\bbeta^{(2)}_{1I}$ as given by
\begin{equation}
  \begin{split}
    & \sigma_2=
    \sum_{n=0}^\infty  {1\over n!}
    \int {d^3q_1\over q_1^0} {d^3q_2\over q_2^0}
    \int \prod_{j=1}^n  {d^3k_j\over k^0_j}
                   \tilde{S}_I(p_1,p_2;k_j)
                   (1-\Theta(\Omega_I;k_j))
    e^{Y(\Omega_I)}
\\&\;\;
   \sum_{n \geq j>k \geq 1}\;\;
   {\bbeta^{(2)}_{2II}(X,p_1,p_2,q_1,q_2,k_j,k_k)
                   \over \tilde{S}_I(k_j)\tilde{S}_I(k_k)}\;
   \delta^{(4)}\bigg(
               p_1+p_2 -q_1-q_2 -\sum_{j=1}^n k_j
            \bigg)
  \end{split}
\end{equation}
This contribution is in a sense more trivial than the previous two
because it is pure \Order{\alpha^2}, it does not have
any  infrared singularity in the two-photon phase-space integral.

We can calculate the contribution from $\bbeta_2$ 
with the same methods as in the case of $\bbeta_0$ or $\bbeta_1$.
The integral is  reorganized easily such that  the integration over 
photon momenta in the $\bbeta^{(2)}_{2II}$ is isolated
and we are able to integrate over final-state fermion momenta
bringing the integral to the standard form 
\begin{equation}
  \sigma_2 = \int\limits_0^1 dv\; \rho_2^{(2)}(v) \sigma^\born\big(s(1-v)\big).
\end{equation}

%/////////////////////////////////////////////////////////////////////////////////////////
%-----------------------------------------------------------------------------------------
\begin{figure}[!ht]
\centering
\setlength{\unitlength}{0.1mm}
\begin{picture}(1600,650)
%%\put( 0,0){\framebox( 1600,600){ }}
\put( 250,605){\makebox(0,0)[b]{\large (a)}}
\put( 800,605){\makebox(0,0)[b]{\large (b)}}
\put(1350,605){\makebox(0,0)[b]{\large (c)}}
\put( 250,500){\makebox(0,0)[t]{\large $\bbeta^{(2)}_{2I}\otimes\bbeta^{(2)}_{0F}$}}
\put( 800,500){\makebox(0,0)[t]{\large $\bbeta^{(2)}_{0I}\otimes\bbeta^{(2)}_{0F}$}}
\put(1350,500){\makebox(0,0)[t]{\large $\bbeta^{(2)}_{1I}\otimes\bbeta^{(2)}_{1F}$}}
\put(  -50, 0){\makebox(0,0)[lb]{\epsfig{file=flat_bt2xbt0.eps,width=55mm,height=60mm}}}
\put(  500, 0){\makebox(0,0)[lb]{\epsfig{file=flat_bt0xbt2.eps,width=55mm,height=60mm}}}
\put( 1050, 0){\makebox(0,0)[lb]{\epsfig{file=flat_bt1xbt1.eps,width=55mm,height=60mm}}}
\end{picture}
%----------------------------------------
\caption{\small\sf 
 The comparison between \KK MC and IEX \Ordpr{\alpha^2} formulas for the
 integrated cross section as a function of the cut-off parameter $v_{\max}$
 on ISR and FSR photons.
 Presented are \KK MC (solid line) and IEX (line with open circles) results for:
 (a) ISR $\bbeta^{(2)}_{2I}$ and the FSR $\bbeta^{(2)}_{0F}$,
 (b) FSR $\bbeta^{(2)}_{2F}$ and the ISR $\bbeta^{(2)}_{0I}$
 (c) ISR $\bbeta^{(2)}_{1I}$ and the FSR $\bbeta^{(2)}_{1F}$.
 The difference between \KK MC and IEX is also included (dots).
 Center of the mass energy is 189GeV. Final state fermion is muon.
}
\label{fig:flat-bt2}
\end{figure}
%----------------------------------------------------------------------------------------

The function $\rho_2^{(2)}(v)$ can be calculated in the LL
approximation with either of the two methods
(keeping an additional spectator photon or not)
and after integration over photon angles the integral
boils down to the following integral over longitudinal
photon momenta, separately for the case
with two collinear and two anticollinear photons
\begin{equation}
  \begin{split}
    \rho_2^{(2)}(v)
    =& \int\limits_0^1 dv_-  dv_+\;
    {\gamma^2\over 4}\; \delta(v-v_+-v_-)\;
    \Bigg[ 
    {1\over 2v_+v_-} \chi(v_+)\;\chi\left( {v_-\over 1-v_+} \right)
\\
&   \qquad\qquad
    +{1\over 2v_+v_-} \chi(v_-)\;\chi\left( {v_+\over 1-v_-} \right)
    -{1\over v_+}\omega(v_-)
    -{1\over v_-}\omega(v_+)
    -{1\over v_+v_-}
   \Bigg]
\\
 +& \int\limits_0^1 dv_-  dv_+
   {\gamma^2\over 4}\; \delta(v- 1 +(1-v_+) (1-v_-) )\;
\\
& \qquad\qquad
   \Bigg[ 
     {1\over v_+v_-} \chi( v_+ ) \chi( v_- )
     -{1\over v_+}\omega(v_-)
     -{1\over v_-}\omega(v_+)
     -{1\over v_+v_-}
   \Bigg]
\\
  =& \gamma^2\; {1\over 4} v,
  \end{split}
\end{equation}
where $\chi(x)=(1+(1-x)^2)/2$ and $\omega(x)= -1+x/2$.

%%%%%%%%%%%%%%%%%%%%%%%%%%%%%%%%%%%%%%%%%%%%%%%%%%%%%%%%%%%%%%%%%%%%%%%%%
%%%%%%%%%%%%%%%%%%%%%%%%%%%%%%%%%%%%%%%%%%%%%%%%%%%%%%%%%%%%%%%%%%%%%%%%%
%%%%%%%%%%%%%%%%%%%%%%%%%%%%%%%%%%%%%%%%%%%%%%%%%%%%%%%%%%%%%%%%%%%%%%%%%
\begin{table}[!ht]
\centering
%%%----------------- initial state -------------------------
\def\mystrut{\rule{0pt}{6mm}}
\begin{tabular}{||l|l||c|c||}
\hline\hline
\mystrut
   &  
      & $d_S$ 
         & $\Delta_H(v)$
\\[1mm] 
\hline\hline
\Order{\alpha^0} \mystrut 
   & $\bbeta_0$ 
      &  $1 $  
         & $ -{1\over 4}\gamma\ln(1-v) 
             -{1\over 2} {\alpha\over\pi} \ln^2(1-v) +0\;\gamma^2 $
\\ [1mm]
\hline \hline
%%%----------------- first order -------------------------
\Order{\alpha^1} \mystrut
   & $\bbeta_0$ 
      &  $1+{\gamma\over 2} $  
         & $ -{1\over 4}\gamma\ln(1-v)$
\\ [1mm]
\mystrut
   & $\bbeta_1$ 
      & $0 $  
         & $v\left(-1+{v\over 2}\right) 
              +\gamma\left[
                  -{v^2\over 2} -{v(2-v)\over 4}\ln(1-v)
                    \right]$
\\ [1mm]
\cline{2-4}
\mystrut
   & All
      &  $1+{\gamma\over 2} $
         & $v\left(-1+{v\over 2}\right) 
              +\gamma\left[
                  -{v^2\over 2} -{(1-v)^2\over 4}\ln(1-v)
                    \right]$
\\[1mm] 
\hline \hline
%%%----------------- second order -------------------------
\Order{\alpha^2} \mystrut
   & $\bbeta_0$ 
      &  $1+{\gamma\over 2}+{\gamma^2\over 8} $  
         & $ -{1\over 4}\gamma\ln(1-v)$
\\ [1mm]
\mystrut
   & $\bbeta_1$ 
      & $0 $  
         &$v\left(-1+{v\over 2}\right) 
          +\gamma\left[
            -{v\over 2}-{v^2\over 4}-{-1+3(1-v)^2\over 8}\ln(1-v)
                \right]$
\\ [1mm]
\mystrut
   & $\bbeta_2$ 
      & $0 $  
         &  $+\gamma {v^2\over 4}$
\\ [1mm]
\cline{2-4}
\mystrut
   & All
      &  $1+{\gamma\over 2}+{\gamma^2\over 8} $
         & $v\left(-1+{v\over 2}\right) 
              +\gamma\left[
                  -{v\over 2} -{1+3(1-v)^2\over 4}\ln(1-v)
                    \right]$
\\[1mm] 
\hline \hline
\multicolumn{2}{||c||}{\mystrut \Order{\alpha^2}-\Order{\alpha^1} }
      &  ${\gamma^2\over 8} $
         & $  +\gamma\left[
                  -{1+(1-v)^2\over 8}\ln(1-v)
                    \right]$
\\[1mm] 
\hline \hline
\end{tabular}
\caption{\sf
  Contributions to the function $\rho_I(v)= d_S+\Delta_H(v)$ from  $\bbeta_k,k=0,1,2$.
  The ISR matrix element is at \Ordpr{\alpha^r}
  with YFS/EEX exponentiation; $r=0,1,2$ is marked in first column.
  Phase-space integration is done analytically always within \Ordpr{\alpha^2},
  except of the \Ordpr{\alpha^0} case in the first row,
  where the phase-space integration is done in \Ordpr{\alpha^3}.
}
\label{table-initial}
\end{table}
%%%%%%%%%%%%%%%%%%%%%%%%%%%%%%%%%%%%%%%%%%%%%%%%%%%%%%%%%%%%%%%%%%%%%%%%%

By eventually keeping additional soft photons in the 
calculation  we get our final result for 
the initial-state \Ordpr{\alpha^2} contribution from $\bbeta_2$  in a more elegant form
\begin{equation}
  \rho_2^{(2)} (v) =  e^{\delta_{YFS}}\; F(\gamma)\; \gamma v^{\gamma-1}
  \left\{   {1\over 4}  \gamma v^2  \right\} +{\cal O}(\gamma^3)
\end{equation}

We have compared numerically the above formula
with the \KK MC in the case of FSR switched off and found an agreement better than 0.1\%.
In Figure~\ref{fig:flat-bt2}(a) we present the comparison in which,
as in the case of the previous $\bbeta$'s,
FSR is switched on.
In Figure~\ref{fig:flat-bt2}(a) we compare the convolution
of the ISR $\bbeta^{(2)}_{2I}$ and the FSR $\bbeta^{(2)}_{0F}$:
%/////////////////////////////////////////////////////////////////////////
\begin{equation}
\label{eq:IEX-bt2xbt0}
    \sigma^{(2)}_{\bbeta^{(2)}_2\otimes\bbeta^{(2)}_0}
   = \int_0^{v_{\max}} dv\;
     \int_0^{v/(1-u_{\max})} du\;
     \sigma^f_{\rm Born}\big(s(1-u)(1-v)\big)\;
     \rho^{(2)}_{I\bbeta^{(2)}_2}(v)\; 
     \rho^{(2)}_{F\bbeta^{(2)}_0}(u)
\end{equation}
The above IEX result is compared with the \KK MC results, and they agree within 0.2\%.
In Figure~\ref{fig:flat-bt2}(b) we show the analogous comparison
for the convolution of the FSR $\bbeta^{(2)}_{2F}$ and the ISR $\bbeta^{(2)}_{0I}$
(anticipating IEX results for FSR $\bbeta^{(2)}_{2F}$ in the next section)
and we find the similar agreement.
Finally, there is another more trivial contribution in the $\bbeta^{(2)}$ family
which correspond to the case with one real photon emitted in the initial state
and one in the final state.
This case does not require a separate analytical phase space integration effort
because the relevant IEX formula involves the convoluton of the already
known expression for the ISR $\bbeta^{(2)}_{1I}$ and the FSR $\bbeta^{(2)}_{1F}$.
The corresponding numerical comparison of the IEX and EEX MC is shown
in Figure Figure~\ref{fig:flat-bt2}(c).
In fact the IEX matrix element was deliberatery constructed 
in such a way (factorizing virtual corrections)
that the above convolution-type IEX formula results.

%%%%%%%%%%%%%%%%%%%%%%%%%%%%%%%%%%%%%%%%%%%%%%%%%%%%%%%%%%
\subsubsection{Summary on IEX for ISR}
%%%%%%%%%%%%%%%%%%%%%%%%%%%%%%%%%%%%%%%%%%%%%%%%%%%%%%%%%%
The entire initial-state \Ordpr{\alpha^2} integrated
cross section is obtained by combining contributions
from all three $\bbeta$'s and it reads
\begin{equation}
  \begin{split}
 & \sigma^{(2)}_I = \int\limits_0^1 dv\; \rho_I^{(2)}(v) \sigma^\born(s(1-v)),
\\
&  \rho_I^{(2)} (v) =   e^{\delta_{YFS}}\; F(\gamma)\; \gamma v^{\gamma-1} 
     \Bigg\{ 1+{\gamma\over 2} +{\gamma \over 8}
\\
&\qquad\qquad
        +v\left(-1 +{1\over 2} \right)
        +\gamma\left[ -{v\over 2} - {1+3(1-v)^2 \over 8} \ln(1-v)  \right]
     \Bigg\}
     +{\cal O}(\gamma^3) +{\cal O}(\gamma\alpha)
 \end{split}
\end{equation}
This above ISR formula has been obtained as a result of ad-hoc exponentiation
(interpolation) in ref.~\cite{yfs2:1990} and was used there
as a numerical parametrization/testing of the cross section from
the Monte Carlo program YFS2. 
It is is now {\em derived} starting from YFS exclusive exponentiation
by means of direct phase-space integration%
\footnote{
  Ad-hoc exponentiation is of course easier to do
  and in ref.~\cite{third-order:1991} even the \Ordpr{\alpha^3}
  formula for the initial-state bremsstrahlung was given
  but the derivation method presented here is much
  better founded and the result does not depend on any kind of
  interpolation or guesswork.}!

Summarizing our IEX calculations for ISR, 
we have obtained through the analytical integration
over the ISR multiphoton phase space the inclusive exponentiated 
cross section for the IEX matrix elements in the
\Ordpr{\alpha^0}, \Ordpr{\alpha^1} and \Ordpr{\alpha^2}
for each $\bbeta_i$ $i=0,1,2$ separately.
The phase-space integration was always done analytically within the \Ordpr{\alpha^2}.
All results from the above extensive study are summarized in
Table \ref{table-initial} where we list the two functions
$d_S$ and $\Delta_H(v)$ in the formula
\begin{equation}
  \begin{split}
  & \sigma_I = \int\limits_0^1 dv\; \rho_I(v) \sigma^\born(s(1-v)),\qquad
    \rho_I(v) =  e^{\delta_{YFS}}\;
    F(\gamma)\; \gamma v^{\gamma-1} \Big( d_S + \Delta_H(v) \Big)
\\&
    \delta_{YFS}= {\gamma\over 4}
                  +{\alpha \over \pi} \left( -{1\over 2} +{\pi^2\over 3}\right),\quad
    \gamma =  2{\alpha\over\pi }\left( \ln{s\over m_e^2} -1 \right),\quad
             F(\gamma)= {e^{C\gamma} \over \Gamma (1+\gamma) }.
 \end{split}
\end{equation}
All notation is recalled for the convenience of the reader.

%%%%%%%%%%%%%%%%%%%%%%%%%%%%%%%%%%%%%%%%%%%%%%%%%%%%%%%%%%%%%%%%%%%%%%%%%
%%%%%%%%%%%%%%%%%%%%%%%%%%%%%%%%%%%%%%%%%%%%%%%%%%%%%%%%%%%%%%%%%%%%%%%%%
%%%%%%%%%%%%%%%%%%%%%%%%%%%%%%%%%%%%%%%%%%%%%%%%%%%%%%%%%%%%%%%%%%%%%%%%%
%%%%\begin{table}[!ht]
\begin{table}[t]
\centering
%%%----------------- final state -------------------------
\def\mystrut{\rule{0pt}{6mm}}
\begin{tabular}{||l|l||c|c||}
\hline\hline
\mystrut
   &  
      & $d'_S$ 
         & $\Delta'_H(u)$
\\[1mm] 
\hline\hline
\Order{\alpha^0} \mystrut 
   & $\bbeta_0$ 
      &  $1 $  
         & $ -{1\over 4}\gamma_f\ln(1-u)$
\\ [1mm]
\hline \hline
%%%----------------- first order -------------------------
\Order{\alpha^1} \mystrut
   & $\bbeta_0$ 
      &  $1+{\gamma_f\over 2} $  
         & $ -{1\over 4}\gamma_f\ln(1-u)$
\\ [1mm]
\mystrut
   & $\bbeta_1$ 
      & $0 $  
         & $u\left(-1+{u\over 2}\right) 
              +\gamma_f\left[
                  -{u^2\over 2} +{u(2-u)\over 2}\ln(1-u)
                    \right]$
\\ [1mm]
\cline{2-4}
\mystrut
   & All
      &  $1+{\gamma_f\over 2} $
         & $u\left(-1+{u\over 2}\right) 
              +\gamma_f\left[
                  -{u^2\over 2} +{-1+4u-2u^2\over 4}\ln(1-v)
                    \right]$
\\[1mm] 
\hline \hline
%%%----------------- second order -------------------------
\Order{\alpha^2} \mystrut
   & $\bbeta_0$ 
      &  $1+{\gamma_f\over 2}+{\gamma_f^2\over 8} $  
         & $ -{1\over 4}\gamma_f\ln(1-u)$
\\ [1mm]
\mystrut
   & $\bbeta_1$ 
      & $0 $  
         &$u\left(-1+{u\over 2}\right) 
          +\gamma_f\left[
            -{u\over 2}-{u^2\over 4} +{2+6u-3u^2\over 8}\ln(1-u)
                \right]$
\\ [1mm]
\mystrut
   & $\bbeta_2$ 
      & $0 $  
         &  $+\gamma_f\left( 
               {u^2\over 4} -{u(2-u)\over 4}\ln(1-u)
                   \right)$
\\ [1mm]
\cline{2-4}
\mystrut
   & All
      &  $1+{\gamma_f\over 2}+{\gamma_f^2\over 8} $
         & $u\left(-1+{u\over 2}\right) 
              +\gamma_f\left[
                  -{u\over 2} +{u(2-u)\over 8}\ln(1-u)
                    \right]$
\\[1mm] 
\hline \hline
\multicolumn{2}{||c||}{\mystrut \Order{\alpha^2}-\Order{\alpha^1} }
      &  ${\gamma_f^2\over 8} $
         & $  +\gamma_f\left[
                  {2-6u+3u^2\over 8}\ln(1-u)
                    \right]$
\\[1mm] 
\hline \hline
\end{tabular}
\caption{\sf
  Contributions to function $\rho_F(u)= d'_S+\Delta'_H(u)$ from  $\bbeta_k,k=0,1,2$.
  The FSR matrix element is \Ordpr{\alpha^r}with YFS/EEX exponentiation, 
  $r=0,1,2$ is marked in the first column.
  Phase-space integration is done analytically always in \Ordpr{\alpha^2}.
}
\label{table-final}
\end{table}
%%%%%%%%%%%%%%%%%%%%%%%%%%%%%%%%%%%%%%%%%%%%%%%%%%%%%%%%%%%%%%%%%%%%%%%%%
%%%%%%%%%%%%%%%%%%%%%%%%%%%%%%%%%%%%%%%%%%%%%%%%%%%%%%%%%%%%%%%%%%%%%%%%%

%/////////////////////////////////////////////////////////////////////////////////////////
%-----------------------------------------------------------------------------------------
\begin{table}[!ht]
\centering
\setlength{\unitlength}{0.1mm}
\begin{picture}(800,800)
\put(   400,  0){\makebox(0,0)[b]{\epsfig{file=flat_total.eps,width=80mm,height=80mm}}}
\end{picture}
%%%%%%%%%%%%%%%%%%%%%%%%%%%%%%%%%%%%%%%%%%%%%%%%
\caption{\small\sf
 The comparison between \KK MC and IEX \Ordpr{\alpha^2} formula.
 The difference between \KK MC in EEX mode and the IEX formula divided by Born
 is plotted with the dotted line, 
 as a function of the cutoff $v_{\max}$ on the total energy ISR and FSR photons.
 Include is also the $10^{-2} \times \sigma(v_{\max})/\sigma_{\rm Born}$,
 as dotted line for IEX and solid line for MC.
}
\label{fig:flat-total}
\end{table}
%-----------------------------------------------------------------------------------------

%%%%%%%%%%%%%%%%%%%%%%%%%%%%%%%%%%%%%%%%%%%%%%%%%%%%%
\subsection{Semi-analytical formulas for FSR}
%%%%%%%%%%%%%%%%%%%%%%%%%%%%%%%%%%%%%%%%%%%%%%%%%%%%%
The calculation of the \Ordpr{\alpha^2} IEX formula for the FSR,
with the $u_{\max}$ cutoff, 
that is $u=1-s'/s<u_{\max}$ is quite similar to one in the ISR case
and we do not enter into details.
We only discuss the basic differences between the ISR and FSR cases
and present the final result.

If we switch off the ISR completely then the FSR integrated cross section 
for the \Ordpr{\alpha^r} $r=0,1,2$ EEX matrix element
reads
\begin{equation}
  \begin{split}
    \label{eq:IEX-FSR}
  & \sigma_F(u_{\max}) 
  = \sigma_{\rm Born}
    \int\limits_0^{u_{\max}} du\;  \rho_{F}(u),\quad
    \rho_F(u)  = 
    e^{\delta_{YFS}}\; F(\gamma_f)\; \gamma_f u^{\gamma_f-1}
    \Big( d'_S + \Delta'_H(u) \Big)
\\& \delta'_{YFS}= 
    {\gamma_f\over 4} -{1\over 2}\gamma_f \ln(1-u)
    +{\alpha \over \pi}
    \left( -{1\over 2} +{\pi^2\over 3}\right),\quad
    \gamma_f= 
    2{\alpha\over\pi }\left( \ln{s\over m_f^2} -1 \right),
  \end{split}
\end{equation}
where the functions $d'_S$ and $\Delta'_H(u)$ obtained with  analytical integration
of the phase space using the \Ordpr{\alpha^2} approximation,
are listed in Table \ref{table-final}.

The main difference and a complication in the phase space analytical integration
with respect to the case of ISR is that
the YFS formfactor $\delta'_{YFS}$ depends in the case of FSR
on the integration variable $u$.
This is why terms of \Order{L^2\alpha^2} are different in the two cases.
In Table \ref{table-final} we show separately the contributions
from each $\bbeta$.
Note that in the case of FSR we did not integrate analytically the phase space for 
$\bbeta_0$ at the \Ordpr{\alpha^3}, like in the case of ISR.
(It was not necessary in order to reach the precision level of 0.2\%.)
We have checked numerically the agreement of the \KK MC
with the eq.~(\ref{eq:IEX-FSR}) separately for each type of the $\bbeta$,
with the ISR switched off (plots are not shown).
We have already presented the complete set of numerical results 
in the case of the ISR switched off in this section, 
for each combination of the ISR and FSR $\bbeta$'s.

%%%%%%%%%%%%%%%%%%%%%%%%%%%%%%%%%%%%%%%%%%%%%%%%%%%%%
\subsection{Semi-analytical IEX for  ISR and FSR}
%%%%%%%%%%%%%%%%%%%%%%%%%%%%%%%%%%%%%%%%%%%%%%%%%%%%%
The last numerical test which we show in Figure~\ref{fig:flat-total}
is the case in which we switch on all ISR and FSR $\bbeta$'s listed in both 
tables \ref{table-initial} and \ref{table-final}.
\begin{equation}
  \sigma_{\rm tot.}
  =\int\limits_0^{v_{\max}} dv\;  
    \int\limits_0^{v/(1-u_{\max})}  du\;\;
     \sigma^f_{\rm Born}\big(s(1-u)(1-v)\big)\;
    \rho_{F}(u)
    \rho_{F}(v).
\end{equation}
It is done for the constant Born cross section, the case with the variable
cross section will be shown in the next section.
We use the IEX formula of the pure \Ordpr{\alpha^2} type (without \Ordpr{\alpha^3}
improvements for ISR).
The everall agreements between IEX and \KK MC is within the advertised 0.2\%.
By looking into all previous figures in this and the previous subsection
it is interesting to note that this difference does not come from
one particular combination of the ISR and FSR $\bbeta$'s, but from
several ones.

The reader may wonder why we elaborate so much in this section for the IEX
semianalytical formula which are related to the EEX type of the matrix element in \KK MC
if in fact the main  matrix element in \KK MC is now CEEX.
The main reason is that historically the EEX was the first available example
of the exclusive exponentiation, and the IEX semianalytical formula were developed
in parallel, providing the valuable cross-check of the MC.
At this stage, as we see in the next section, both IEX and EEX provide the reference
calculation and valuable test for the CEEX.
The precision of the present \Ordpr{\alpha^2} IEX is limited, but it could be improved
to the full \Ordpr{\alpha^3} if necessary.
More important limitation in the present \Ordpr{\alpha^2} IEX as a test of the CEEX
model is the lack of the ISR$\otimes$FSR interference.
We believe that this effect can be included in the semianalytical IEX if necessary.
The ad-hoc variant of the \Order{\alpha^1} exponentiation 
including the ISR$\otimes$FSR interference
is already available in refs.~\cite{greco:1975,greco:1980}.

%%%%%%%%%%%%%%%%%%%%%%%%%%%%%%%%%%%%%%%%%%%%%%%%%%%%%%%%%%%%%%%%%%%%%%%%%%%%%%%%%%%%%%%%%%%
%%%%%%%%%%%%%%%%%%%%%%%%%%%%%%%%%%%%%%%%%%%%%%%%%%%%%%%%%%%%%%%%%%%%%%%%%%%%%%%%%%%%%%%%%%%
%%%%%%%%%%%%%%%%%%%%%%%%%%%%%%%%%%%%%%%%%%%%%%%%%%%%%%%%%%%%%%%%%%%%%%%%%%%%%%%%%%%%%%%%%%%

%%%%%%%%%%%%%%%%%%%%%%%%%
% make ceex2-all-ps
%%%%%%%%%%%%%%%%%%%%%%%%%

\newpage
%%%%%%%%%%%%%%%%%%%%%%%%%%%%%%%%%%%%%%%%%%%%%%%%%%%%%%%%%%%%%%%%%%%%%%%%%%%%%%%%%%%%%%%%%%%%%
\section{Numerical results and tests}
%%%%%%%%%%%%%%%%%%%%%%%%%%%%%%%%%%%%%%%%%%%%%%%%%%%%%%%%%%%%%%%%%%%%%%%%%%%%%%%%%%%%%%%%%%%%%
In this section we shall mainly present the numerical results from the \KK MC
in which the Standard Model amplitudes for the process $e^-e^+\to f\bar{f}+n\gamma$
of the previous Section~\ref{sec:eex} (EEX) and Section \ref{sec:ceex} (CEEX)
are implemented.
The analytical results of the Section \ref{sec:semi} will be also exploited
to obtain numerical results from the semianalytical program \KK sem.
These results are mainly for the $\mu^-\mu^+$ final state.
For more results on the quark final states and other interesting numerical results
from \KK MC we refer the reader to forthcoming proceedings
of the LEP2 Monte Carlo Workshop\cite{lep2mc2f:2000}.

%/////////////////////////////////////////////////////////////////////////////////////////
%-----------------------------------------------------------------------------------------
\begin{figure}[!ht]
\centering
\setlength{\unitlength}{0.1mm}
\begin{picture}(800,1000)
\put(  0, 0){\makebox(0,0)[lb]{\epsfig{file=flow-mcgen2.eps,width=80mm,height=100mm}}}
\end{picture}
%----------------------------------------
\caption{\small\sf
  General structure of the \KK Monte Carlo program.
}
\label{fig:kkmc}
\end{figure}
%----------------------------------------------------------------------------------------
The general structure of the \KK MC code is depicted in the Figure~\ref{fig:kkmc}.
The program is divided in the two distinct parts (levels):
\begin{itemize}
  \item[(a)] Phase-space Monte-Carlo integration engine with the common importance sampling
    for the entire family of QED distributions (EEX and CEEX)
  \item[(a)] Collection (library) of programs for the SM/QED spin amplitudes
    and differential distributions, at various orders, with various styles of exponentiation.
\end{itemize}
In this work we do not enter into the description of the MC integration algorithm
in the universal MC integration engine.
The Monte Carlo method of the phase space integration
is fully documented (for the first time) in ref.~\cite{kkcpc:1999}
and some aspects of the phase-space parametrization are documented
in the forthcoming ref.~\cite{kinematicon:1999}.
Here we regard this low-level MC program as a black box capable to integrate
the phase space exactly (up to a statistical error).

The life is however not that simple and the numerical program which ``in principle''
is doing something ``exactly/rigorously'' may still give imprecise results due
to programming bugs and numerical instabilities, 
especially in a program as complicated as \KK MC is.
This is why we always introduce a concept of the {\em technical precision} of the given
program/calculation, see below.
The basic aim of our numerical exercises presented in this Section is the determination
of the total {\em theoretical precision} associated with our calculation of 
Standard Model predictions for experimental observables
(although we limit the discussion to QED part of SM for most of our discussion).
As for observables we shall concentrate mainly on the total cross section and
charge asymmetry at LEP1, LEP2 and Linear Collider energies.

What are the technical and physical precisions?
The {\em technical precision} we define as all uncertainties related to pure numerical problems
like programming bugs, numerical instabilities, numerical approximation, etc.
In our case the question of the technical precision will mainly concern the MC integration engine.
It is important to determine it at the early stage of the work and it should be generally
much better than the physical precision.
On the other hand, the physical precision is the total uncertainty related 
to neglected higher orders in coupling constant $\alpha$ or in other expansion parameters like
the inverse of the big-log $1/L$, or the ratio of the width to mass $\Gamma/M$ for a narrow resonance.
For physical precision we understand that the above truncations are done in the spin amplitudes
and/or differential cross section.
If some of them are done in the phase space integration then we tend to associate them with
the technical precision (as phase space integration is a technical problem).

We shall start this section with the basic discussion of the technical precision, 
then we proceed to a subsection elaborating on the physical precision for the EEX matrix element,
based on  comparisons of the \KK MC and semi-analytical results,
and later we discuss the physical precision for the case of the full CEEX matrix element.
In this Section we also present numerical results 
and a rather complete discussion of the effects due to the ISR-FSR interference
in the fermion pair production process.

We note that it would be good to include also more numerical tests and lower energies $\sim 10GeV$,
and for very high energies $\sim 1TeV$, and some more tests specific to spin effects.
However, the basic pattern of the spin correlations in double $\tau$ decay
was already cross-checked in ref.~\cite{gps:1998}.

%%%%%%%%%%%%%%%%%%%%%%%%%%%%%%%%%%%%%%%%%%%%%%%%%%%%%%%%%%%%%%%%%
\subsection{Basic test of technical precision}
%%%%%%%%%%%%%%%%%%%%%%%%%%%%%%%%%%%%%%%%%%%%%%%%%%%%%%%%%%%%%%%%%

%/////////////////////////////////////////////////////////////////////////////////////////
%-----------------------------------------------------------------------------------------
\begin{figure}[!ht]
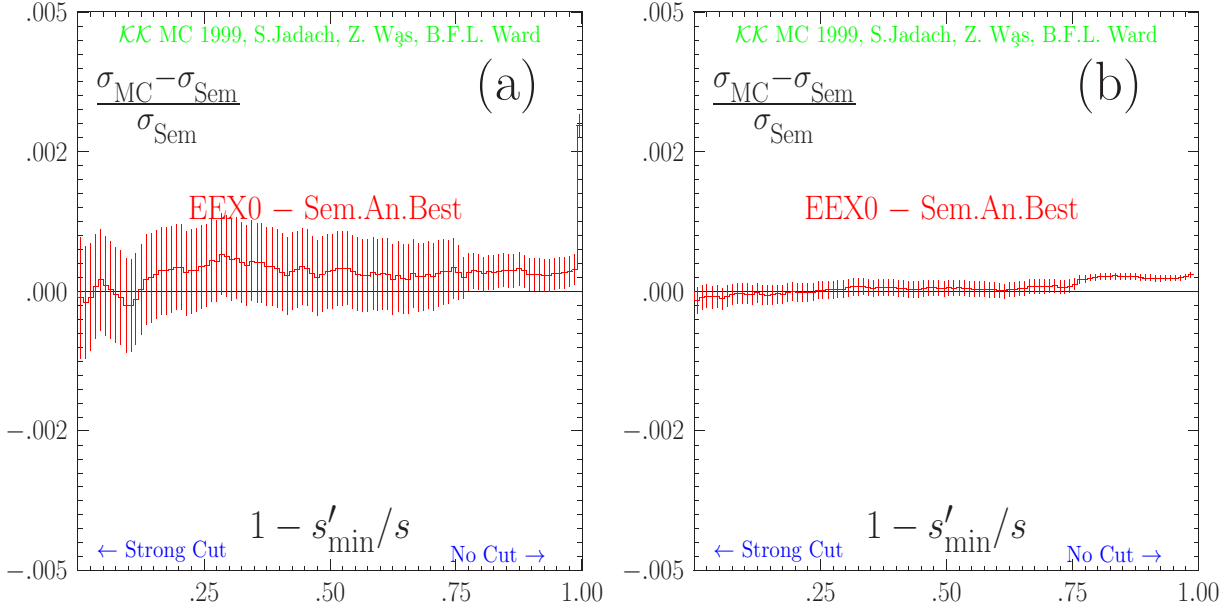

\centering
\setlength{\unitlength}{0.1mm}
\begin{picture}(1600,800)
%\put( 0,0){\framebox( 1600,850){ }}
\put( 650,650){\makebox(0,0)[b]{\LARGE (a)}}
\put(1450,650){\makebox(0,0)[b]{\LARGE (b)}}
\put(  -20, 0){\makebox(0,0)[lb]{\epsfig{file=chi-mcan-O0tech.eps,width=80mm,height=80mm}}}
\put(  800, 0){\makebox(0,0)[lb]{\epsfig{file=chi-mcan-O0tech.eps.999,width=80mm,height=80mm}}}
\end{picture}
%%%%%%%%%%%%%%%%%%%%%%%%%%%%%%%%%%%%%%%%%%%%%%%%
\caption{\small\sf
  Evaluation of the technical precision of the \KK\ Monte Carlo
  using simplified QED multiphoton distribution.
  The difference of the \KK\ MC result and semianalytical result divided by semianalytical
  is plotted as a function of $v_{\max}=1-s'_{\min}/s$.
  results are shown for $\mu^+\mu^-$ final state at $\protect\sqrt{s}=189$GeV.
  In the case (a) the phase-space limit $v_{\max}=1-4m_\mu^2/s$ is taken;
  the last bin represent the entire phase phase space.
  In the case (b) $v_{\max}=0.999$.
}
\label{fig:BasicTech}
\end{figure}
%-----------------------------------------------------------------------------------------

The best way to determine technical precision is to compare results of the two or even more
independent calculations which implement the same physics model but differ in technical
details of the actual implementation like the method of phase space integration,
independent coding, etc.
The best two possible methods are:
(a) to compare two independent Monte Carlo calculations or 
(b) to compare Monte Carlo results with the results of a semi-analytical calculation.
The method (a) is generally better because it can be done for arbitrary kinematical selections (cuts)
and for the simplified QED matrix element,
while method (b) is limited to a simple or absent kinematical selections.
In the following we shall use method (b).

For our basic test of the technical precision we use the simplest possible
variant of the QED model, that is of the type \Oeex{\alpha^0}
defined in Section~\ref{sec:eex}.
For this type of QED matrix element we were able to integrate analytically the total cross section
in the Section~\ref{sec:semi}.
The relevant formula can be read from the first row in Tables~\ref{table-initial} and \ref{table-final}.
For the sake of completeness we write down the complete expression explicitly:
%/////////////////////////////////////////////////////////////////////////
\begin{equation}
  \label{eq:eex0-best}
  \begin{split}
    &\sigma^f_{\rm SAN}
   = \int_0^{v_{\max}} dv\;
     \sigma^f_{\rm Born}(s(1-u)(1-v))\;
     \rho^{(0)}_{I}(v)\; \rho^{(0)}_{F}(u),\\
    &\rho^{(0)}_{I}(v) = F(\gamma_e)\;
                 e^{ {1\over 4} \gamma_e +{\alpha\over\pi}\big({1\over 2} +{\pi^2\over 3}\big)}\;
                 \gamma_e v^{\gamma_e-1}\;
                 \bigg(1 -{1\over 4}\gamma_e \ln(1-v) 
                         -{1\over 2} {\alpha\over\pi} \ln^2(1-v) 
                         +0\;\gamma_e^2
                 \bigg),\\
    &\rho^{(0)}_{F}(u) = F(\gamma_f)\;
                 e^{ {1\over 4} \gamma_f -{1\over 2} \gamma_f \ln(1-u)
                     +{\alpha\over\pi}\big({1\over 2} +{\pi^2\over 3}\big)}\;
                 \gamma_f u^{\gamma_f-1}\;
                 \bigg(1 -{1\over 4}\gamma_f \ln(1-u) \bigg),
  \end{split}
\end{equation}
As we remember the coefficient in front of the \Order{L^3\alpha^3} term is zero, as marked explicitly.
It was essential to calculate analytically and introduce the ISR term of \Order{L^1\alpha^2}
because it amounts numerically to several per cent for the cross section located close to $v=1$.

In fig.~\ref{fig:BasicTech} we present the comparison of the \KK\ MC 
with the semianalytical formula of eq.~(\ref{eq:eex0-best}).
The difference between the MC result and the semianalytical result 
is divided by the semianalytical result
and as we see it is remarkably small!
The comparison is done for the $\mu^+\mu^-$ final state at $\sqrt{s}=189$ GeV,
as a function of $v_{\max}$.
In the last point (bin)  the entire phase space is covered, $v_{\max}=1-4m_\mu^2/s$.

The  conclusion from the above exercise
is that we control the phase-space integration at the level of 
$2\times 10^{-4}$ for $v_{\max}< 0.999$,
including the $Z$ radiative return,
and at the level of $3\times 10^{-3}$ for no cuts at~all.

The possible loophole in this estimate of precision is that it may break
down when we cut the transverse momenta of the real photons, or switch to
a more sophisticated QED model.  The second is very unlikely as the
phase space and the actual SM model matrix element are separated into
completely separate modules in the program.  The question of the cut
transverse momenta of the real photons requires further discussion.
Here, it has to be stressed that in our MC the so-called big-logarithm
\begin{equation}
  L= \ln\Big( {s\over m_f^2}\Big) - 1
\end{equation}
is the  {\em result of the phase space integration} and if this
integration were not correct then we would witness the breakdown of the
infrared (IR) cancellation and the fermion mass cancellation for FSR.
We do not see anything like that at the 0.02\% precision level.  In
addition there is a wealth of comparisons with many {\em independent
codes} of the phase space integration for $n_\gamma=1,2,3$ real
photons, with and without cuts on photon $p_T$.  It should be
remembered that the multiphoton phase space integration module/code in
${\cal KK}$MC is unchanged since last 10 years.  For ISR it is based
on YFS2 algorithm of ref.~\cite{yfs2:1990} and for FSR on YFS3
algorithm of ref.~\cite{yfs3-pl:1992}, these modules/codes were part of
the KORALZ~\cite{koralz4:1994} multiphoton MC from the very
beginning, already at the time of the LEP1 1989
workshop~\cite{Z-physics-at-lep-1:89}, and they were continuously
tested since then.  The phase space integration for $n_\gamma=1$ was
tested very early by the authors of YFS2/YFS3 against the older MC
programs MUSTRAAL~\cite{mustraal-cpc:1983} and KORALB~\cite{koralb:1985} and with
analytical calculations, at the precision level $<0.1\%$, with and
without cuts on photon $p_T$.  The phase space integration for
$n_\gamma=2,3$ with cuts on photon $p_T$ was tested very many times
over the years by the authors of the YFS2/YFS3/KORALZ and
independently by all four LEP collaborations, using other
integration programs like COMPHEP, GRACE and other ones, in the context of the
search of the anomalous $2\gamma$ and $3\gamma$ events.  Another
important series of tests was done in ref.~\cite{colas:1990} for ISR
$n_\gamma=1,2$ photons (with cuts sensitive to $p_T$ of photons),
comparing KORALZ/YFS2 with the other independent 
MC's for the $\nu\bar{\nu}\gamma(\gamma)$ final states.  
Typically, these tests, in which QED matrix element was programmed in several
independent ways, showed agreement at the level of 10\% for the cross
section for $n_\gamma=2$ which was of order 0.1\% of the Born, or
0.2-0.5\% for $n_\gamma=1$ which was of order 1\% of the Born, so they
never invalidated our present technical precision of 0.02\% in terms
of Born cross section (or total cross section in terms of Z-inclusive cut).

We conclude therefore that the technical precision of ${\cal KK}$MC
due to phase space integration is 0.02\% of the integrated cross
section, for any cuts on photon energies Z-inclusive and Z-exclusive,
stronger than%
\footnote{It downgrades to 0.5\% for $M_{inv}(\mu\bar{\mu}) \leq
   2m_\mu$, i.e. full phase space.}
$M_{inv}(f\bar{f})>0.1\sqrt{s}$ and any mild cut on the
transverse photon energies due to any typical realistic experimental
cuts.  For the cross sections with a single photon tagged it is about
0.2-0.5\% and with two photons tagged it is $\sim 10\%$ of the
corresponding integrated cross section.  These conclusions are based on
the comparisons with at least six other independent codes.

%%%%%%%%%%%%%%%%%%%%%%%%%%%%%%%%%%%%%%%%%%%%%%%%%%%%%%%%%%%%%%%%%
\subsection{Physical precision, the case of EEX}
%%%%%%%%%%%%%%%%%%%%%%%%%%%%%%%%%%%%%%%%%%%%%%%%%%%%%%%%%%%%%%%%%
\label{sec:eex3best}

%/////////////////////////////////////////////////////////////////////////////////////////
%-----------------------------------------------------------------------------------------
\begin{figure}[!ht]
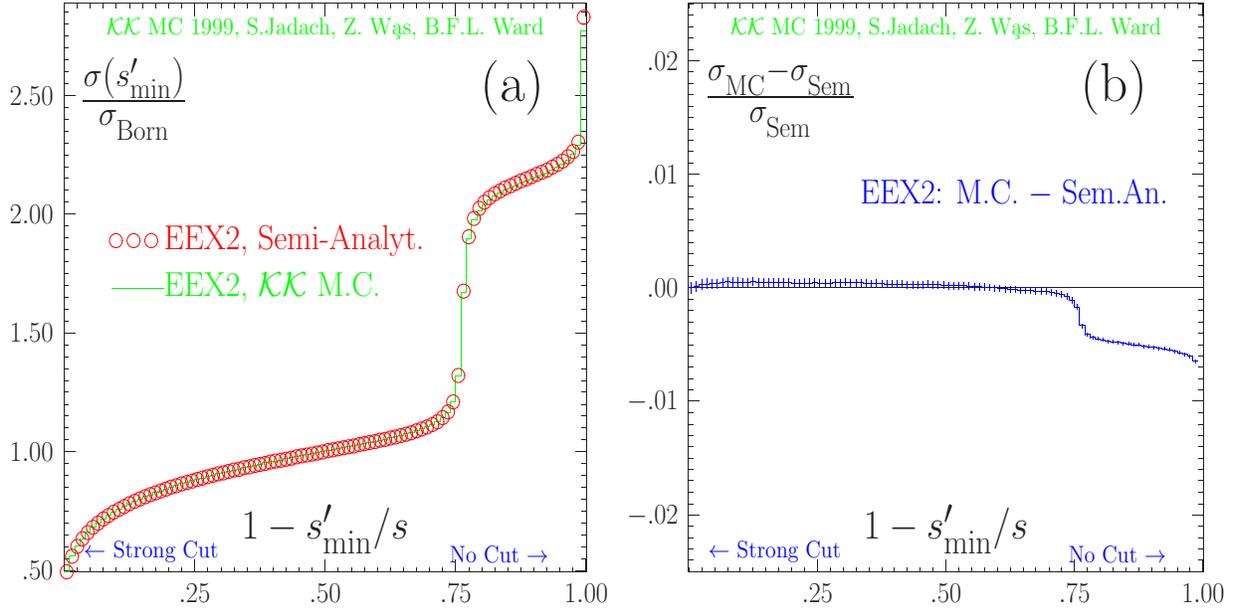

\centering
\setlength{\unitlength}{0.1mm}
\begin{picture}(1600,800)
%\put( 0,0){\framebox( 1600,800){ }}
\put( 650,650){\makebox(0,0)[b]{\LARGE (a)}}
\put(1450,650){\makebox(0,0)[b]{\LARGE (b)}}
\put(  -20, 0){\makebox(0,0)[lb]{\epsfig{file=chi-mcan-O2mca.eps,width=80mm,height=80mm}}}
\put(  800, 0){\makebox(0,0)[lb]{\epsfig{file=chi-mcan-O2dif.eps,width=80mm,height=80mm}}}
\end{picture}
%----------------------------------------
\caption{\small\sf
  Process: $e^-e^+\to f\bar{f}$, for $f=\mu^-$, at $\sqrt{s}=$189 GeV.
  ISR and FSR, IFI=ISR*FSR interf. is off, EW corr. are off.
  Total cross-section $\sigma(s'>s'_{\min})$ where $s'=m^2_{f\bar{f}}$.
}
\label{fig:EEX2}
\end{figure}
%----------------------------------------------------------------------------------------
We now start the presentation of the  numerical results from \KK MC run in the EEX mode
with semianalytical calculations based of the results in Section~\ref{sec:semi}.
Note that the EEX matrix element of Section~\ref{sec:eex}
is very similar (basically the same) to that implemented since
many years in KORALZ program ~\cite{koralz4:1994}.
We do for two reasons: (a) these tests were historically the first,
(they existed unpublished for many years giving us confidence that the KORALZ/YFS3 program
provides correct results) and 
(b) they are now still useful as a reference calculation for the newer CEEX scheme.
They will also allow us to introduce some notations and gradually introduce the reader to the
subject of the discussion of the theoretical precision of our results.
Of course, we shall remember that in the case of EEX we do not include 
the ISR-FSR interferences (IFI).

In Figure~\ref{fig:EEX2} we show the dependence of the total cross section on the cut
on the total photon energy (ISR+FSR).
The comparison is done for the $\mu^+\mu^-$ final state at $\sqrt{s}=189$ GeV,
as a function of $v_{\max}$.
In the last point (bin)  the {\em entire phase space} is covered, i.e. $v_{\max}=1-4m_\mu^2/s$.
The very striking (and well known) phenomenon is that the total cross section
due to huge ISR correction is almost 3 times the Born cross section, 
in the absence of any kinematical cuts.
Part of this ISR contribution is located close to $v=1$, $s'\sim 4m_\mu^2/s$,
let us call it the $\gamma\gamma^*$ process, it amounts to as much as the Born cross section itself,
$\sigma_{\gamma\gamma^*} \sim \sigma_{\rm Born}$,
while dominant part of the cross section $\sigma_{\rm ZRR} \sim 2\sigma_{\rm Born}$
is concentrated close to $v=1-M^2_Z/s \sim 0.75$, and is associated with the so called
``$Z$ radiative return'' (ZRR) process, that is the resonant production of Z,
after emission of rather hard ISR photon, usually lost in the beam pipe.
In the experiment the $\gamma\gamma^*$ process is almost always eliminated
from the data, and the ZRR process is also not very often included in the data sample.
The typical experimental cut is situated somewhere in the range $0.1<v_{\max}<0.3$.
As we see in Figure~\ref{fig:EEX2} (a), the total QED corrections 
$(\sigma(v_{\max})-\sigma_{\rm Born})/\sigma_{\rm Born}$ 
is in this case quite close to zero, in fact slightly negative.

In  Figure~\ref{fig:EEX2}(b) we compare the \KK MC with the semianalytical expression based
on the phase-space integration in Section~\ref{sec:semi}.
In the MC calculation we use the second order EEX type of the QED model
EEX2$\equiv$\Oeex{1,\alpha,\alpha L,\alpha^2 L^2}, defined in Section~\ref{sec:eex}.
The semianalytical formula used in Figure~\ref{fig:EEX2}(b) is 
also in the class EEX2. It is defined as follows
%/////////////////////////////////////////////////////////////////////////
\begin{equation}
\label{eq:KKsem}
  \sigma^f_{\rm SAN} = \int_0^{v_{\max}} dv\;
     \sigma^f_{\rm Born}(s(1-u)(1-v))\;
     \rho^{(2)}_{I}(v)\; \rho^{(2)}_{F}(u),
\end{equation}
where the distributions  $\rho^{(2)}_{I}$ and  $\rho^{(2)}_{F}$ are from
the Tables~\ref{table-initial} and \ref{table-final} in Section~\ref{sec:semi}.

What kind of lesson can we draw from Figure~\ref{fig:EEX2}(b)?
We treat the result in Figure~\ref{fig:EEX2}(b) as an indication that,
the contribution from QED (non-IFI) photonic corrections to
combined physical and technical precision in the  EEX2-class integrated cross section 
for the standard cut $v_{\max}\sim 0.2$ is about 0.2\%, 
for the ZRR process it is 0.7\% and 
for the $\gamma\gamma^*$ process it is 3\%.
We are here talking about the technical precision
of the coding of the EEX2 matrix element, not associated with the phase space integration
(covered in the previous section).

%/////////////////////////////////////////////////////////////////////////////////////////
%-----------------------------------------------------------------------------------------
\begin{figure}[!ht]
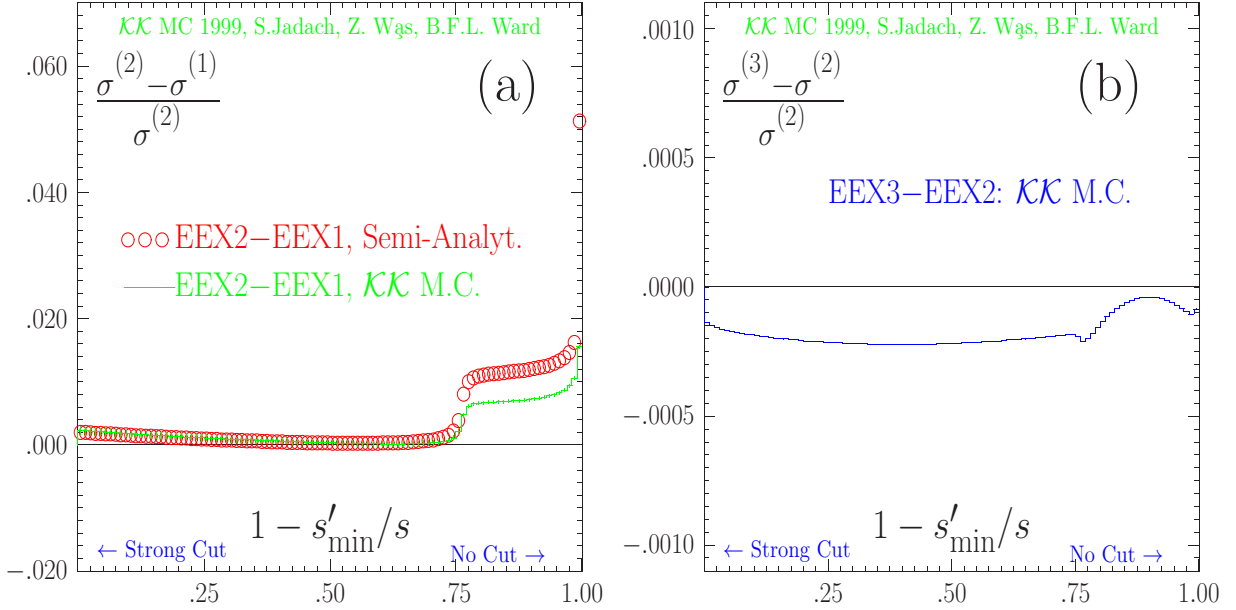

\centering
\setlength{\unitlength}{0.1mm}
\begin{picture}(1600,800)
%\put( 0,0){\framebox( 1600,800){ }}
\put( 650,650){\makebox(0,0)[b]{\LARGE (a)}}
\put(1450,650){\makebox(0,0)[b]{\LARGE (b)}}
\put(  -20, 0){\makebox(0,0)[lb]{\epsfig{file=chi-mcan-O2mO1.eps,width=80mm,height=80mm}}}
\put(  800, 0){\makebox(0,0)[lb]{\epsfig{file=chi-mcan-O3dO2.eps,width=80mm,height=80mm}}}
\end{picture}
%----------------------------------------
\caption{\small\sf
  An attempt to estimate physical precision for EEX: \Order{\alpha^2} and \Order{\alpha^3}
  Process, energy, definition of cuts the same as in Fig.~\protect\ref{fig:EEX2}.
}
\label{fig:EEXdif}
\end{figure}
%----------------------------------------------------------------------------------------
In the next Figure~\ref{fig:EEXdif} we make an attempt to estimate the physical precision
of the QED model in the EEX class.
Specifically, we look into difference between EEX2 (as defined above) and EEX1,
with the EEX1 being the
\Oeex{\alpha^1} of Section~\ref{sec:eex}, EEX1$\equiv$\Oeex{1,\alpha,\alpha L}.
It is plotted in Figure~\ref{fig:EEXdif}(a)
both from \KK MC and semianalytical formula.
Taking conservatively (see the discussion below) half of the difference between 
EEX2 and EEX1 as an estimate of the physical precision of EEX2 we arrive to a similar
estimate of about 0.2\% for the standard cut $v_{\max}\sim 0.2$,
0.7\% for the ZRR process and up to 3\% for the $\gamma\gamma^*$ process.

The other useful piece of information comes from
Figure~\ref{fig:EEXdif}(b), where we plot the difference EEX3$-$EEX2, with
EEX3$\equiv$\Oeex{1,\alpha,\alpha L,\alpha^2 L^2,\alpha^3 L^3},
provides direct insight into the neglected third order LL contributions.
As we see it is always below $3\cdot 10^{-4}$,
(This estimate will be also useful for the case of CEEX.)
If the \Order{L^3 \alpha^3} corrections is of this size, then necessarily
the main contribution to the above estimate of the theoretical error is coming
from the \Order{L^1 \alpha^2} corrections!

In fact the lack of the \Order{\alpha^2 L^1} corrections in both EEX2 and EEX1 is
the main deficiency of the above tests, so they cannot pin down
directly the size of this contribution.
Keeping this limitation in mind, from the above test we nevertheless estimate
tentatively the combined physical and technical precision in the integrated EEX3-class
cross section of the \KK MC to be 0.2\% for the standard cut $v_{\max}\sim 0.2$,
0.7\% for the ZRR process and about to 1.5\% for the $\gamma\gamma^*$ process.
The caveat of this exercise is that we know retrospectively the 
QED non-IFI component of the precision on the KORALZ/YFS3
Monte Carlo at LEP2 energies because the EEX of KORALZ and the EEX of \KK MC are practically the same%
\footnote{KORALZ version 4.02 and earlier have EEX implemented differently from \KK MC.}.
The above does help indeed, in spite of the fact that the neglected IFI contribution to
integrated cross section is of order 1\%, because KORALZ in the non-exponentiated 
\Order{\alpha} mode can calculate IFI separately, see discussion in the following subsections.

%/////////////////////////////////////////////////////////////////////////////////////////
%-----------------------------------------------------------------------------------------
\begin{figure}[!ht]
\centering
\setlength{\unitlength}{0.1mm}
\begin{picture}(800,800)
\put(  0, 0){\makebox(0,0)[lb]{\epsfig{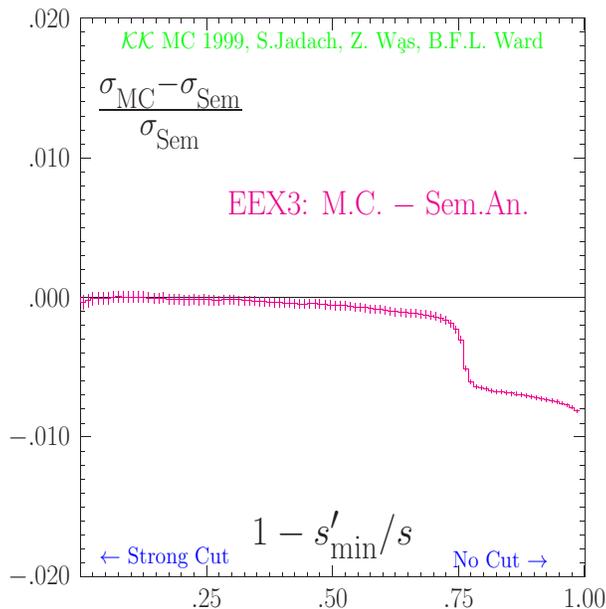}}}
\end{picture}
%----------------------------------------
\caption{\small\sf
  Final attempt to estimate physical precision for EEX3:
  the difference between EEX3 from \KK MC and semianalytical
  EEX3best, see definition in the text.
  Process, energy, definition of cuts the same as in Fig.~\protect\ref{fig:EEX2}.
}
\label{fig:EEXulti}
\end{figure}
%----------------------------------------------------------------------------------------
Let us finally make an ultimate effort to estimate the total precision,
staying all the time within the EEX model.
As we have already noted the most important missing contribution seems to be the \Order{L^1 \alpha^2},
most probably the ISR part of it.
In the the semianalytical formula for the total cross section we are able to add it,
since it is known from ref.~\cite{berends-neerver-burgers:1988}.
The \Order{L^3 \alpha^3} corrections we may add as well and in this way
we replace $\rho^{(2)}_{I}$ by the $\rho^{(3)}_{I}$
of ref.~\cite{third-order:1991} which is the true \Ordpr{\alpha^3} for ISR
(according to the terminology in the Introduction)
and \Ordpr{\alpha^2} for FSR (no IFI). Let us call it
EEX3best$\equiv$\Oeex{1,\alpha,\alpha L,\alpha^2 L^2,\alpha^2 L^1,\alpha^3 L^3}.
The difference between semianalytical EEX3best and EEX3 from \KK MC
is plotted in Figure~\ref{fig:EEXulti}.
As we see this final test confirms the previous estimate of the physical
precision of the EEX type of the matrix element.

%%%%%%%%%%%%%%%%%%%%%%%%%%%%%%%%%%%%%%%%%%%%%%%%%%%%%%%%%%%%%%%%%
\subsection{Physical precision, the case of CEEX}
%%%%%%%%%%%%%%%%%%%%%%%%%%%%%%%%%%%%%%%%%%%%%%%%%%%%%%%%%%%%%%%%%

%/////////////////////////////////////////////////////////////////////////////////////////
%-----------------------------------------------------------------------------------------
\begin{figure}[!ht]
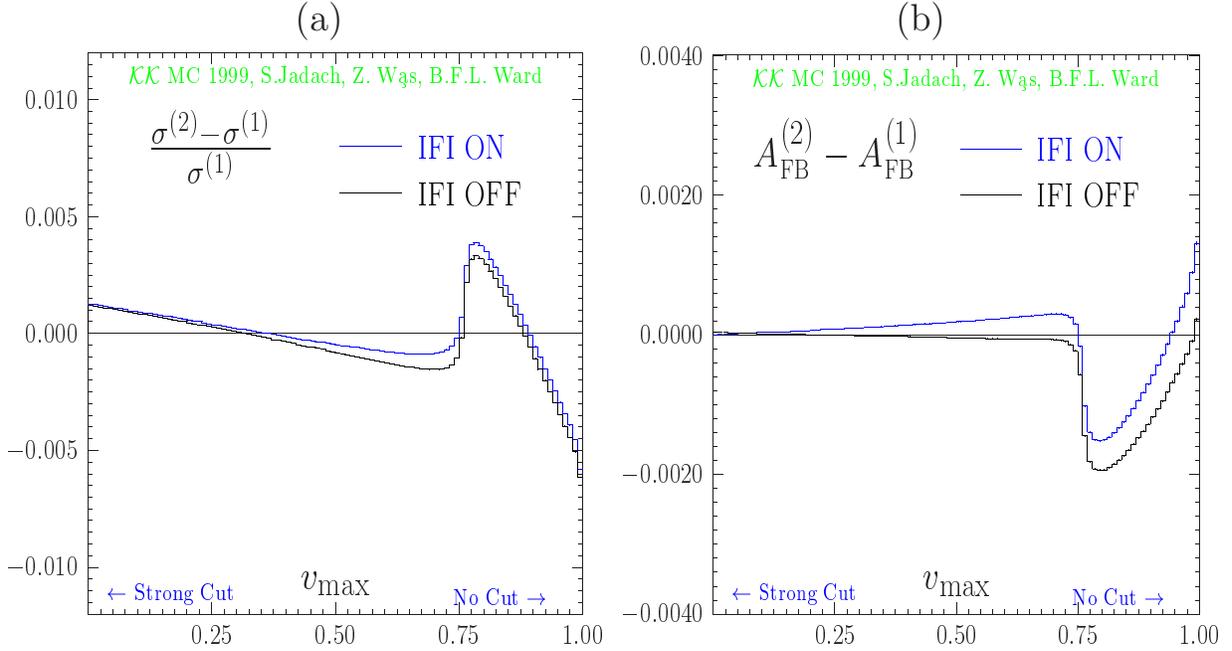

\centering
%\htmlimage{scale=1.6}
%----------------------------
\setlength{\unitlength}{0.1mm}
\begin{picture}(1600,850)
%\put( 0,0){\framebox( 1600,850){ }}
\put( 400,805){\makebox(0,0)[b]{\large (a)}}
\put(1200,805){\makebox(0,0)[b]{\large (b)}}
\put(  -20, 0){\makebox(0,0)[lb]{\epsfig{file=afb-int-sigHO.eps,width=80mm,height=79mm}}}
\put(  800, 0){\makebox(0,0)[lb]{\epsfig{file=afb-int-afbHO.eps,width=80mm,height=80mm}}}
\end{picture}
%%%%%%%%%%%%%%%%%%%%%%%%%%%%%%%%%%%%%%%%%%%%%%%%
\caption{\small\sf
  Evaluation of the physical precision for total cross section and charge asymmetry.
  The difference between \Oceex{\alpha^2} and \Oceex{\alpha^1}
  is plotted as a function of $v_{\max}=1-s'_{\min}/s$.
  Results are shown for the $\mu^+\mu^-$ final state at $\protect\sqrt{s}=189$GeV.
% end-of-caption
}
\label{fig:BasicPhys}
\end{figure}
%/////////////////////////////////////////////////////////////////////////////////////////

Quantitative determination of the {\em physical precision} should be based on the
comparison of the calculations in two consecutive orders
in the expansion parameters, 
for instance by comparing results from \Order{\alpha^r} and \Order{\alpha^{r-1}} calculation,
or \Order{L^r\alpha^n} versus \Order{L^{r-1}\alpha^n}, etc.
For example when only the Born and \Order{\alpha^1} results are available one should take
the difference between the two (or some fraction of it) as an estimate of the physical precision.
The above conservative recipe gives solid estimate of the physical precision and we shall
employ it as our basic method in the following.
In most cases in the literature, however, 
authors try to {\em estimate} the {\em uncalculated} higher order effects
with some ``rule of thumb''.
For instance in the case when  Born and \Order{\alpha^1} results are known they
take ${1\over 2} L {\alpha\over\pi}$ as an estimate of the missing/uncalculated \Order{\alpha^2}
corrections.
This has to be done with care because one may easily overlook some ``enhancement factor''.
For example the cross section close to a resonance can be modified by additional powers of
the big logarithm $\ln{\Gamma\over M}$.
In most cases these ``enhancement factors'' are already seen in the \Order{\alpha^1} calculation
so it is not difficult to trace them.

We are in rather comfortable situation because for QED ``photonic'' corrections we have
at our disposal \Order{\alpha^0}, \Order{\alpha^1} and \Order{\alpha^2} calculations
(at least for ISR, where they are the biggest).
We can therefore afford to take half of the difference between 
the \Order{\alpha^1} and \Order{\alpha^2} calculations as a conservative estimate of the
physical precision due to QED ``photonic'' corrections.
We also profit from the fact that the exponentiation speeds up considerably
the convergence of the perturbative series by ``advanced  summation'' of certain class of corrections
to infinite order,
and by not introducing additional spurious cut-off parameters dividing real emissions into soft and hard ones
which are typical for the calculations without exponentiation 
(see discussion on the famous $k_0$ parameter in the 1989 LEP workshop~\cite{Z-physics-at-lep-1:89}).

Let us mention that in our estimates of the physical precision we omit from the discussion
the \Order{\alpha^2} effects due to an additional fermion pair, either real or virtual.
We do it because:
(a) there are many MC programs which implement production of the four fermion final
states (often with additional ISR) and
(b) in the experiment this contribution can be eliminated at the early stage from the data
in the experimental data analysis aimed at single fermion pair production,
see for example ref.~\cite{DELPUB172}.
In fact this point is still under debate, see forthcoming proceedings of the LEP2 Monte Carlo
workshop~\cite{lep2mc2f:2000}.
It was proposed that in the final combined LEP2
data certain the so called non-singlet initial and final state secondary
pair contribution will be kept in the data, 
as done by OPAL, see refs.~\cite{OPAL2F172,OPAL2F183,OPAL2F189}.
We have recently included the virtual corrections of the 
``vacuum polarization'' type with the fermionic bubble 
in the \Order{\alpha^2} photonic contributions like vertex corrections
in yet unpublished version of 4.14 of \KK MC.
This is done having in mind combining results of \KK MC with the other MC program for four-fermion
production process like KORALW~\cite{koralw:1998}.
The tandem of \KK MC and KORALW programs will be able to realize any possible
scenario of the treatment of the soft/light pair corrections in the LEP2 data.

In Fig.~\ref{fig:BasicPhys} we present the numerical results
on which we base our quantitative estimate of the physical precision due to photonic QED corrections.
In this figure we plot the difference between \Oceex{\alpha^2} and \Oceex{\alpha^1}
for the total cross section and charge asymmetry at 189GeV as a function of the cut on the total energy
emitted by all ISR and FSR photons for $\mu^+\mu^-$ final state.
The cut is formulated with the $s'>s'_{\min}$ or equivalently $v<v_{\max}$ condition,
where $s'$ is the effective mass squared of the $\mu^+\mu^-$ pair and $v=1-s'/s$, as usual.
One should remember that the actual experimental cut is around $v_{\max}\sim 0.2$
(eliminating $Z$ radiative return) in the case of the standard data analysis, 
and sometimes around $v_{\max}\sim 0.9$ in the case when  $Z$ radiative return is admitted in the data.
The ``kink'' around $v_{\max}\sim 0.75$ is at the position of the $Z$ radiative return.
In either case, whether we admit or eliminate the $Z$ radiative return, that is for $v_{\max}\sim 0.9$,
the difference between \Oceex{\alpha^2} and \Oceex{\alpha^1} in total cross section is below 0.4\% and
for the charge asymmetry it is below 0.002.

%-----------------------------------------------------------------------------------------
%/////////////////////////////////////////////////////////////////////////////////////////
%-----------------------------------------------------------------------------------------
\begin{figure}[!ht]
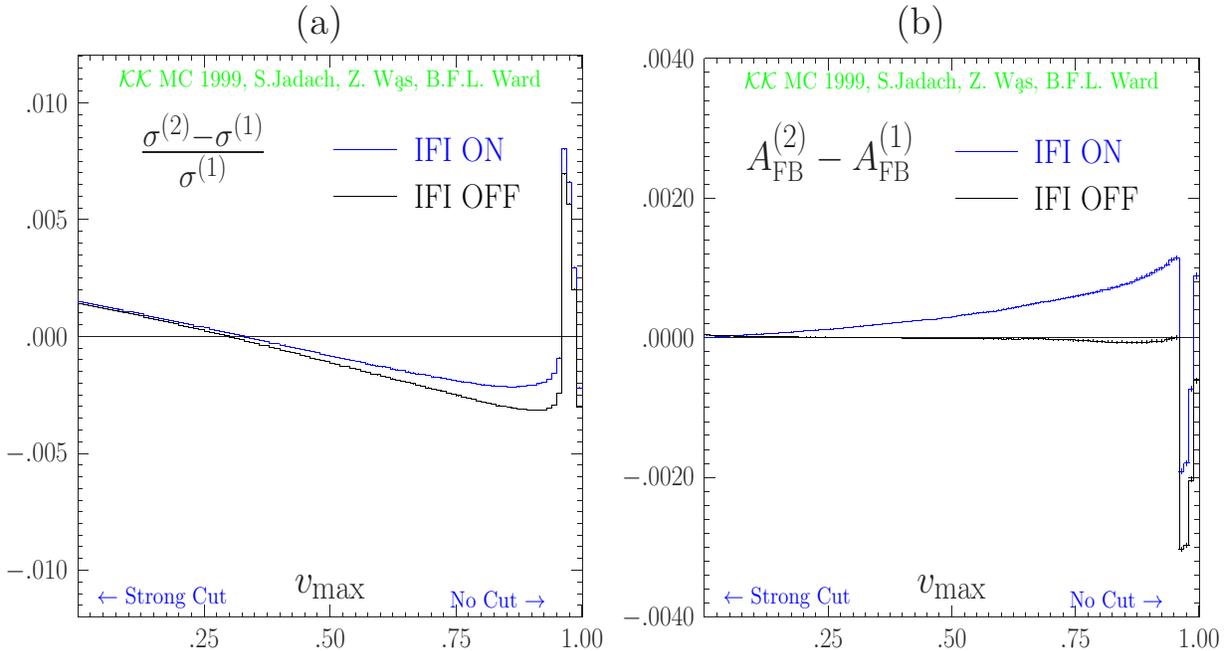

\centering
%\htmlimage{scale=1.6}
%----------------------------
\setlength{\unitlength}{0.1mm}
\begin{picture}(1600,850)
%\put( 0,0){\framebox( 1600,850){ }}
\put( 400,805){\makebox(0,0)[b]{\large (a)}}
\put(1200,805){\makebox(0,0)[b]{\large (b)}}
\put(  -20, 0){\makebox(0,0)[lb]{\epsfig{file=afb-int-sigHO-500GeV.eps,width=80mm,height=79mm}}}
\put(  800, 0){\makebox(0,0)[lb]{\epsfig{file=afb-int-afbHO-500GeV.eps,width=80mm,height=80mm}}}
\end{picture}
%%%%%%%%%%%%%%%%%%%%%%%%%%%%%%%%%%%%%%%%%%%%%%%%
\caption{\small\sf
  Evaluation of the physical precision for total cross section and charge asymmetry.
  at $\protect\sqrt{s}=500$GeV.
  The difference between \Oceex{\alpha^2} and \Oceex{\alpha^1}
  is plotted as a function of $v_{\max}=1-s'_{\min}/s$ for $\mu^+\mu^-$ final state.
}
\label{fig:BasicPhys500GeV}
\end{figure}
%/////////////////////////////////////////////////////////////////////////////////////////
%-----------------------------------------------------------------------------------------
Taking {\em conservatively} half of this difference among \Oceex{\alpha^2} and \Oceex{\alpha^1}
as an estimate of the neglected \Oceex{\alpha^3} and higher orders we conclude that
the physical precision due to photonic QED corrections of our \Oceex{\alpha^2} calculation
for all possible cutoffs within $0<v_{\max}<0.9$ range
is 0.2\% in the total cross reaction and 0.001 in the charge asymmetry.
This estimate would be even a factor of two better,
if we restricted ourselves to the most typical cut-off range $0.1<v_{\max}<0.3$.
The above estimate will be confirmed by more auxiliary tests in the following.

As we see we have improved on the physical precision estimate with respect to the previous
estimates for the EEX model -- in addition we do include IFI all the time.
For the ZRR process we now quote for the integrated cross-section 0.2\% instead of the previous 0.7\%
and for $\gamma\gamma^*$ we have something like 0.3\% instead of the previous 1.5\%.
This we interpret as a result of inclusion of the \Order{L \alpha^2} ISR correction
in our CEEX spin amplitudes.

We have to stress very strongly that the estimate of the physical precision
depends on the type of the observable (we took $\sigma$ and $A_{\rm FB}$),
the type of the final state
(we took $\mu$ pair final state; for the quark final state the QED FSR effects 
are smaller due to the smaller electric charges of quarks)
and on many other input parameters, for example the total CMS energy.
The great thing about the Monte Carlo is that the type of the evaluation
we proposed and implemented in this Section 
(half of difference \Order{\alpha^2}$-$\Order{\alpha^1})
can be repeated for any observable, any final state and any energy.
For example in Figure~\ref{fig:BasicPhys500GeV} we repeat our evaluation
of the physical precision for $\sigma$ and $A_{\rm FB}$ at the Linear Collider
energy 500GeV.
As we see the resulting precision is worse, the worsening is negligible
for a mild cut of order $v_{\max}<0.5$ and almost factor two for the $Z$ radiative return,
which is now placed close the $v=0.95$.

%%%%%%%%%%%%%%%%%%%%%%%%%%%%%%%%%%%%%%%%%%%%%%%%%%%%%%%%%%%%%%%%%%%%%%%%%%%%%%%%%
\subsection{Absolute predictions, more on physical/technical precision}
%%%%%%%%%%%%%%%%%%%%%%%%%%%%%%%%%%%%%%%%%%%%%%%%%%%%%%%%%%%%%%%%%%%%%%%%%%%%%%%%%
In this Section we shall present the SM absolute predictions for the total cross section
and charge asymmetry at LEP2 (189GeV) and at the Linear Collider (500GeV).
We compare them with our own semianalytical program \KK sem,
with KORALZ~\cite{koralz4:1994} and in some cases with ZFITTER~\cite{zfitter6:1999}.
They may not improve our basic estimates of the technical and physical precision
from the previous sections,
but they can confirm them (or disprove them!).

%/////////////////////////////////////////////////////////////////////////////////////////
%-----------------------------------------------------------------------------------------
\begin{table}[!ht]
\centering
\setlength{\unitlength}{0.1mm}
\begin{picture}(1600,800)
%\put( 0,0){\framebox( 1600,800){ }}
\put(  -20, 0){\makebox(0,0)[lb]{\epsfig{file=afb-int-tab1.eps,width=164mm,height=80mm}}}
\end{picture}
%%%%%%%%%%%%%%%%%%%%%%%%%%%%%%%%%%%%%%%%%%%%%%%%
\caption{\small\sf
  Absolute prediction for total cross section an charge asymmetry.
  They are for $\mu^+\mu^-$ final state at $\protect\sqrt{s}=189$GeV.
  Results are plotted as a function of the cut-off on the total photon energy $v_{\max}=1-s'_{\min}/s$.
  The ``reference'' $\sigma$ and $A_{\rm FB}$ in first column are from \KK sem semi-analytical program.
  We have used Higgs boson mass 100GeV and top mass 175GeV as input parameters.
}
\label{tab:absolut}
\end{table}
%-----------------------------------------------------------------------------------------

In Table~\ref{tab:absolut} we show numerical results for the total cross-section
$\sigma(v_{\max})$ and charge asymmetry $A_{\rm FB}(v_{\max})$ as a function of the cut
$v_{\max}$ on the total photon energy 
(the cut-off parameter $v_{\max}$ is defined as in the previous subsection).
Generally, in Table~\ref{tab:absolut} we show results
with the ISR-FSR interference  (IFI) switched on and off.
The \KK sem semianalytical program
(part of the \KK MC package) provides {\em ``reference results''} 
for $\sigma$ and $A_{\rm FB}$, see the first column in Table~\ref{tab:absolut}, 
which are without IFI,
obtained from using EEX3best formula defined in the previous
Section~\ref{sec:eex3best}.
For the charge asymmetry we use the convolution-type semianalytical formula like
that of eq.~(\ref{eq:KKsem}).
(In fact we use this formula separately for the cross section in the forward and backward
hemispheres and then we calculate $A_{\rm FB}$ from these partial integrals.)
Results from the \KK MC in Table~\ref{tab:absolut}  are shown for two types of QED matrix element:
\Oceex{\alpha^2} with and without IFI.
In addition we include results from KORALZ are for the \Order{\alpha^1} 
matrix element with and without IFI which will be discussed in the next Section.

%/////////////////////////////////////////////////////////////////////////////////////////
%-----------------------------------------------------------------------------------------
\begin{figure}[!ht]
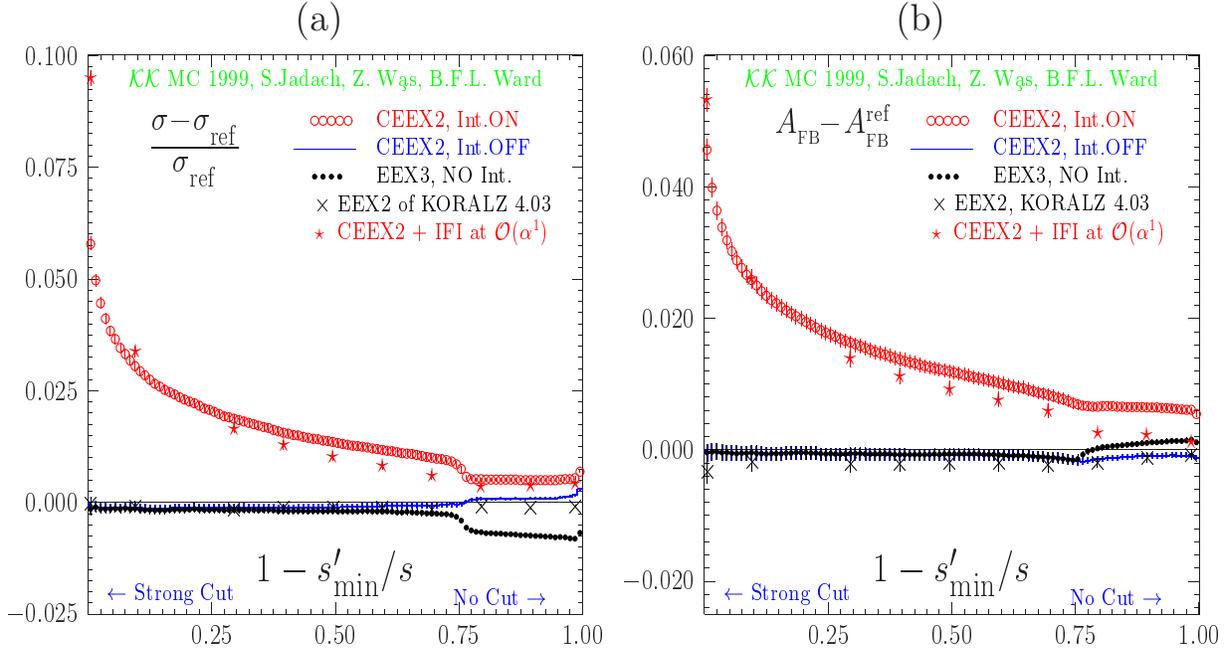

\centering
\setlength{\unitlength}{0.1mm}
\begin{picture}(1600,850)
%\put( 0,0){\framebox( 1600,850){ }}
\put( 400,805){\makebox(0,0)[b]{\large (a)}}
\put(1200,805){\makebox(0,0)[b]{\large (b)}}
\put(  -20, 0){\makebox(0,0)[lb]{\epsfig{file=afb-int-Gsig.eps,width=80mm,height=80mm}}}
\put(  800, 0){\makebox(0,0)[lb]{\epsfig{file=afb-int-Gafb.eps,width=80mm,height=80mm}}}
\end{picture}
%%%%%%%%%%%%%%%%%%%%%%%%%%%%%%%%%%%%%%%%%%%%%%%%
\caption{\small\sf
  Absolute prediction for total cross section an charge asymmetry.
  They are for $\mu^+\mu^-$ final state at $\protect\sqrt{s}=189$GeV.
  Results are plotted as a function of the cut-off on the total photon energy $v_{\max}=1-s'_{\min}/s$.
  The ``reference'' $\sigma$ and $A_{\rm FB}$ in the first column are from \KK sem semi-analytical program.
}
\label{fig:absolut}
\end{figure}
%-----------------------------------------------------------------------------------------

As tables with list of numbers are difficult to comprehend, 
we present the essential
results of the Table~\ref{tab:absolut} in Figure~\ref{fig:absolut},
where they are all plotted as a difference 
with the {\em reference} results of our semianalytical program \KK sem.
(In other words the results from \KK sem are exactly on the $x$-axis.)

In the case of IFI switched on \KK sem cannot be used as a cross-check of the \KK MC.
Remembering that IFI in KORALZ in the \Order{\alpha^1} mode (without exponentiation)
is very well tested we combine the \Order{\alpha^1} IFI contribution with
the CEEX result without IFI.
Such a hybrid solution denoted in Fig.~\ref{fig:absolut} as ``CEEX2+IFI at \Order{\alpha^1}''
us used as our primary test of the full CEEX matrix element with IFI switched on.
The above procedure is done separately for cross sections
in the forward and backward hemispheres such that the prediction for charge asymetry
is also available.

It is worth to mention that the above hybrid solution
was already successfully used in refs.\cite{holt:1996,holt:1997} 
for the study of the IFI contribution at Z peak, imposing strong acollinearity cut.
It is also implemented in a semianalytical form in ZFITTER 6.x.
On general ground we expect this recipe to be rather good, because
IFI correction itself at \Order{\alpha^1} does not contain any large mass logarithm
and is relatively small and can be handled additively.

In  Figure~\ref{fig:absolut} we also show the numerical results from 
\KK MC in the EEX3 mode (no IFI), from KORALZ in the EEX2 mode (no IFI),
which are not included in Table~\ref{tab:absolut}.

Let us now comment on the results in Figure~\ref{fig:absolut}.
The EEX3 from \KK MC differs from EEX3best of \KK sem (no IFI in both)
by about 0.7\% for the ZRR process, as we have already seen,
and we interpret this difference as the result of the missing \Order{L^1 \alpha^2}.
The EEX2 of KORALZ 4.03 is closer to the EEX3best of \KK sem for ZRR process
-- we do not see any contradiction in this since the implementation of EEX
in KORALZ and \KK MC differs in the details 
(causing a difference of \Order{L^1 \alpha^2} in the integrated cross section.)

In the case of IFI switched off,
the CEEX2 result, corresponding exactly to \Oceex{\alpha^2}, 
defined in Section~\ref{sec:ceex}, as implemented in \KK MC 4.13,
agrees very well with the EEX3best of \KK sem.
This result is compatible with the total theoretical precision of 0.2\%
for the integrated cross-section, even including the ZRR process.

In the case of IFI switched on, the hybrid solution ``CEEX2+IFI at \Order{\alpha^1}''
also agrees with the full CEEX2 result confirming the total theoretical precision of 0.2\%
for the integrated cross-section, including the ZRR process.

For the charge asymmetry in Figure~\ref{fig:absolut} situation is quite similar.
The IFI effect is up to 4\% for strong cuts.
In the case of IFI switched off,
the CEEX2 result agrees with the EEX3best of \KK sem to within 0.2\%.
For IFI included, the CEEX2 agrees with the hybrid solution rather well,
to within 0.4\%.
Note that in the above Monte Carlo exercise we have
used the symmetric definition of the scattering angle
$\theta^\bullet$ of ref.~\cite{afb-prd:1991}
(which is close to what is used in the LEP experiments).

Summarising, the numerical results in Figure~\ref{fig:absolut} establish
our basic estimate of the theoretical precision  of the \KK MC, due to QED effects,
at LEP2 energies of about 0.2\% for total cross section and 0.2-0.4\% (depending on cut-offs)
for charge asymmetry.

%/////////////////////////////////////////////////////////////////////////////////////////
%-----------------------------------------------------------------------------------------
\begin{figure}[!ht]
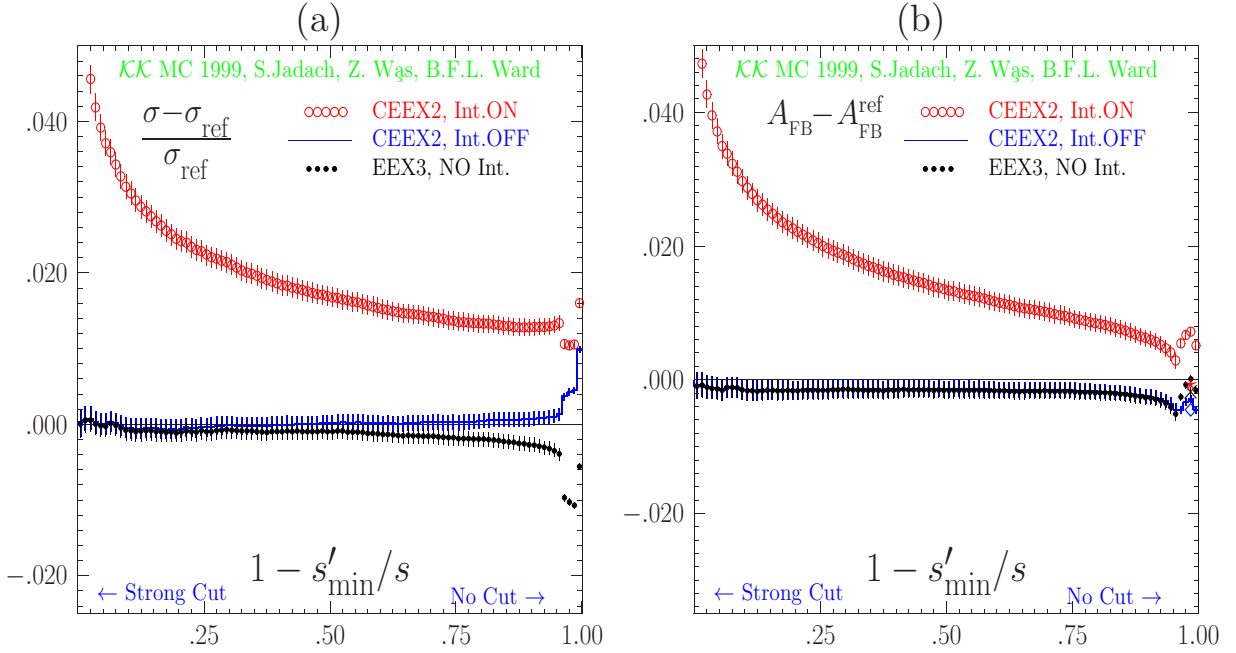

\centering
\setlength{\unitlength}{0.1mm}
\begin{picture}(1600,850)
%\put( 0,0){\framebox( 1600,850){ }}
\put( 400,805){\makebox(0,0)[b]{\large (a)}}
\put(1200,805){\makebox(0,0)[b]{\large (b)}}
\put(  -20, 0){\makebox(0,0)[lb]{\epsfig{file=afb-int-Gsig-500GeV.eps,width=80mm,height=80mm}}}
\put(  800, 0){\makebox(0,0)[lb]{\epsfig{file=afb-int-Gafb-500GeV.eps,width=80mm,height=80mm}}}
\end{picture}
%%%%%%%%%%%%%%%%%%%%%%%%%%%%%%%%%%%%%%%%%%%%%%%%
\caption{\small\sf
  Total cross section an charge asymmetry
  for $\mu^+\mu^-$ final state at $\protect\sqrt{s}=500$GeV.
  Results analogous as in Fig.~\protect\ref{fig:absolut}.
}
\label{fig:absolut2}
\end{figure}
%-----------------------------------------------------------------------------------------
Finally, we examine the analogous results from the \KK MC at 500GeV in  Figure~\ref{fig:absolut2}.
In this case we include only results from the \KK MC and \KK sem.
The pattern of agreement is up to a factor two the same as at 189GeV.

%%%%%%%%%%%%%%%%%%%%%%%%%%%%%%%%%%%%%%%%%%%%%%%%%%%%%%%%%%%%%%%%%
\subsection{Initial-final state interference}
%%%%%%%%%%%%%%%%%%%%%%%%%%%%%%%%%%%%%%%%%%%%%%%%%%%%%%%%%%%%%%%%%
The control of the initial-final state interference correction 
down to the precision of 0.2\% in the integrated cross-section 
and in the charge asymmetry is rather important -- 
this is why we dedicate this section to a more detailed
study of this QED correction.
In particular we would like to answer the following questions:
\begin{itemize}
    \item How big is the ISR$\otimes$FSR interference in $\sigma_{tot}$, $A_{FB}$?
    \item Do we know ISR$\otimes$FSR at \Order{\alpha^1}?
    \item Do we know ISR$\otimes$FSR beyond \Order{\alpha^1}?
    \item How sensitive is ISR$\otimes$FSR to cut-off changes?
\end{itemize}

%/////////////////////////////////////////////////////////////////////////////////////////
%-----------------------------------------------------------------------------------------
\begin{figure}[!ht]
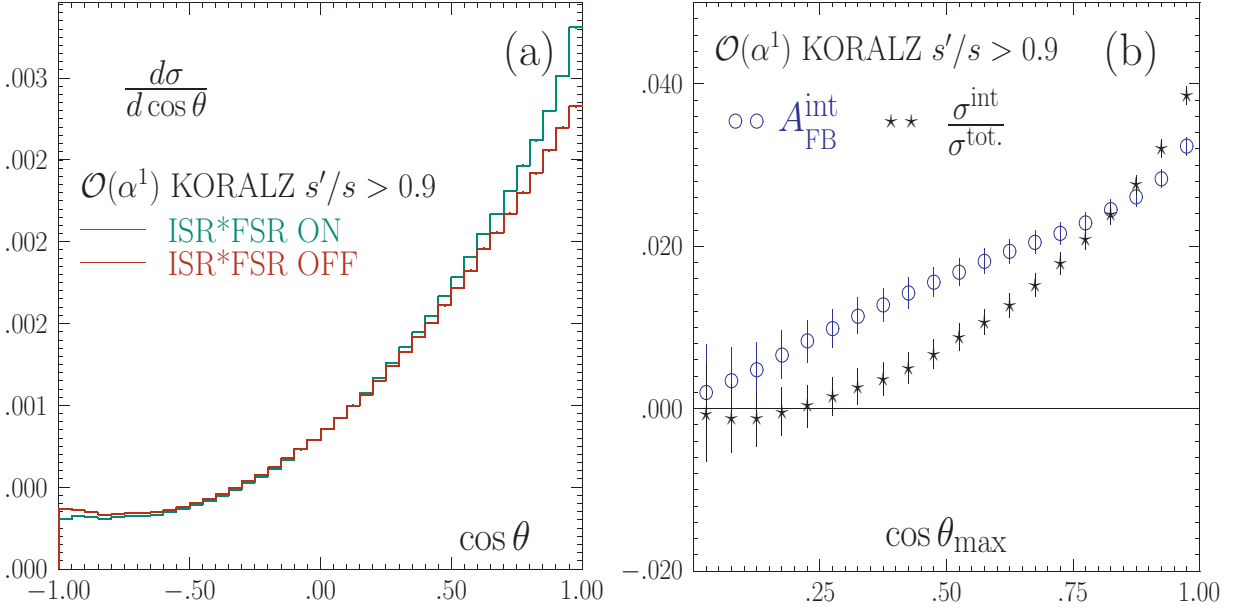

\centering
\setlength{\unitlength}{0.1mm}
\begin{picture}(1600,800)
\put( 680,700){\makebox(0,0)[b]{\Large (a)}}
\put(1480,700){\makebox(0,0)[b]{\Large (b)}}
\put(  -20, 0){\makebox(0,0)[lb]{\epsfig{file=afb-int-AngMx.eps,width=80mm,height=80mm}}}
\put(  800, 0){\makebox(0,0)[lb]{\epsfig{file=afb-int-comMx.eps,width=80mm,height=80mm}}}
\end{picture}
%%%%%%%%%%%%%%%%%%%%%%%%%%%%%%%%%%%%%%%%%%%%%%%%
\caption{\small\sf
  Results from \Order{\alpha^1} KORALZ (no exponentiation)
  for $\mu^+\mu^-$ final state at $\sqrt{s}$=189GeV.
  The energy cut is on $s'/s$, where $s'=m^2_{f\bar{f}}$.
  The angular cut is $|\cos\theta|<\cos\theta_{\max}$.
  The scattering angle is $\theta=\theta^{\bullet}$ of ref.~\protect\cite{afb-prd:1991}.
}
\label{fig:ifi-koralz}
\end{figure}
%-----------------------------------------------------------------------------------------
%/////////////////////////////////////////////////////////////////////////////////////////
%-----------------------------------------------------------------------------------------
\begin{figure}[!ht]
\centering
\setlength{\unitlength}{0.1mm}
\begin{picture}(1600,1630)
\put( 650,1500){\makebox(0,0)[tr]{\large (a)}}
\put(1500,1500){\makebox(0,0)[tr]{\large (b)}}
\put( 650, 700){\makebox(0,0)[tr]{\large (c)}}
\put(1500, 700){\makebox(0,0)[tr]{\large (d)}}
\put(  -20, 830){\makebox(0,0)[lb]{\epsfig{file=afb-int-G1x.eps,  width=80mm,height=79mm}}}
\put(  800, 830){\makebox(0,0)[lb]{\epsfig{file=afb-int-com1x.eps,width=80mm,height=80mm}}}
\put(  -20,   0){\makebox(0,0)[lb]{\epsfig{file=afb-int-G1.eps,   width=80mm,height=79mm}}}
\put(  800,   0){\makebox(0,0)[lb]{\epsfig{file=afb-int-com1.eps, width=80mm,height=80mm}}}
\end{picture}
%%%%%%%%%%%%%%%%%%%%%%%%%%%%%%%%%%%%%%%%%%%%%%%%
\caption{\small\sf
  Results from \Order{\alpha^2} \KK MC.
  for $\mu^+\mu^-$ final state at $\sqrt{s}$=189GeV.
  The energy cut is on $s'/s$, where $s'=m^2_{f\bar{f}}$.
  The angular cut is $|\cos\theta|<\cos\theta_{\max}$.
  Scattering angle is $\theta=\theta^{\bullet}$ of ref.~\protect\cite{afb-prd:1991}.
}
\label{fig:ifi-kkmc}
\end{figure}
%-----------------------------------------------------------------------------------------
KORALZ is the best starting point and reference for the problem of calculating the ISR$\otimes$FSR.
In Figure~\ref{fig:ifi-koralz} we show results from the \Order{\alpha^1} KORALZ (no exponentiation)
for the $\mu^+\mu^-$ final state at $\sqrt{s}$=189GeV.
Angular distributions from KORALZ, pure \Order{\alpha^1} (without exponentiation),
were verified very precisely at the level of $\sim 0.01\%$ using a special analytical calculation,
see ref.~\cite{afb-prd:1991},
so we know ISR$\otimes$FSR at \Order{\alpha^1} very precisely.
As we see the ISR$\otimes$FSR contribution to the integrated cross-section is about 3\% and
about 0.03 to $A_{\rm FB}$.
This is definitely above the ultimate experimental error tag for the combined
LEP2 data at the end of LEP2 operation.
The energy cut on the total photon energy is fixed in the results of Figure~\ref{fig:ifi-koralz}
to just one value $v<v_{\max}=0.1$ (where $v_{\max}=1-s'/s$ is defined as usual).
This is close to the usual value in the experimental LEP2 data analysis.
We introduce also the angular cut $|\cos\theta|<\cos\theta_{\max}$ and vary the value
of $\cos\theta_{\max}$, see Figure~\ref{fig:ifi-koralz}(b),
the value used in the experimental LEP2 data analysis
is around $\cos\theta_{\max}=0.9$; this corresponds to two bins before
the last one in Figure~\ref{fig:ifi-koralz}(b)
(the last point in the plot is for $\cos\theta_{\max}=1$).
In this way already we have answered the first two questions from the list above.

In Figure~\ref{fig:ifi-kkmc} we present similar results from the \KK MC which will help
us to answer whether we know ISR$\otimes$FSR beyond \Order{\alpha^1} and inspect
in a more detail the dependence on cut-offs.
In Figure~\ref{fig:ifi-kkmc}(a-b) we essentially repeat the exercise of Figure~\ref{fig:ifi-koralz}
finding out the ISR$\otimes$FSR contribution to the angular distribution and $A_{\rm FB}$
for the same energy-cut using \KK MC instead of KORALZ.
As we see the results change slightly, the ISR$\otimes$FSR effect is about 20\%-30\% smaller.
We attribute it mainly to 
(a) different (better) treatment of the ISR in \KK MC
(b) exponentiation of the ISR$\otimes$FSR effect is in \KK MC.
As it is well known, in \Order{\alpha^1}, the ISR$\otimes$FSR
contributes like $4Q_eQ_f{\alpha\over\pi} \ln {1-\cos\theta \over 1+\cos\theta}$
to the angular distribution 
-- this even causes the angular distribution to be negative close to $\cos\theta=-1$.
In the CEEX exponentiation the above singularity is summed up to infinite order and
the angular distribution near $|\cos\theta|=1$ is not singular any more.
(This kind of exponentiation will be implemented in 
the next version of ZFITTER, see~\cite{lep2mc2f:2000} for first numerical results.)
The typical experimental cut $|\cos\theta|<0.9$ eliminates most of the above trouble
anyway -- what is probably more important is the correct ``convolution'' of the
IR-finite \Order{\alpha^1} ISR$\otimes$FSR with the \Order{\alpha^2} ISR.
In the \KK MC this is done in a maximally clean way from the theoretical/physical point of view
(at the amplitude level) while in the semianalytical programs like ZFITTER~\cite{zfitter6:1999}
this is done in a more ``ad hoc'' manner.
Let us remind the reader that we still lack the genuine IR-finite \Order{\alpha^2}
corrections in the ISR$\otimes$FSR class from diagrams like 2-boxes and 5-boxes,
see Section~\ref{sec:ceex}.
These contributions are most likely negligible, of order \Order{L^1\alpha^2} at most.

%/////////////////////////////////////////////////////////////////////////////////////////
%-----------------------------------------------------------------------------------------
\begin{figure}[!ht]
\centering
\setlength{\unitlength}{0.1mm}
\begin{picture}(1000,800)
\put(  -20, 0){\makebox(0,0)[lb]{\epsfig{file=afb-int-afb2.eps,width=100mm,height=80mm}}}
\end{picture}
%%%%%%%%%%%%%%%%%%%%%%%%%%%%%%%%%%%%%%%%%%%%%%%%
\caption{\small\sf
 ISR$\otimes$FSR contribution to $A_{\rm FB}$ ``bin-per-bin''.
}
\label{fig:AFBinBins}
\end{figure}
%-----------------------------------------------------------------------------------------
In Figure~\ref{fig:ifi-kkmc}(c-d) we make the energy cut more loose, $v_{\max}=0.9$,
thus admitting the ZRR into the available phase-space.
As a result the relative ISR$\otimes$FSR decreases by a factor 3, simply because
it gets ``diluted'' in the factor 3 bigger integrated cross section,
while ZRR does not contribute to ISR$\otimes$FSR
because of its narrow-resonance character, 
already discussed in Section~\ref{sec:ceex} at length.
The fact that the ZRR does not contribute to the ISR$\otimes$FSR can be seen explicitly in 
Figure~\ref{fig:AFBinBins} where we plot the ISR$\otimes$FSR contribution
to $A_{\rm FB}$ ``bin-per-bin'', that is calculated in each bin separately. 
As we see the contribution from the ZRR which at this energy (189GeV) is located
at $v=0.75$ is very small, smaller than from all other $v$'s where there is no Z resonance.

%/////////////////////////////////////////////////////////////////////////////////////////
%-----------------------------------------------------------------------------------------
\begin{figure}[!ht]
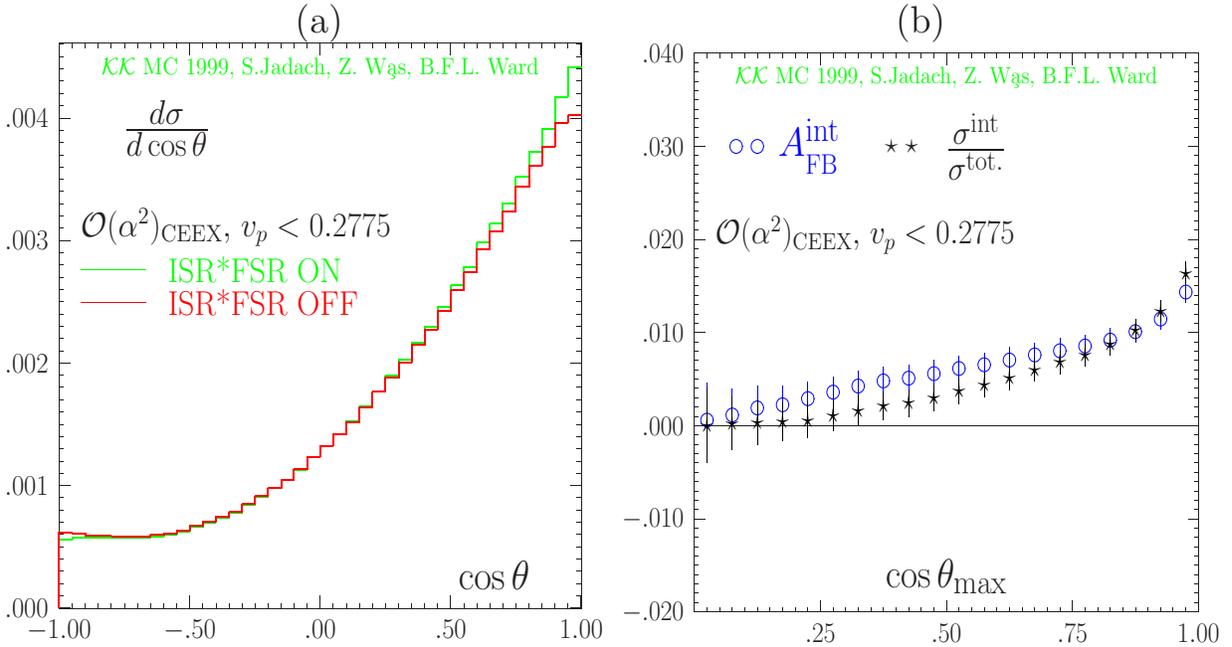

\centering
\setlength{\unitlength}{0.1mm}
\begin{picture}(1600,820)
\put( 400,800){\makebox(0,0)[b]{\large (a)}}
\put(1200,800){\makebox(0,0)[b]{\large (b)}}
\put(  -20, 0){\makebox(0,0)[lb]{\epsfig{file=afb-int-G1xxx.eps,width=80mm,height=80mm}}}
\put(  800, 0){\makebox(0,0)[lb]{\epsfig{file=afb-int-com1xxx.eps,width=80mm,height=80mm}}}
\end{picture}
%%%%%%%%%%%%%%%%%%%%%%%%%%%%%%%%%%%%%%%%%%%%%%%%
\caption{\small\sf
  Results from \Order{\alpha^1} from \KK MC
  for $\mu^+\mu^-$ final state at $\sqrt{s}$=189GeV.
  The energy cut is on $v_p=1-s'/s$, where $s'$ is estimate of the ``propagator eff. mass''
  as defined by ALEPH.
  The angular cut is $|\cos\theta|<\cos\theta_{\max}$.
  Scattering angle is $\theta=\theta^{\bullet}$ of ref.~\protect\cite{afb-prd:1991}.
}
\label{fig:ifi-mix-vprop}
\end{figure}
\begin{figure}[!ht]
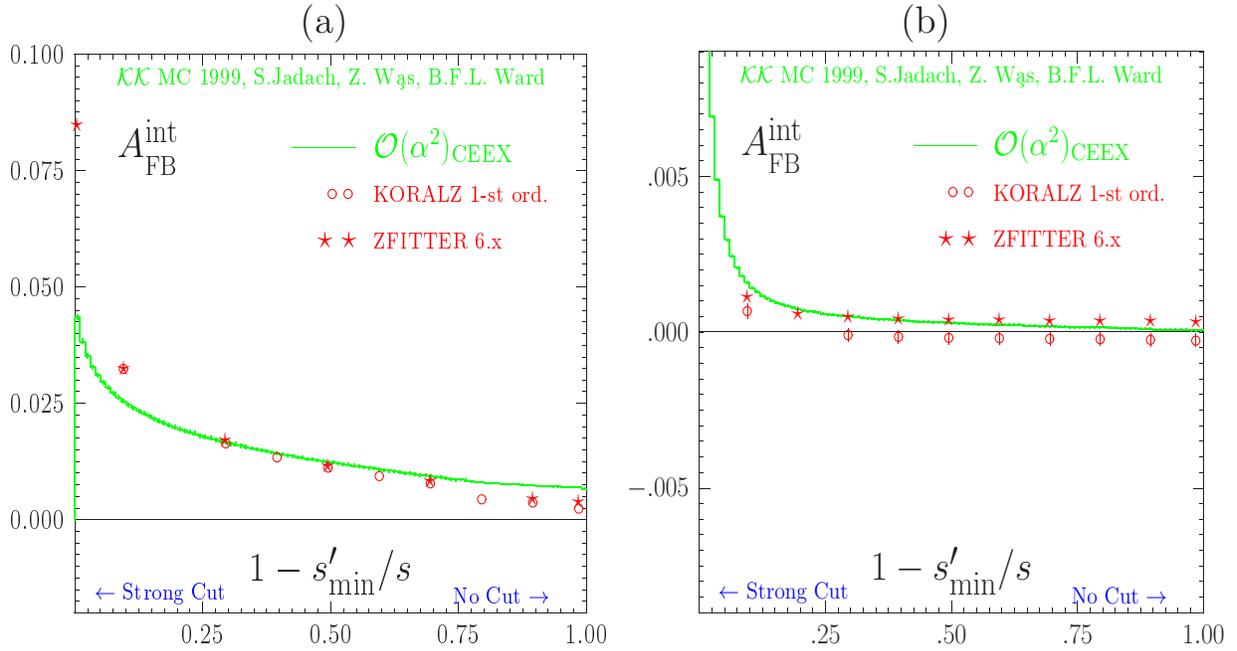

\centering
\setlength{\unitlength}{0.1mm}
\begin{picture}(1600,820)
\put( 400,800){\makebox(0,0)[b]{\large (a)}}
\put(1200,800){\makebox(0,0)[b]{\large (b)}}
\put(  -20, 0){\makebox(0,0)[lb]{\epsfig{file=afb-int-afb1.eps,width=80mm,height=80mm}}}
\put(  800, 0){\makebox(0,0)[lb]{\epsfig{file=afb-int-afb1-91GeV.eps,width=80mm,height=79mm}}}
\end{picture}
%%%%%%%%%%%%%%%%%%%%%%%%%%%%%%%%%%%%%%%%%%%%%%%%
\caption{\small\sf
  $s'$-cut dependence of $\delta A_{FB}$. No $\theta$-cut
}
\label{fig:ifi-afb-vcut}
\end{figure}

%-----------------------------------------------------------------------------------------
In the above exercises and also in the following we use always the energy cut on the $v=1-s'/s$ variable
defined in terms of the effective mass of the ``bare'' final fermions, that is without
any attempt of combining them with the collinear FSR photons.
This is experimentally well justified for the $\mu$-pair final states but not for $\tau$-pairs
or quarks.
It is possible and in fact rather easy to define a ``propagator'' or ``reduced'' $s'_p$ which
takes into account the loss of energy due to ISR but not FSR.
In other words the  $s'_p$ effective mass squared sums up FSR photons.
One can ask a legitimate question:
If we would cut not on the ``bare'' final fermion variable $s'$, but instead on the ``propagator''  $s'_p$
then perhaps the estimate of the ISR$\otimes$FSR contribution would be  dramatically different,
for instance it would be much smaller?
In Figure~\ref{fig:ifi-afb-vcut} we show a numerical exercise in which we employ the energy cut
in terms of $v_p=1-s'_p/s$.
One can construct such a $s'_p$ looking into angles of the outgoing fermions.
This type of variable was used in ref~\cite{afb-prd:1991}.
In Figure~\ref{fig:ifi-afb-vcut} we use the definition of $s'_p$ of ALEPH~\cite{aleph-angle:1996}.
As we see in this Figure, the result is not dramatically different from what we have seen
in Figure~\ref{fig:ifi-kkmc}.
The magnitude of ISR$\otimes$FSR contribution is close to what we could see if we applied
the same value of the energy cut for the ``bare'' $s'$ (as we have checked independently).

%/////////////////////////////////////////////////////////////////////////////////////////
%-----------------------------------------------------------------------------------------
%-----------------------------------------------------------------------------------------
We shall now examine the dependence of the ISR$\otimes$FSR contribution
on the energy-cut $v_{\max}$ in a more detail.
In Figure~\ref{fig:ifi-afb-vcut} we show
the ISR$\otimes$FSR contribution to $A_{\rm FB}$ as a function of energy-cut $v_{\max}$
at two energies (a) 189GeV and (b) at the $Z$ peak $\sqrt{s}=M_Z$.
No cut on $\cos\theta$ is applied.
In addition to \KK MC results we show the results from \Order{\alpha^1} mode of KORALZ
and from ZFITTER.
At 189GeV and for the typical energy cut $0.2<v_{\max}<0.3$ all three programs agree very well.
This cut is relatively ``inclusive'' such that exponentiation effects are not so important
and ISR is eliminated in a ``gentle'' way (the total cross section is close to the Born value).
For stronger cuts $v_{\max}<0.2$ we see a large (factor 2) discrepancy among 
the results from \KK MC and both KORALZ and ZFITTER,
because of the lack of exponentiation in KORALZ and ZFITTER.
(in ZFITTER ISR$\otimes$FSR is taken without exponentiation and combined with ISR ``additively'').
We also observe the discrepancy of about 0.2\% among \KK MC on one hand and both KORALZ and ZFITTER
on the other hand for the ZRR.
Our guess is that it is due to the difference in the method of combining ISR$\otimes$FSR
with the second order ISR
(of course, we believe that the CEEX method of doing it at the amplitude level
is the best one can do).
In Figure~\ref{fig:ifi-afb-vcut}(b) we see, first of all,  the well known phenomenon
of the strong suppression of the ISR$\otimes$FSR contribution at the resonance,
especially for the loose cut-off.
Even for a strong cut, $v_{\max}=0.1$, the ISR$\otimes$FSR contribution is about 0.01,
about factor 30 smaller than in the off-resonance case.
Here, \KK MC agrees rather well with KORALZ and ZFITTER.
The differences are generally%
\footnote{
  The difference between KORALZ and ZFITTER should be perhaps smaller because both are
  \Order{\alpha^1}? May be the difference is due to the angle definition?}
up to 0.0015.
%/////////////////////////////////////////////////////////////////////////////////////////
%-----------------------------------------------------------------------------------------
\begin{figure}[!ht]
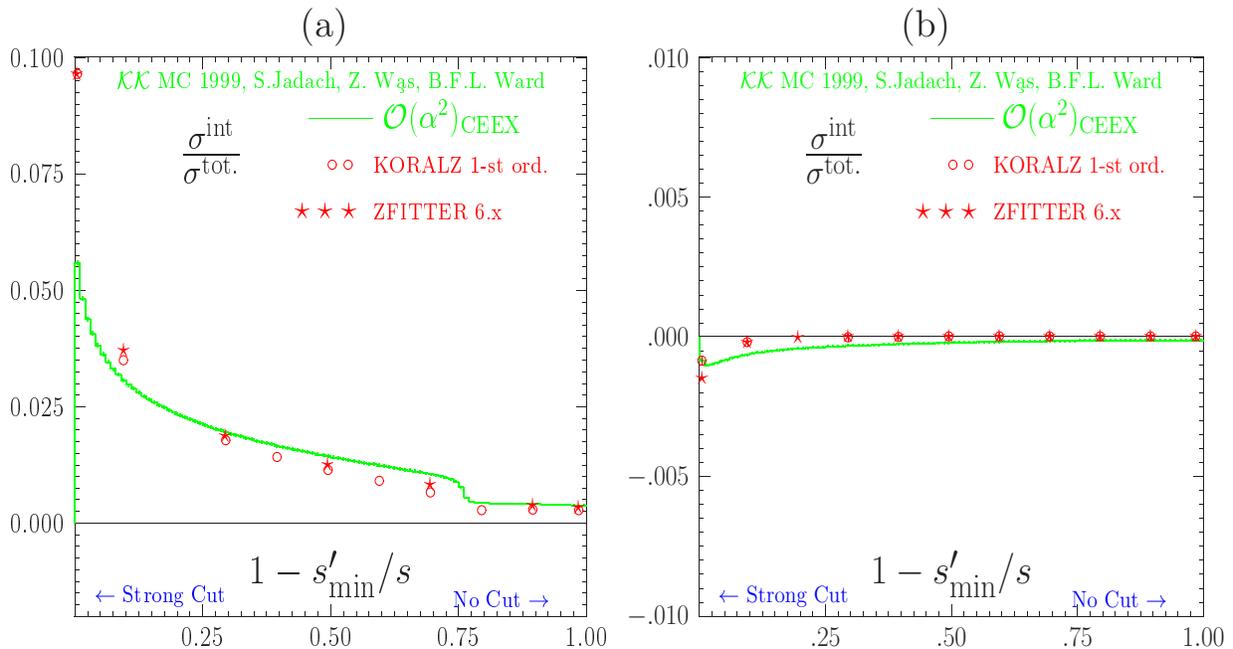

\centering
\setlength{\unitlength}{0.1mm}
\begin{picture}(1600,820)
\put( 400,800){\makebox(0,0)[b]{\large (a)}}
\put(1200,800){\makebox(0,0)[b]{\large (b)}}
\put(  -20, 0){\makebox(0,0)[lb]{\epsfig{file=afb-int-sig1.eps,width=80mm,height=80mm}}}
\put(  800, 0){\makebox(0,0)[lb]{\epsfig{file=afb-int-sig1-91GeV.eps,width=80mm,height=80mm}}}
\end{picture}
%%%%%%%%%%%%%%%%%%%%%%%%%%%%%%%%%%%%%%%%%%%%%%%%
\caption{\small\sf
  $s'$-cut dependence of $\delta\sigma$, No $\theta$-cut.
}
\label{fig:ifi-sig-vcut}
\end{figure}
%-----------------------------------------------------------------------------------------

In Figure~\ref{fig:ifi-sig-vcut}(a) we examine the ISR$\otimes$FSR contribution
to integrated cross section as a function of the energy cut $v_{\max}$.
At 189GeV and for the typical energy cut $0.1<v_{\max}<0.6$ all three programs agree reasonably well,
KORALZ and ZFITTER are generally closer to each other than to \KK MC.
After admitting the ZRR, $v_{\max}>0.8$ all three programs agree even better.
For a very strong cut, $v_{\max}<0.1$ KORALZ and ZFITTER disagree dramatically with \KK MC
because of the lack of exponentiation in KORALZ and ZFITTER.
In Figure~\ref{fig:ifi-sig-vcut}(b) we see, again  the strong suppression of the 
ISR$\otimes$FSR contribution at the resonance, especially for the loose cut-off.
Suppression is cut-off dependent and generally stronger for KORALZ and ZFITTER than for \KK MC.
Most of the comments which we made on ISR$\otimes$FSR contribution to $A_{\rm FB}$
apply also here.

%/////////////////////////////////////////////////////////////////////////////////////////
%-----------------------------------------------------------------------------------------
\begin{figure}[!ht]
\centering
\setlength{\unitlength}{0.1mm}
\begin{picture}(1200,1200)
\put(  -20, 0){\makebox(0,0)[lb]{ \epsfig{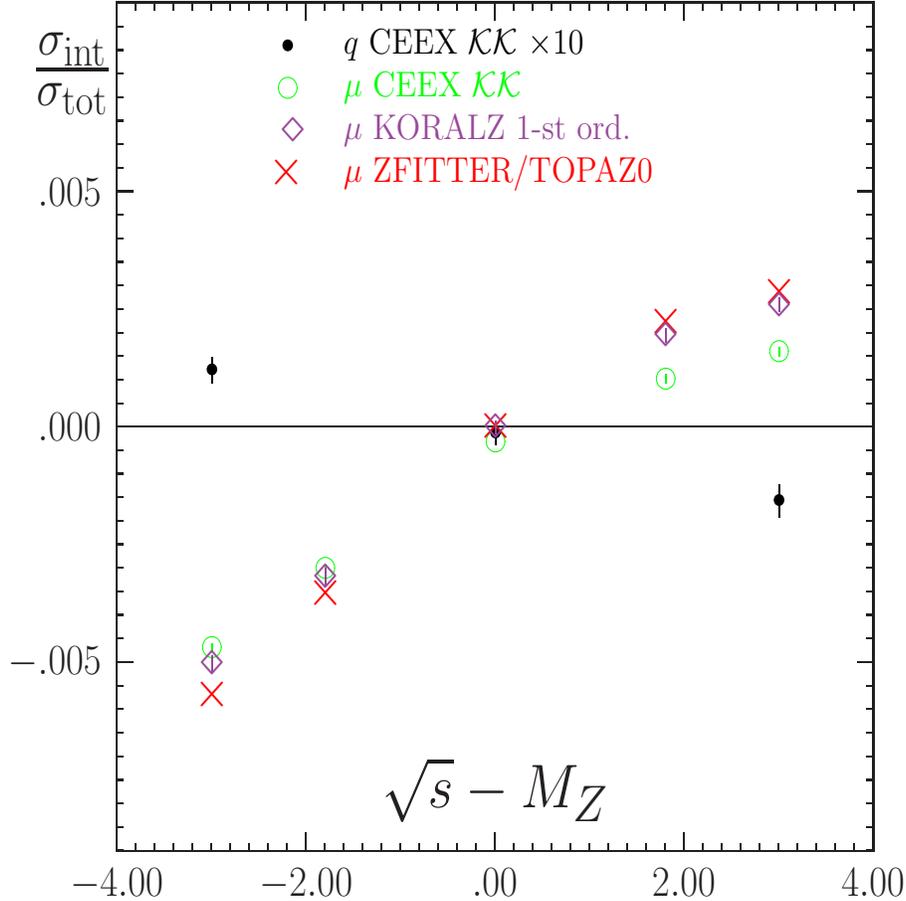}}}
\end{picture}
%%%%%%%%%%%%%%%%%%%%%%%%%%%%%%%%%%%%%%%%%%%%%%%%
\caption{\small\sf
  Back on Z peak.
}
\label{fig:ifi-Z-peak}
\end{figure}
%-----------------------------------------------------------------------------------------

Finally in Figure~\ref{fig:ifi-Z-peak} we go back to a vicinity of the $Z$ peak (LEP1)
and we show  the magnitude of the ISR$\otimes$FSR contribution
to the integrated cross section as a function of the CMS energy, for the $\mu^-\mu^+$
final state and for all five quark final states taken together 
(the so-called hadronic cross section)
from the \KK Monte Carlo.
No angular cut or energy cut is applied (full phase space).
For the $\mu^-\mu^+$ final state we also 
include results from the KORALZ \Order{\alpha^1} and ZFITTER/TOPAZ0~\cite{zfitter6:1999,topaz0:1998}.
Results from quarks are multiplied by a factor 10 to be visible, because ISR$\otimes$FSR contribution
in this case is small. 
It is not only suppressed by the smallness of the quark charge, but we also have partial
cancellation among up- and down-type quarks, see ref~\cite{afb-prd:1991}.
However, the ISR$\otimes$FSR contribution to hadronic cross section has to be known
much more precisely (factor $\sim 3$) because it is measured much more precisely, due to higher statistics.
In Fig.~\ref{fig:ifi-Z-peak} we see that the suppression of ISR$\otimes$FSR is much weaker
as we go away from the centre of the resonance, and it changes the resonance curve
in such a way that it affects the fitted mass of the $Z$.
The actual size of the shift of $M_Z$ was studied in ref. \cite{IFI-with-bolek:1999}
and it was found to be 0.15MeV.
Results of the \KK MC are smaller about 10-20\% than
the \Order{\alpha^1} estimates of KORALZ \Order{\alpha^1} and ZFITTER,
away from the $Z$ peak.
This is compatible with the 10-20\% size of the \Order{L^2\alpha^2} ISR corrections
with respect to \Order{L^1\alpha^1} corrections, which are included in \KK MC and are not
included in KORALZ and apparently also not included in ZFITTER/TOPAZ0
(which agree very well with KORALZ).
Our last comment concerns the reliability of our estimate for the ISR$\otimes$FSR contribution
in the absence of the correct implementation of the simultaneous emission of the FSR photon
and FSR gluon.
We think that through the usual arguments, see ref. \cite{IFI-with-bolek:1999},
we can neglect from consideration emission of the FSR single gluon, as long as we stick
to very a inclusive cross section, like the total cross section in Figure~\ref{fig:ifi-Z-peak}.
For stronger angular cuts, or events with definite jet multiplicity, we would need to
improve our calculation.

We summarize  the results of this section on ISR$\otimes$FSR  as follows:
\begin{itemize}
\item
  For a typical expt. energy cut of 0.3 ISR$\otimes$FSR int. is about 1.5\% in $\sigma_{tot}$ 
  and $A_{FB}$.
\item
  For the energy cut 0.1 it is a factor 2  bigger.
\item
  The cut $|\cos\theta|<0.9$ makes it 25\% smaller.
\item
  The \Order{\alpha^1} ISR$\otimes$FSR int. is under total control using KORALZ and \KK\ Monte Carlo
  for arbitrary cuts.
\item
  Effects beyond \Order{\alpha^1} are negligible, ($<$20\% of \Order{\alpha^1}),
  except when the energy cut is stronger than 0.1.
\item
  ISR$\otimes$FSR int. at the Z radiative return is very small, as expected.
\item
  Changing from $s'$ to $Q^2$-propagator in the energy cut has no effect.
\end{itemize}

%%%%%%%%%%%%%%%%%%%%%%%%%%%%%%%%%%%%%%%%%%%%%%%%%%%%%%%%%%%%%%%%%
\subsection{Total theoretical precision}
%%%%%%%%%%%%%%%%%%%%%%%%%%%%%%%%%%%%%%%%%%%%%%%%%%%%%%%%%%%%%%%%%

Let us summarize the total theoretical precision:
\begin{itemize}
  \item 
    For the most typical cut-off range $0.1<v_{\max}<0.3$ excluding the Z radiative return
    we quote for CEEX the total precision 0.2\%
    for LEP2 and for the LC at 0.5TeV.
  \item
    For a cut-off including ZRR we quote  0.2\% total precision for LEP2 
    and 0.4\% total precision for the LC at 0.5TeV.
  \item
    For $\gamma\gamma^*$ we quote 0.3\% at LEP2 (no firm result for LC).
\end{itemize}
In the above estimates the technical component was significantly below the physical one.
Restrictions apply: No light fermion pairs (pure photonic QED), no EW component.

%%%%%%%%%%%%%%%%%%%%%%%%%%%%%%%%%%%%%%%%%%%%%%%%%%%%%%%%%%%%%%%%%%%%%%%%%%%%%%%%%%%%%%%%%%%%%
%%%%%%%%%%%%%%%%%%%%%%%%%%%%%%%%%%%%%%%%%%%%%%%%%%%%%%%%%%%%%%%%%%%%%%%%%%%%%%%%%%%%%%%%%%%%%
\section{Outlook and summary}
%%%%%%%%%%%%%%%%%%%%%%%%%%%%%%%%%%%%%%%%%%%%%%%%%%%%%%%%%%%%%%%%%%%%%%%%%%%%%%%%%%%%%%%%%%%%%
The most important new features in the present CEEX are the
ISR-FSR interference, the second-order subleading corrections, and the exact matrix
element for two hard photons.
This makes CEEX already a unique source of SM predictions for the LEP2 physics program
and for the LC physics program.
Note that for these the electroweak correction library has to be reexamined at LC energies.
The most important omission in the present version is the lack of neutrino and electron
channels.
Let us stress that the present program is an excellent starting platform
for the construction of the second-order Bhabha MC generator based on CEEX exponentiation.
We hope to be able to include the Bhabha and neutrino channels soon, 
possibly in the next version.
The other important directions for the development are 
the inclusion of the exact matrix element for three hard photons, together
with virtual corrections up to \Order{\alpha^3L^3} and the
emission of the light fermion pairs.
The inclusion of the $W^+W^-$ and $t\bar{t}$ final states is still in a farther perspective.

%%%%%%%%%%%%%%%%%%%%%%%%%%%%%%%%%%%%%%%%%%%%%%%%%%%%%%%%%%%%%%%%%%%%%%%%%%%%%%%%%%%%%%%%%%%%%
%%%%%%%%%%%%%%%%%%%%%%%%%%%%%%%%%%%%%%%%%%%%%%%%%%%%%%%%%%%%%%%%%%%%%%%%%%%%%%%%%%%%%%%%%%%%%
\section*{Acknowledgements}
%%%%%%%%%%%%%%%%%%%%%%%%%%%%%%%%%%%%%%%%%%%%%%%%%%%%%%%%%%%%%%%%%%%%%%%%%%%%%%%%%%%%%%%%%%%%%
Two of us (SJ and BFLW) would like to thank the  CERN EP and TH Divisions. We are grateful
to all four LEP Collaborations and their members for support. 
In particular we would like to thank Dr. D. Schlatter of ALEPH for continuous support and help.
One of us (S.J.) would like to thank the DESY Directorate for its generous support
in the critical stage of the beginning of this project.
We would like to express our gratitude to W.~P\l{}aczek, E.~Richter-W\c{a}s, M.~Skrzypek and 
S.~Yost for valuable comments.

\newpage
%%%%%%%%%%%%%%%%%%%%%%%%%%%%%%%%%%%%%%%%%%%%%%%%%%%%%%%%%%%%%%%%%%%%%%%%%%%%%%%%%%%%%%%%%%%%%
%%%%%%%%%%%%%%%%%%%%%%%%%%%%%%%%%%%%%%%%%%%%%%%%%%%%%%%%%%%%%%%%%%%%%%%%%%%%%%%%%%%%%%%%%%%%%
%%%%%%%%%%%%%%%%%%%%%%%%%%%%%%%%%%%%%%%%%%%%%%%%%%%%%%%%%%%%%%%%%%%%%%%%%%%%%%%%%%%%%%%%%%%%%
%\section{\normalsize Appendix A: Basic KS/GPS spinors and photon polarizations}
\section{Appendix A: Basic KS/GPS spinors and photon polarizations}
%%%%%%%%%%%%%%%%%%%%%%%%%%%%%%%%%%%%%%%%%%%%%%%%%%%%%%%%%%%%%%%%%%%

The arbitrary massless spinor $u_\lambda(p)$
of momentum $p$ and chirality $\lambda$ is defined according to KS methods
\cite{kleiss-stirling:1985,kleiss-stirling:1986}.
In the following we follow closely the notation of ref.~\cite{gps:1998}
(in particular we also use $\zeta=\zeta_{\downarrow}$).
In the above framework every spinor is transformed out of the two 
{\em constant basic} spinors
$\umf_\lambda(\zeta)$, of opposite chirality $\lambda=\pm$, as follows
%/////////////////////////////////////////////////////
\begin{equation}
\label{def-massless}
  u_\lambda(p) 
     = {1\over \sqrt{ 2p\cdot \zeta}} \not\!p  \umf_{-\lambda}(\zeta),\quad
     \umf_+(\zeta)=\not\!\eta  \umf_-(\zeta),\quad
     \eta^2=-1,\quad
     (\eta\zeta)=0.
\end{equation}
The usual relations hold:
 $\not\!\zeta \umf_\lambda(\zeta) =0$,
 $\omega_\lambda \umf_\lambda(\zeta) =\umf_\lambda(\zeta)$,
 $\umf_\lambda(\zeta) \bar{\umf}_\lambda(\zeta) = \not\!\zeta \omega_\lambda $,
 $\not\!p u_\lambda(p) =0$,
 $\omega_\lambda u_\lambda(p) =u_\lambda(p)$,
 $u_\lambda(p) \bar{u}_\lambda(p) = \not\!\!p \omega_\lambda$, where
 $\omega_\lambda = {1\over 2} (1+\lambda \gamma_5)$.
Spinors for the massive particle with four-momentum $p$ (with $p^2=m^2$)
and spin projection $\lambda/2$ are defined similarly
%%%%/////////////////////////////////////////////////////
\begin{equation}
\label{def-massive}
  u(p,\lambda) 
    = {1\over \sqrt{2p\cdot \zeta}}\; (\not\!p +m)\; u_{-\lambda}(\zeta),\qquad
  v(p,\lambda) 
    = {1\over \sqrt{2p\cdot \zeta}}\; (\not\!p -m)\; u_{ \lambda}(\zeta).
\end{equation}
or, equivalently, in terms of massless spinors
%/////////////////////////////////////////////////////
\begin{equation}
\label{def-massive2}
u(p,\lambda)
        = u_\lambda(p_\zeta)    +{m\over \sqrt{2p\zeta}} \umf_{-\lambda}(\zeta),\qquad
v(p,\lambda)
        = u_{-\lambda}(p_\zeta) -{m\over \sqrt{2p\zeta}} \umf_{\lambda}(\zeta),
\end{equation}
where 
$  p_\zeta  \equiv \hat{p}  \equiv p - \zeta\; m^2/(2\zeta p)$
is the light-cone projection ($ p_\zeta^2 =0 $) 
of the $p$ obtained with the help of the constant auxiliary vector $\zeta$.

The above definition is supplemented in ref.~\cite{gps:1998} with the precise prescription
on spin quantization axes, translation from spin amplitudes to density matrices
(also in vector notation) and the methodology of connecting production and decay
for unstable fermions.
We collectively call these rules global positioning of spin (GPS).
Thanks to these we are able to easily introduce polarizations for beams
and implement polarization effects for final fermion decays 
(of $\tau$-leptons, $t$-quarks), for the first time
also in the presence of emission of many ISR and FSR photons!

%%%%%%%%%%%%%%%%%%%% added paragraph defining GPS rules %%%%%%%%%%%%%%%%%%%%%%%%%%%%
The GPS rules determining the spin quantization frame
for $u(p,\pm)$ and $v(p,\pm)$ of eq.~(\ref{def-massive2})
are summarized as follows:
\begin{itemize}
\item[(a)]
  In the rest frame of the fermion, take the $z$-axis along $-\vec{\zeta}$.
\item[(b)]
  Place the $x$-axis in the plane defined by the $z$-axis from the previous point
  and the vector $\vec{\eta}$, in the same half-plane as $\vec{\eta}$.
\item[(c)]
  With the $y$-axis, complete the right-handed system of coordinates.
  The rest frame defined in this way we call the GPS frame of the particular fermion.
\end{itemize}
See ref.~\cite{gps:1998} for more details.
In the following we shall assume that polarization vectors of beams
and of outgoing fermions are defined in their corresponding GPS frames.
%%%%%%%%%%%%%%%%%%%%%%%%%%%%%%%%%%%%%%%%%%%%%%%%%%%%%%%%%%%%%%%%%%%%%%%%%%%%%%%%%%%%

The inner product of the two massless spinors is defined as follows
%//////////////////////////////////////////////////
%    Inner product
%//////////////////////////////////////////////////
  \begin{equation}
        s_{+}(p_1,p_2) \equiv \bar{u}_{+}(p_1) u_{-}(p_2),\quad
        s_{-}(p_1,p_2) \equiv \bar{u}_{-}(p_1) u_{+}(p_2) 
                       = -( s_{+}(p_1,p_2))^*.
  \end{equation}
%%------------
The above inner product can be evaluated
using the Kleiss-Stirling expression
%//////////////////////////////////////////////////
\begin{equation}
  s_+(p,q) = 2\; (2p\zeta)^{-1/2}\; (2q\zeta)^{-1/2}\; 
  \left[ (p\zeta)(q\eta) - (p\eta)(q\zeta) 
    -i\epsilon_{\mu\nu\rho\sigma} \zeta^\mu \eta^\nu p^\rho q^\sigma 
  \right]
\end{equation}
in any reference frame.
In particular, in the laboratory frame we typically use
$\zeta=(1,1,0,0)$ and $\eta=(0,0,1,0)$, which leads to the following ``massless''
inner product
%//////////////////////////////////////////////////
\begin{equation}
  s_+(p,q) = -(q^2 + iq^3) \sqrt{ (p^0 -p^1)/( q^0-q^1) }
             +(p^2 + ip^3) \sqrt{ (q^0 -q^1)/( p^0-p^1) }.
\end{equation}
Equation~(\ref{def-massive2}) immediately
provides us also with the {\em inner product} for massive spinors
%%------------
\begin{equation}
  \begin{split}
    & \bar{u}(p_1,\lambda_1)  u(p_2,\lambda_2)
                =S(p_1,m_1,\lambda_1,  p_2,m_2,\lambda_2), \\
    & \bar{u}(p_1,\lambda_1)  v(p_2,\lambda_2)
                =S(p_1,m_1,\lambda_1,  p_2,-m_2,-\lambda_2), \\
    & \bar{v}(p_1,\lambda_1)  u(p_2,\lambda_2)
                =S(p_1,-m_1,-\lambda_1,  p_2,m_2,\lambda_2), \\
    & \bar{v}(p_1,\lambda_1)  v(p_2,\lambda_2)
                =S(p_1,-m_1,-\lambda_1,  p_2,-m_2,-\lambda_2), \\
  \end{split}
\end{equation}
%%------------
where 
%%------------
\begin{equation}
\label{inner-massive}
  S(p_1,m_1,\lambda_1,  p_2,m_2,\lambda_2)
  = \delta_{\lambda_1,-\lambda_2} s_{\lambda_1}({p_1}_\zeta, {p_2}_\zeta)
   +\delta_{\lambda_1, \lambda_2} 
  \left(
     m_1 \sqrt{ {2\zeta p_2 \over 2\zeta p_1}  }
    +m_2 \sqrt{ {2\zeta p_1 \over 2\zeta p_2}  }
  \right).
\end{equation}
%%------------
In our spinor algebra we shall exploit the completeness relations
%//////////////////////////////////////////////////
%                Completness
%//////////////////////////////////////////////////
\begin{equation}
  \begin{split}
    &\not\!{p}+m = \sum_\lambda u(p,\lambda) \bar{u}(p,\lambda),\quad
     \not\!{p}-m = \sum_\lambda v(p,\lambda) \bar{v}(p,\lambda),\\
    &\not\!{k}   = \sum_\lambda u(k,\lambda) \bar{u}(k,\lambda),\quad k^2=0.
  \end{split}
\end{equation}
%%------

For a circularly polarized photon with four-momentum $k$
and helicity $\sigma=\pm 1$ we adopt the KS choice
(see also ref.~\cite{beijing:1987}) of polarization vector%
\footnote{ 
  Contrary to other papers on Weyl spinor techniques
  \protect\cite{kleiss-stirling:1985,calkul}
  we keep here the explicitly complex conjugation in $\epsilon$. This conjugation
  is cancelled by another conjugation following from Feynman rules, but only
  for outgoing photons, not for a beam photon, as in the Compton process, 
  see ref.~\protect\cite{stuart:1989}.
  }
%//////////////////////////////////////////////////
%                Photon polarizations
%//////////////////////////////////////////////////
\begin{equation}
\label{phot-pol}
  (\epsilon^\mu_\sigma(\beta))^*
     ={\bar{u}_\sigma(k) \gamma^\mu u_\sigma(\beta)
       \over \sqrt{2}\; \bar{u}_{-\sigma}(k) u_\sigma(\beta)},\quad
  (\epsilon^\mu_\sigma(\zeta))^*
     ={\bar{u}_\sigma(k) \gamma^\mu \umf_\sigma(\zeta)
       \over \sqrt{2}\; \bar{u}_{-\sigma}(k) \umf_\sigma(\zeta)},
\end{equation}
%%------
where $\beta$ is an arbitrary light-like four-vector $\beta^2=0$.
The second choice with $\umf_\sigma(\zeta)$ 
(not exploited in \cite{kleiss-stirling:1985})
often leads  to simplifications in the resulting photon emission amplitudes.
Using the Chisholm identity%
\footnote{ For $\beta=\zeta$ the identity is slightly different 
  because of the additional minus sign in the ``line-reversal'' rule, i.e. 
  $ \bar{u}_\sigma(k) \gamma^\mu \umf_\sigma(\zeta) = 
  - \bar{\umf}_{-\sigma}(\zeta) \gamma^\mu u_{-\sigma}(k)$,
  in contrast to the usual
  $ \bar{u}_\sigma(k) \gamma^\mu u_\sigma(\beta) = 
  + \bar{u}_{-\sigma}(\beta) \gamma^\mu u_{-\sigma}(k).$}
%//////////////////////////////////////////////////
%           Chisholm
%//////////////////////////////////////////////////
\begin{align}
  \label{Chisholm}
    \bar{u}_\sigma(k) \gamma_\mu u_\sigma(\beta)\; \gamma^\mu
   &= 2    u_\sigma(\beta)\;   \bar{u}_\sigma(k)
    + 2    u_{-\sigma}(k)\;    \bar{u}_{-\sigma}(\beta),\\
  \label{Chisholm2}
    \bar{u}_\sigma(k) \gamma_\mu \umf_\sigma(\zeta)\; \gamma^\mu
   &= 2 \umf_\sigma(\zeta)\;    \bar{u}_\sigma(k)      
    - 2 u_{-\sigma}(k)\;    \bar{\umf}_{-\sigma}(\zeta),
\end{align}
we get two useful expressions, equivalent to eq.~(\ref{phot-pol}):
%//////////////////////////////////////////////////
\begin{equation}
  \begin{split}
   &({\not\!\epsilon}_\sigma (k,\beta) )^*
    = {\sqrt{2} \over \bar{u}_{-\sigma}(k) u_\sigma(\beta)}
    \left[ u_\sigma(\beta)  \bar{u}_\sigma(k) 
      +u_{-\sigma}(k)   \bar{u}_{-\sigma}(\beta)
    \right]\\
   &({\not\!\epsilon}_\sigma (k,\zeta) )^*
    = {\sqrt{2} \over \sqrt{2\zeta k} }
    \left[ \umf_\sigma(\zeta)  \bar{u}_\sigma(k) 
      -u_{-\sigma}(k)   \bar{\umf}_{-\sigma}(\zeta)
    \right].\\
  \end{split}
\end{equation}
%%------

In the evaluation of photon emission
spin amplitudes we shall use the following important building block --
the elements of the ``transition matrices''  $U$ and $V$  defined as follows
%//////////////////////////////////////////////////
%          Transition matrix definitions
%//////////////////////////////////////////////////
\begin{equation}
  \begin{split}
    \label{transition-defs}
   &\bar{u}(p_1,\lambda_1) 
         \not\!{\epsilon}^\star_\sigma(k,\beta)\;
    u(p_2,\lambda_2)
  = U\left( \st^k_\sigma \right)\!
     \left[ \st^{p_1}_{\lambda_1} \st^{p_2}_{\lambda_2} \right]
  = U^\sigma_{\lambda_1,\lambda_2} (k,p_1,m_1,p_2,m_2),\\
   &\bar{v}(p_1,\lambda_1) 
         \not\!{\epsilon}^\star_\sigma(k,\zeta)\;
    v(p_2,\lambda_2)
  = V\left( \st^k_\sigma \right)\!
     \left[ \st^{p_1}_{\lambda_1} \st^{p_2}_{\lambda_2} \right]
  = V^\sigma_{\lambda_1,\lambda_2} (k,p_1,m_1,p_2,m_2).\\
  \end{split}
\end{equation}
%%------
In the case of $\umf_\sigma(\zeta)$ 
the above transition matrices are rather simple%
\footnote{
  Our $U$ and $V$ matrices are not the same as
  the $M$-matrices of ref.~\protect\cite{kleiss-stirling:1986}, but rather
  products of several of those.
  }:
%//////////////////////////////////////////////////
%         U and V beta=zeta case
%//////////////////////////////////////////////////
\begin{equation}
  \begin{align}
  \label{UVsimple}
   &U^+(k,p_1,m_1,p_2,m_2) =\sqrt{2} 
   \begin{bmatrix} 
        \sqrt{ {2\zeta p_2 \over 2\zeta k} } s_+(k,\hat{p_1}),
      & 0 \\ 
        m_2\sqrt{ {2\zeta p_1 \over 2\zeta p_2} }
       -m_1\sqrt{ {2\zeta p_2 \over 2\zeta p_1} },
      & \sqrt{ {2\zeta p_1 \over 2\zeta k} } s_+(k,\hat{p_2})\\
    \end{bmatrix},\\
    \label{UVsimple2}
   &      U^-_{ \lambda_1, \lambda_2}(k,p_1, m_1,p_2, m_2) 
  =\left[-U^+_{ \lambda_2, \lambda_1}(k,p_2, m_2,p_1, m_1)  \right]^*,
\\
\label{UVsimple3}
   &V^\sigma_{ \lambda_1, \lambda_2}(k,p_1, m_1,p_2, m_2) 
   =U^\sigma_{-\lambda_1,-\lambda_2}(k,p_1,-m_1,p_2,-m_2).
  \end{align}
\end{equation}
%%------
The more general case with $u_\sigma(\beta)$ looks a little bit more complicated:
%//////////////////////////////////////////////////
%         U and V general case
%//////////////////////////////////////////////////
\begin{equation}
  \label{UVgeneral}
  \begin{split}
   &U^+(k,p_1,m_1,p_2,m_2) =
   {\sqrt{2} \over s_-(k,\beta)}\times \\
   &
    \begin{bmatrix} 
%%(++)
         s_+(\hat{p}_1,k) s_-(\beta,\hat{p}_2)
        +m_1m_2\sqrt{ {2\zeta \beta \over 2\zeta p_1}
                      {2\zeta k     \over 2\zeta p_2} },
%%(+-)
      &  m_1 \sqrt{ {2\zeta \beta \over 2\zeta p_1} } s_+(k,\hat{p}_2)
        +m_2 \sqrt{ {2\zeta \beta \over 2\zeta p_2} } s_+(\hat{p}_1,k)\\ 
%%(-+)
         m_1 \sqrt{ {2\zeta k \over 2\zeta p_1} } s_-(\beta,\hat{p}_2)
        +m_2 \sqrt{ {2\zeta k \over 2\zeta p_2} } s_-(\hat{p}_1,\beta),
%%(--)
      &  s_-(\hat{p}_1,\beta) s_+(k,\hat{p}_2)
        +m_1m_2\sqrt{ {2\zeta \beta \over 2\zeta p_1}
                      {2\zeta k     \over 2\zeta p_2} }\\
    \end{bmatrix},\\
  \end{split}
\end{equation}
with the same relations (\ref{UVsimple2}) and (\ref{UVsimple3}).

In the above the following numbering of elements in matrices 
$U$ and $V$ was adopted
%//////////////////////////////////////////////////
\begin{equation}
  \{ (\lambda_1,\lambda_2) \} = 
    \begin{bmatrix}
      (++) & (+-) \\
      (-+) & (--) \\
    \end{bmatrix}.
\end{equation}
When analysing (multi-) bremsstrahlung amplitudes we shall 
also often employ the following compact notation
%//////////////////////////////////////////////////
\begin{equation}
U\left[  \st^{p}_{\lambda_1}
         \st^{k}_{\sigma}\;
         \st^{p}_{\lambda_2}
 \right]
=U^\sigma_{\lambda_1,\lambda_2}(k,p_1,m_1,p_2,m_2),\quad
V\left[  \st^{p}_{\lambda_1}
         \st^{k}_{\sigma}\;
         \st^{p}_{\lambda_2}
 \right]
=V^\sigma_{\lambda_1,\lambda_2}(k,p_1,m_1,p_2,m_2),\quad
\end{equation}

When analysing the soft real photon limit
we shall exploit the following important {\em diagonality} property%
\footnote{
  Let us also keep in mind the relation
  $ b_{-\sigma}(k,p)= -(b_\sigma(k,p))^* $, which can
  save time in the numerical calculations.}
%//////////////////////////////////////////////////
%         Diagonality
%//////////////////////////////////////////////////
\begin{equation}
  \begin{align}
    \label{diagonality}
&   U\left[  \st^{p}_{\lambda_1}
             \st^k_\sigma\;
             \st^{p}_{\lambda_2}
           \right]
    = 
    V\left[  \st^{p}_{\lambda_1}
             \st^k_\sigma\;
             \st^{p}_{\lambda_2}
           \right]
    = b_\sigma(k,p)\; \delta_{\lambda_1 \lambda_2},\\
&
    b_\sigma(k,p)
    =  \sqrt{2}\; { \bar{u}_\sigma(k) \not\!p \; \umf_\sigma(\zeta)
                    \over      \bar{u}_{-\sigma}(k) \umf_\sigma(\zeta) }
    =  \sqrt{2}\; \sqrt{ {2\zeta p \over 2\zeta k} } s_\sigma(k,\hat{p}),
  \end{align}
\end{equation}
which also holds
in the general case of $u_\sigma(\beta)$, where
%//////////////////////////////////////////////////
\begin{equation}
    b_\sigma(k,p)=
        {\sqrt{2} \over s_{-\sigma}(k,\beta)}
        \left( s_{-\sigma}(\beta,\hat{p}) s_\sigma(\hat{p},k) 
          +{m^2\over 2\zeta\hat{p}}\; \sqrt{ (2\beta\zeta)\; (2\zeta k) }
        \right).
\end{equation}

%%%%%%%%%%%%%%%%%%%%%%%%%%%%%%%%%%%%%%%%%%%%%%%%%%%%%%%%%%%%%%%%%%%%%%%%%%%%%%%%%%%%%%%%%%%
%%%%%%%%%%%%%%%%%%%%%%%%%%%%%%%%%%%%%%%%%%%%%%%%%%%%%%%%%%%%%%%%%%%%%%%%%%%%%%%%%%%%%%%%%%%
%%%%%%%%%%%%%%%%%%%%%%%%%%%%%%%%%%%%%%%%%%%%%%%%%%%%%%%%%%%%%%%%%%%%%%%%%%%%%%%%%%%%%%%%%%%

%%%%%%%%%%%%%%%%%%%%%%%%%%%%%%%%%%%%%%%%%%%%%%%%%%%%%%%%%%%%%%%%%%%%%%%%%%%%
%%%%%%%%%%%%%%%%%%%%%%%%%%%%%%%%%%%%%%%%%%%%%%%%%%%%%%%%%%%%%%%%%%%%%%%%%%%%
%%%\bibliographystyle{prsty}
%%%%\bibliographystyle{plain}
%%%\bibliography{KK}

%%%%%%%%%%%%%%%%%%%%%%%%%%%%%%%%%%%%%%%%%%%%%%%%%%%%%%%%%%%%%%%%%%%%%%%%%%%%
%%%%%%%%%%%%%%%%%%%%%%%%%%%%%%%%%%%%%%%%%%%%%%%%%%%%%%%%%%%%%%%%%%%%%%%%%%%%

                                     %%%%%%%%%%%%%%%%%%
\end{document}